%% file: OSA-journal-template.tex
\documentclass[9pt,twocolumn,twoside]{osajnl}
\usepackage[utf8]{inputenc}
\usepackage{amsmath}
\usepackage{bm}
\usepackage{pgfplots}
\usepackage{fancyhdr}
\DeclareUnicodeCharacter{2212}{−}
\usepgfplotslibrary{groupplots,dateplot}
\usetikzlibrary{patterns,shapes,arrows}
\pgfplotsset{compat=newest}
\usetikzlibrary{positioning}
\graphicspath{{figures/}}

\DeclareMathOperator*{\argmax}{\arg\!\max}
\DeclareMathOperator*{\argmin}{\arg\!\min}
\newcommand{\mbeq}{\overset{!}{=}}
\newcommand{\LinRegA}{a}
\newcommand{\LinRegB}{b}
\newcommand{\BSGuided}{B_\text{g}}
\newcommand{\BSRecon}{B_\text{r}}
\newcommand{\SmoothingMatrixScale}{d}
\newcommand{\SmoothingMatrixScaleEst}{\hat{d}}
\newcommand{\GuidedFilteringLoss}{E}
\newcommand{\GuideWeightsScale}{g}
\newcommand{\IntensityLevel}{l}
\newcommand{\NumMSChannels}{M}
\newcommand{\NumHSChannels}{N}
\newcommand{\LensSpectrum}{o}
\newcommand{\CameraSpectrum}{m}
\newcommand{\SpatialDecay}{p}
\newcommand{\LightSourceSpectrum}{q}
\newcommand{\ReflectanceSpectrum}{r}
\newcommand{\SpectrumFunction}{s}
\newcommand{\WindowGuide}{w}
\newcommand{\SmoothingMatrixInvFactor}{\alpha}
\newcommand{\Wavelength}{\lambda}
\newcommand{\WindowPixelSet}{\Xi}
\newcommand{\GuidedFilteringParam}{\theta}
\newcommand{\SmoothingVar}{\sigma_{\text{s}}}

\newcommand{\MSChannels}{\mathbf{c}}
\newcommand{\MSChannelsRandom}{\mathbf{\tilde{c}}}
\newcommand{\DotScaleSpectral}{\mathbf{d}_{\text{m}}}
\newcommand{\DotScaleMS}{\mathbf{d}_{\text{c}}}
\newcommand{\Eigenvalues}{\mathbf{e}}
\newcommand{\SingleFilter}{\mathbf{f}}
\newcommand{\Noise}{\mathbf{n}}
\newcommand{\HSChannels}{\mathbf{s}}
\newcommand{\HSChannelsEst}{\mathbf{\hat{s}}}
\newcommand{\HSChannelsRandom}{\mathbf{\tilde{s}}}
\newcommand{\Eigenvector}{\mathbf{v}}
\newcommand{\GuideWeights}{\mathbf{w}}
\newcommand{\MixingWeights}{\mathbf{z}}
\newcommand{\NoiseVar}{\bm{\sigma}}

\newcommand{\PlaceholderMatrixA}{\mathbf{A}}
\newcommand{\PlaceholderMatrixB}{\mathbf{B}}
\newcommand{\MSImage}{\mathbf{C}}
\newcommand{\MSImageMean}{\mathbf{\bar{C}}}
\newcommand{\MSImageNoiseless}{\mathbf{\check{C}}}
\newcommand{\VecMSImageBlock}{\mathbf{C}_{\text{b}}}
\newcommand{\DifferenceMatrix}{\mathbf{D}}
\newcommand{\FilterMatrix}{\mathbf{F}}
\newcommand{\BigFilterMatrix}{\mathbf{\hat{F}}}
\newcommand{\Guide}{\mathbf{G}}
\newcommand{\GuideMean}{\mathbf{\bar{G}}}
\newcommand{\Identity}{\mathbf{I}}
\newcommand{\CorrelationSpatioSpectral}{\mathbf{K}}
\newcommand{\CorrelationMS}{\mathbf{K}_{\text{c}}}
\newcommand{\CorrelationHS}{\mathbf{K}_{\text{r}}}
\newcommand{\CorrelationHSMS}{\mathbf{K}_{\text{rc}}}
\newcommand{\CorrelationSpatial}{\mathbf{K}_{\text{s}}}
\newcommand{\CorrelationMSImageNoiseless}{\mathbf{K}_{\bm{\check{c}}}}
\newcommand{\CorrelationMSImageNoise}{\mathbf{K}_{\text{m}}}
\newcommand{\SmoothingMatrix}{\mathbf{M}}
\newcommand{\ScaledSmoothingMatrix}{\mathbf{\hat{M}}}
\newcommand{\MSNoiseVarMatrix}{\mathbf{N}}
\newcommand{\HSNoiseVarMatrix}{\mathbf{N}^{\text{HS}}}
\newcommand{\BigMSNoiseVarMatrix}{\mathbf{\hat{N}}}
\newcommand{\PickerSpatioSpectral}{\mathbf{P}}
\newcommand{\SpatialMarkovMatrix}{\mathbf{R}}
\newcommand{\VecHSImageBlock}{\mathbf{S}_{\text{b}}}
\newcommand{\HSImage}{\mathbf{S}}
\newcommand{\HSImageEst}{\mathbf{\hat{S}}}
\newcommand{\HSImageNoisy}{\mathbf{\hat{S}}^{\text{SSW}}}
\newcommand{\HSImageLinReg}{\mathbf{\check{S}}}
\newcommand{\HSImageMean}{\mathbf{\bar{S}}}
\newcommand{\HSImageFiltered}{\mathbf{\hat{S}}^{\text{GF}}}
\newcommand{\HSImageDiff}{\mathbf{S}^{\Delta}}
\newcommand{\NoiseEstFilterMatrix}{\mathbf{V}}
\newcommand{\WienerDenoi}{\mathbf{W}_{\text{d}}}
\newcommand{\WienerRecon}{\mathbf{W}_{\text{r}}}
\newcommand{\WienerSSW}{\mathbf{W}_{\text{SSW}}}

\fancypagestyle{firststyle}
{
	\fancyhf{}
	\fancyfoot[L]{\copyright 2020 Optica Publishing Group. One print or electronic copy may be made for personal use only. Systematic reproduction and distribution, duplication of any material in this paper for a fee or for commercial purposes, or modifications of the content of this paper are prohibited.}
}

\newcommand{\AllImageCoord}{\bm{\Omega}}

\journal{josaa}

\setboolean{shortarticle}{false}

\title{Structure-Preserving Spectral Reflectance Estimation using Guided Filtering}

\author[1,*]{Frank Sippel}
\author[1]{Jürgen Seiler}
\author[1]{Nils Genser}
\author[1]{André Kaup}

\affil[1]{Chair of Multimedia Communications and Signal Processing, Friedrich-Alexander University Erlangen-Nuremberg, 91058 Erlangen, Germany}

\affil[*]{Corresponding author: frank.sippel@fau.de}

\dates{Received 16 June 2020; revised 7 September 2020; accepted 7 September 2020; posted 10 September 2020 (Doc. ID 400485); \\ published 7 October 2020}
\doi{\url{https://doi.org/10.1364/JOSAA.400485}}

\begin{abstract}
Light spectra are a very important source of information for diverse classification problems, e.g., for discrimination of materials. To lower the cost for acquiring this information, multispectral cameras are used. Several techniques exist for estimating light spectra out of multispectral images by exploiting properties about the spectrum. Unfortunately, especially when capturing multispectral videos, the images are heavily affected by noise due to the nature of limited exposure times in videos. Therefore, models that explicitly try to lower the influence of noise on the reconstructed spectrum are highly desirable. Hence, a novel reconstruction algorithm is presented. This novel estimation method is based on the guided filtering technique which preserves basic structures, while using spatial information to reduce the influence of noise. The evaluation based on spectra of natural images reveals that this new technique yields better quantitative and subjective results in noisy scenarios than other state-of-the-art spatial reconstruction methods. Specifically, the proposed algorithm lowers the mean squared error and the spectral angle up to 46\% and 35\% in noisy scenarios, respectively. Furthermore, it is shown that the proposed reconstruction technique works out-of-the-box and does not need any calibration or training by reconstructing spectra from a real-world multispectral camera with nine channels.
\end{abstract}

\setboolean{displaycopyright}{true}

\begin{document}

\maketitle
\thispagestyle{firststyle}
\section{Introduction}
Typically, hyperspectral cameras are used to record light spectra. These cameras usually capture around 25 to 200 different spectral bands using narrowband filters. The result is a fine detailed spectrum for each pixel. By using the result of these cameras, a lot of classification tasks can be solved. For example, material classification, specifically the discrimination of different types of plastic \cite{moroni-pet-2015} can be solved as well as determining the degree of burn \cite{sowa-classification-2006}, detecting drug counterfeit \cite{degardin-near-2016}, classifying different wheat kernels \cite{dowell-automated-1998}, or gasoline discrimination \cite{balabin-gasoline-2010}. A big advantage is that using hyperspectral imaging in the context of these applications is a non-destructive method.

Unfortunately, there are multiple problems with these systems. First of all, they are usually extremely expensive, since a lot of different filters are needed. Consequently, multispectral cameras, which have much less different filters, are used to estimate the hyperspectral images. Then, there are other disadvantages dependent on the camera design. Using a multispectral filter array approach \cite{lapray-multispectral-2014} leads to low resolution images. Here, the number of bands represents the pixels used for different bands. In the end, the information gained for such a block of pixels is equivalent to one big pixel with the given amount of bands, which results in a low spatial resolution image. The advantage of such a system is that it is capable of capturing video sequences. On the other hand, approaches based on line scan cameras \cite{lambrechts-cmos-compatible-2014} or filter wheels \cite{brauers-multispectral-2008} are not able to capture video sequences properly, since the rotating or scanning procedure takes too much time to yields proper video sequences. However, when there is enough time to capture the scene, these imaging techniques yield high resolution images.

\begin{figure}[t]
	\centering
	\includegraphics[]{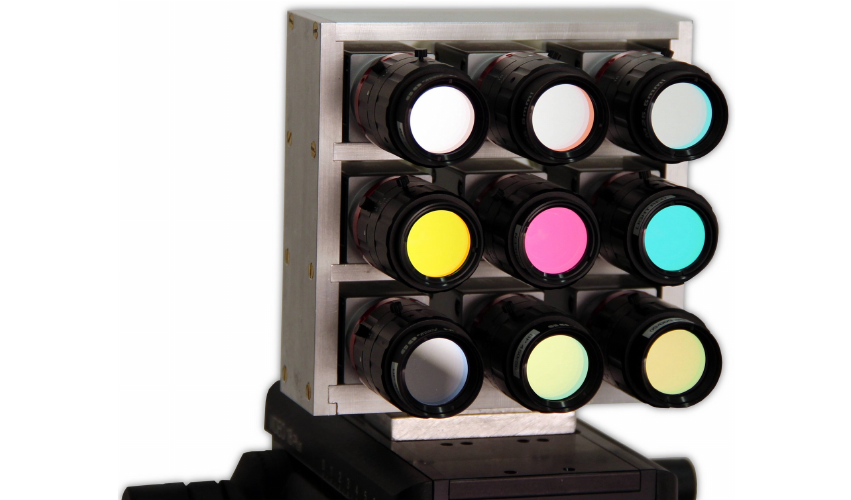}
	\caption{\label{fig:camsi}  The Camera Array for Multi-Spectral Imaging (CAMSI) \cite{genser-multispectral-2019} has nine channels and is able to record high-resolution multispectral video sequences. Note that by using the algorithms of \cite{genser-deep-2020} and \cite{genser-joint-2020}, all views are reconstructed to the center view.}
\end{figure}

The ability of capturing proper video sequences is an important property for a lot of classification algorithms. For example, imagine a drone flying over agriculture fields and estimating the plant health in real-time based on the provided spectra. Therefore, the desired multispectral camera should  be able to capture high-resolution multispectral video sequences. This is possible using an approach which uses multiple cameras, where each of them is equipped with just one single filter. This setup is shown exemplarily in Fig. \ref{fig:camsi}. After combining the resulting images of this setup properly, this system can yield extremely high resolution images and video sequences, while the cost for such a system stays quite low, since it uses classical cameras and fairly common filters. Since the goal is to deliver images, which yield a fine-detailed spectrum for every pixel, the task is to reconstruct hyperspectral images from this multispectral camera. The reconstruction of spectra exploits the information, which is given by the overlapping filters.

Unfortunately, when capturing video sequences or fast moving objects, the exposure time of each image is very limited. Furthermore, the bandwidth of multispectral filters is fairly low. Both of these circumstances lead to the situation that the amount of light that falls onto the sensor is limited. Consequently, the resulting images are typically very noisy. This paper presents a novel spectral reflectance estimation algorithm, which produces high quality results, even if the environment yields extremely noisy images. Furthermore, our novel technique works out of the box and does not need any training database.

Section \ref{sec:basic} establishes the basic relationship between multispectral images and hyperspectral images, formulates the problem and introduces a noise model. Afterwards, Section \ref{sec:sota} presents the state of the art. Our novel reconstruction algorithm is described in Section \ref{sec:spre}. The superiority of our new method is shown in Section \ref{sec:evaluation}. Finally, Section \ref{sec:applications} presents the power of our technique using CAMSI \cite{genser-multispectral-2019} and Section \ref{sec:conclusion} summarizes the results.

\section{Problem formulation}
\label{sec:basic}

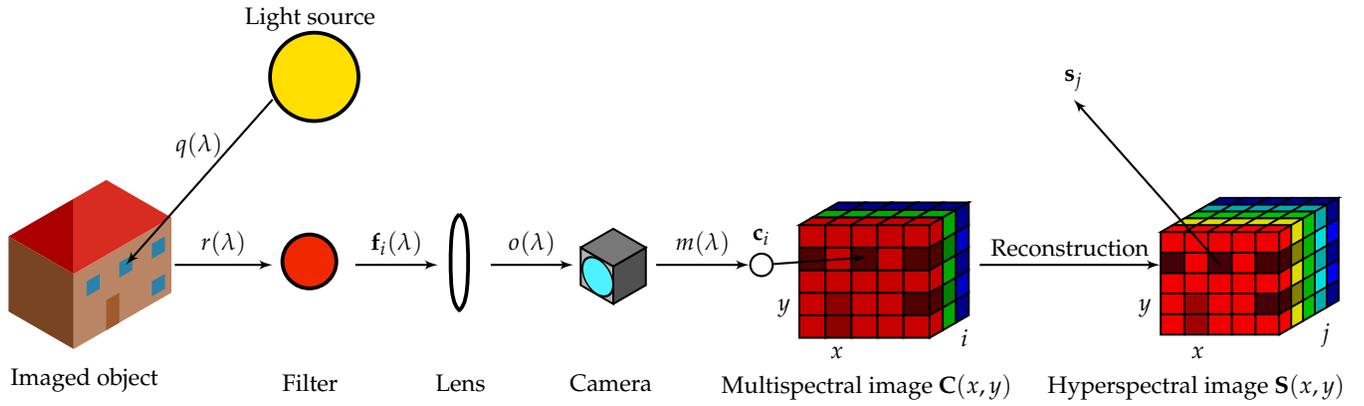
\begin{figure*}
	\centering
	\begin{tikzpicture}
		\tikzstyle{l} = [draw, -latex',thick]
		\tikzstyle{c} = [circle, draw, node distance=1.1cm, thick, minimum size=0.25cm]
		
		\node[inner sep=0pt] (sun) at (3,2.5) {\includegraphics[width=1in]{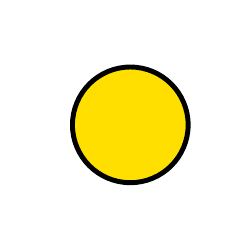}};
		\node[inner sep=0pt] (cam) at (7,0) {\includegraphics[width=1in]{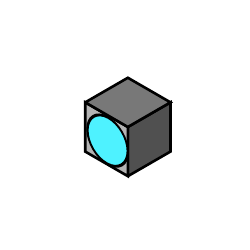}};
		\node[inner sep=0pt] (lens) at (5,0) {\includegraphics[width=1in]{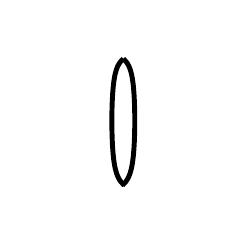}};
		\node[inner sep=0pt] (filter) at (3,0) {\includegraphics[width=1in]{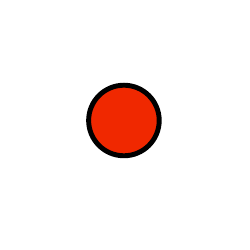}};
		\node[inner sep=0pt] (house) at (0,0) {\includegraphics[width=1in]{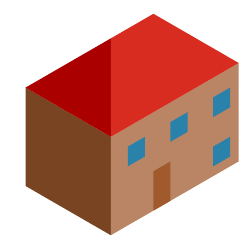}};
		\node[inner sep=0pt] (ms) at (10.6,0) {\includegraphics[width=1in]{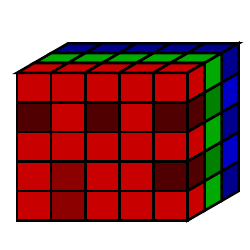}};
		\node[inner sep=0pt] (hs) at (15.5,0) {\includegraphics[width=1in]{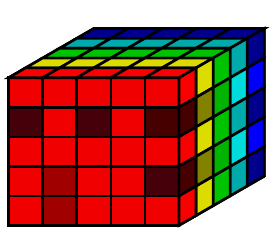}};
		\node[c] (rc) at (9, 0) {};
		
		\coordinate[label=above:Light source] () at (3, 3);
		\coordinate[label=above:Imaged object] () at (0, -1.8);
		\coordinate[label=above:Filter] () at (3, -1.8);
		\coordinate[label=above:Lens] () at (5, -1.8);
		\coordinate[label=above:Camera] () at (7, -1.8);
		\coordinate[label=above:{Multispectral image $\MSImage(x, y)$}] () at (10.4, -1.9);
		\coordinate[label=above:{Hyperspectral image $\HSImage(x, y)$}] () at (14.78, -1.9);
		\coordinate[label=above:{$\MSChannels_i$}] () at (9, 0.15);
		\coordinate[label=above:{$\HSChannels_j$}] () at (13.15, 2.2);
		
		\path[l] (2.5, 2.2) -- (0.55, 0) node[midway, above=0.2] {$\LightSourceSpectrum(\Wavelength)$};
		\path[l] (1.2, 0) -- (2.5, 0) node[midway, above] {$\ReflectanceSpectrum(\Wavelength)$};
		\path[l] (3.6, 0) -- (4.7, 0) node[midway, above] {$\SingleFilter_i(\Wavelength)$};
		\path[l] (5.4, 0) -- (6.5, 0) node[midway, above] {$\LensSpectrum(\Wavelength)$};
		\path[l] (7.6, 0) -- (rc) node[midway, above] {$\CameraSpectrum(\Wavelength)$};
		\path[l] (rc) -- (10.4, 0.1);
		\path[l] (11.9, 0) -- (14.3, 0) node[midway, above] {Reconstruction};
		\path[l] (15.1, 0) -- (13.15, 2.2);
		
		\coordinate[label=above:$x$] () at (10, -1.35);
		\coordinate[label=above:$y$] () at (9.3, -0.8);
		\coordinate[label=above:$i$] () at (11.7, -1.2);
		
		\coordinate[label=above:$x$] () at (14.8, -1.35);
		\coordinate[label=above:$y$] () at (14.1, -0.8);
		\coordinate[label=above:$j$] () at (16.5, -1.2);
	\end{tikzpicture}
	\caption{\label{fig:basic_pipeline} The basic pipeline from the emitting light source to a reconstructed hyperspectral datacube. Note that this figure only shows one out of $M$ filters. First, the light is emitted from the light source and reflected by the imaged object. Thereafter, the reflected light passes through one of filter, the lens and the camera. In the end, the camera counts the remaining number of photons and creates one $x-y$-plane of the multispectral datacube. Of course, other light rays will pass through the other filters such that the full multispectral datacube is created. The purpose of this paper is to reconstruct a hyperspectral datacube out of this multispectral datacube under the influence of strong noise. Finally, a spectrum of a single pixel can be analyzed.}
\end{figure*}
The basic physical relationship between the multispectral channels $\MSChannels$ and the spectrum is given by counting photons, which passed through the corresponding filters $\SingleFilter$ \cite{cortes-multipectral-2003}
\begin{equation}
	\MSChannels_i = \int_{\Wavelength_\text{min}}^{\Wavelength_\text{max}} \LightSourceSpectrum(\Wavelength)\ReflectanceSpectrum(\Wavelength)\SingleFilter_i(\Wavelength)\LensSpectrum(\Wavelength)\CameraSpectrum(\Wavelength) \ \text{d}\Wavelength,
\end{equation}
where $\MSChannels_i$ is the $i$-th multispectral channel of a single pixel, $\LightSourceSpectrum(\Wavelength)$ is the spectrum of the light source, $\ReflectanceSpectrum(\Wavelength)$ is the spectral reflectance of the imaged object, $\SingleFilter_i(\Wavelength)$ is the transfer function of the $i$-th filter, $\LensSpectrum(\Wavelength)$ is the spectral permeability of the camera lens and $\CameraSpectrum(\Wavelength)$ is the spectral response of the camera. Since $\LightSourceSpectrum(\Wavelength)$, $\ReflectanceSpectrum\left(\Wavelength\right)$, $\LensSpectrum(\Wavelength)$ and $\CameraSpectrum(\Wavelength)$ are all unknown, the spectrum which can be estimated reads as ${\SpectrumFunction(\Wavelength) = \LightSourceSpectrum(\Wavelength)\ReflectanceSpectrum\left(\Wavelength\right)\LensSpectrum(\Wavelength)\CameraSpectrum(\Wavelength)}$. Note that the spectral reflectance $\ReflectanceSpectrum\left(\Wavelength\right)$ could be estimated, if $\LightSourceSpectrum(\Wavelength)$, $\LensSpectrum(\Wavelength)$ and $\CameraSpectrum(\Wavelength)$ are known, since they could be multiplied to the transfer functions of the filter.

As estimating a continuous spectrum is cumbersome, the spectrum and the filters are sampled, and the integration is replaced by a sum. Sampling the spectrum and the filters with $N$ sampling points leads to ${\HSChannels = \left[\SpectrumFunction(\Wavelength_\text{1}), \SpectrumFunction(\Wavelength_\text{2}), \cdots, \SpectrumFunction(\Wavelength_\NumHSChannels)\right] \in \mathcal{R}^\NumHSChannels}$ and ${\FilterMatrix_i=\left[\SingleFilter_i(\Wavelength_\text{1}), \SingleFilter_i(\Wavelength_\text{2}), \cdots, \SingleFilter_i(\Wavelength_\NumHSChannels)\right]}$, respectively, where $\FilterMatrix_i$ denotes the $i$-th row of the filter matrix ${\FilterMatrix \in \mathcal{R}^{\NumMSChannels \times \NumHSChannels}}$. The sampled spectrum $\HSChannels$ corresponds to one pixel of a hyperspectral image. Therefore, the adjective \emph{hyperspectral} is identical to a fine sampling of a spectrum, while \emph{multispectral} corresponds to less channels, which result from a weighted integration over a spectrum area. Stacking all $\NumMSChannels$ multispectral channels results for the noiseless case in the underdetermined system
\begin{equation}
	\MSChannels = \FilterMatrix \HSChannels,
\end{equation}
where $\MSChannels$ are the recorded multispectral channels and $\FilterMatrix$ is the filter matrix, which contains the relationships between the combined reflectance spectrum $\HSChannels$ and the recorded channels.

Later on, it is necessary to take neighboring pixels into account. Since $\MSChannels$ and $\HSChannels$ only contain the information about one single pixel, they need to be extended. First of all, multispectral images and hyperspectral images need to be defined. Therefore, $\MSImage$ and $\HSImage$ are introduced, where $\MSImage$ is the multispectral image and $\HSImage$ is the hyperspectral image. To get the value of the $i$-th multispectral channel or $j$-th hyperspectral channel at pixel position $(x, y)$, $\MSImage_i(x, y)$ and $\HSImage_j(x, y)$ are used, respectively.
Fig. \ref{fig:basic_pipeline} shows these different images and the capturing process.

A very important property of images is the spatial correlation between different pixels in uniform regions. To exploit that information, blocks of images are processed. Moreover, for an optimal support area, each pixel has its own block, where this specific pixel is the center of the block. Therefore, more definitions are necessary. First of all, the blocksize for reconstruction purposes is $\BSRecon$. Consequently, blocks with $\BSRecon \times \BSRecon$ pixels are reconstructed when using spatial information. Furthermore, since there should be a center pixel, $\BSRecon$ shall be odd. Moreover, these blocks are vectorized, which means that all multispectral channels and spectra within one block are stacked on top of each other. This is done by traversing the image in the $x$-$y$ plane row by row. In the end, the first elements of this vector contain all multispectral or hyperspectral channels of the first pixel in this block, respectively. Thereafter, all channels of the pixel right to the first pixel follow. Consequently, the multispectral and hyperspectral block variables are ${\VecMSImageBlock^{x, y} \in \mathcal{R}^{\BSRecon^2 \NumMSChannels}}$ and ${\VecHSImageBlock^{x, y} \in \mathcal{R}^{\BSRecon^2 \NumHSChannels}}$, respectively, where $\VecMSImageBlock^{x, y}$ returns the $\BSRecon \times \BSRecon$ multispectral block around center pixel $(x, y)$ in vectorized form.

When reconstructing whole blocks at once, the filter matrix $\FilterMatrix$ has to be expanded such that each pixel in the vectorized multispectral block $\VecMSImageBlock^{x, y}$ can be reconstructed with the filter matrix. This results in a block-diagonal matrix with the filter matrix $\FilterMatrix$ on its main block diagonal. The expanded filter matrix can be written as
\begin{equation}
	\label{eq:expanded_F}
	\BigFilterMatrix = \Identity^{\BSRecon^2 \times \BSRecon^2} \otimes \FilterMatrix,
\end{equation}
where $\otimes$ indicates the Kronecker product and $\Identity^{\BSRecon^2 \times \BSRecon^2}$ is the identity matrix with size $\BSRecon^2 \times \BSRecon^2$. Therefore, the expanded multispectral imaging equation for the noiseless case reads as
\begin{equation}
	\VecMSImageBlock^{x, y} = \BigFilterMatrix \VecHSImageBlock^{x, y}.
\end{equation}
The first challenge is to employ prior knowledge about the spectrum to solve this underdetermined linear system of equations.

Unfortunately, not only the undetermined linear system of equations causes problems, but also noise. There are multiple noise sources within the imaging pipeline of multispectral cameras. The most important noise source is shot noise \cite{schoberl-photometric-2012}, which is caused by miscounting photons. This noise source is signal-dependent, which results in a higher absolute deviation for images in a bright environment. However, the relative deviation, which compares the variance to the mean, is much lower, when a lot of light falls into the sensor compared to a dark imaging environment. Furthermore, since the number of photons is counted, the distribution for the multispectral channels is non-negative and discrete. These observations result in the distribution \cite{gow-comprehensive-2007}
\begin{equation}
	\MSChannels_i \sim \mathcal{P}(\FilterMatrix_i \HSChannels),
\end{equation}
where $\mathcal{P}(\FilterMatrix_i \HSChannels)$ is a Poisson distribution with mean $\FilterMatrix_i \HSChannels$. For a Poisson distribution the mean equals the variance. Therefore, the variance of the noise is higher, if more photons are counted. There are a couple of problems using a Poisson distribution to model the spectral reconstruction problem. Firstly, it is cumbersome to derive solutions based on the Poisson distribution, since it often ends in non-closed-form solutions. Secondly, since the noise distribution is signal dependent, the statistical moments would be different for every pixel. Consequently, it would be much more difficult to estimate noise statistics. The estimation is much easier when assuming a constant noise variance for a single multispectral image. Furthermore, the Poisson distribution can be nicely modeled by a Gaussian distribution for a high number of photons. Consequently, in the following it is assumed that the noise is Gaussian distributed and not signal-dependent. Other noise sources are dark current noise, amplifier noise, and reset noise. Usually, these noise sources are modeled using additive white Gaussian noise \cite{gow-comprehensive-2007}. Consequently, all independent noise sources can be summarized into one single additive white Gaussian noise source yielding the relationship \cite{pratt-spectral-1976}
\begin{equation}
	\MSChannels = \FilterMatrix \HSChannels + \Noise,
\end{equation}
where $\Noise$ is the additive white Gaussian noise with zero mean and different variances for each multispectral image. This is necessary, since, for example, different bandwidths for the filters are allowed. This leads to a different number of photons counted by the corresponding channel, which in the end yields different noise variances. In multispectral video sequences, different noise variances in the multispectral channels also occur fairly often, since the illumination varies in its intensity over different wavelength areas. Then, for less illuminated wavelength areas, the corresponding multispectral channel contains much more noise in comparison to the other channels.

Some of the reconstruction methods need an estimate of the noise variance. For a single multispectral image, the noise variance can be estimated using \cite{immerkaer-fast-1996}. Here, the filter kernel
\begin{equation}
	\NoiseEstFilterMatrix = \begin{pmatrix}
		1 & -2 & 1\\
		-2 & 4 & -2\\
		1 & -2 & 1
	\end{pmatrix},
\end{equation}
is used to estimate the noise standard deviation for the $i$-th multispectral image
\begin{equation}
	\label{eq:noise_estimation}
	\NoiseVar_{\text{n},i} = \frac{1}{6 |\AllImageCoord|} \sqrt{\frac{\pi}{2}} \sum_{(x, y) \in \AllImageCoord}|\MSImage_i(x,y) \ast \NoiseEstFilterMatrix|,
\end{equation}
where $\AllImageCoord$ contains all pixel coordinates in the image and the operator $\ast$ denotes a convolution.

Often a noise matrix is necessary for the upcoming reconstructing techniques. Since the noise of different channels is independent from each other, this noise matrix $\MSNoiseVarMatrix$ just contains the noise variances of the individual channels $\NoiseVar_{\text{n},i}^2$ on its main diagonal. When reconstructing whole blocks, the noise matrix has to be expanded in a similar form to \eqref{eq:expanded_F}, which reads as
\begin{equation}
	\BigMSNoiseVarMatrix = \Identity^{\BSRecon^2 \times \BSRecon^2} \otimes \MSNoiseVarMatrix.
\end{equation}

\section{State of the art}
\label{sec:sota}

In the following section, different reconstruction techniques are presented. Some of them do no take the spatial correlation between pixels into account. These are the so-called single-pixel reconstruction techniques, which are deployed pixel by pixel. These reconstruction methods miss out essential information to reduce the influence of noise on the reconstruction. Nevertheless, they are the state-of-the-art non-learning-based methods for the noiseless case. Therefore it is essential to describe them before building the spatial reconstruction algorithms on top of these. There are diverse ideas to tackle the single-pixel problem. For example, one could use compressed sensing \cite{zhang-study-2015}, a basis parameter estimation \cite{mansouri-representation-2008}, or machine-learning based approaches (e.g. \cite{eckhard-evaluating-2014, shen-spectral-2006, zhang-spectral-2008, shen-reflectance-2007, mansouri-representation-2008}). When using machine-learning, a training database for representative spectra or statistical evaluations is necessary. The training-based reconstruction methods can lead to superior results. However, there is a high risk that such a system runs into overfitting to a specific environment. Consequently, approaches, which do not require a training database, are used to deliver a general purpose spectral estimation system. First, two other very popular single-pixel reconstruction are presented in the following, which are actually tightly related. Afterwards, two state-of-the-art spatial reconstruction methods are presented.
\subsection{Smoothed pseudoinverse (SP)}
One way to tackle the reconstruction problem is by assuming the spectrum to be smooth. One measure to define smoothness is by assuming small derivatives in the continuous spectrum, thus differences in the sampled spectrum \cite{pratt-spectral-1976}. With this information, one can setup an optimization problem, which tries to find the spectrum with the smallest differences within the solution set of the underdetermined linear system of equations, which has infinitely many solutions as long as there is no contradiction in the system. This results in the optimization problem of the smoothed pseudoinverse (SP)
\begin{equation}
	\begin{aligned}
		\HSChannelsEst^{\text{SP}} = \ & \argmin_{\HSChannels} & & ||\DifferenceMatrix\HSChannels||_2^2 \\
		\ & \text{s.t.} & & \MSChannels = \FilterMatrix \HSChannels,
	\end{aligned}
\end{equation}
where $\DifferenceMatrix$ is the smoothing matrix. Using the $\text{l}_2$-norm for this optimization problem leads to a closed-form solution. The order of the difference matrix $\DifferenceMatrix$ can be chosen arbitrarily. The higher the order of the differences used, the more flexibility is allowed. Usually, the first-order or the second-order differences are used, since the procedure rather should produce less oscillating results. For the first differences, the matrix $\DifferenceMatrix$ looks like
\begin{equation}
	\begin{aligned}
		\DifferenceMatrix_1 &= \begin{pmatrix}
			1 & -1 & 0 & \cdots & 0 \\
			0 & 1 & -1 & \cdots & 0 \\
			\vdots & \ddots & \ddots & \ddots & \vdots  \\
			0 & \cdots & 0 & 1 & -1
		\end{pmatrix}.
	\end{aligned}
\end{equation}
The first-order differences just calculate the delta between neighboring spectral entries, for example the first difference is $\DifferenceMatrix_{1, 1}\HSChannels = \HSChannels_1 - \HSChannels_2$, where $\DifferenceMatrix_{1, 1}$ indicates the first row of $\DifferenceMatrix_{1}$. Then, the second derivative is calculated by just applying the derivative again to the first-order derivative. This can be expressed by $\DifferenceMatrix_2 = \DifferenceMatrix_1^{\NumHSChannels-2 \times \NumHSChannels-1} \DifferenceMatrix_1^{\NumHSChannels-1 \times \NumHSChannels}$. The superscripts denote the matrix size.

By applying the derivative to the Lagrangian function, the result can be written in closed-form as
\begin{equation}
	\HSChannelsEst^{\text{SP}} = \SmoothingMatrix^{-1}\FilterMatrix^{\text{T}}(\FilterMatrix\SmoothingMatrix^{-1}\FilterMatrix^{\text{T}})^{-1}\MSChannels,
\end{equation}
where $\SmoothingMatrix = \DifferenceMatrix^{\text{T}}\DifferenceMatrix + \SmoothingMatrixInvFactor\Identity$. The additional scaled identity matrix is necessary because $\DifferenceMatrix^{\text{T}}\DifferenceMatrix$ is not invertible.

\subsection{Single-pixel Wiener filter (WF)}

Another approach to tackle this inversion problem statistically is by minimizing the mean squared error between the estimation $\HSChannelsEst$ and the random variable  $\HSChannelsRandom$, which represents the spectrum statistically. The underlying system yielding the observations $\MSChannels$ is known, in particular $\MSChannelsRandom = \FilterMatrix\HSChannelsRandom + \Noise$, where $\MSChannelsRandom$ is the random variable for the observed multispectral channels and $\Noise$ is the statistical independent Gaussian noise vector with zero mean. Using the covariance matrices $\CorrelationMS = \FilterMatrix\CorrelationHS\FilterMatrix^{\text{T}} + \MSNoiseVarMatrix$, where $\CorrelationMS$ is the covariance matrix for the multispectral channels and $\CorrelationHS$ is the covariance matrix for the spectral reflectances, and the cross-covariance matrix $\CorrelationHSMS = \CorrelationHS\FilterMatrix^{\text{T}}$ yields the single-pixel Wiener filter (WF) \cite{pratt-spectral-1976}
\begin{equation}
	\label{eq:single_pixel_wiener}
	\HSChannelsEst^{\text{WF}} = \CorrelationHSMS \CorrelationMS^{-1} \MSChannels = \CorrelationHS\FilterMatrix^{\text{T}}(\FilterMatrix\CorrelationHS\FilterMatrix^{\text{T}} + \MSNoiseVarMatrix)^{-1}\MSChannels.
\end{equation}
Using this formula assumes the spectral reflectances to have zero mean. This is obviously not ideal, however, giving a reasonable estimate is difficult. Firstly, a training database for the mean would be necessary. Secondly, the multispectral images would have to stay within the same bounds as the training images. Thirdly, a different multispectral channels to hyperspectral channels ratio would lead to a different mean value.

A novel connection between the SP and the WF can be found in the appendix. This insight is used for all reconstruction methods based on the Wiener filter, namely $\CorrelationHS = \SmoothingMatrixScale \ \SmoothingMatrix^{-1}$.

\subsection{Spatio-spectral Wiener filter (SSW)}

The big problem of single-pixel reconstruction methods is, that they are limited in handling noise, since they are missing spatial information. The spatio-spectral Wiener filter (SSW), described in \cite{murakami-color-2008}, tries to tackle that problem by exploiting the spatial correlation between neighboring pixels. It is an extension to the single-pixel Wiener filter. This method replaces the covariance matrix by a combined spatio-spectral covariance matrix $\CorrelationSpatioSpectral$. The spectral reflectances of pixel $(x, y)$ are estimated by
\begin{equation}
	\HSImageEst^{\text{SSW}}(x, y) = \PickerSpatioSpectral\CorrelationSpatioSpectral\BigFilterMatrix^{\text{T}}(\BigFilterMatrix\CorrelationSpatioSpectral\BigFilterMatrix^{\text{T}} + \BigMSNoiseVarMatrix)^{-1}\VecMSImageBlock^{x, y},
\end{equation}
where $\PickerSpatioSpectral$ is a $\NumHSChannels \times \BSRecon^2 \NumHSChannels$ matrix which picks out the spectrum of the middle pixel. The covariance matrix $\CorrelationSpatioSpectral$ is formed by assuming it to be separable into the elements $\CorrelationHS$ and $\CorrelationSpatial$, where $\CorrelationHS$ is the spectral covariance matrix known from above and $\CorrelationSpatial$ is the spatial covariance matrix. Then, $\CorrelationSpatioSpectral$ can be calculated by
\begin{equation}
	\CorrelationSpatioSpectral = \CorrelationSpatial \otimes \CorrelationHS.
\end{equation}
Furthermore, the spatial covariance $\CorrelationSpatial$ is obtained by assuming it to be separable into vertical and horizontal direction. This results in
\begin{equation}
	\CorrelationSpatial = \SpatialMarkovMatrix(\SpatialDecay) \otimes \SpatialMarkovMatrix(\SpatialDecay),
\end{equation}
where $\SpatialMarkovMatrix(\SpatialDecay)$ is the covariance in either horizontal or vertical direction. Moreover, it is assumed that the covariance in horizontal and vertical direction can be modeled by a first order Markov sequence. Consequently, the matrix $\SpatialMarkovMatrix(\SpatialDecay)$ is formed by
\begin{equation}
	\SpatialMarkovMatrix(\SpatialDecay) = \begin{pmatrix}
		\SpatialDecay^0 & \SpatialDecay^1 & \cdots & \SpatialDecay^{\BSRecon - 1} \\
		\SpatialDecay^1 & \SpatialDecay^0 & \cdots & \SpatialDecay^{\BSRecon - 2} \\
		\vdots  & \vdots  & \ddots & \vdots  \\
		\SpatialDecay^{\BSRecon - 1} &\SpatialDecay^{\BSRecon - 2} & \cdots & \SpatialDecay^0 
	\end{pmatrix},
\end{equation}
where $\SpatialDecay$ is the correlation coefficient or the decay factor.

\subsection{Edge-preserving spatio-spectral Wiener filter (EPSSW)}
\begin{figure*}[t!]
	\centering
	\begin{tikzpicture}
		\tikzstyle{b} = [rectangle, draw, node distance=1.3cm, text width=10em, text centered, rounded corners, minimum height=1.2cm, thick]
		\tikzstyle{c} = [circle, draw, node distance=1.1cm, minimum size=0.15cm, inner sep=0pt, outer sep=0pt]
		\tikzstyle{l} = [draw, -latex',thick]
		\tikzstyle{l2} = [draw, thick]
		\node[draw=none, fill=none] (ms) {$\MSImage(x, y)$};
		\node[b, right=of ms] (ne) {Noise estimation};
		\node[b, below=of ne] (re) {Reconstruction};
		\node[b, above=of ne] (gg) {Guide generation};
		\node[b, right=of ne] (gf) {Guided filtering};
		\node[b, right=of gf] (mx) {Mixing};
		\node[draw=none, fill=none, right=of mx] (hs) {$\HSImageEst^{\text{SPRE}}(x, y)$};
		\node[c, fill=black](rc) at (gf |- re) {};
		
		\path[l] (ms) |- (re);
		\path[l] (ms) |- (ne);
		\path[l] (ms) |- (gg);
		\path[l] (gf) -- (mx) node[midway, above] {$\HSImageFiltered(x, y)$};
		\path[l] (ne) -- (re) node[midway, left] {$\MSNoiseVarMatrix$};
		\path[l] (ne) -- (gg) node[midway, left] {$\MSNoiseVarMatrix$};
		\path[l] (mx) |- (hs);
		\path[l] (gg) -| (gf) node[near end, left] {$\Guide(x, y)$};
		\path[l2] (re) -- (rc) node[midway , above] {$\HSImageNoisy(x, y)$};
		\path[l] (rc) -- (gf);
		\path[l] (rc) -| (mx);
		\path[l] (re.350) -| (mx.320) node[near end, right] {$\HSNoiseVarMatrix$};
	\end{tikzpicture}
	\caption{\label{fig:spre_flow} Signal flow diagram of our novel SPRE algorithm. The noise estimation is done using \eqref{eq:noise_estimation}. The basic idea is to generate a guide image from the multispectral images, which is used by the guided filtering to improve the reconstruction result. Due to unfavourable properties of the guided filtering, a mixing step is necessary in the end.}
\end{figure*}
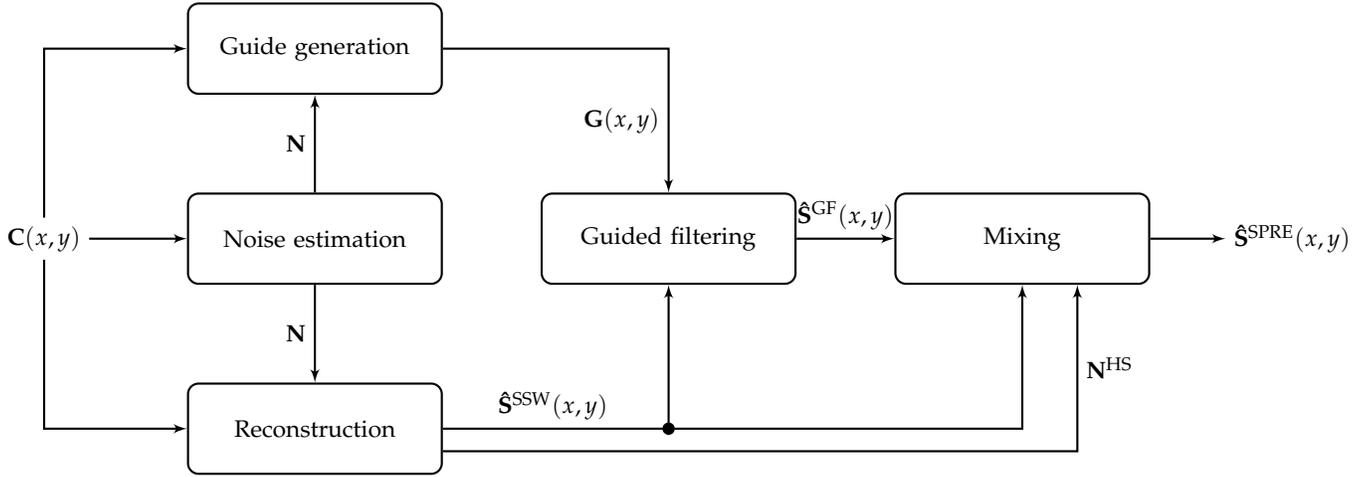

The idea of \cite{urban-spectral-2009}, the edge-preserving spatio-spectral Wiener filter (EPSSW), is to combine a Wiener filter for denoising the multispectral images first, which uses weights similar to bilateral filter approach, and afterwards apply a spectral reconstruction Wiener filter using the resulting noise covariance of the denoising filter. This results in
\begin{equation}
	\begin{aligned}
		\HSImageEst^{\text{EPSSW}}(x, y) = &\WienerRecon(x, y)\bigl(\WienerDenoi(x, y)\bigl[\MSImage(x, y) -\\
		&- \MSImageMean(x, y)\bigr] + \MSImageMean(x, y)\bigr),
	\end{aligned}
\end{equation}
where $\WienerDenoi(x, y)$ is the denoising Wiener filter, $\WienerRecon(x, y)$ is the Wiener filter for spectral reflectance estimation and $\MSImageMean(x, y)$ is a weighted average of the multispectral channels within a block using bilateral weights. These weights have a spatial and a range component. The spatial component is responsible for allocating higher weights for closer pixels. The speed of this decay towards distant pixels is described by a spatial variance. The range component is responsible for weighting similar pixels higher. This is done by using the squared $l_2$-norm of the difference between the multispectral channels of two pixels, namely, $||\MSImage(x, y) - \MSImage(u, v)||_2^2$. Afterwards, this term is converted into an affinity measure by an exponential term. Furthermore, the decay of the similarity measure is controlled by a range variance. In the end, both components are combined by a multiplication and are scaled, such that the sum of all weights within one block equals one. For a detailed description of these weights and the Wiener filters please refer to \cite{urban-spectral-2009}. Moreover, the mean of the spectral reflectances is omitted again to have a fair comparison. Otherwise, the mean could be also easily used in the WF and the SSW. Originally, this paper formulated the problem by assuming a constant noise variance over all multispectral images.

\section{Structure-preserving reflectance estimation (SPRE)}
\label{sec:spre}

The SSW and the EPSSW try to embed spatial information to reduce the influence of noise on the reflectance estimation. However, the spatio spectral Wiener weights each pixel in a window for every position in the same way, or in other words content-independent. Thus, this algorithm does not include the structure of the image into the reconstruction. On the other hand, the EPSSW tries to embed this structure by applying affinity weights in the denoising step. This affinity is calculated by a decaying exponential term and the $l_2$-norm of the difference between neighboring multispectral pixels and the middle pixel. Even if this distance gets higher in scenarios with some noise, the affinity measure still works, since the weights are normalized to sum up to one. However, the more noise an image has, the higher the range distance to other pixels than the middle pixel. Therefore, for low SNR images, this middle pixel gets a lot of weight, while the surrounding pixels approximately get the same range affinity. In the end, this procedure cannot reconstruct the structure for low SNR images properly anymore.

This novel reconstruction algorithm, called structure-preserving reflectance estimation (SPRE), is designed to simultaneously preserve structure, while using spatial information to tackle the influence of noise on the reconstruction in uniform regions. The idea of this novel technique is to use the independence of the noise across different multispectral images to create an image that contains the structure of the scene. This combined image contains much less noise and serves as a guide for the reconstruction. Therefore, the idea of guided filtering, described in \cite{he-guided-2013}, is embedded into the spectral reconstruction problem.

The basic signal flow is shown in Fig. \ref{fig:spre_flow}. The foundation for the SPRE is the SSW. Then, the noisy hyperspectral images $\HSImageNoisy_i$, and a guide image $\Guide$, which contains the basic structure of the image, are combined to form a denoised structure-preserving hyperspectral image $\HSImageFiltered$. A proper guide generation procedure, which effectively reduces the influence of noise, is extremely important. In the end, $\HSImageNoisy$ and $\HSImageFiltered$ are mixed for the high SNR case.

\subsection{Guided filtering}

This technique assumes that the noiseless hyperspectral image $\HSImageLinReg_i$ is a linear combination of the guide in a window $\WindowGuide_{x,y}$, which is centered around pixel $(x, y)$ and has size $\BSGuided \times \BSGuided$. This can be formulated as
\begin{equation}
	\HSImageLinReg_i(u, v) = \LinRegA_{x,y} \Guide(u, v) + \LinRegB_{x, y} \quad \forall (u, v) \in \WindowGuide_{x,y},
\end{equation}
where $\LinRegA_{x,y}$ and $\LinRegB_{x,y}$ are the linear regression coefficients for window $\WindowGuide_{x,y}$. Within this window the basic structure is assumed to be the same, however, scale factors, flips and offsets are allowed. Here, one can see why a proper guide is essential for this technique to work. If the guide $\Guide$ is too noisy and edges are not shown properly, the resulting hyperspectral image will be disturbed heavily, since the noise will take direct influence through the parameter $\LinRegA_{x,y}$. The loss function
\begin{equation}
	\begin{aligned}
		&\argmin_{\LinRegA_{x,y}, \LinRegB_{x,y}} \GuidedFilteringLoss(\LinRegA_{x,y}, \LinRegB_{x,y}) =\\
		&\argmin_{\LinRegA_{x,y}, \LinRegB_{x,y}} \sum_{(u, v) \in \WindowPixelSet} \Bigl( \bigl( \LinRegA_{x,y} \Guide(u, v) + \LinRegB_{x, y}	- \HSImageNoisy_i(u, v) \bigr)^2 + \GuidedFilteringParam \LinRegA_{x,y}^2 \Bigr)
	\end{aligned}
\end{equation}
is needed to estimate these coefficients $\LinRegA_{x,y}$ and $\LinRegB_{x, y}$, where ${\WindowPixelSet = \{(u, v) | (u, v) \in \WindowGuide_{x,y}\} }$ and $|\WindowPixelSet| = \BSGuided^2$, or in other words, $\WindowPixelSet$ contains all pixel positions within window $\WindowGuide_{x,y}$. This minimization problem tries to simultaneously minimize the difference between the noisy hyperspectral image and the constructed image out of the guide and the coefficients $\LinRegA_{x,y}$. The intuition for this is that in homogeneous regions the denoised hyperspectral image should rely more on averaging the noisy hyperspectral image to reconstruct proper spectra, while in regions with a lot of structure the guide contains the valuable information of the position of these structures. The balance between these extreme cases is kept with the hyperparameter $\GuidedFilteringParam$. If $\GuidedFilteringParam$ is high, the parameters $\LinRegA_{x,y}$ have to be low. Thus, the reconstruction relies more on using spatial information and the image is smoother. This might end in an image that is too smooth. If $\GuidedFilteringParam$ is low, more structures will be adopted. This might end in an image, that adopted some noisy parts of the guided image.

Taking the derivative with respect to $\LinRegA_{x,y}$ and $\LinRegB_{x, y}$ leads to
\begin{equation}
	\begin{aligned}
		\frac{\partial \GuidedFilteringLoss}{\partial \LinRegA_{x,y}} = \LinRegA_{x,y} \sum_{(u, v) \in \WindowPixelSet} \Guide(u, v)^2 + \LinRegB_{x,y} \sum_{(u, v) \in \WindowPixelSet} \Guide(u, v) -\\
		-\sum_{(u, v) \in \WindowPixelSet} \Guide(u, v)\HSImageNoisy_i(u, v) + \GuidedFilteringParam \BSGuided^2 \LinRegA_{x,y} \mbeq 0,
	\end{aligned}
\end{equation}
where $\mbeq$ is an operator which enforces equality, and
\begin{equation}
	\frac{\partial \GuidedFilteringLoss}{\partial \LinRegB_{x,y}} = \LinRegA_{x,y} \sum_{(u, v) \in \WindowPixelSet} \Guide(u, v) + \LinRegB_{x,y} \BSGuided^2 - \sum_{(u, v) \in \WindowPixelSet} \HSImageNoisy_i(u, v)\mbeq 0,
\end{equation}
respectively. Using the second equation, one can solve for the offset coefficient $\LinRegB_{x,y}$
\begin{equation}
	\LinRegB_{x,y} = \HSImageMean_{i, \WindowGuide} - \LinRegA_{x,y}\GuideMean_{\WindowGuide},
\end{equation}
where $\GuideMean_\WindowGuide$ and $\HSImageMean_{i, \WindowGuide}$ is the mean of $\Guide(u, v)$ and $\HSImageNoisy_i(u, v)$ in the window $\WindowGuide_{x,y}$, respectively. Then one can use this result and plug $\LinRegB_{x,y}$ into the derivative for scale coefficient $\LinRegA_{x,y}$, which leads to
\begin{equation}
	\LinRegA_{x,y} = \frac{\frac{1}{\BSGuided^2} \sum_{(u, v) \in \WindowPixelSet} \Guide(u, v) \HSImageNoisy_i(u, v) - \GuideMean_\WindowGuide \HSImageMean_{i,\WindowGuide}}{\text{var}_{\WindowGuide_{x,y}}\left(\Guide(u, v)\right) + \GuidedFilteringParam},
\end{equation}
where $\text{var}_{\WindowGuide_{x,y}}\left(\Guide(u, v)\right)$ calculates the variance of $\Guide(u, v)$ in the window $\WindowGuide_{x,y}$.

Since every pixel is the center of one window, there are as many windows as noisy pixels, assuming the image was padded beforehand. This is also referred to as a sliding window. Then, the above presented calculations are performed for every window. This results in two parameters for every window, namely $\LinRegA_{x,y}$ and  $\LinRegB_{x,y}$. Therefore, the result of these calculations are basically two coefficient images, where one contains all $\LinRegA_{x,y}$, and the other carries all $\LinRegB_{x,y}$. Consequently, these images have the same width and height as the noisy image. 
Now one could calculate one denoised hyperspectral pixel $(x, y)$ by taking the coefficients $\LinRegA_{x,y}$ and $\LinRegB_{x,y}$ and plug them into the linear regression equation. However, the pixel is part of much more windows, and thus much more regression coefficients could be used. Therefore, to get the denoised image pixel $(x, y)$, the parameters $\LinRegA_{x,y}$ and $\LinRegB_{x,y}$ are averaged for every window, that contains pixel $(x, y)$
\begin{equation}
	\HSImageFiltered_i(x, y) = \frac{1}{\BSGuided^2} \sum_{(u, v) \in \WindowPixelSet} \LinRegA_{u,v} \Guide(x, y) + \frac{1}{\BSGuided^2} \sum_{(u, v) \in \WindowPixelSet} \LinRegB_{u,v}.
\end{equation}
This can be implemented again using a sliding window over the coefficient images, which has the purpose of averaging all coefficients within that window. Since every pixel has its own window again, every pixel has its own two averaged coefficients for the linear regression. This sliding window must have the same block size $\BSGuided$. Assuming being able to generate a perfect guide, which indicates smooth region and edges, this method nicely can rely on neighboring pixels with the same spectral properties while preserving edges to other spectra. Consequently, the block size should be chosen relatively small, because the individual images of a three-dimensional hyperspectral block, that contains one or two different reflectance spectra, can be nicely mapped by the linear regression. On the other hand, the relationship between more spectra might not be projected by a linear regression of the guide properly, because the guide image is fixed.

\subsection{Guide generation}
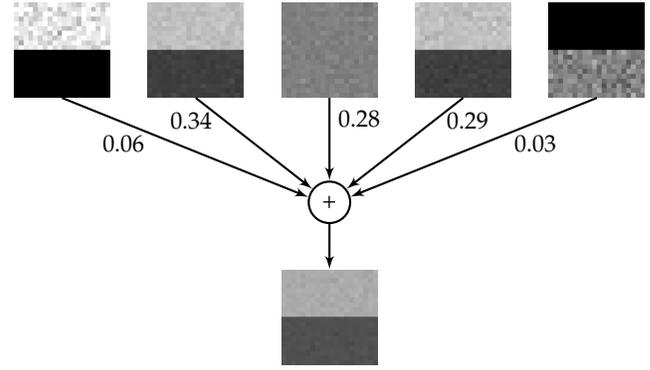
\begin{figure}[t!]
	\centering
	\begin{tikzpicture}
		\tikzstyle{l} = [draw, -latex',thick]
		\tikzstyle{c} = [circle, draw, node distance=1.1cm, thick, minimum size=0.25cm]
		
		\node[inner sep=0pt] (i1) at (0.0in, 0) {\includegraphics[width=0.5in]{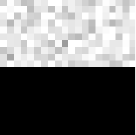}};
		\node[inner sep=0pt] (i2) at (0.7in, 0) {\includegraphics[width=0.5in]{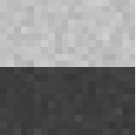}};
		\node[inner sep=0pt] (i3) at (1.4in, 0) {\includegraphics[width=0.5in]{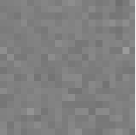}};
		\node[inner sep=0pt] (i4) at (2.1in, 0) {\includegraphics[width=0.5in]{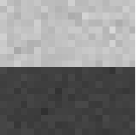}};
		\node[inner sep=0pt] (i5) at (2.8in, 0) {\includegraphics[width=0.5in]{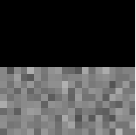}};
		
		\node[c, below=of i3](rc) {+};
		
		\node[inner sep=0pt] (gg) at (1.4in, -1.4in) {\includegraphics[width=0.5in]{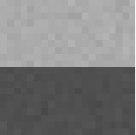}};
		
		\path[l] (i1.270) -- (rc.166) node[near start, below=0.02in] {$0.06$};
		\path[l] (i2.270) --(rc) node[near start, left=0.02in] {$0.34$};
		\path[l] (i3.270) -- (rc) node[near start, right] {$0.28$};
		\path[l] (i4.270) -- (rc) node[near start, right=0.02in] {$0.29$};
		\path[l] (i5.270) -- (rc.15) node[near start, below=0.02in] {$0.03$};
		\path[l] (rc) -- (gg);
	\end{tikzpicture}
	\caption{\label{fig:guide_generation} Example guide generation. The amount of noise in the guide image is strongly reduced due to a smart mixing procedure. Note that it does not matter, that the middle image does not show the structure of the image, or that the last image is flipped.}
\end{figure}
As already mentioned, for this technique to work properly, it is crucial to generate a guide image, which has much less noise than the multispectral images. The idea to generate a proper guidance image $\Guide$ is to use the multispectral images, which roughly indicate differences in spectra, while taking advantage of the image-independent noise. Consequently, a weighted average of multispectral images is formed
\begin{equation}
	\Guide(x, y) = \sum_{i=1}^{\NumMSChannels} \GuideWeights_i \MSImage_i(x, y),
\end{equation}
where $\GuideWeights_i$ indicates the weight for the $i$-th multispectral image. The weights should take the amount of noise as well as the image signal power and correlation into account. This concept is shown in Fig. \ref{fig:guide_generation}. Therefore, the signal-to-noise ratio (SNR) is optimized with respect to the weights $\GuideWeights$. It is assumed that the guide at pixel $(x, y)$ can be split into an unknown signal part and an independent unknown noise part
\begin{equation}
	\Guide(x, y) = \sum_{i=1}^{\NumMSChannels} \GuideWeights_i \MSImageNoiseless_i(x, y) + \sum_{i=1}^{\NumMSChannels} \GuideWeights_i \Noise_i = \GuideWeights^{\text{T}}\MSImageNoiseless(x, y) + \GuideWeights^{\text{T}}\Noise,
\end{equation}
where $\MSImageNoiseless(x, y)$ are the noiseless multispectral images. Treating $\MSImageNoiseless(x, y)$ and $\Noise$ as random variables, the SNR is optimized by
\begin{equation}
	\label{eq:SNR}
	\argmax_{\GuideWeights} \frac{\mathcal{E}\left[\GuideWeights^{\text{T}}\MSImageNoiseless(x, y)\MSImageNoiseless(x, y)^{\text{T}}\GuideWeights\right]}{\mathcal{E}\left[\GuideWeights^{\text{T}}\Noise\Noise^{\text{T}}\GuideWeights\right]} = \argmax_{\GuideWeights} \frac{\GuideWeights^{\text{T}}\CorrelationMSImageNoiseless\GuideWeights}{\GuideWeights^{\text{T}}\MSNoiseVarMatrix\GuideWeights},
\end{equation}
where $\mathcal{E}\left[\cdot\right]$ is the expectation operator and $\CorrelationMSImageNoiseless$ is the second-order moment matrix of the denoised multispectral images. This matrix can be estimated using $\CorrelationMSImageNoise = \CorrelationMSImageNoiseless + \MSNoiseVarMatrix$, where $\CorrelationMSImageNoise$ is the second-order moment matrix of the observed multispectral images and therefore can be measured. Here, the independence between the noise and the noise-free multispectral image is used to yield the formula for $\CorrelationMSImageNoise$. The generalized Rayleigh quotient gives an upper bound for \eqref{eq:SNR} and reads as
\begin{equation}
	\frac{\GuideWeights^{\text{T}}\CorrelationMSImageNoiseless\GuideWeights}{\GuideWeights^{\text{T}}\MSNoiseVarMatrix\GuideWeights} \leq \Eigenvalues^{\text{max}}\left(\MSNoiseVarMatrix^{-1}\CorrelationMSImageNoiseless\right),
\end{equation}
where $\Eigenvalues^{\text{max}}(\PlaceholderMatrixA)$ is the maximum eigenvalue of matrix $\PlaceholderMatrixA$. This equation holds with equality if \cite{therrien-discrete-1992}
\begin{equation}
	\GuideWeights = \GuideWeightsScale \ \Eigenvector^{\text{max}}\left(\MSNoiseVarMatrix^{-1}\CorrelationMSImageNoiseless\right),
\end{equation}
where $\Eigenvector^{\text{max}}(\PlaceholderMatrixA)$ is the eigenvector which corresponds to the maximum eigenvalue of matrix $\PlaceholderMatrixA$ and $\GuideWeightsScale$ is an arbitrary scale factor, since a constant scale factor on the weights does not influence the SNR. Therefore, the weights are normalized such, that the guide image $\Guide$ is in the same value bounds as the multispectral images. Consequently, the weights should sum up to one, which reads as
\begin{equation}
	\GuideWeightsScale = \frac{1}{\sum_{i=1}^{\NumMSChannels}\Eigenvector^{\text{max}}\left(\MSNoiseVarMatrix^{-1}\CorrelationMSImageNoiseless\right)}.
\end{equation}
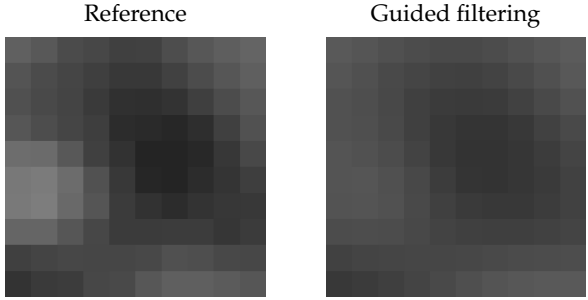
\begin{figure}[t!]
	\begin{center}
		\input{figures/raw_guided_filtering.pgf}
	\end{center}
	\caption{\label{fig:raw_guided_output} Raw guided filtering output for noiseless image. Unfortunately, the guided image filtering tends to smooth out more than it should.}
\end{figure}

\subsection{Mixing}
This guided filtering tends to smooth out more than it should for high SNR images as shown in Fig. \ref{fig:raw_guided_output}. Hence, it is necessary to lower the influence in these situations. Therefore, the noisy hyperspectral images, which are not heavily affected by noise for high SNR images anymore, and the result of the guided filtering are mixed to form the final result
\begin{equation}
	\begin{aligned}
		\HSImageEst^{\text{SPRE}}_i(x, y) &= \MixingWeights_i \HSImageNoisy_i(x, y) + (1 - \MixingWeights_i) \HSImageFiltered_i(x, y)\\
		&= \HSImageFiltered_i(x, y) + \MixingWeights_i (\HSImageNoisy_i(x, y) - \HSImageFiltered_i(x, y))\\
		&= \HSImageFiltered_i(x, y) + \MixingWeights_i \HSImageDiff_{i}(x, y),
	\end{aligned}
\end{equation}
where $\MixingWeights_i$ is the mixing parameter for hyperspectral image $i$ and bounded between zero and one, and $\HSImageDiff_i(x, y)$ is the difference image between the $i$-th noisy hyperspectral image and the $i$-th filtered hyperspectral image.

Now, the weight $\MixingWeights_i$ should be chosen such that the variance of the difference between the final result and the noisy hyperspectral image should be equal to the diagonal of the hyperspectral noise matrix, since the influence of the guided filtering should be lowered if it modifies the noisy hyperspectral image more than the hyperspectral noise variance allows. The noise variance for each hyperspectral image resulting from the multispectral noise can be calculated by
\begin{equation}
	\HSNoiseVarMatrix = \mathcal{E}\left[\WienerSSW\Noise\Noise^{\text{T}}\WienerSSW^{\text{T}}\right] = \WienerSSW \MSNoiseVarMatrix \WienerSSW^{\text{T}},
\end{equation}
where $\WienerSSW = \PickerSpatioSpectral\CorrelationSpatioSpectral\BigFilterMatrix^{\text{T}}(\BigFilterMatrix\CorrelationSpatioSpectral\BigFilterMatrix^{\text{T}} + \MSNoiseVarMatrix)^{-1}$ is the spatio-spectral Wiener filter and $\HSNoiseVarMatrix$ is the hyperspectral noise variance matrix. Subsequently, the condition
\begin{equation}
	\begin{aligned}
		\mathcal{E}\left[\left(\HSImageFiltered_i(x, y) + \MixingWeights_i \HSImageDiff_{i}(x, y) - \HSImageNoisy_i(x, y)\right)^2\right] &\mbeq \HSNoiseVarMatrix_{ii}\\
		\mathcal{E}\left[\left(\MixingWeights_i \HSImageDiff_{i}(x, y) - \HSImageDiff_{i}(x, y)\right)^2\right] &\mbeq \HSNoiseVarMatrix_{ii}\\
		(\MixingWeights_i - 1)^2\mathcal{E}\left[\left(\HSImageDiff_{i}(x, y)\right)^2\right] &\mbeq \HSNoiseVarMatrix_{ii},
	\end{aligned}
\end{equation}
leads to the optimal coefficients
\begin{equation}
	\MixingWeights_i = \left[1 - \sqrt{\frac{\HSNoiseVarMatrix_{ii}}{\mathcal{E}\left[\left(\HSImageDiff_{i}(x, y)\right)^2\right]}}\right]_{+},
\end{equation}
where $\mathcal{E}\left[\left(\HSImageDiff_{i}(x, y)\right)^2\right]$ can be measured by the actual results and $\left[\cdot\right]_{+}$ sets negative values to zero to fulfill the boundary condition on $\MixingWeights_i$. Negative values would appear if the guided filtering modified the noisy hyperspectral image less than the hyperspectral noise variance allows. Using these coefficients, one can calculate the final result $\HSImageEst^{\text{SPRE}}(x, y)$.

\section{Simulation on synthetic data}
\label{sec:evaluation}

\begin{figure}[t]
	\begin{center}
		\input{figures/filters.pgf}
	\end{center}
	\caption{\label{fig:filters} Filters used to create multispectral data. The different colors indicate different filters.}
\end{figure}
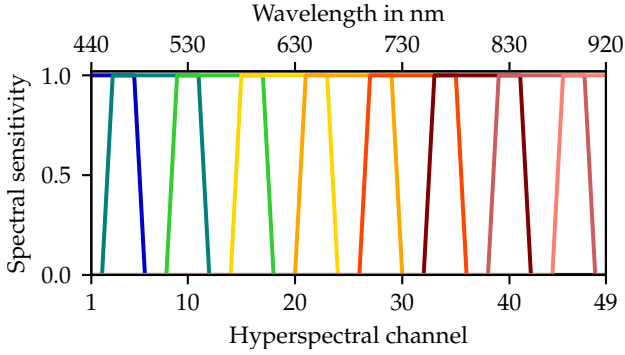
To test the performance of the presented methods, an evaluation of these algorithms is essential. In this paper, two different simulation scenarios are evaluated. First, the illumination intensity level is varied but constant for each multispectral channel. Example images and spectra are shown for this scenario. Afterwards, the illumination intensity level is varied across different multispectral channels, which is mimicking the behaviour of infrared and ultraviolet channels, which usually do not get as much light as the channels in the visible light area.

The multispectral data is created using a synthetic filter matrix. The filters used here are constant band-pass filters, which let a specific spectrum area pass and block all other frequencies. Fig. \ref{fig:filters} shows these filters. They are picked because they are mimicking the filters used for the multispectral camera in Section \ref{sec:applications} fairly well.

To evaluate the noise behaviour of the reconstruction techniques, signal-dependent Poisson noise is added to the multispectral images, which are scaled to fit between zero and one. To simulate different strengths of noise, a sample of 
\begin{equation}
	\label{eq:poisson_noise}
	\MSImage_i(x, y) \sim \frac{\mathcal{P}\left(\IntensityLevel \ \MSImageNoiseless_i(x, y) \right)}{\IntensityLevel}
\end{equation}
is generated, where $\IntensityLevel$ represents the illumination intensity level. A low $\IntensityLevel$ represents a lot of noise, while a high $\IntensityLevel$ is used to simulate a good lighting environment. An illumination intensity level of infinity (inf) represents no added noise at all. In general, the intensity level is closely related to the maximum amount of photons, that are counted by the image sensor.

For the evaluation, two metrics are used. First, the well known mean squared error (MSE) is used to calculate the difference between the original spectrum $\HSChannels$ and the reconstructed spectrum $\HSChannelsEst$
\begin{equation}
	\text{MSE}\left(\HSChannels, \HSChannelsEst\right) = \frac{1}{\NumHSChannels} \sum_{i=1}^{\NumHSChannels} (\HSChannels_i - \HSChannelsEst_i)^2.
\end{equation}
For a whole image, the average of the pixel-wise MSEs is calculated. Since the MSE does not capture how well the shape of the spectrum is reconstructed, another evaluation metric is necessary. Furthermore, an algorithm that reconstructs high energy spectra accurately is preferred by the MSE, since for high energy spectra, the shape is better preserved with the same absolute error in comparison to a low energy spectrum. Therefore, another metric, called spectral angle (SA) \cite{kruse-spectral-1993}, is used to quantify the quality of the shape reconstruction result with
\begin{equation}
	\text{SA}(\HSChannels, \HSChannelsEst) = \arccos{\left(\frac{\HSChannels^{\text{T}}}{||\HSChannels||_2}\frac{\HSChannelsEst}{||\HSChannelsEst||_2}\right)}.
\end{equation}
The spectral angle calculates the angle between two hyperspectral data vectors by applying the inverse cosine to the dot product of the normalized spectra. For a whole image the SA is averaged. Moreover, if the original spectrum is all zero, and therewith the denominator, for a single pixel, this pixel does not contribute to the SA. If the reconstructed spectrum is all zero and the original spectrum is not, the SA for this pixel is $\pi$, which is the highest possible value. The best SA a spectrum can have is zero, however, the reconstructed spectrum can be a scaled version of the original version to reach this value.
\begin{figure}[t]
	\begin{center}
		\input{figures/example_rgb_images.pgf}
	\end{center}
	\caption{\label{fig:rgb_examples} Four example false color images of the database used for the evaluation. All images from the evaluation database are recorded in an urban scene, while maintaining a good mix between plants, houses, objects and public places. Furthermore, the textures of the houses and the underground vary a lot.}
\end{figure}
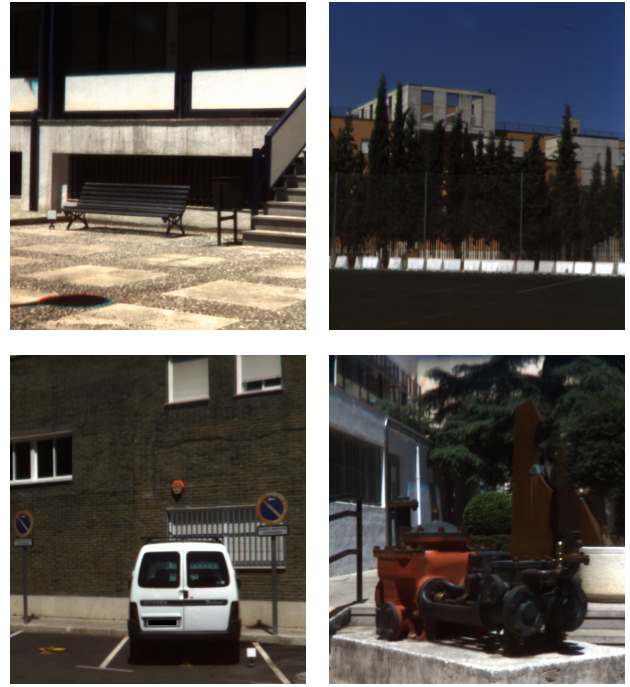
\begin{figure*}[!p]
	\begin{center}
		\input{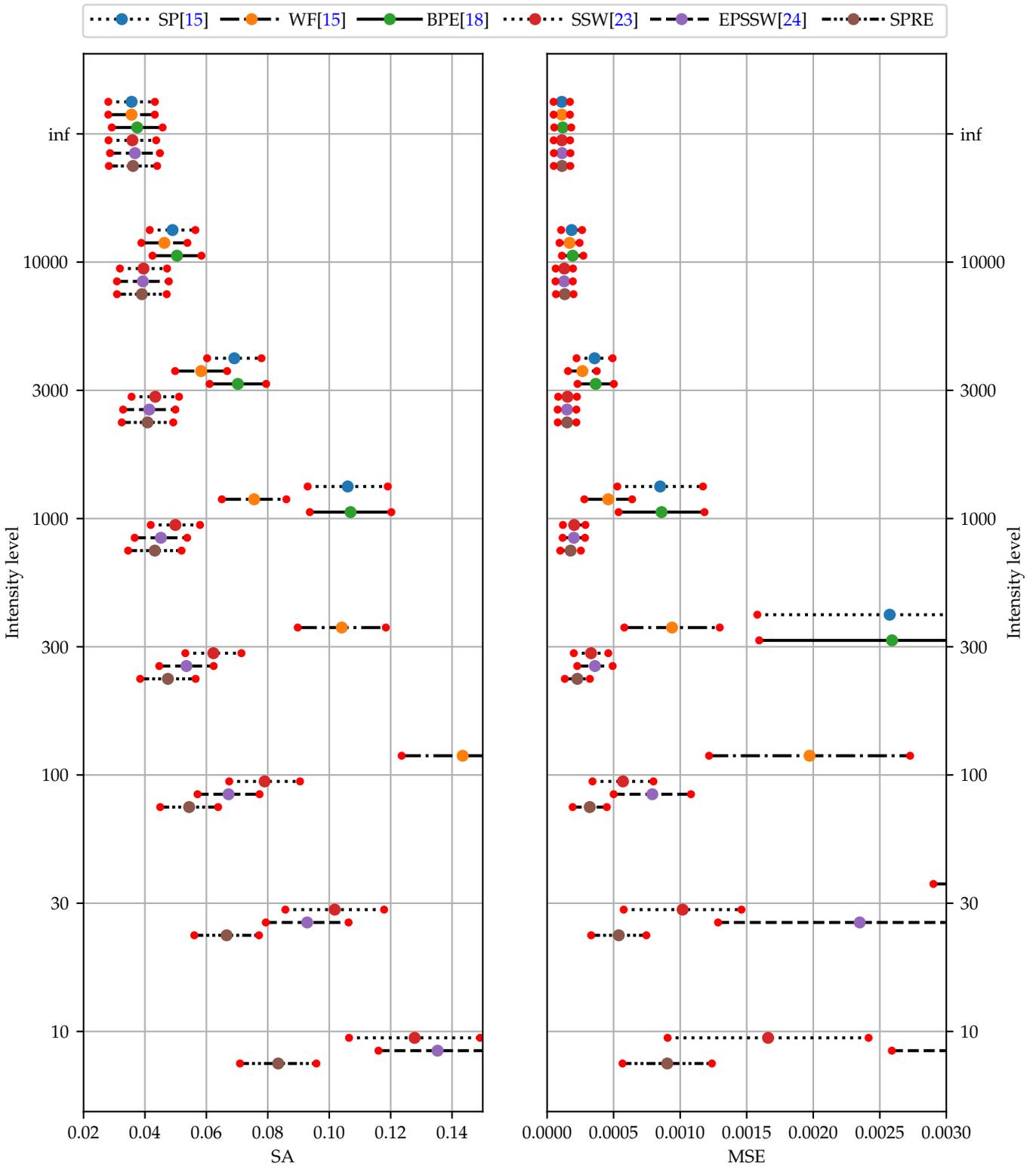}
	\end{center}
	\caption{\label{fig:evaluation_same_noise}Evaluation of the SA and the MSE. Lower is better for both metrics. The big circles indicate the mean while the horizontal lines with the small red circles at the end show plus/minus the standard deviation. The mean and the standard deviation are calculated over 14 images of the evaluation database. Simulated Poisson noise with different intensity levels is applied to every multispectral image uniformly. Note that our novel SPRE method outperforms all other methods under the influence of noise.}
\end{figure*}
\begin{figure*}[!p]
	\begin{center}
		\input{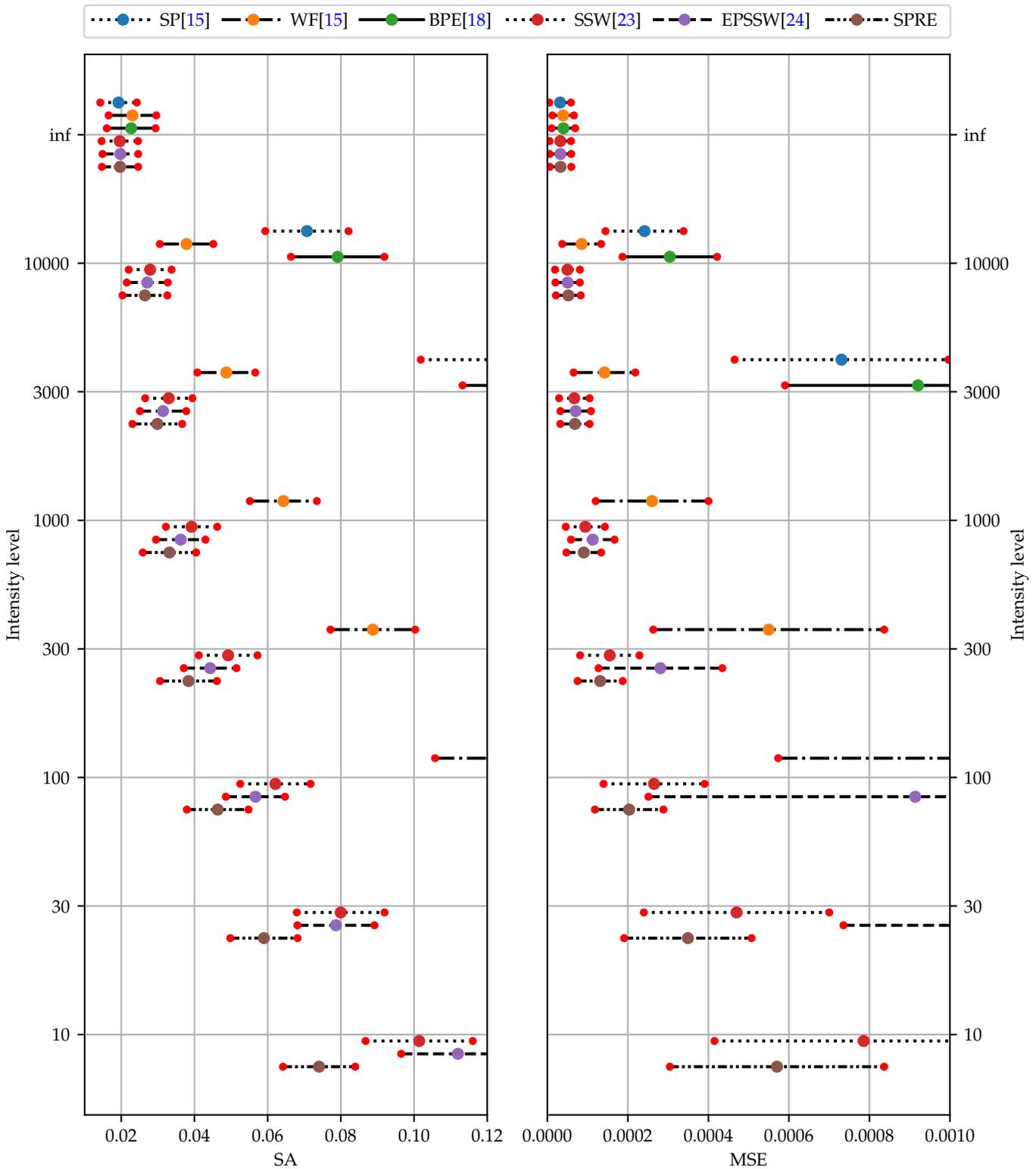}
	\end{center}
	\caption{\label{fig:evaluation_same_noise_database_6}Evaluation with different intensity levels applied to every multispectral image uniformly on a representative subset of images of the ARAD database \cite{leibe-sparse-2016}, namely \textit{4cam\_0411-1648}, \textit{BGU\_0403-1419-1}, \textit{eve\_0331-1656}, \textit{gavyam\_0823-0944}, \textit{Labtest\_0910-1513}, \textit{Lehavim\_0910-1630}, \textit{lst\_0408-0950}, \textit{nachal\_0823-1127}, \textit{objects\_0924-1634}, \textit{omer\_0331-1130}, \textit{peppers\_0503-1315}, \textit{plt\_0411-1200-1}, \textit{prk\_0328-1025}, \textit{rsh\_0406-1441-1}, \textit{sat\_0406-1107}, \textit{selfie\_0822-0906}. The subset is chosen, since the database contains a lot of repetitive hyperspectral images. Exactly the same parameter for the algorithms were used as described in the evaluations. Note that the hyperspectral images of the ARAD database only have 31 channels. Again, the big circles represent the mean while the horizontal lines with the small red circles at the end show plus/minus the standard deviation. These results are close to the results using the evaluation database.}
\end{figure*}
\begin{figure}[htp]
	\begin{center}
		\input{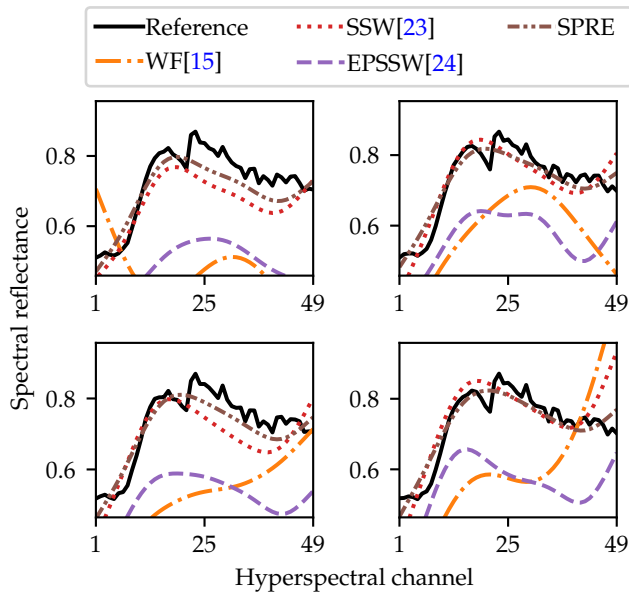}
	\end{center}
	\caption{\label{fig:example_spectra}Example spectra of a $2 \times 2$ uniform region with intensity level 10. The EPSSW has a problem reconstructing the offset properly, while the SSW shows a strong peak in the bottom right, that is not present at all. The WF results are not usuable at all.}
\end{figure}
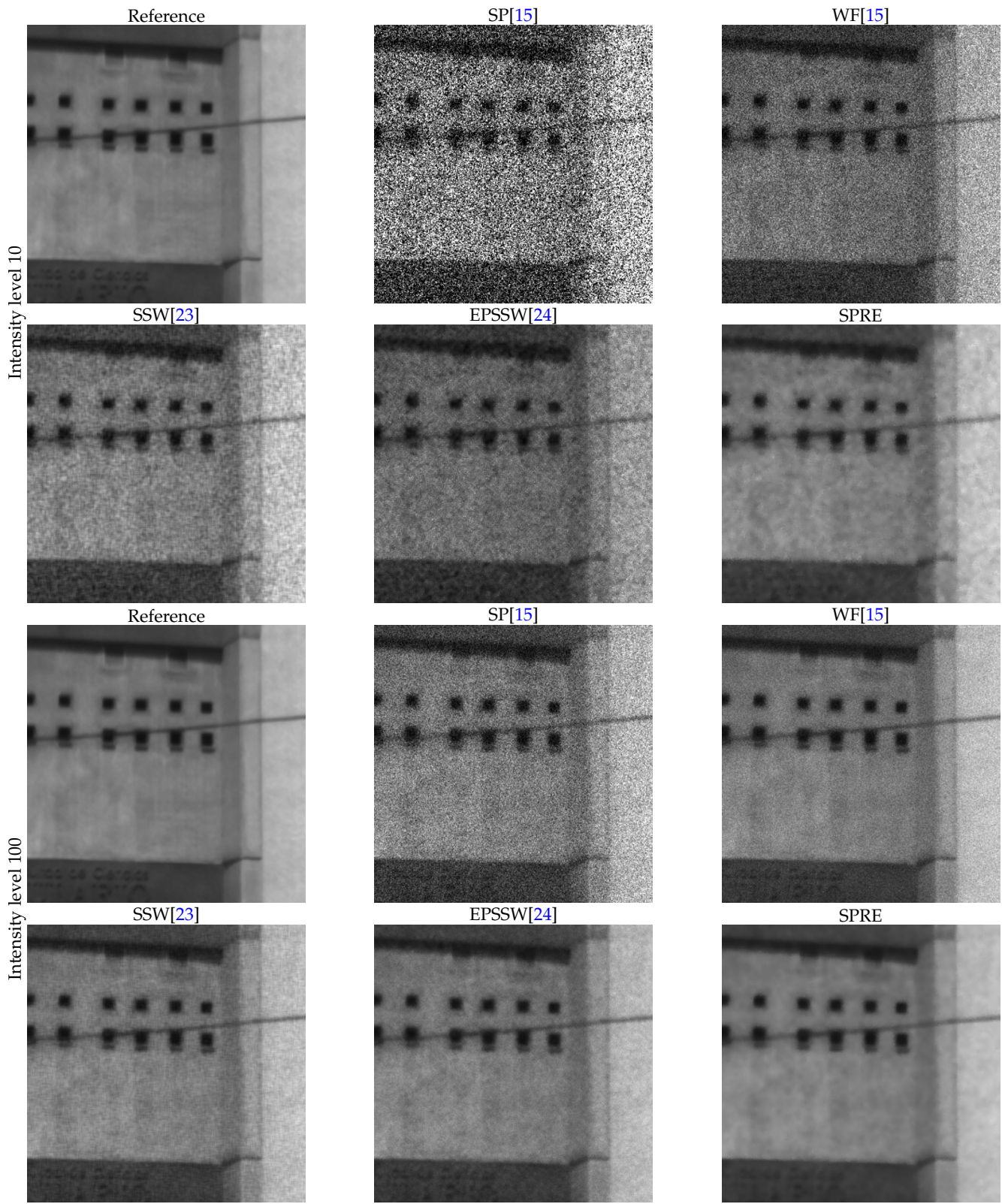
\begin{figure*}[!p]
	\begin{center}
		\input{figures/example_reconstruction_images_2.pgf}
	\end{center}
	\caption{\label{fig:example_images}Example reconstruction images of channel 7 with intensity level 10 and intensity level 100. The images are generated by first generating multispectral images out of the reference hyperspectral image by applying synthetic filters. Then, Poisson noise is added to the multispectral channels. In the end, the reconstructed hyperspectral images are created using the different algorithms. Here, the resulting images are cropped to a $300\times300$ area to get a better impression of the artefacts. Channel 7 is picked exemplarily and due to its diversity in grayscale. Note that the \textit{reference} image is the ground truth.}
\end{figure*}
\begin{figure*}[t]
	\begin{center}
		\input{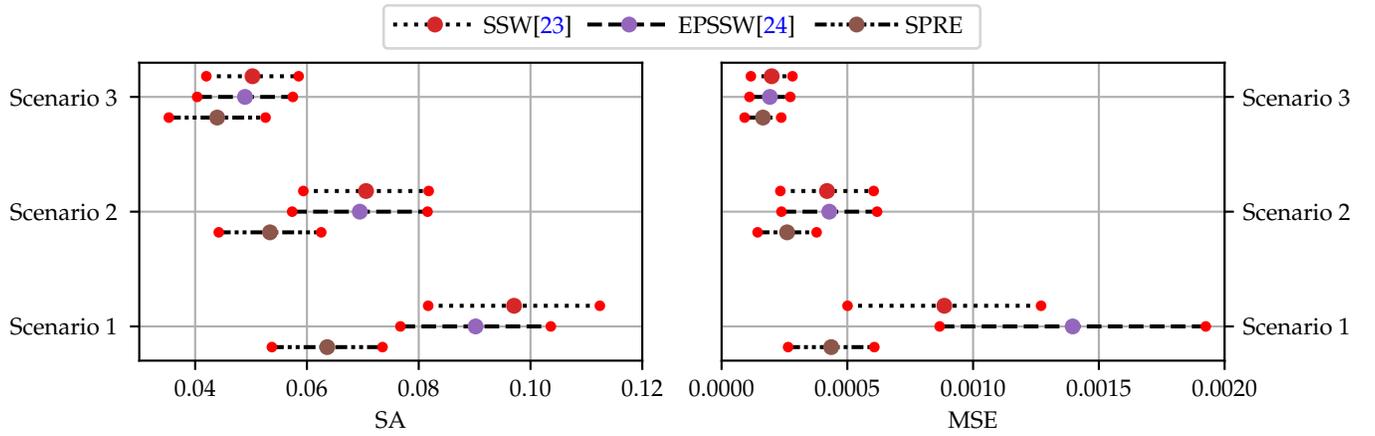}
	\end{center}
	\caption{\label{fig:evaluation_different_noise}Evaluation of different scenarios, where the multispectral channels on the border are more heavily corrupted by noise than the middle ones. The amount of noise on the border and in the middle is varied between each scenario according to Table \ref{tab:intensity}. The big circles indicate the mean, while the vertical bars with small red circles at the end show plus/minus the standard deviation over the 14 images. Other algorithms are omitted, since they perform even worse and would decrease the readability of this plot. Note that our novel SPRE algorithm is better than the SSW and EPSSW.}
\end{figure*}
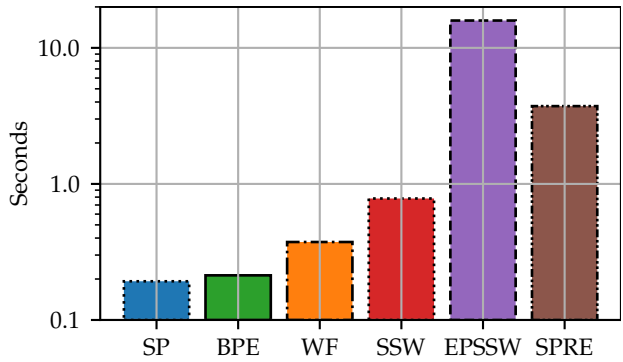
\begin{figure}[t]
	\begin{center}
		\input{figures/runtime_evaluation.pgf}
	\end{center}
	\caption{\label{fig:evaluation_runtime} Average runtimes for the different algorithms for an image of size $1000 \times 900$ at intensity level 10.}
\end{figure}

To create multispectral data and evaluate reconstructed spectra, a hyperspectral database is necessary. The database for the evaluations shown here was created by \cite{eckhard-outdoor-2015}. In total there are 14 hyperspectral pictures showing different urban scenes. Some of these scenes are shown in Fig. \ref{fig:rgb_examples}. Unfortunately, several images are heavily corrupted by noise, specifically the first four and last eight bands. Therefore, these bands are omitted to obtain a meaningful result. This results in 49 hyperspectral bands uniformly distributed in the spectrum area of 440nm to 920nm (both inclusive).

Moreover, hyperparameters of the individual methods need to be fixed. First of all, all methods are using the second-order differences as a-priori information about the spectrum. Therefore, the correlation matrix $\CorrelationHS$ is set to $\SmoothingMatrixScale \left(\DifferenceMatrix_2^{\text{T}}\DifferenceMatrix_2 + \SmoothingMatrixInvFactor\Identity\right)^{-1}$. The technique to get a proper scale factor $\SmoothingMatrixScale$ for each reconstruction method is described in Appendix \ref{subsec:connection}. Parameter $\SmoothingMatrixInvFactor$ is set to $0.0001$ to avoid inversion problems. The block size for the SSW\cite{murakami-color-2008}, the EPSSW\cite{urban-spectral-2009} and our proposed SPRE is set to $\BSRecon=5$ and $\BSGuided=5$, since this blocksize is able to use enough spatial information, while still yielding acceptable runtimes. Furthermore, the spatial weight variance is set to 16, while the range weight variance is set to 0.4 for the edge-preserving reconstruction technique. The decay factor of the SSW is set to $\SpatialDecay=0.97$. Parameter $\GuidedFilteringParam$ is set to $0.001$. All of these hyperparameters are found by a manual optimization using the CAVE hyperspectral database, described in \cite{yasuma-generalized-2010}. The blocksize for denoising and reconstruction is the same as in the paper for the SSW and the EPSSW.

Fig. \ref{fig:evaluation_same_noise} shows the mean and the standard deviation over 14 images in a graphical manner, such that one can easily compare different methods. This statistical measure indicates that both metrics are robust regarding the evaluation of different images, since the standard deviation is often only a fraction of the mean. In general, the SA seems to be more robust than the MSE due to the lower standard deviation to mean ratio. Note that all standard deviations for intensity level infinity are approximately the same, which serves as plausibility check.

In environments with low light, and thus images with heavy noise, the novel structure-preserving algorithm outperforms all other methods. The algorithm BPE\cite{mansouri-representation-2008} is based on the discrete cosine transform (DCT). This reconstruction technique calculates the optimal weights for the first $\NumMSChannels$ DCT functions by a matrix inversion. Therefore, this method also works out of the box, but has an entirely different approach to solve the underdetermined system. This method was added to the evaluation, because it is an entirely different type of algorithm. Thus, if the BPE would perform much better than the SP, one should rather try to embed the spatial correlation into this method. However, this is not the case, since the single-pixel SP\cite{pratt-spectral-1976} without noise consideration as well as the DCT-based method BPE output very similar results and even struggle with very light noise. This gets better when using the WF\cite{pratt-spectral-1976}, which models the noise, but still does not use any spatial information, which ends in a bad performance for darker environments. Then, the EPSSW and the SSW share the second place. Depending on the metric, the one or the other method performs better for different environment.

The higher MSE produced by the EPSSW for heavy noise environments can be explained by Fig. \ref{fig:example_spectra}. One can see that this method has a problem with reconstructing the offset of the spectrum properly, while the estimation for the shape is much better. Looking at more reconstructed spectra examples, the pattern, that the reconstructed spectrum is below the actual spectrum, is repeated. This has several reasons. First of all, a zero mean of the spectrum is implicitly assumed for all reconstruction Wiener filters. This impacts the WF and EPSSW more heavily than the SSW. Consequently, the EPSSW heavily depends on the trained spectrum mean for reconstructing the offset. Furthermore, this problem may be amplified by the bilateral Wiener denoising prior to reconstruction. The Poisson noise affects pixel with a high intensity more. Therefore, the difference vector to higher intensity might be higher, compared to the difference vector with lower intensity pixels than the middle pixel. Thus, the similarity to darker pixels will be higher, which results in an offset towards zero for the reconstructed spectra. However, this offset problem is not extremely important, since most classification algorithms rely on the shape of the spectrum. Therefore, the SA metric is more meaningful than the MSE. For the intensity levels up to 1000 the structure-preserving reconstruction method outperforms the other methods. For very low SNR environments, the proposed SPRE lowers the MSE up to 46\% and the SA up to 35\%. For low SNR multispectral images (intensity level 100) our SPRE still outperforms the EPSSW by 19\% in the SA and decreases the MSE by approximately 44\% compared to the SSW. For good lighting environments, the spatial-considering methods roughly perform on the same level, while the novel SPRE algorithm is still a tiny bit better in the noisy scenarios. This is an expected behaviour since the methods are all based on the same Wiener filter, which assumes spatial smoothness. If there is no noise at all (intensity level infinity), all methods produce more or less the same result. Therefore the metrics are almost identical for this case.

Note that very similar results are obtained when evaluating these algorithms with the same parameters on another database \cite{leibe-sparse-2016}, that contains natural urban scenes, see Fig. \ref{fig:evaluation_same_noise_database_6}. Both metrics, namely the MSE and the SA, still can be decreased up to 27\% compared to the SSW with intensity level 10. The novel algorithm SPRE outperforms all other methods in the noisy scenarios. The SP and BPE perform even worse using this database, which emphasizes the need for methods which reduce the influence of noise effectively. Interestingly, the shape of the curves for the SSW and the EPSSW are very similar. Especially, the SA of the EPSSW is always better than the SA of the SSW except for intensity level 10, where the SSW wins.

Fig. \ref{fig:example_spectra} also reveals some other properties. While the spectra of our structure-preserving reconstruction is pretty consistent and fairly usable throughout all pixels, the spectra of the other spatio-spectral algorithms differ a lot in comparison. Furthermore, some of these spectra do not represent the spectrum well at all. The single-pixel WF is expected to perform bad in this environment.

Fig. \ref{fig:example_images} depicts visual examples from reconstructed hyperspectral images of channel 7. This channel is picked exemplarily and because it has a wide range of different grayscale values. While all other methods struggle to compensate the noise, the proposed SPRE manages to eliminate a lot of artifacts through its structure-adaptive smoothing. This leads to the insight that this new method produces much better visual results in noisy environments. Unfortunately, this figure also reveals that if the guide does not contain the structure, the spatial resolution will decrease slightly, since edges are smoothed out. Further work will tackle this problem by an advanced guide generation technique.

Fig. \ref{fig:evaluation_different_noise} shows another evaluation of noisy environments. However, this time the multispectral images are affected by different intensity levels. Especially in environments with a limited exposure time, this situation is very common, since the amount of photons captured by the individual multispectral channels varies a lot. Therefore, it is crucial to measure the performance of the different algorithms in this circumstance, especially regarding the real-world experiments in Section \ref{sec:applications}. Typically, the amount of photons captured in the visible light area is much higher, and therefore the noise variance much lower than in the infrared or ultraviolet area. The outer channels of the database used for the evaluation are at least close to or even in these areas and the multispectral channels also can be located to these parts of the spectrum due to the bandpass filters. Consequently, this behaviour is modeled by a $\cap$-shape of the intensity levels. This corresponds to more noise in the outer channels, because the amount of photons (intensity) that reach the sensor is much lower than in the middle channels. Generally, different functions could be used to model this bell shape. In this case, a sinusoidal is used due to its simple formulation, but of course  a Gaussian could also be used. Specifically, the intensity levels $\IntensityLevel(i)$ used for the Poisson noise in \eqref{eq:poisson_noise} are
\begin{equation}
	\IntensityLevel(i) = \sin(\frac{i}{\NumMSChannels - 1} \pi) \ (\IntensityLevel_\text{max} - \IntensityLevel_\text{min}) + \IntensityLevel_\text{min},
\end{equation}
where $\IntensityLevel_\text{max}$ and $\IntensityLevel_\text{min}$ are the maximum (middle of $\cap$-shape) and minimum (borders of $\cap$-shape) intensity levels, respectively. Three different scenarios are evaluated defined by Table \ref{tab:intensity}.
\begin{table}
	\begin{center}
		\begin{tabular}{ c c c }
			Scenario & $\IntensityLevel_\text{min}$ & $\IntensityLevel_\text{max}$ \\ \hline
			1 & 10 & 100 \\
			2 & 10 & 1000 \\
			3 & 100 & 10000.
		\end{tabular}
	\end{center}
	\caption{Definition of intensity levels for different evaluation scenarios.}
	\label{tab:intensity}
\end{table}
The result of this evaluation is shown in Fig. \ref{fig:evaluation_different_noise}. The vertical bars are showing the standard deviation. Again, the standard deviation leads to the insight that the SA and the MSE are robust metrics. The SP, WF and BPE are omitted, since they even perform worse than the spatial methods for high intensity levels in the previous evaluation, which draws a similar picture as this evaluation. The SSW and the EPSSW are alternating in outperforming the other one. Again, the proposed novel method performs best in all three scenarios for both metrics. Speaking in quantitative words, the proposed SPRE decreases the SA by 29\%, 23\% and 10\% compared to the second place in the three scenarios, respectively. The MSE can be decreased by approximately 51\%, 38\% and 14\% compared to the second best method, respectively. Especially, for the high noise scenario 1 our novel algorithm produces much better results than the other methods.

Fig. \ref{fig:evaluation_runtime} shows a short evaluation of the runtimes of different algorithms on a single thread. The runtime of all images (size $1000\times900$) of the evaluation database, corrupted by noise with intensity level 10, is measured. Note that the algorithms are not optimized and the algorithms are implemented in Python, mainly using numpy. However, this evaluation still indicates, in which order of magnitudes the algorithms stay. If one needs a real-time reconstruction of a multispectral video, embedding spatial correlation will be difficult, but not impossible, since there are a lot of possibilities for parallelization for all spatial algorithms. However, if one wants to postprocess a multispectral video, all algorithms are suitable.

\section{Real-world data experiment}
\label{sec:applications}

\begin{figure}[t]
	\centering
	\includegraphics[width=\linewidth]{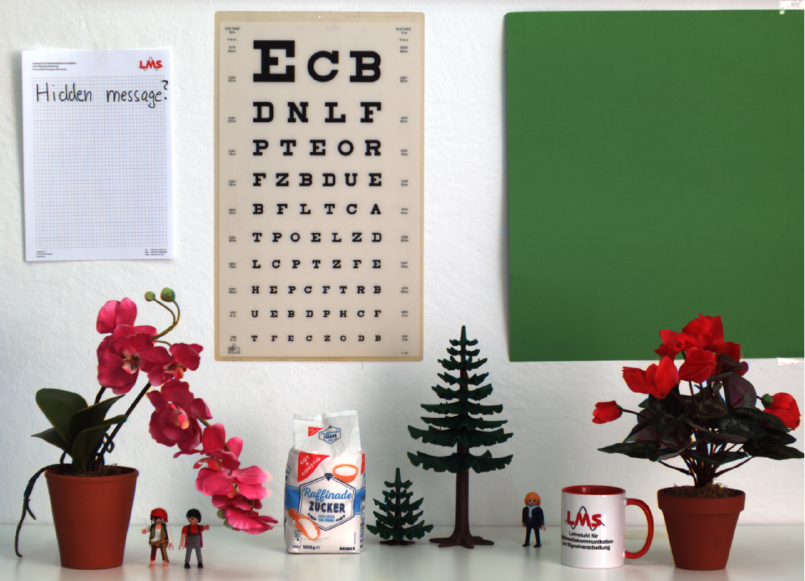}
	\caption{\label{fig:real_rgb} The scene used for evaluating the algorithms with a real-world multispectral camera. The scene shows a good mixture between uniform regions and areas with a high frequency.}
\end{figure}
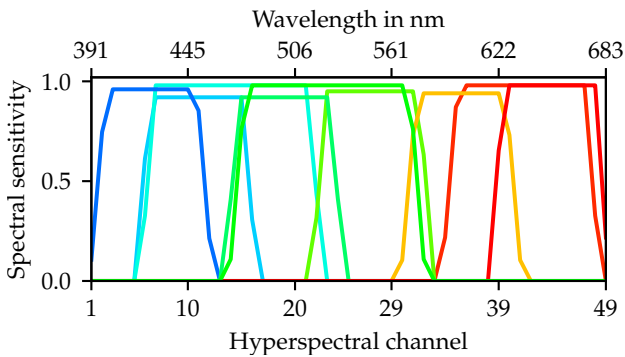
\begin{figure}[t]
	\begin{center}
		\input{figures/real_filters.pgf}
	\end{center}
	\caption{\label{fig:real_filters} The filters of the multispectral camera. Note that these filter are quite similar to the ones used in the simulation of the synthetic data.}
\end{figure}
Though synthetic filters and synthetic noise allow for flexible evaluations, real-world data may differ from the simulated environment. Therefore, an evaluation based on this data is essential. To overcome the absence of a hyperspectral camera, which could produce reference data, the following procedure is used. Firstly, the scene shown in Fig. \ref{fig:real_rgb} is recorded with an exposure time of 1 ms, which imitates a frame in a video sequence heavily affected by noise. Secondly, the same scene is captured multiple times (in this case 1000) and averaged to reduce shot noise to an extent where it is barely noticeable. Furthermore, a scene with no light at all is captured multiple times to extract static noise, which is removed from the averaged scene image. This leads to multispectral images, which contain barely any noise. This near noiseless image is then reconstructed by the single-pixel WF, which serves as reference to evaluate the spatial reconstruction methods in the noisy scenario. The single-pixel WF is chosen over the spatial reconstruction algorithms, because this procedure eliminates the possibility of a good evaluation of the chosen spatial reconstruction procedure due to reconstructing the same artifacts. For example, a hypothetical reference algorithm, that always produces white hyperspectral images, will score a perfect evaluation in the noisy scenario.

The scene is captured using the multispectral camera shown in Fig. \ref{fig:camsi}. Unfortunately, the transfer functions of the filters are not exactly known. Thus, they are approximated by trapezoids. Consequently, parameter $\SmoothingMatrixInvFactor$ is set to $0.00001$ (from $0.0001$ in the evaluations), which leads to smoother spectra, since a lower $\SmoothingMatrixInvFactor$ increases the correlation between different spectrum entries. Afterwards, these filter curves are sampled to yield 49 hyperspectral channels, which is the same amount of hyperspectral channels as for the evaluation. The resulting filter curves are shown in Fig. \ref{fig:real_filters}. These values are embedded into the filter matrix $\FilterMatrix$, which has a size of $9 \times 49$ in this case. Other parameters are equivalent to the ones used in the evaluation.
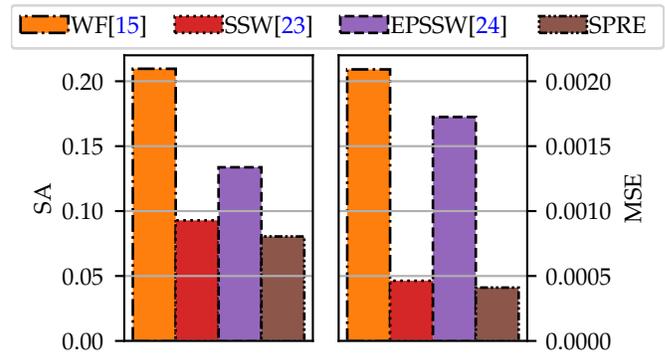
\begin{figure}[t]
	\begin{center}
		\input{figures/real_evaluation.pgf}
	\end{center}
	\caption{\label{fig:real_evaluation} Evaluation of different algorithm tested against the reconstruction results of the Wiener filter of the clean image. Again, the novel SPRE method wins this evaluation.}
\end{figure}

The results of the evaluation are shown in Fig. \ref{fig:real_evaluation}. While the EPSSW outperforms the reference algorithm in the noisy case, the SSW and our novel SPRE algorithm outperform the reference algorithm even more. Furthermore, the SSW and our SPRE deliver similar results, however, our novel method still performs better. This similarity is confirmed by evaluating the SA between the results of the SSW and the results of our SPRE, which results in 0.0284. The guided filtering still has a lot of influence on the result of the SPRE method, since the mixing vector $\MixingWeights$ values vary between 0.35 and 0.5.

Since the reference spectra are not known, this evaluation does not give a final evaluation on the real-world performance, and therefore, the results of this specific evaluation should not be weighted too much. Nevertheless, this evaluation still gives a hint, that all spatial reconstruction methods are capable of better managing noise than the single-pixel WF, while our novel SPRE algorithm tends to reduce the influence of noise the most.

\section{Conclusion}
\label{sec:conclusion}

This paper introduced a novel spectral reconstruction method, that focuses on preserving structure in noisy environments like multispectral video sequences. To preserve structure, a guide image is used. The guide image was created exploiting the independence of the noise in different multispectral channels. This technique yields a guide image which contains the basic structure, while the noise variance is much lower compared to a single multispectral image. Afterwards, a technique was shown how to embed this nearly noiseless guide into the spectral reconstruction. It was presented, that this novel reflectance estimation algorithm performs much better in noisy environments, specifically, it lowers the mean squared error and the spectral angle up to 46\% and 35\%, respectively. Moreover, this also applies to noisy environments, where the noise strength varies between different multispectral images. Furthermore, this paper showed how the spectral correlation matrix used by the Wiener filter can be replaced with the inverse squared difference matrix, which is often used by the smoothed pseudoinverse. Using this insight leads to an out-of-the-box design, that is now possible using the Wiener filter, since the correlation matrix does not need to be estimated anymore. The experiment showed that our novel reconstruction technique also works outside a simulated environment in reality.

\section{Appendix}
\label{sec:appendix}

\subsection{Smoothed pseudoinverse statistical derivation}

A novel way to derive the optimization problem of the SP from a statistical point of view is by assuming, that the current spectrum sample is Gaussian distributed around the previous one. This leads to $\text{P}(\HSChannels_j) = \mathcal{N}(\HSChannels_{j - 1}, \SmoothingVar^2)$. By assuming independence, a maximum likelihood estimation can be performed
\begin{equation}
	\begin{aligned}
		\HSChannelsEst^{\text{SP}} = \ & \argmax_{\HSChannels} \prod_{j=2}^{\NumHSChannels} \mathcal{N}(\HSChannels_{j - 1}, \SmoothingVar^2)
		= && \argmin_{\HSChannels} ||\DifferenceMatrix_1\HSChannels||_2^2\\
		& \text{s.t.} \quad \MSChannels = \FilterMatrix \HSChannels
		&& \text{s.t.} \quad \MSChannels = \FilterMatrix \HSChannels.
	\end{aligned}
\end{equation}
Interestingly, this results in the SP optimization problem using first order differences. A similar procedure can be performed resulting in the second-order difference matrix.\\
In the noisy case, this novel derivation is modified, such that the multispectral channels are assumed to be Gaussian-distributed around the filtering result $\text{P}(\MSChannels_i|\HSChannels) = \mathcal{N}(\FilterMatrix_i\HSChannels, \NoiseVar_{\text{n},i}^2)$. With this likelihood and the prior information, a maximum-a-posteriori estimation can be performed to calculate the noise-considering smoothed pseudoinverse (NSP)
\begin{equation}
	\begin{aligned}
		\HSChannelsEst^{\text{NSP}} &= \ \argmin_{\HSChannels} -\log \text{P}(\MSChannels|\HSChannels) - \log\text{P}(\HSChannels)\\
		&= \ \argmin_{\HSChannels} \frac{1}{2}||\MSNoiseVarMatrix^{-\frac{1}{2}}(\FilterMatrix\HSChannels - \MSChannels)||^2_2 + \frac{1}{2\SmoothingVar^2}||\DifferenceMatrix_1\HSChannels||_2^2,\\
	\end{aligned}
\end{equation}
where $\MSNoiseVarMatrix = \text{diag}\left(\NoiseVar_{\text{n}}^2\right)$ is a diagonal matrix with the noise variances of the multispectral channels on its main diagonal. Taking the gradient and solving for $\HSChannels$ leads to
\begin{equation}
	\HSChannelsEst^{\text{NSP}} =  (\FilterMatrix^\text{T}\MSNoiseVarMatrix^{-\frac{1}{2}}\MSNoiseVarMatrix^{-\frac{1}{2}}\FilterMatrix + \ScaledSmoothingMatrix)^{-1}\FilterMatrix^{\text{T}}\MSNoiseVarMatrix^{-\frac{1}{2}}\MSNoiseVarMatrix^{-\frac{1}{2}}\MSChannels,
\end{equation}
where $\ScaledSmoothingMatrix$ is a scaled version of $\SmoothingMatrix$ embedding the scale factor $\frac{1}{\sigma_s^2}$. Using the searle set of identities \cite{petersen-matrix}, specifically $\left(\PlaceholderMatrixA + \PlaceholderMatrixB\PlaceholderMatrixB^{\text{T}}\right)^{-1}\PlaceholderMatrixB = \PlaceholderMatrixA^{-1}\PlaceholderMatrixB\left(\PlaceholderMatrixB^{\text{T}}\PlaceholderMatrixA^{-1}\PlaceholderMatrixB + \Identity\right)^{-1}$, where $\Identity$ indicates an identity matrix with suitable dimensions, and rearranging the inverse standard deviation matrix $\MSNoiseVarMatrix^{-\frac{1}{2}}$ results in
\begin{equation}
	\label{eq:ssp_noise}
	\HSChannelsEst^{\text{NSP}} = \ScaledSmoothingMatrix^{-1}\FilterMatrix^{\text{T}}(\FilterMatrix\ScaledSmoothingMatrix^{-1}\FilterMatrix^{\text{T}} + \MSNoiseVarMatrix)^{-1}\MSChannels.
\end{equation}

\subsection{Relationship between Wiener filter and smoothed pseudoinverse}
\label{subsec:connection}
Comparing \eqref{eq:single_pixel_wiener} and \eqref{eq:ssp_noise} results in the novel observation that $\CorrelationHS = \ScaledSmoothingMatrix^{-1}$. It is interesting that these two methods, widely used in literature, are closely related. Often $\CorrelationHS$ is estimated using a training set in literature. Assuming the spectrum to be smooth via the definition of differences is a much more general assumption than estimating the covariance matrix from a database. Therefore, all methods, which are based on the Wiener filter are using the inverse squared smoothing matrix as their covariance matrix. Consequently, the single-pixel WF works out of the box without estimating the covariance matrix for a specific purpose.\\
Furthermore, by using the WF derivation a proper scale factor for $\SmoothingMatrix^{-1}$ can be found. This is necessary, because the smoothing matrix itself has a constant scale, while the noise matrix $\MSNoiseVarMatrix$ depends on the scale of the image. Therefore, a proper scale factor is essential. This can be done by using the relationship $\CorrelationMS = \FilterMatrix\ScaledSmoothingMatrix^{-1}\FilterMatrix^{\text{T}} + \MSNoiseVarMatrix$, where $\ScaledSmoothingMatrix^{-1} = \SmoothingMatrixScale \ \SmoothingMatrix^{-1}$. $\CorrelationMS$ can be measured using the observations and $\MSNoiseVarMatrix$ is estimated using a noise variance estimator. Then, since $\MSNoiseVarMatrix$ just contains non-zero values on the main diagonal, the most important area is the main diagonal of these matrices. This would still result in $M$ different scale factors. Consequently, a least squares estimator
\begin{equation}
	\SmoothingMatrixScaleEst = \argmin_{\SmoothingMatrixScale}||\SmoothingMatrixScale \ \DotScaleSpectral - \DotScaleMS||_2^2
\end{equation}
is used, where $\DotScaleSpectral = \text{diag}(\FilterMatrix\ScaledSmoothingMatrix^{-1}\FilterMatrix^{\text{T}})$ and $\DotScaleMS = \text{diag}(\CorrelationMS - \MSNoiseVarMatrix)$. This optimization problem results in the scale factor $\SmoothingMatrixScaleEst = \frac{\DotScaleSpectral^{\text{T}}\DotScaleMS}{\DotScaleSpectral^{\text{T}}\DotScaleSpectral}$.

\medskip
\noindent\textbf{Disclosures.} The authors declare no conflicts of interest.

% Bibliography
\bibliography{sample}

\end{document}

%% file: figures/raw_guided_filtering.pgf
%% Creator: Matplotlib, PGF backend
%%
%% To include the figure in your LaTeX document, write
%%   \input{<filename>.pgf}
%%
%% Make sure the required packages are loaded in your preamble
%%   \usepackage{pgf}
%%
%% and, on pdftex
%%   \usepackage[utf8]{inputenc}\DeclareUnicodeCharacter{2212}{-}
%%
%% or, on luatex and xetex
%%   \usepackage{unicode-math}
%%
%% Figures using additional raster images can only be included by \input if
%% they are in the same directory as the main LaTeX file. For loading figures
%% from other directories you can use the `import` package
%%   \usepackage{import}
%%
%% and then include the figures with
%%   \import{<path to file>}{<filename>.pgf}
%%
%% Matplotlib used the following preamble
%%   \usepackage{fontspec}
%%
\begingroup%
\makeatletter%
\begin{pgfpicture}%
\pgfpathrectangle{\pgfpointorigin}{\pgfqpoint{3.033227in}{1.528685in}}%
\pgfusepath{use as bounding box, clip}%
\begin{pgfscope}%
\pgfsetbuttcap%
\pgfsetmiterjoin%
\definecolor{currentfill}{rgb}{1.000000,1.000000,1.000000}%
\pgfsetfillcolor{currentfill}%
\pgfsetlinewidth{0.000000pt}%
\definecolor{currentstroke}{rgb}{1.000000,1.000000,1.000000}%
\pgfsetstrokecolor{currentstroke}%
\pgfsetdash{}{0pt}%
\pgfpathmoveto{\pgfqpoint{0.000000in}{0.000000in}}%
\pgfpathlineto{\pgfqpoint{3.033227in}{0.000000in}}%
\pgfpathlineto{\pgfqpoint{3.033227in}{1.528685in}}%
\pgfpathlineto{\pgfqpoint{0.000000in}{1.528685in}}%
\pgfpathclose%
\pgfusepath{fill}%
\end{pgfscope}%
\begin{pgfscope}%
\pgfpathrectangle{\pgfqpoint{0.000000in}{-0.000000in}}{\pgfqpoint{1.357227in}{1.357227in}}%
\pgfusepath{clip}%
\pgfsys@transformshift{0.000000in}{0.000000in}%
\pgftext[left,bottom]{\includegraphics[interpolate=true,width=1.357500in,height=1.357500in]{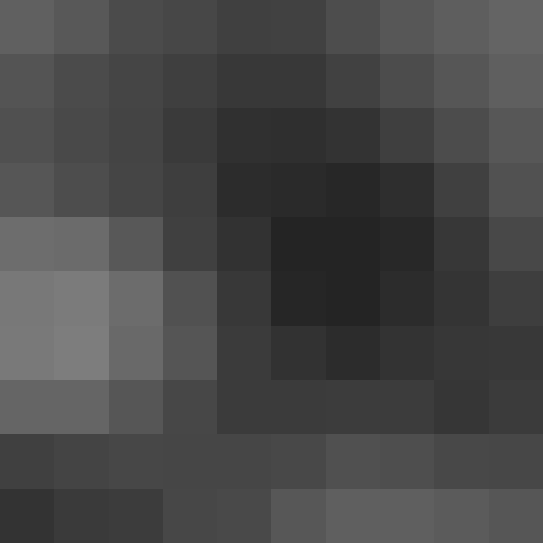}}%
\end{pgfscope}%
\begin{pgfscope}%
\definecolor{textcolor}{rgb}{0.000000,0.000000,0.000000}%
\pgfsetstrokecolor{textcolor}%
\pgfsetfillcolor{textcolor}%
\pgftext[x=0.678613in,y=1.440560in,,base]{\color{textcolor}\rmfamily\fontsize{9.000000}{10.800000}\selectfont Reference}%
\end{pgfscope}%
\begin{pgfscope}%
\pgfpathrectangle{\pgfqpoint{1.676000in}{-0.000000in}}{\pgfqpoint{1.357227in}{1.357227in}}%
\pgfusepath{clip}%
\pgfsys@transformshift{1.676000in}{0.000000in}%
\pgftext[left,bottom]{\includegraphics[interpolate=true,width=1.357500in,height=1.357500in]{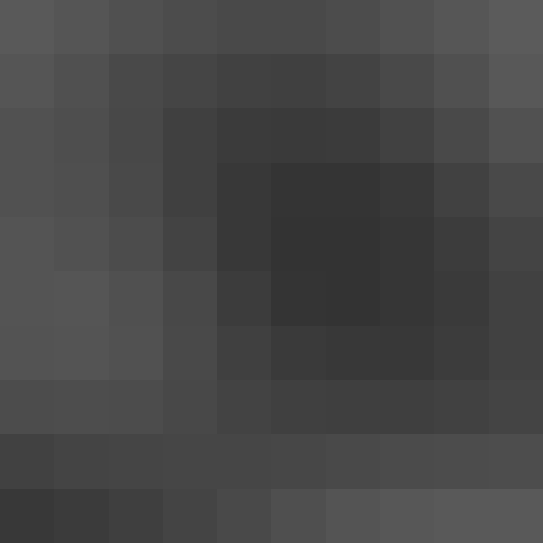}}%
\end{pgfscope}%
\begin{pgfscope}%
\definecolor{textcolor}{rgb}{0.000000,0.000000,0.000000}%
\pgfsetstrokecolor{textcolor}%
\pgfsetfillcolor{textcolor}%
\pgftext[x=2.354613in,y=1.440560in,,base]{\color{textcolor}\rmfamily\fontsize{9.000000}{10.800000}\selectfont Guided filtering}%
\end{pgfscope}%
\end{pgfpicture}%
\makeatother%
\endgroup%

%% file: figures/filters.pgf
%% Creator: Matplotlib, PGF backend
%%
%% To include the figure in your LaTeX document, write
%%   \input{<filename>.pgf}
%%
%% Make sure the required packages are loaded in your preamble
%%   \usepackage{pgf}
%%
%% and, on pdftex
%%   \usepackage[utf8]{inputenc}\DeclareUnicodeCharacter{2212}{-}
%%
%% or, on luatex and xetex
%%   \usepackage{unicode-math}
%%
%% Figures using additional raster images can only be included by \input if
%% they are in the same directory as the main LaTeX file. For loading figures
%% from other directories you can use the `import` package
%%   \usepackage{import}
%%
%% and then include the figures with
%%   \import{<path to file>}{<filename>.pgf}
%%
%% Matplotlib used the following preamble
%%   \usepackage{fontspec}
%%
\begingroup%
\makeatletter%
\begin{pgfpicture}%
\pgfpathrectangle{\pgfpointorigin}{\pgfqpoint{3.240750in}{1.813325in}}%
\pgfusepath{use as bounding box, clip}%
\begin{pgfscope}%
\pgfsetbuttcap%
\pgfsetmiterjoin%
\definecolor{currentfill}{rgb}{1.000000,1.000000,1.000000}%
\pgfsetfillcolor{currentfill}%
\pgfsetlinewidth{0.000000pt}%
\definecolor{currentstroke}{rgb}{1.000000,1.000000,1.000000}%
\pgfsetstrokecolor{currentstroke}%
\pgfsetdash{}{0pt}%
\pgfpathmoveto{\pgfqpoint{0.000000in}{0.000000in}}%
\pgfpathlineto{\pgfqpoint{3.240750in}{0.000000in}}%
\pgfpathlineto{\pgfqpoint{3.240750in}{1.813325in}}%
\pgfpathlineto{\pgfqpoint{0.000000in}{1.813325in}}%
\pgfpathclose%
\pgfusepath{fill}%
\end{pgfscope}%
\begin{pgfscope}%
\pgfsetbuttcap%
\pgfsetmiterjoin%
\definecolor{currentfill}{rgb}{1.000000,1.000000,1.000000}%
\pgfsetfillcolor{currentfill}%
\pgfsetlinewidth{0.000000pt}%
\definecolor{currentstroke}{rgb}{0.000000,0.000000,0.000000}%
\pgfsetstrokecolor{currentstroke}%
\pgfsetstrokeopacity{0.000000}%
\pgfsetdash{}{0pt}%
\pgfpathmoveto{\pgfqpoint{0.440778in}{0.386152in}}%
\pgfpathlineto{\pgfqpoint{3.134375in}{0.386152in}}%
\pgfpathlineto{\pgfqpoint{3.134375in}{1.452797in}}%
\pgfpathlineto{\pgfqpoint{0.440778in}{1.452797in}}%
\pgfpathclose%
\pgfusepath{fill}%
\end{pgfscope}%
\begin{pgfscope}%
\pgfsetbuttcap%
\pgfsetroundjoin%
\definecolor{currentfill}{rgb}{0.000000,0.000000,0.000000}%
\pgfsetfillcolor{currentfill}%
\pgfsetlinewidth{0.803000pt}%
\definecolor{currentstroke}{rgb}{0.000000,0.000000,0.000000}%
\pgfsetstrokecolor{currentstroke}%
\pgfsetdash{}{0pt}%
\pgfsys@defobject{currentmarker}{\pgfqpoint{0.000000in}{-0.048611in}}{\pgfqpoint{0.000000in}{0.000000in}}{%
\pgfpathmoveto{\pgfqpoint{0.000000in}{0.000000in}}%
\pgfpathlineto{\pgfqpoint{0.000000in}{-0.048611in}}%
\pgfusepath{stroke,fill}%
}%
\begin{pgfscope}%
\pgfsys@transformshift{0.440778in}{0.386152in}%
\pgfsys@useobject{currentmarker}{}%
\end{pgfscope}%
\end{pgfscope}%
\begin{pgfscope}%
\definecolor{textcolor}{rgb}{0.000000,0.000000,0.000000}%
\pgfsetstrokecolor{textcolor}%
\pgfsetfillcolor{textcolor}%
\pgftext[x=0.440778in,y=0.288930in,,top]{\color{textcolor}\rmfamily\fontsize{9.000000}{10.800000}\selectfont 1}%
\end{pgfscope}%
\begin{pgfscope}%
\pgfsetbuttcap%
\pgfsetroundjoin%
\definecolor{currentfill}{rgb}{0.000000,0.000000,0.000000}%
\pgfsetfillcolor{currentfill}%
\pgfsetlinewidth{0.803000pt}%
\definecolor{currentstroke}{rgb}{0.000000,0.000000,0.000000}%
\pgfsetstrokecolor{currentstroke}%
\pgfsetdash{}{0pt}%
\pgfsys@defobject{currentmarker}{\pgfqpoint{0.000000in}{-0.048611in}}{\pgfqpoint{0.000000in}{0.000000in}}{%
\pgfpathmoveto{\pgfqpoint{0.000000in}{0.000000in}}%
\pgfpathlineto{\pgfqpoint{0.000000in}{-0.048611in}}%
\pgfusepath{stroke,fill}%
}%
\begin{pgfscope}%
\pgfsys@transformshift{0.945827in}{0.386152in}%
\pgfsys@useobject{currentmarker}{}%
\end{pgfscope}%
\end{pgfscope}%
\begin{pgfscope}%
\definecolor{textcolor}{rgb}{0.000000,0.000000,0.000000}%
\pgfsetstrokecolor{textcolor}%
\pgfsetfillcolor{textcolor}%
\pgftext[x=0.945827in,y=0.288930in,,top]{\color{textcolor}\rmfamily\fontsize{9.000000}{10.800000}\selectfont 10}%
\end{pgfscope}%
\begin{pgfscope}%
\pgfsetbuttcap%
\pgfsetroundjoin%
\definecolor{currentfill}{rgb}{0.000000,0.000000,0.000000}%
\pgfsetfillcolor{currentfill}%
\pgfsetlinewidth{0.803000pt}%
\definecolor{currentstroke}{rgb}{0.000000,0.000000,0.000000}%
\pgfsetstrokecolor{currentstroke}%
\pgfsetdash{}{0pt}%
\pgfsys@defobject{currentmarker}{\pgfqpoint{0.000000in}{-0.048611in}}{\pgfqpoint{0.000000in}{0.000000in}}{%
\pgfpathmoveto{\pgfqpoint{0.000000in}{0.000000in}}%
\pgfpathlineto{\pgfqpoint{0.000000in}{-0.048611in}}%
\pgfusepath{stroke,fill}%
}%
\begin{pgfscope}%
\pgfsys@transformshift{1.506993in}{0.386152in}%
\pgfsys@useobject{currentmarker}{}%
\end{pgfscope}%
\end{pgfscope}%
\begin{pgfscope}%
\definecolor{textcolor}{rgb}{0.000000,0.000000,0.000000}%
\pgfsetstrokecolor{textcolor}%
\pgfsetfillcolor{textcolor}%
\pgftext[x=1.506993in,y=0.288930in,,top]{\color{textcolor}\rmfamily\fontsize{9.000000}{10.800000}\selectfont 20}%
\end{pgfscope}%
\begin{pgfscope}%
\pgfsetbuttcap%
\pgfsetroundjoin%
\definecolor{currentfill}{rgb}{0.000000,0.000000,0.000000}%
\pgfsetfillcolor{currentfill}%
\pgfsetlinewidth{0.803000pt}%
\definecolor{currentstroke}{rgb}{0.000000,0.000000,0.000000}%
\pgfsetstrokecolor{currentstroke}%
\pgfsetdash{}{0pt}%
\pgfsys@defobject{currentmarker}{\pgfqpoint{0.000000in}{-0.048611in}}{\pgfqpoint{0.000000in}{0.000000in}}{%
\pgfpathmoveto{\pgfqpoint{0.000000in}{0.000000in}}%
\pgfpathlineto{\pgfqpoint{0.000000in}{-0.048611in}}%
\pgfusepath{stroke,fill}%
}%
\begin{pgfscope}%
\pgfsys@transformshift{2.068159in}{0.386152in}%
\pgfsys@useobject{currentmarker}{}%
\end{pgfscope}%
\end{pgfscope}%
\begin{pgfscope}%
\definecolor{textcolor}{rgb}{0.000000,0.000000,0.000000}%
\pgfsetstrokecolor{textcolor}%
\pgfsetfillcolor{textcolor}%
\pgftext[x=2.068159in,y=0.288930in,,top]{\color{textcolor}\rmfamily\fontsize{9.000000}{10.800000}\selectfont 30}%
\end{pgfscope}%
\begin{pgfscope}%
\pgfsetbuttcap%
\pgfsetroundjoin%
\definecolor{currentfill}{rgb}{0.000000,0.000000,0.000000}%
\pgfsetfillcolor{currentfill}%
\pgfsetlinewidth{0.803000pt}%
\definecolor{currentstroke}{rgb}{0.000000,0.000000,0.000000}%
\pgfsetstrokecolor{currentstroke}%
\pgfsetdash{}{0pt}%
\pgfsys@defobject{currentmarker}{\pgfqpoint{0.000000in}{-0.048611in}}{\pgfqpoint{0.000000in}{0.000000in}}{%
\pgfpathmoveto{\pgfqpoint{0.000000in}{0.000000in}}%
\pgfpathlineto{\pgfqpoint{0.000000in}{-0.048611in}}%
\pgfusepath{stroke,fill}%
}%
\begin{pgfscope}%
\pgfsys@transformshift{2.629325in}{0.386152in}%
\pgfsys@useobject{currentmarker}{}%
\end{pgfscope}%
\end{pgfscope}%
\begin{pgfscope}%
\definecolor{textcolor}{rgb}{0.000000,0.000000,0.000000}%
\pgfsetstrokecolor{textcolor}%
\pgfsetfillcolor{textcolor}%
\pgftext[x=2.629325in,y=0.288930in,,top]{\color{textcolor}\rmfamily\fontsize{9.000000}{10.800000}\selectfont 40}%
\end{pgfscope}%
\begin{pgfscope}%
\pgfsetbuttcap%
\pgfsetroundjoin%
\definecolor{currentfill}{rgb}{0.000000,0.000000,0.000000}%
\pgfsetfillcolor{currentfill}%
\pgfsetlinewidth{0.803000pt}%
\definecolor{currentstroke}{rgb}{0.000000,0.000000,0.000000}%
\pgfsetstrokecolor{currentstroke}%
\pgfsetdash{}{0pt}%
\pgfsys@defobject{currentmarker}{\pgfqpoint{0.000000in}{-0.048611in}}{\pgfqpoint{0.000000in}{0.000000in}}{%
\pgfpathmoveto{\pgfqpoint{0.000000in}{0.000000in}}%
\pgfpathlineto{\pgfqpoint{0.000000in}{-0.048611in}}%
\pgfusepath{stroke,fill}%
}%
\begin{pgfscope}%
\pgfsys@transformshift{3.134375in}{0.386152in}%
\pgfsys@useobject{currentmarker}{}%
\end{pgfscope}%
\end{pgfscope}%
\begin{pgfscope}%
\definecolor{textcolor}{rgb}{0.000000,0.000000,0.000000}%
\pgfsetstrokecolor{textcolor}%
\pgfsetfillcolor{textcolor}%
\pgftext[x=3.134375in,y=0.288930in,,top]{\color{textcolor}\rmfamily\fontsize{9.000000}{10.800000}\selectfont 49}%
\end{pgfscope}%
\begin{pgfscope}%
\definecolor{textcolor}{rgb}{0.000000,0.000000,0.000000}%
\pgfsetstrokecolor{textcolor}%
\pgfsetfillcolor{textcolor}%
\pgftext[x=1.787576in,y=0.122375in,,top]{\color{textcolor}\rmfamily\fontsize{9.000000}{10.800000}\selectfont Hyperspectral channel}%
\end{pgfscope}%
\begin{pgfscope}%
\pgfsetbuttcap%
\pgfsetroundjoin%
\definecolor{currentfill}{rgb}{0.000000,0.000000,0.000000}%
\pgfsetfillcolor{currentfill}%
\pgfsetlinewidth{0.803000pt}%
\definecolor{currentstroke}{rgb}{0.000000,0.000000,0.000000}%
\pgfsetstrokecolor{currentstroke}%
\pgfsetdash{}{0pt}%
\pgfsys@defobject{currentmarker}{\pgfqpoint{-0.048611in}{0.000000in}}{\pgfqpoint{0.000000in}{0.000000in}}{%
\pgfpathmoveto{\pgfqpoint{0.000000in}{0.000000in}}%
\pgfpathlineto{\pgfqpoint{-0.048611in}{0.000000in}}%
\pgfusepath{stroke,fill}%
}%
\begin{pgfscope}%
\pgfsys@transformshift{0.440778in}{0.386152in}%
\pgfsys@useobject{currentmarker}{}%
\end{pgfscope}%
\end{pgfscope}%
\begin{pgfscope}%
\definecolor{textcolor}{rgb}{0.000000,0.000000,0.000000}%
\pgfsetstrokecolor{textcolor}%
\pgfsetfillcolor{textcolor}%
\pgftext[x=0.179305in, y=0.342777in, left, base]{\color{textcolor}\rmfamily\fontsize{9.000000}{10.800000}\selectfont 0.0}%
\end{pgfscope}%
\begin{pgfscope}%
\pgfsetbuttcap%
\pgfsetroundjoin%
\definecolor{currentfill}{rgb}{0.000000,0.000000,0.000000}%
\pgfsetfillcolor{currentfill}%
\pgfsetlinewidth{0.803000pt}%
\definecolor{currentstroke}{rgb}{0.000000,0.000000,0.000000}%
\pgfsetstrokecolor{currentstroke}%
\pgfsetdash{}{0pt}%
\pgfsys@defobject{currentmarker}{\pgfqpoint{-0.048611in}{0.000000in}}{\pgfqpoint{0.000000in}{0.000000in}}{%
\pgfpathmoveto{\pgfqpoint{0.000000in}{0.000000in}}%
\pgfpathlineto{\pgfqpoint{-0.048611in}{0.000000in}}%
\pgfusepath{stroke,fill}%
}%
\begin{pgfscope}%
\pgfsys@transformshift{0.440778in}{0.909017in}%
\pgfsys@useobject{currentmarker}{}%
\end{pgfscope}%
\end{pgfscope}%
\begin{pgfscope}%
\definecolor{textcolor}{rgb}{0.000000,0.000000,0.000000}%
\pgfsetstrokecolor{textcolor}%
\pgfsetfillcolor{textcolor}%
\pgftext[x=0.179305in, y=0.865642in, left, base]{\color{textcolor}\rmfamily\fontsize{9.000000}{10.800000}\selectfont 0.5}%
\end{pgfscope}%
\begin{pgfscope}%
\pgfsetbuttcap%
\pgfsetroundjoin%
\definecolor{currentfill}{rgb}{0.000000,0.000000,0.000000}%
\pgfsetfillcolor{currentfill}%
\pgfsetlinewidth{0.803000pt}%
\definecolor{currentstroke}{rgb}{0.000000,0.000000,0.000000}%
\pgfsetstrokecolor{currentstroke}%
\pgfsetdash{}{0pt}%
\pgfsys@defobject{currentmarker}{\pgfqpoint{-0.048611in}{0.000000in}}{\pgfqpoint{0.000000in}{0.000000in}}{%
\pgfpathmoveto{\pgfqpoint{0.000000in}{0.000000in}}%
\pgfpathlineto{\pgfqpoint{-0.048611in}{0.000000in}}%
\pgfusepath{stroke,fill}%
}%
\begin{pgfscope}%
\pgfsys@transformshift{0.440778in}{1.431882in}%
\pgfsys@useobject{currentmarker}{}%
\end{pgfscope}%
\end{pgfscope}%
\begin{pgfscope}%
\definecolor{textcolor}{rgb}{0.000000,0.000000,0.000000}%
\pgfsetstrokecolor{textcolor}%
\pgfsetfillcolor{textcolor}%
\pgftext[x=0.179305in, y=1.388507in, left, base]{\color{textcolor}\rmfamily\fontsize{9.000000}{10.800000}\selectfont 1.0}%
\end{pgfscope}%
\begin{pgfscope}%
\definecolor{textcolor}{rgb}{0.000000,0.000000,0.000000}%
\pgfsetstrokecolor{textcolor}%
\pgfsetfillcolor{textcolor}%
\pgftext[x=0.123750in,y=0.919475in,,bottom,rotate=90.000000]{\color{textcolor}\rmfamily\fontsize{9.000000}{10.800000}\selectfont Spectral sensitivity}%
\end{pgfscope}%
\begin{pgfscope}%
\pgfpathrectangle{\pgfqpoint{0.440778in}{0.386152in}}{\pgfqpoint{2.693597in}{1.066644in}}%
\pgfusepath{clip}%
\pgfsetrectcap%
\pgfsetroundjoin%
\pgfsetlinewidth{1.505625pt}%
\definecolor{currentstroke}{rgb}{0.000000,0.000000,0.803922}%
\pgfsetstrokecolor{currentstroke}%
\pgfsetdash{}{0pt}%
\pgfpathmoveto{\pgfqpoint{0.440778in}{1.431882in}}%
\pgfpathlineto{\pgfqpoint{0.496894in}{1.431882in}}%
\pgfpathlineto{\pgfqpoint{0.553011in}{1.431882in}}%
\pgfpathlineto{\pgfqpoint{0.609127in}{1.431882in}}%
\pgfpathlineto{\pgfqpoint{0.665244in}{1.431882in}}%
\pgfpathlineto{\pgfqpoint{0.721361in}{0.386152in}}%
\pgfpathlineto{\pgfqpoint{0.777477in}{0.386152in}}%
\pgfpathlineto{\pgfqpoint{0.833594in}{0.386152in}}%
\pgfpathlineto{\pgfqpoint{0.889711in}{0.386152in}}%
\pgfpathlineto{\pgfqpoint{0.945827in}{0.386152in}}%
\pgfpathlineto{\pgfqpoint{1.001944in}{0.386152in}}%
\pgfpathlineto{\pgfqpoint{1.058060in}{0.386152in}}%
\pgfpathlineto{\pgfqpoint{1.114177in}{0.386152in}}%
\pgfpathlineto{\pgfqpoint{1.170294in}{0.386152in}}%
\pgfpathlineto{\pgfqpoint{1.226410in}{0.386152in}}%
\pgfpathlineto{\pgfqpoint{1.282527in}{0.386152in}}%
\pgfpathlineto{\pgfqpoint{1.338643in}{0.386152in}}%
\pgfpathlineto{\pgfqpoint{1.394760in}{0.386152in}}%
\pgfpathlineto{\pgfqpoint{1.450877in}{0.386152in}}%
\pgfpathlineto{\pgfqpoint{1.506993in}{0.386152in}}%
\pgfpathlineto{\pgfqpoint{1.563110in}{0.386152in}}%
\pgfpathlineto{\pgfqpoint{1.619226in}{0.386152in}}%
\pgfpathlineto{\pgfqpoint{1.675343in}{0.386152in}}%
\pgfpathlineto{\pgfqpoint{1.731460in}{0.386152in}}%
\pgfpathlineto{\pgfqpoint{1.787576in}{0.386152in}}%
\pgfpathlineto{\pgfqpoint{1.843693in}{0.386152in}}%
\pgfpathlineto{\pgfqpoint{1.899809in}{0.386152in}}%
\pgfpathlineto{\pgfqpoint{1.955926in}{0.386152in}}%
\pgfpathlineto{\pgfqpoint{2.012043in}{0.386152in}}%
\pgfpathlineto{\pgfqpoint{2.068159in}{0.386152in}}%
\pgfpathlineto{\pgfqpoint{2.124276in}{0.386152in}}%
\pgfpathlineto{\pgfqpoint{2.180393in}{0.386152in}}%
\pgfpathlineto{\pgfqpoint{2.236509in}{0.386152in}}%
\pgfpathlineto{\pgfqpoint{2.292626in}{0.386152in}}%
\pgfpathlineto{\pgfqpoint{2.348742in}{0.386152in}}%
\pgfpathlineto{\pgfqpoint{2.404859in}{0.386152in}}%
\pgfpathlineto{\pgfqpoint{2.460976in}{0.386152in}}%
\pgfpathlineto{\pgfqpoint{2.517092in}{0.386152in}}%
\pgfpathlineto{\pgfqpoint{2.573209in}{0.386152in}}%
\pgfpathlineto{\pgfqpoint{2.629325in}{0.386152in}}%
\pgfpathlineto{\pgfqpoint{2.685442in}{0.386152in}}%
\pgfpathlineto{\pgfqpoint{2.741559in}{0.386152in}}%
\pgfpathlineto{\pgfqpoint{2.797675in}{0.386152in}}%
\pgfpathlineto{\pgfqpoint{2.853792in}{0.386152in}}%
\pgfpathlineto{\pgfqpoint{2.909908in}{0.386152in}}%
\pgfpathlineto{\pgfqpoint{2.966025in}{0.386152in}}%
\pgfpathlineto{\pgfqpoint{3.022142in}{0.386152in}}%
\pgfpathlineto{\pgfqpoint{3.078258in}{0.386152in}}%
\pgfpathlineto{\pgfqpoint{3.134375in}{0.386152in}}%
\pgfusepath{stroke}%
\end{pgfscope}%
\begin{pgfscope}%
\pgfpathrectangle{\pgfqpoint{0.440778in}{0.386152in}}{\pgfqpoint{2.693597in}{1.066644in}}%
\pgfusepath{clip}%
\pgfsetrectcap%
\pgfsetroundjoin%
\pgfsetlinewidth{1.505625pt}%
\definecolor{currentstroke}{rgb}{0.000000,0.501961,0.501961}%
\pgfsetstrokecolor{currentstroke}%
\pgfsetdash{}{0pt}%
\pgfpathmoveto{\pgfqpoint{0.440778in}{0.386152in}}%
\pgfpathlineto{\pgfqpoint{0.496894in}{0.386152in}}%
\pgfpathlineto{\pgfqpoint{0.553011in}{1.431882in}}%
\pgfpathlineto{\pgfqpoint{0.609127in}{1.431882in}}%
\pgfpathlineto{\pgfqpoint{0.665244in}{1.431882in}}%
\pgfpathlineto{\pgfqpoint{0.721361in}{1.431882in}}%
\pgfpathlineto{\pgfqpoint{0.777477in}{1.431882in}}%
\pgfpathlineto{\pgfqpoint{0.833594in}{1.431882in}}%
\pgfpathlineto{\pgfqpoint{0.889711in}{1.431882in}}%
\pgfpathlineto{\pgfqpoint{0.945827in}{1.431882in}}%
\pgfpathlineto{\pgfqpoint{1.001944in}{1.431882in}}%
\pgfpathlineto{\pgfqpoint{1.058060in}{0.386152in}}%
\pgfpathlineto{\pgfqpoint{1.114177in}{0.386152in}}%
\pgfpathlineto{\pgfqpoint{1.170294in}{0.386152in}}%
\pgfpathlineto{\pgfqpoint{1.226410in}{0.386152in}}%
\pgfpathlineto{\pgfqpoint{1.282527in}{0.386152in}}%
\pgfpathlineto{\pgfqpoint{1.338643in}{0.386152in}}%
\pgfpathlineto{\pgfqpoint{1.394760in}{0.386152in}}%
\pgfpathlineto{\pgfqpoint{1.450877in}{0.386152in}}%
\pgfpathlineto{\pgfqpoint{1.506993in}{0.386152in}}%
\pgfpathlineto{\pgfqpoint{1.563110in}{0.386152in}}%
\pgfpathlineto{\pgfqpoint{1.619226in}{0.386152in}}%
\pgfpathlineto{\pgfqpoint{1.675343in}{0.386152in}}%
\pgfpathlineto{\pgfqpoint{1.731460in}{0.386152in}}%
\pgfpathlineto{\pgfqpoint{1.787576in}{0.386152in}}%
\pgfpathlineto{\pgfqpoint{1.843693in}{0.386152in}}%
\pgfpathlineto{\pgfqpoint{1.899809in}{0.386152in}}%
\pgfpathlineto{\pgfqpoint{1.955926in}{0.386152in}}%
\pgfpathlineto{\pgfqpoint{2.012043in}{0.386152in}}%
\pgfpathlineto{\pgfqpoint{2.068159in}{0.386152in}}%
\pgfpathlineto{\pgfqpoint{2.124276in}{0.386152in}}%
\pgfpathlineto{\pgfqpoint{2.180393in}{0.386152in}}%
\pgfpathlineto{\pgfqpoint{2.236509in}{0.386152in}}%
\pgfpathlineto{\pgfqpoint{2.292626in}{0.386152in}}%
\pgfpathlineto{\pgfqpoint{2.348742in}{0.386152in}}%
\pgfpathlineto{\pgfqpoint{2.404859in}{0.386152in}}%
\pgfpathlineto{\pgfqpoint{2.460976in}{0.386152in}}%
\pgfpathlineto{\pgfqpoint{2.517092in}{0.386152in}}%
\pgfpathlineto{\pgfqpoint{2.573209in}{0.386152in}}%
\pgfpathlineto{\pgfqpoint{2.629325in}{0.386152in}}%
\pgfpathlineto{\pgfqpoint{2.685442in}{0.386152in}}%
\pgfpathlineto{\pgfqpoint{2.741559in}{0.386152in}}%
\pgfpathlineto{\pgfqpoint{2.797675in}{0.386152in}}%
\pgfpathlineto{\pgfqpoint{2.853792in}{0.386152in}}%
\pgfpathlineto{\pgfqpoint{2.909908in}{0.386152in}}%
\pgfpathlineto{\pgfqpoint{2.966025in}{0.386152in}}%
\pgfpathlineto{\pgfqpoint{3.022142in}{0.386152in}}%
\pgfpathlineto{\pgfqpoint{3.078258in}{0.386152in}}%
\pgfpathlineto{\pgfqpoint{3.134375in}{0.386152in}}%
\pgfusepath{stroke}%
\end{pgfscope}%
\begin{pgfscope}%
\pgfpathrectangle{\pgfqpoint{0.440778in}{0.386152in}}{\pgfqpoint{2.693597in}{1.066644in}}%
\pgfusepath{clip}%
\pgfsetrectcap%
\pgfsetroundjoin%
\pgfsetlinewidth{1.505625pt}%
\definecolor{currentstroke}{rgb}{0.196078,0.803922,0.196078}%
\pgfsetstrokecolor{currentstroke}%
\pgfsetdash{}{0pt}%
\pgfpathmoveto{\pgfqpoint{0.440778in}{0.386152in}}%
\pgfpathlineto{\pgfqpoint{0.496894in}{0.386152in}}%
\pgfpathlineto{\pgfqpoint{0.553011in}{0.386152in}}%
\pgfpathlineto{\pgfqpoint{0.609127in}{0.386152in}}%
\pgfpathlineto{\pgfqpoint{0.665244in}{0.386152in}}%
\pgfpathlineto{\pgfqpoint{0.721361in}{0.386152in}}%
\pgfpathlineto{\pgfqpoint{0.777477in}{0.386152in}}%
\pgfpathlineto{\pgfqpoint{0.833594in}{0.386152in}}%
\pgfpathlineto{\pgfqpoint{0.889711in}{1.431882in}}%
\pgfpathlineto{\pgfqpoint{0.945827in}{1.431882in}}%
\pgfpathlineto{\pgfqpoint{1.001944in}{1.431882in}}%
\pgfpathlineto{\pgfqpoint{1.058060in}{1.431882in}}%
\pgfpathlineto{\pgfqpoint{1.114177in}{1.431882in}}%
\pgfpathlineto{\pgfqpoint{1.170294in}{1.431882in}}%
\pgfpathlineto{\pgfqpoint{1.226410in}{1.431882in}}%
\pgfpathlineto{\pgfqpoint{1.282527in}{1.431882in}}%
\pgfpathlineto{\pgfqpoint{1.338643in}{1.431882in}}%
\pgfpathlineto{\pgfqpoint{1.394760in}{0.386152in}}%
\pgfpathlineto{\pgfqpoint{1.450877in}{0.386152in}}%
\pgfpathlineto{\pgfqpoint{1.506993in}{0.386152in}}%
\pgfpathlineto{\pgfqpoint{1.563110in}{0.386152in}}%
\pgfpathlineto{\pgfqpoint{1.619226in}{0.386152in}}%
\pgfpathlineto{\pgfqpoint{1.675343in}{0.386152in}}%
\pgfpathlineto{\pgfqpoint{1.731460in}{0.386152in}}%
\pgfpathlineto{\pgfqpoint{1.787576in}{0.386152in}}%
\pgfpathlineto{\pgfqpoint{1.843693in}{0.386152in}}%
\pgfpathlineto{\pgfqpoint{1.899809in}{0.386152in}}%
\pgfpathlineto{\pgfqpoint{1.955926in}{0.386152in}}%
\pgfpathlineto{\pgfqpoint{2.012043in}{0.386152in}}%
\pgfpathlineto{\pgfqpoint{2.068159in}{0.386152in}}%
\pgfpathlineto{\pgfqpoint{2.124276in}{0.386152in}}%
\pgfpathlineto{\pgfqpoint{2.180393in}{0.386152in}}%
\pgfpathlineto{\pgfqpoint{2.236509in}{0.386152in}}%
\pgfpathlineto{\pgfqpoint{2.292626in}{0.386152in}}%
\pgfpathlineto{\pgfqpoint{2.348742in}{0.386152in}}%
\pgfpathlineto{\pgfqpoint{2.404859in}{0.386152in}}%
\pgfpathlineto{\pgfqpoint{2.460976in}{0.386152in}}%
\pgfpathlineto{\pgfqpoint{2.517092in}{0.386152in}}%
\pgfpathlineto{\pgfqpoint{2.573209in}{0.386152in}}%
\pgfpathlineto{\pgfqpoint{2.629325in}{0.386152in}}%
\pgfpathlineto{\pgfqpoint{2.685442in}{0.386152in}}%
\pgfpathlineto{\pgfqpoint{2.741559in}{0.386152in}}%
\pgfpathlineto{\pgfqpoint{2.797675in}{0.386152in}}%
\pgfpathlineto{\pgfqpoint{2.853792in}{0.386152in}}%
\pgfpathlineto{\pgfqpoint{2.909908in}{0.386152in}}%
\pgfpathlineto{\pgfqpoint{2.966025in}{0.386152in}}%
\pgfpathlineto{\pgfqpoint{3.022142in}{0.386152in}}%
\pgfpathlineto{\pgfqpoint{3.078258in}{0.386152in}}%
\pgfpathlineto{\pgfqpoint{3.134375in}{0.386152in}}%
\pgfusepath{stroke}%
\end{pgfscope}%
\begin{pgfscope}%
\pgfpathrectangle{\pgfqpoint{0.440778in}{0.386152in}}{\pgfqpoint{2.693597in}{1.066644in}}%
\pgfusepath{clip}%
\pgfsetrectcap%
\pgfsetroundjoin%
\pgfsetlinewidth{1.505625pt}%
\definecolor{currentstroke}{rgb}{1.000000,0.843137,0.000000}%
\pgfsetstrokecolor{currentstroke}%
\pgfsetdash{}{0pt}%
\pgfpathmoveto{\pgfqpoint{0.440778in}{0.386152in}}%
\pgfpathlineto{\pgfqpoint{0.496894in}{0.386152in}}%
\pgfpathlineto{\pgfqpoint{0.553011in}{0.386152in}}%
\pgfpathlineto{\pgfqpoint{0.609127in}{0.386152in}}%
\pgfpathlineto{\pgfqpoint{0.665244in}{0.386152in}}%
\pgfpathlineto{\pgfqpoint{0.721361in}{0.386152in}}%
\pgfpathlineto{\pgfqpoint{0.777477in}{0.386152in}}%
\pgfpathlineto{\pgfqpoint{0.833594in}{0.386152in}}%
\pgfpathlineto{\pgfqpoint{0.889711in}{0.386152in}}%
\pgfpathlineto{\pgfqpoint{0.945827in}{0.386152in}}%
\pgfpathlineto{\pgfqpoint{1.001944in}{0.386152in}}%
\pgfpathlineto{\pgfqpoint{1.058060in}{0.386152in}}%
\pgfpathlineto{\pgfqpoint{1.114177in}{0.386152in}}%
\pgfpathlineto{\pgfqpoint{1.170294in}{0.386152in}}%
\pgfpathlineto{\pgfqpoint{1.226410in}{1.431882in}}%
\pgfpathlineto{\pgfqpoint{1.282527in}{1.431882in}}%
\pgfpathlineto{\pgfqpoint{1.338643in}{1.431882in}}%
\pgfpathlineto{\pgfqpoint{1.394760in}{1.431882in}}%
\pgfpathlineto{\pgfqpoint{1.450877in}{1.431882in}}%
\pgfpathlineto{\pgfqpoint{1.506993in}{1.431882in}}%
\pgfpathlineto{\pgfqpoint{1.563110in}{1.431882in}}%
\pgfpathlineto{\pgfqpoint{1.619226in}{1.431882in}}%
\pgfpathlineto{\pgfqpoint{1.675343in}{1.431882in}}%
\pgfpathlineto{\pgfqpoint{1.731460in}{0.386152in}}%
\pgfpathlineto{\pgfqpoint{1.787576in}{0.386152in}}%
\pgfpathlineto{\pgfqpoint{1.843693in}{0.386152in}}%
\pgfpathlineto{\pgfqpoint{1.899809in}{0.386152in}}%
\pgfpathlineto{\pgfqpoint{1.955926in}{0.386152in}}%
\pgfpathlineto{\pgfqpoint{2.012043in}{0.386152in}}%
\pgfpathlineto{\pgfqpoint{2.068159in}{0.386152in}}%
\pgfpathlineto{\pgfqpoint{2.124276in}{0.386152in}}%
\pgfpathlineto{\pgfqpoint{2.180393in}{0.386152in}}%
\pgfpathlineto{\pgfqpoint{2.236509in}{0.386152in}}%
\pgfpathlineto{\pgfqpoint{2.292626in}{0.386152in}}%
\pgfpathlineto{\pgfqpoint{2.348742in}{0.386152in}}%
\pgfpathlineto{\pgfqpoint{2.404859in}{0.386152in}}%
\pgfpathlineto{\pgfqpoint{2.460976in}{0.386152in}}%
\pgfpathlineto{\pgfqpoint{2.517092in}{0.386152in}}%
\pgfpathlineto{\pgfqpoint{2.573209in}{0.386152in}}%
\pgfpathlineto{\pgfqpoint{2.629325in}{0.386152in}}%
\pgfpathlineto{\pgfqpoint{2.685442in}{0.386152in}}%
\pgfpathlineto{\pgfqpoint{2.741559in}{0.386152in}}%
\pgfpathlineto{\pgfqpoint{2.797675in}{0.386152in}}%
\pgfpathlineto{\pgfqpoint{2.853792in}{0.386152in}}%
\pgfpathlineto{\pgfqpoint{2.909908in}{0.386152in}}%
\pgfpathlineto{\pgfqpoint{2.966025in}{0.386152in}}%
\pgfpathlineto{\pgfqpoint{3.022142in}{0.386152in}}%
\pgfpathlineto{\pgfqpoint{3.078258in}{0.386152in}}%
\pgfpathlineto{\pgfqpoint{3.134375in}{0.386152in}}%
\pgfusepath{stroke}%
\end{pgfscope}%
\begin{pgfscope}%
\pgfpathrectangle{\pgfqpoint{0.440778in}{0.386152in}}{\pgfqpoint{2.693597in}{1.066644in}}%
\pgfusepath{clip}%
\pgfsetrectcap%
\pgfsetroundjoin%
\pgfsetlinewidth{1.505625pt}%
\definecolor{currentstroke}{rgb}{1.000000,0.647059,0.000000}%
\pgfsetstrokecolor{currentstroke}%
\pgfsetdash{}{0pt}%
\pgfpathmoveto{\pgfqpoint{0.440778in}{0.386152in}}%
\pgfpathlineto{\pgfqpoint{0.496894in}{0.386152in}}%
\pgfpathlineto{\pgfqpoint{0.553011in}{0.386152in}}%
\pgfpathlineto{\pgfqpoint{0.609127in}{0.386152in}}%
\pgfpathlineto{\pgfqpoint{0.665244in}{0.386152in}}%
\pgfpathlineto{\pgfqpoint{0.721361in}{0.386152in}}%
\pgfpathlineto{\pgfqpoint{0.777477in}{0.386152in}}%
\pgfpathlineto{\pgfqpoint{0.833594in}{0.386152in}}%
\pgfpathlineto{\pgfqpoint{0.889711in}{0.386152in}}%
\pgfpathlineto{\pgfqpoint{0.945827in}{0.386152in}}%
\pgfpathlineto{\pgfqpoint{1.001944in}{0.386152in}}%
\pgfpathlineto{\pgfqpoint{1.058060in}{0.386152in}}%
\pgfpathlineto{\pgfqpoint{1.114177in}{0.386152in}}%
\pgfpathlineto{\pgfqpoint{1.170294in}{0.386152in}}%
\pgfpathlineto{\pgfqpoint{1.226410in}{0.386152in}}%
\pgfpathlineto{\pgfqpoint{1.282527in}{0.386152in}}%
\pgfpathlineto{\pgfqpoint{1.338643in}{0.386152in}}%
\pgfpathlineto{\pgfqpoint{1.394760in}{0.386152in}}%
\pgfpathlineto{\pgfqpoint{1.450877in}{0.386152in}}%
\pgfpathlineto{\pgfqpoint{1.506993in}{0.386152in}}%
\pgfpathlineto{\pgfqpoint{1.563110in}{1.431882in}}%
\pgfpathlineto{\pgfqpoint{1.619226in}{1.431882in}}%
\pgfpathlineto{\pgfqpoint{1.675343in}{1.431882in}}%
\pgfpathlineto{\pgfqpoint{1.731460in}{1.431882in}}%
\pgfpathlineto{\pgfqpoint{1.787576in}{1.431882in}}%
\pgfpathlineto{\pgfqpoint{1.843693in}{1.431882in}}%
\pgfpathlineto{\pgfqpoint{1.899809in}{1.431882in}}%
\pgfpathlineto{\pgfqpoint{1.955926in}{1.431882in}}%
\pgfpathlineto{\pgfqpoint{2.012043in}{1.431882in}}%
\pgfpathlineto{\pgfqpoint{2.068159in}{0.386152in}}%
\pgfpathlineto{\pgfqpoint{2.124276in}{0.386152in}}%
\pgfpathlineto{\pgfqpoint{2.180393in}{0.386152in}}%
\pgfpathlineto{\pgfqpoint{2.236509in}{0.386152in}}%
\pgfpathlineto{\pgfqpoint{2.292626in}{0.386152in}}%
\pgfpathlineto{\pgfqpoint{2.348742in}{0.386152in}}%
\pgfpathlineto{\pgfqpoint{2.404859in}{0.386152in}}%
\pgfpathlineto{\pgfqpoint{2.460976in}{0.386152in}}%
\pgfpathlineto{\pgfqpoint{2.517092in}{0.386152in}}%
\pgfpathlineto{\pgfqpoint{2.573209in}{0.386152in}}%
\pgfpathlineto{\pgfqpoint{2.629325in}{0.386152in}}%
\pgfpathlineto{\pgfqpoint{2.685442in}{0.386152in}}%
\pgfpathlineto{\pgfqpoint{2.741559in}{0.386152in}}%
\pgfpathlineto{\pgfqpoint{2.797675in}{0.386152in}}%
\pgfpathlineto{\pgfqpoint{2.853792in}{0.386152in}}%
\pgfpathlineto{\pgfqpoint{2.909908in}{0.386152in}}%
\pgfpathlineto{\pgfqpoint{2.966025in}{0.386152in}}%
\pgfpathlineto{\pgfqpoint{3.022142in}{0.386152in}}%
\pgfpathlineto{\pgfqpoint{3.078258in}{0.386152in}}%
\pgfpathlineto{\pgfqpoint{3.134375in}{0.386152in}}%
\pgfusepath{stroke}%
\end{pgfscope}%
\begin{pgfscope}%
\pgfpathrectangle{\pgfqpoint{0.440778in}{0.386152in}}{\pgfqpoint{2.693597in}{1.066644in}}%
\pgfusepath{clip}%
\pgfsetrectcap%
\pgfsetroundjoin%
\pgfsetlinewidth{1.505625pt}%
\definecolor{currentstroke}{rgb}{1.000000,0.270588,0.000000}%
\pgfsetstrokecolor{currentstroke}%
\pgfsetdash{}{0pt}%
\pgfpathmoveto{\pgfqpoint{0.440778in}{0.386152in}}%
\pgfpathlineto{\pgfqpoint{0.496894in}{0.386152in}}%
\pgfpathlineto{\pgfqpoint{0.553011in}{0.386152in}}%
\pgfpathlineto{\pgfqpoint{0.609127in}{0.386152in}}%
\pgfpathlineto{\pgfqpoint{0.665244in}{0.386152in}}%
\pgfpathlineto{\pgfqpoint{0.721361in}{0.386152in}}%
\pgfpathlineto{\pgfqpoint{0.777477in}{0.386152in}}%
\pgfpathlineto{\pgfqpoint{0.833594in}{0.386152in}}%
\pgfpathlineto{\pgfqpoint{0.889711in}{0.386152in}}%
\pgfpathlineto{\pgfqpoint{0.945827in}{0.386152in}}%
\pgfpathlineto{\pgfqpoint{1.001944in}{0.386152in}}%
\pgfpathlineto{\pgfqpoint{1.058060in}{0.386152in}}%
\pgfpathlineto{\pgfqpoint{1.114177in}{0.386152in}}%
\pgfpathlineto{\pgfqpoint{1.170294in}{0.386152in}}%
\pgfpathlineto{\pgfqpoint{1.226410in}{0.386152in}}%
\pgfpathlineto{\pgfqpoint{1.282527in}{0.386152in}}%
\pgfpathlineto{\pgfqpoint{1.338643in}{0.386152in}}%
\pgfpathlineto{\pgfqpoint{1.394760in}{0.386152in}}%
\pgfpathlineto{\pgfqpoint{1.450877in}{0.386152in}}%
\pgfpathlineto{\pgfqpoint{1.506993in}{0.386152in}}%
\pgfpathlineto{\pgfqpoint{1.563110in}{0.386152in}}%
\pgfpathlineto{\pgfqpoint{1.619226in}{0.386152in}}%
\pgfpathlineto{\pgfqpoint{1.675343in}{0.386152in}}%
\pgfpathlineto{\pgfqpoint{1.731460in}{0.386152in}}%
\pgfpathlineto{\pgfqpoint{1.787576in}{0.386152in}}%
\pgfpathlineto{\pgfqpoint{1.843693in}{0.386152in}}%
\pgfpathlineto{\pgfqpoint{1.899809in}{1.431882in}}%
\pgfpathlineto{\pgfqpoint{1.955926in}{1.431882in}}%
\pgfpathlineto{\pgfqpoint{2.012043in}{1.431882in}}%
\pgfpathlineto{\pgfqpoint{2.068159in}{1.431882in}}%
\pgfpathlineto{\pgfqpoint{2.124276in}{1.431882in}}%
\pgfpathlineto{\pgfqpoint{2.180393in}{1.431882in}}%
\pgfpathlineto{\pgfqpoint{2.236509in}{1.431882in}}%
\pgfpathlineto{\pgfqpoint{2.292626in}{1.431882in}}%
\pgfpathlineto{\pgfqpoint{2.348742in}{1.431882in}}%
\pgfpathlineto{\pgfqpoint{2.404859in}{0.386152in}}%
\pgfpathlineto{\pgfqpoint{2.460976in}{0.386152in}}%
\pgfpathlineto{\pgfqpoint{2.517092in}{0.386152in}}%
\pgfpathlineto{\pgfqpoint{2.573209in}{0.386152in}}%
\pgfpathlineto{\pgfqpoint{2.629325in}{0.386152in}}%
\pgfpathlineto{\pgfqpoint{2.685442in}{0.386152in}}%
\pgfpathlineto{\pgfqpoint{2.741559in}{0.386152in}}%
\pgfpathlineto{\pgfqpoint{2.797675in}{0.386152in}}%
\pgfpathlineto{\pgfqpoint{2.853792in}{0.386152in}}%
\pgfpathlineto{\pgfqpoint{2.909908in}{0.386152in}}%
\pgfpathlineto{\pgfqpoint{2.966025in}{0.386152in}}%
\pgfpathlineto{\pgfqpoint{3.022142in}{0.386152in}}%
\pgfpathlineto{\pgfqpoint{3.078258in}{0.386152in}}%
\pgfpathlineto{\pgfqpoint{3.134375in}{0.386152in}}%
\pgfusepath{stroke}%
\end{pgfscope}%
\begin{pgfscope}%
\pgfpathrectangle{\pgfqpoint{0.440778in}{0.386152in}}{\pgfqpoint{2.693597in}{1.066644in}}%
\pgfusepath{clip}%
\pgfsetrectcap%
\pgfsetroundjoin%
\pgfsetlinewidth{1.505625pt}%
\definecolor{currentstroke}{rgb}{0.501961,0.000000,0.000000}%
\pgfsetstrokecolor{currentstroke}%
\pgfsetdash{}{0pt}%
\pgfpathmoveto{\pgfqpoint{0.440778in}{0.386152in}}%
\pgfpathlineto{\pgfqpoint{0.496894in}{0.386152in}}%
\pgfpathlineto{\pgfqpoint{0.553011in}{0.386152in}}%
\pgfpathlineto{\pgfqpoint{0.609127in}{0.386152in}}%
\pgfpathlineto{\pgfqpoint{0.665244in}{0.386152in}}%
\pgfpathlineto{\pgfqpoint{0.721361in}{0.386152in}}%
\pgfpathlineto{\pgfqpoint{0.777477in}{0.386152in}}%
\pgfpathlineto{\pgfqpoint{0.833594in}{0.386152in}}%
\pgfpathlineto{\pgfqpoint{0.889711in}{0.386152in}}%
\pgfpathlineto{\pgfqpoint{0.945827in}{0.386152in}}%
\pgfpathlineto{\pgfqpoint{1.001944in}{0.386152in}}%
\pgfpathlineto{\pgfqpoint{1.058060in}{0.386152in}}%
\pgfpathlineto{\pgfqpoint{1.114177in}{0.386152in}}%
\pgfpathlineto{\pgfqpoint{1.170294in}{0.386152in}}%
\pgfpathlineto{\pgfqpoint{1.226410in}{0.386152in}}%
\pgfpathlineto{\pgfqpoint{1.282527in}{0.386152in}}%
\pgfpathlineto{\pgfqpoint{1.338643in}{0.386152in}}%
\pgfpathlineto{\pgfqpoint{1.394760in}{0.386152in}}%
\pgfpathlineto{\pgfqpoint{1.450877in}{0.386152in}}%
\pgfpathlineto{\pgfqpoint{1.506993in}{0.386152in}}%
\pgfpathlineto{\pgfqpoint{1.563110in}{0.386152in}}%
\pgfpathlineto{\pgfqpoint{1.619226in}{0.386152in}}%
\pgfpathlineto{\pgfqpoint{1.675343in}{0.386152in}}%
\pgfpathlineto{\pgfqpoint{1.731460in}{0.386152in}}%
\pgfpathlineto{\pgfqpoint{1.787576in}{0.386152in}}%
\pgfpathlineto{\pgfqpoint{1.843693in}{0.386152in}}%
\pgfpathlineto{\pgfqpoint{1.899809in}{0.386152in}}%
\pgfpathlineto{\pgfqpoint{1.955926in}{0.386152in}}%
\pgfpathlineto{\pgfqpoint{2.012043in}{0.386152in}}%
\pgfpathlineto{\pgfqpoint{2.068159in}{0.386152in}}%
\pgfpathlineto{\pgfqpoint{2.124276in}{0.386152in}}%
\pgfpathlineto{\pgfqpoint{2.180393in}{0.386152in}}%
\pgfpathlineto{\pgfqpoint{2.236509in}{1.431882in}}%
\pgfpathlineto{\pgfqpoint{2.292626in}{1.431882in}}%
\pgfpathlineto{\pgfqpoint{2.348742in}{1.431882in}}%
\pgfpathlineto{\pgfqpoint{2.404859in}{1.431882in}}%
\pgfpathlineto{\pgfqpoint{2.460976in}{1.431882in}}%
\pgfpathlineto{\pgfqpoint{2.517092in}{1.431882in}}%
\pgfpathlineto{\pgfqpoint{2.573209in}{1.431882in}}%
\pgfpathlineto{\pgfqpoint{2.629325in}{1.431882in}}%
\pgfpathlineto{\pgfqpoint{2.685442in}{1.431882in}}%
\pgfpathlineto{\pgfqpoint{2.741559in}{0.386152in}}%
\pgfpathlineto{\pgfqpoint{2.797675in}{0.386152in}}%
\pgfpathlineto{\pgfqpoint{2.853792in}{0.386152in}}%
\pgfpathlineto{\pgfqpoint{2.909908in}{0.386152in}}%
\pgfpathlineto{\pgfqpoint{2.966025in}{0.386152in}}%
\pgfpathlineto{\pgfqpoint{3.022142in}{0.386152in}}%
\pgfpathlineto{\pgfqpoint{3.078258in}{0.386152in}}%
\pgfpathlineto{\pgfqpoint{3.134375in}{0.386152in}}%
\pgfusepath{stroke}%
\end{pgfscope}%
\begin{pgfscope}%
\pgfpathrectangle{\pgfqpoint{0.440778in}{0.386152in}}{\pgfqpoint{2.693597in}{1.066644in}}%
\pgfusepath{clip}%
\pgfsetrectcap%
\pgfsetroundjoin%
\pgfsetlinewidth{1.505625pt}%
\definecolor{currentstroke}{rgb}{0.803922,0.360784,0.360784}%
\pgfsetstrokecolor{currentstroke}%
\pgfsetdash{}{0pt}%
\pgfpathmoveto{\pgfqpoint{0.440778in}{0.386152in}}%
\pgfpathlineto{\pgfqpoint{0.496894in}{0.386152in}}%
\pgfpathlineto{\pgfqpoint{0.553011in}{0.386152in}}%
\pgfpathlineto{\pgfqpoint{0.609127in}{0.386152in}}%
\pgfpathlineto{\pgfqpoint{0.665244in}{0.386152in}}%
\pgfpathlineto{\pgfqpoint{0.721361in}{0.386152in}}%
\pgfpathlineto{\pgfqpoint{0.777477in}{0.386152in}}%
\pgfpathlineto{\pgfqpoint{0.833594in}{0.386152in}}%
\pgfpathlineto{\pgfqpoint{0.889711in}{0.386152in}}%
\pgfpathlineto{\pgfqpoint{0.945827in}{0.386152in}}%
\pgfpathlineto{\pgfqpoint{1.001944in}{0.386152in}}%
\pgfpathlineto{\pgfqpoint{1.058060in}{0.386152in}}%
\pgfpathlineto{\pgfqpoint{1.114177in}{0.386152in}}%
\pgfpathlineto{\pgfqpoint{1.170294in}{0.386152in}}%
\pgfpathlineto{\pgfqpoint{1.226410in}{0.386152in}}%
\pgfpathlineto{\pgfqpoint{1.282527in}{0.386152in}}%
\pgfpathlineto{\pgfqpoint{1.338643in}{0.386152in}}%
\pgfpathlineto{\pgfqpoint{1.394760in}{0.386152in}}%
\pgfpathlineto{\pgfqpoint{1.450877in}{0.386152in}}%
\pgfpathlineto{\pgfqpoint{1.506993in}{0.386152in}}%
\pgfpathlineto{\pgfqpoint{1.563110in}{0.386152in}}%
\pgfpathlineto{\pgfqpoint{1.619226in}{0.386152in}}%
\pgfpathlineto{\pgfqpoint{1.675343in}{0.386152in}}%
\pgfpathlineto{\pgfqpoint{1.731460in}{0.386152in}}%
\pgfpathlineto{\pgfqpoint{1.787576in}{0.386152in}}%
\pgfpathlineto{\pgfqpoint{1.843693in}{0.386152in}}%
\pgfpathlineto{\pgfqpoint{1.899809in}{0.386152in}}%
\pgfpathlineto{\pgfqpoint{1.955926in}{0.386152in}}%
\pgfpathlineto{\pgfqpoint{2.012043in}{0.386152in}}%
\pgfpathlineto{\pgfqpoint{2.068159in}{0.386152in}}%
\pgfpathlineto{\pgfqpoint{2.124276in}{0.386152in}}%
\pgfpathlineto{\pgfqpoint{2.180393in}{0.386152in}}%
\pgfpathlineto{\pgfqpoint{2.236509in}{0.386152in}}%
\pgfpathlineto{\pgfqpoint{2.292626in}{0.386152in}}%
\pgfpathlineto{\pgfqpoint{2.348742in}{0.386152in}}%
\pgfpathlineto{\pgfqpoint{2.404859in}{0.386152in}}%
\pgfpathlineto{\pgfqpoint{2.460976in}{0.386152in}}%
\pgfpathlineto{\pgfqpoint{2.517092in}{0.386152in}}%
\pgfpathlineto{\pgfqpoint{2.573209in}{1.431882in}}%
\pgfpathlineto{\pgfqpoint{2.629325in}{1.431882in}}%
\pgfpathlineto{\pgfqpoint{2.685442in}{1.431882in}}%
\pgfpathlineto{\pgfqpoint{2.741559in}{1.431882in}}%
\pgfpathlineto{\pgfqpoint{2.797675in}{1.431882in}}%
\pgfpathlineto{\pgfqpoint{2.853792in}{1.431882in}}%
\pgfpathlineto{\pgfqpoint{2.909908in}{1.431882in}}%
\pgfpathlineto{\pgfqpoint{2.966025in}{1.431882in}}%
\pgfpathlineto{\pgfqpoint{3.022142in}{1.431882in}}%
\pgfpathlineto{\pgfqpoint{3.078258in}{0.386152in}}%
\pgfpathlineto{\pgfqpoint{3.134375in}{0.386152in}}%
\pgfusepath{stroke}%
\end{pgfscope}%
\begin{pgfscope}%
\pgfpathrectangle{\pgfqpoint{0.440778in}{0.386152in}}{\pgfqpoint{2.693597in}{1.066644in}}%
\pgfusepath{clip}%
\pgfsetrectcap%
\pgfsetroundjoin%
\pgfsetlinewidth{1.505625pt}%
\definecolor{currentstroke}{rgb}{0.980392,0.501961,0.447059}%
\pgfsetstrokecolor{currentstroke}%
\pgfsetdash{}{0pt}%
\pgfpathmoveto{\pgfqpoint{0.440778in}{0.386152in}}%
\pgfpathlineto{\pgfqpoint{0.496894in}{0.386152in}}%
\pgfpathlineto{\pgfqpoint{0.553011in}{0.386152in}}%
\pgfpathlineto{\pgfqpoint{0.609127in}{0.386152in}}%
\pgfpathlineto{\pgfqpoint{0.665244in}{0.386152in}}%
\pgfpathlineto{\pgfqpoint{0.721361in}{0.386152in}}%
\pgfpathlineto{\pgfqpoint{0.777477in}{0.386152in}}%
\pgfpathlineto{\pgfqpoint{0.833594in}{0.386152in}}%
\pgfpathlineto{\pgfqpoint{0.889711in}{0.386152in}}%
\pgfpathlineto{\pgfqpoint{0.945827in}{0.386152in}}%
\pgfpathlineto{\pgfqpoint{1.001944in}{0.386152in}}%
\pgfpathlineto{\pgfqpoint{1.058060in}{0.386152in}}%
\pgfpathlineto{\pgfqpoint{1.114177in}{0.386152in}}%
\pgfpathlineto{\pgfqpoint{1.170294in}{0.386152in}}%
\pgfpathlineto{\pgfqpoint{1.226410in}{0.386152in}}%
\pgfpathlineto{\pgfqpoint{1.282527in}{0.386152in}}%
\pgfpathlineto{\pgfqpoint{1.338643in}{0.386152in}}%
\pgfpathlineto{\pgfqpoint{1.394760in}{0.386152in}}%
\pgfpathlineto{\pgfqpoint{1.450877in}{0.386152in}}%
\pgfpathlineto{\pgfqpoint{1.506993in}{0.386152in}}%
\pgfpathlineto{\pgfqpoint{1.563110in}{0.386152in}}%
\pgfpathlineto{\pgfqpoint{1.619226in}{0.386152in}}%
\pgfpathlineto{\pgfqpoint{1.675343in}{0.386152in}}%
\pgfpathlineto{\pgfqpoint{1.731460in}{0.386152in}}%
\pgfpathlineto{\pgfqpoint{1.787576in}{0.386152in}}%
\pgfpathlineto{\pgfqpoint{1.843693in}{0.386152in}}%
\pgfpathlineto{\pgfqpoint{1.899809in}{0.386152in}}%
\pgfpathlineto{\pgfqpoint{1.955926in}{0.386152in}}%
\pgfpathlineto{\pgfqpoint{2.012043in}{0.386152in}}%
\pgfpathlineto{\pgfqpoint{2.068159in}{0.386152in}}%
\pgfpathlineto{\pgfqpoint{2.124276in}{0.386152in}}%
\pgfpathlineto{\pgfqpoint{2.180393in}{0.386152in}}%
\pgfpathlineto{\pgfqpoint{2.236509in}{0.386152in}}%
\pgfpathlineto{\pgfqpoint{2.292626in}{0.386152in}}%
\pgfpathlineto{\pgfqpoint{2.348742in}{0.386152in}}%
\pgfpathlineto{\pgfqpoint{2.404859in}{0.386152in}}%
\pgfpathlineto{\pgfqpoint{2.460976in}{0.386152in}}%
\pgfpathlineto{\pgfqpoint{2.517092in}{0.386152in}}%
\pgfpathlineto{\pgfqpoint{2.573209in}{0.386152in}}%
\pgfpathlineto{\pgfqpoint{2.629325in}{0.386152in}}%
\pgfpathlineto{\pgfqpoint{2.685442in}{0.386152in}}%
\pgfpathlineto{\pgfqpoint{2.741559in}{0.386152in}}%
\pgfpathlineto{\pgfqpoint{2.797675in}{0.386152in}}%
\pgfpathlineto{\pgfqpoint{2.853792in}{0.386152in}}%
\pgfpathlineto{\pgfqpoint{2.909908in}{1.431882in}}%
\pgfpathlineto{\pgfqpoint{2.966025in}{1.431882in}}%
\pgfpathlineto{\pgfqpoint{3.022142in}{1.431882in}}%
\pgfpathlineto{\pgfqpoint{3.078258in}{1.431882in}}%
\pgfpathlineto{\pgfqpoint{3.134375in}{1.431882in}}%
\pgfusepath{stroke}%
\end{pgfscope}%
\begin{pgfscope}%
\pgfsetrectcap%
\pgfsetmiterjoin%
\pgfsetlinewidth{0.803000pt}%
\definecolor{currentstroke}{rgb}{0.000000,0.000000,0.000000}%
\pgfsetstrokecolor{currentstroke}%
\pgfsetdash{}{0pt}%
\pgfpathmoveto{\pgfqpoint{0.440778in}{0.386152in}}%
\pgfpathlineto{\pgfqpoint{0.440778in}{1.452797in}}%
\pgfusepath{stroke}%
\end{pgfscope}%
\begin{pgfscope}%
\pgfsetrectcap%
\pgfsetmiterjoin%
\pgfsetlinewidth{0.803000pt}%
\definecolor{currentstroke}{rgb}{0.000000,0.000000,0.000000}%
\pgfsetstrokecolor{currentstroke}%
\pgfsetdash{}{0pt}%
\pgfpathmoveto{\pgfqpoint{3.134375in}{0.386152in}}%
\pgfpathlineto{\pgfqpoint{3.134375in}{1.452797in}}%
\pgfusepath{stroke}%
\end{pgfscope}%
\begin{pgfscope}%
\pgfsetrectcap%
\pgfsetmiterjoin%
\pgfsetlinewidth{0.803000pt}%
\definecolor{currentstroke}{rgb}{0.000000,0.000000,0.000000}%
\pgfsetstrokecolor{currentstroke}%
\pgfsetdash{}{0pt}%
\pgfpathmoveto{\pgfqpoint{0.440778in}{0.386152in}}%
\pgfpathlineto{\pgfqpoint{3.134375in}{0.386152in}}%
\pgfusepath{stroke}%
\end{pgfscope}%
\begin{pgfscope}%
\pgfsetrectcap%
\pgfsetmiterjoin%
\pgfsetlinewidth{0.803000pt}%
\definecolor{currentstroke}{rgb}{0.000000,0.000000,0.000000}%
\pgfsetstrokecolor{currentstroke}%
\pgfsetdash{}{0pt}%
\pgfpathmoveto{\pgfqpoint{0.440778in}{1.452797in}}%
\pgfpathlineto{\pgfqpoint{3.134375in}{1.452797in}}%
\pgfusepath{stroke}%
\end{pgfscope}%
\begin{pgfscope}%
\pgfsetbuttcap%
\pgfsetroundjoin%
\definecolor{currentfill}{rgb}{0.000000,0.000000,0.000000}%
\pgfsetfillcolor{currentfill}%
\pgfsetlinewidth{0.803000pt}%
\definecolor{currentstroke}{rgb}{0.000000,0.000000,0.000000}%
\pgfsetstrokecolor{currentstroke}%
\pgfsetdash{}{0pt}%
\pgfsys@defobject{currentmarker}{\pgfqpoint{0.000000in}{0.000000in}}{\pgfqpoint{0.000000in}{0.048611in}}{%
\pgfpathmoveto{\pgfqpoint{0.000000in}{0.000000in}}%
\pgfpathlineto{\pgfqpoint{0.000000in}{0.048611in}}%
\pgfusepath{stroke,fill}%
}%
\begin{pgfscope}%
\pgfsys@transformshift{0.440778in}{1.452797in}%
\pgfsys@useobject{currentmarker}{}%
\end{pgfscope}%
\end{pgfscope}%
\begin{pgfscope}%
\definecolor{textcolor}{rgb}{0.000000,0.000000,0.000000}%
\pgfsetstrokecolor{textcolor}%
\pgfsetfillcolor{textcolor}%
\pgftext[x=0.440778in,y=1.550019in,,bottom]{\color{textcolor}\rmfamily\fontsize{9.000000}{10.800000}\selectfont 440}%
\end{pgfscope}%
\begin{pgfscope}%
\pgfsetbuttcap%
\pgfsetroundjoin%
\definecolor{currentfill}{rgb}{0.000000,0.000000,0.000000}%
\pgfsetfillcolor{currentfill}%
\pgfsetlinewidth{0.803000pt}%
\definecolor{currentstroke}{rgb}{0.000000,0.000000,0.000000}%
\pgfsetstrokecolor{currentstroke}%
\pgfsetdash{}{0pt}%
\pgfsys@defobject{currentmarker}{\pgfqpoint{0.000000in}{0.000000in}}{\pgfqpoint{0.000000in}{0.048611in}}{%
\pgfpathmoveto{\pgfqpoint{0.000000in}{0.000000in}}%
\pgfpathlineto{\pgfqpoint{0.000000in}{0.048611in}}%
\pgfusepath{stroke,fill}%
}%
\begin{pgfscope}%
\pgfsys@transformshift{0.945827in}{1.452797in}%
\pgfsys@useobject{currentmarker}{}%
\end{pgfscope}%
\end{pgfscope}%
\begin{pgfscope}%
\definecolor{textcolor}{rgb}{0.000000,0.000000,0.000000}%
\pgfsetstrokecolor{textcolor}%
\pgfsetfillcolor{textcolor}%
\pgftext[x=0.945827in,y=1.550019in,,bottom]{\color{textcolor}\rmfamily\fontsize{9.000000}{10.800000}\selectfont 530}%
\end{pgfscope}%
\begin{pgfscope}%
\pgfsetbuttcap%
\pgfsetroundjoin%
\definecolor{currentfill}{rgb}{0.000000,0.000000,0.000000}%
\pgfsetfillcolor{currentfill}%
\pgfsetlinewidth{0.803000pt}%
\definecolor{currentstroke}{rgb}{0.000000,0.000000,0.000000}%
\pgfsetstrokecolor{currentstroke}%
\pgfsetdash{}{0pt}%
\pgfsys@defobject{currentmarker}{\pgfqpoint{0.000000in}{0.000000in}}{\pgfqpoint{0.000000in}{0.048611in}}{%
\pgfpathmoveto{\pgfqpoint{0.000000in}{0.000000in}}%
\pgfpathlineto{\pgfqpoint{0.000000in}{0.048611in}}%
\pgfusepath{stroke,fill}%
}%
\begin{pgfscope}%
\pgfsys@transformshift{1.506993in}{1.452797in}%
\pgfsys@useobject{currentmarker}{}%
\end{pgfscope}%
\end{pgfscope}%
\begin{pgfscope}%
\definecolor{textcolor}{rgb}{0.000000,0.000000,0.000000}%
\pgfsetstrokecolor{textcolor}%
\pgfsetfillcolor{textcolor}%
\pgftext[x=1.506993in,y=1.550019in,,bottom]{\color{textcolor}\rmfamily\fontsize{9.000000}{10.800000}\selectfont 630}%
\end{pgfscope}%
\begin{pgfscope}%
\pgfsetbuttcap%
\pgfsetroundjoin%
\definecolor{currentfill}{rgb}{0.000000,0.000000,0.000000}%
\pgfsetfillcolor{currentfill}%
\pgfsetlinewidth{0.803000pt}%
\definecolor{currentstroke}{rgb}{0.000000,0.000000,0.000000}%
\pgfsetstrokecolor{currentstroke}%
\pgfsetdash{}{0pt}%
\pgfsys@defobject{currentmarker}{\pgfqpoint{0.000000in}{0.000000in}}{\pgfqpoint{0.000000in}{0.048611in}}{%
\pgfpathmoveto{\pgfqpoint{0.000000in}{0.000000in}}%
\pgfpathlineto{\pgfqpoint{0.000000in}{0.048611in}}%
\pgfusepath{stroke,fill}%
}%
\begin{pgfscope}%
\pgfsys@transformshift{2.068159in}{1.452797in}%
\pgfsys@useobject{currentmarker}{}%
\end{pgfscope}%
\end{pgfscope}%
\begin{pgfscope}%
\definecolor{textcolor}{rgb}{0.000000,0.000000,0.000000}%
\pgfsetstrokecolor{textcolor}%
\pgfsetfillcolor{textcolor}%
\pgftext[x=2.068159in,y=1.550019in,,bottom]{\color{textcolor}\rmfamily\fontsize{9.000000}{10.800000}\selectfont 730}%
\end{pgfscope}%
\begin{pgfscope}%
\pgfsetbuttcap%
\pgfsetroundjoin%
\definecolor{currentfill}{rgb}{0.000000,0.000000,0.000000}%
\pgfsetfillcolor{currentfill}%
\pgfsetlinewidth{0.803000pt}%
\definecolor{currentstroke}{rgb}{0.000000,0.000000,0.000000}%
\pgfsetstrokecolor{currentstroke}%
\pgfsetdash{}{0pt}%
\pgfsys@defobject{currentmarker}{\pgfqpoint{0.000000in}{0.000000in}}{\pgfqpoint{0.000000in}{0.048611in}}{%
\pgfpathmoveto{\pgfqpoint{0.000000in}{0.000000in}}%
\pgfpathlineto{\pgfqpoint{0.000000in}{0.048611in}}%
\pgfusepath{stroke,fill}%
}%
\begin{pgfscope}%
\pgfsys@transformshift{2.629325in}{1.452797in}%
\pgfsys@useobject{currentmarker}{}%
\end{pgfscope}%
\end{pgfscope}%
\begin{pgfscope}%
\definecolor{textcolor}{rgb}{0.000000,0.000000,0.000000}%
\pgfsetstrokecolor{textcolor}%
\pgfsetfillcolor{textcolor}%
\pgftext[x=2.629325in,y=1.550019in,,bottom]{\color{textcolor}\rmfamily\fontsize{9.000000}{10.800000}\selectfont 830}%
\end{pgfscope}%
\begin{pgfscope}%
\pgfsetbuttcap%
\pgfsetroundjoin%
\definecolor{currentfill}{rgb}{0.000000,0.000000,0.000000}%
\pgfsetfillcolor{currentfill}%
\pgfsetlinewidth{0.803000pt}%
\definecolor{currentstroke}{rgb}{0.000000,0.000000,0.000000}%
\pgfsetstrokecolor{currentstroke}%
\pgfsetdash{}{0pt}%
\pgfsys@defobject{currentmarker}{\pgfqpoint{0.000000in}{0.000000in}}{\pgfqpoint{0.000000in}{0.048611in}}{%
\pgfpathmoveto{\pgfqpoint{0.000000in}{0.000000in}}%
\pgfpathlineto{\pgfqpoint{0.000000in}{0.048611in}}%
\pgfusepath{stroke,fill}%
}%
\begin{pgfscope}%
\pgfsys@transformshift{3.134375in}{1.452797in}%
\pgfsys@useobject{currentmarker}{}%
\end{pgfscope}%
\end{pgfscope}%
\begin{pgfscope}%
\definecolor{textcolor}{rgb}{0.000000,0.000000,0.000000}%
\pgfsetstrokecolor{textcolor}%
\pgfsetfillcolor{textcolor}%
\pgftext[x=3.134375in,y=1.550019in,,bottom]{\color{textcolor}\rmfamily\fontsize{9.000000}{10.800000}\selectfont 920}%
\end{pgfscope}%
\begin{pgfscope}%
\definecolor{textcolor}{rgb}{0.000000,0.000000,0.000000}%
\pgfsetstrokecolor{textcolor}%
\pgfsetfillcolor{textcolor}%
\pgftext[x=1.787576in,y=1.716575in,,base]{\color{textcolor}\rmfamily\fontsize{9.000000}{10.800000}\selectfont Wavelength in nm}%
\end{pgfscope}%
\begin{pgfscope}%
\pgfsetrectcap%
\pgfsetmiterjoin%
\pgfsetlinewidth{0.803000pt}%
\definecolor{currentstroke}{rgb}{0.000000,0.000000,0.000000}%
\pgfsetstrokecolor{currentstroke}%
\pgfsetdash{}{0pt}%
\pgfpathmoveto{\pgfqpoint{0.440778in}{0.386152in}}%
\pgfpathlineto{\pgfqpoint{0.440778in}{1.452797in}}%
\pgfusepath{stroke}%
\end{pgfscope}%
\begin{pgfscope}%
\pgfsetrectcap%
\pgfsetmiterjoin%
\pgfsetlinewidth{0.803000pt}%
\definecolor{currentstroke}{rgb}{0.000000,0.000000,0.000000}%
\pgfsetstrokecolor{currentstroke}%
\pgfsetdash{}{0pt}%
\pgfpathmoveto{\pgfqpoint{3.134375in}{0.386152in}}%
\pgfpathlineto{\pgfqpoint{3.134375in}{1.452797in}}%
\pgfusepath{stroke}%
\end{pgfscope}%
\begin{pgfscope}%
\pgfsetrectcap%
\pgfsetmiterjoin%
\pgfsetlinewidth{0.803000pt}%
\definecolor{currentstroke}{rgb}{0.000000,0.000000,0.000000}%
\pgfsetstrokecolor{currentstroke}%
\pgfsetdash{}{0pt}%
\pgfpathmoveto{\pgfqpoint{0.440778in}{0.386152in}}%
\pgfpathlineto{\pgfqpoint{3.134375in}{0.386152in}}%
\pgfusepath{stroke}%
\end{pgfscope}%
\begin{pgfscope}%
\pgfsetrectcap%
\pgfsetmiterjoin%
\pgfsetlinewidth{0.803000pt}%
\definecolor{currentstroke}{rgb}{0.000000,0.000000,0.000000}%
\pgfsetstrokecolor{currentstroke}%
\pgfsetdash{}{0pt}%
\pgfpathmoveto{\pgfqpoint{0.440778in}{1.452797in}}%
\pgfpathlineto{\pgfqpoint{3.134375in}{1.452797in}}%
\pgfusepath{stroke}%
\end{pgfscope}%
\end{pgfpicture}%
\makeatother%
\endgroup%

%% file: figures/example_rgb_images.pgf
%% Creator: Matplotlib, PGF backend
%%
%% To include the figure in your LaTeX document, write
%%   \input{<filename>.pgf}
%%
%% Make sure the required packages are loaded in your preamble
%%   \usepackage{pgf}
%%
%% and, on pdftex
%%   \usepackage[utf8]{inputenc}\DeclareUnicodeCharacter{2212}{-}
%%
%% or, on luatex and xetex
%%   \usepackage{unicode-math}
%%
%% Figures using additional raster images can only be included by \input if
%% they are in the same directory as the main LaTeX file. For loading figures
%% from other directories you can use the `import` package
%%   \usepackage{import}
%%
%% and then include the figures with
%%   \import{<path to file>}{<filename>.pgf}
%%
%% Matplotlib used the following preamble
%%   \usepackage{fontspec}
%%
\begingroup%
\makeatletter%
\begin{pgfpicture}%
\pgfpathrectangle{\pgfpointorigin}{\pgfqpoint{3.209750in}{3.565700in}}%
\pgfusepath{use as bounding box, clip}%
\begin{pgfscope}%
\pgfsetbuttcap%
\pgfsetmiterjoin%
\definecolor{currentfill}{rgb}{1.000000,1.000000,1.000000}%
\pgfsetfillcolor{currentfill}%
\pgfsetlinewidth{0.000000pt}%
\definecolor{currentstroke}{rgb}{1.000000,1.000000,1.000000}%
\pgfsetstrokecolor{currentstroke}%
\pgfsetdash{}{0pt}%
\pgfpathmoveto{\pgfqpoint{0.000000in}{0.000000in}}%
\pgfpathlineto{\pgfqpoint{3.209750in}{0.000000in}}%
\pgfpathlineto{\pgfqpoint{3.209750in}{3.565700in}}%
\pgfpathlineto{\pgfqpoint{0.000000in}{3.565700in}}%
\pgfpathclose%
\pgfusepath{fill}%
\end{pgfscope}%
\begin{pgfscope}%
\pgfpathrectangle{\pgfqpoint{-0.000000in}{1.850350in}}{\pgfqpoint{1.543815in}{1.715350in}}%
\pgfusepath{clip}%
\pgfsys@transformshift{-0.000000in}{1.850350in}%
\pgftext[left,bottom]{\includegraphics[interpolate=true,width=1.545000in,height=1.717500in]{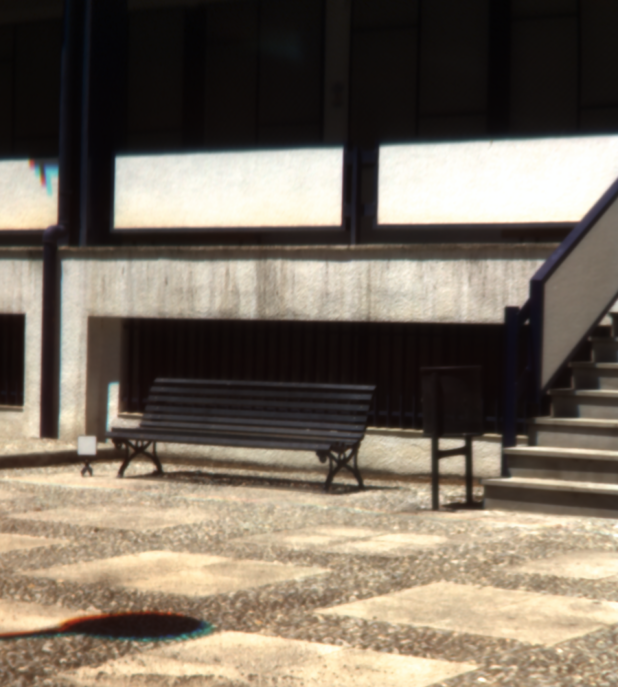}}%
\end{pgfscope}%
\begin{pgfscope}%
\pgfpathrectangle{\pgfqpoint{1.665935in}{1.850350in}}{\pgfqpoint{1.543815in}{1.715350in}}%
\pgfusepath{clip}%
\pgfsys@transformshift{1.665935in}{1.850350in}%
\pgftext[left,bottom]{\includegraphics[interpolate=true,width=1.545000in,height=1.717500in]{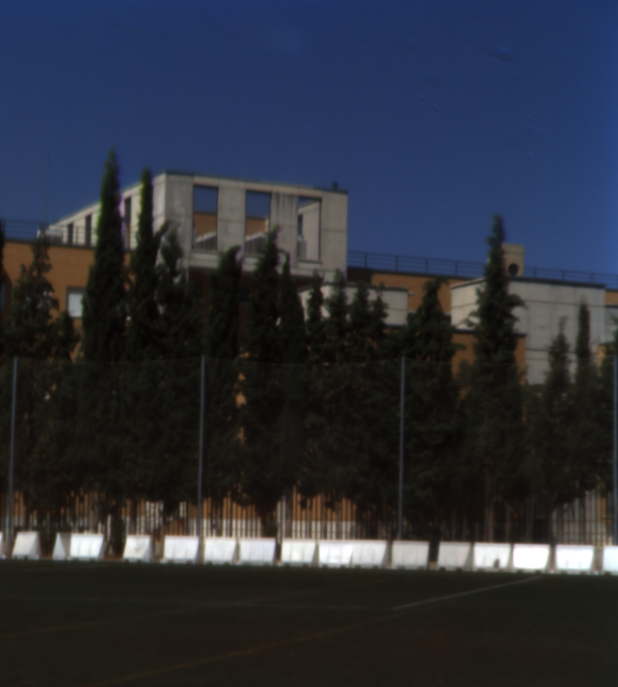}}%
\end{pgfscope}%
\begin{pgfscope}%
\pgfpathrectangle{\pgfqpoint{-0.000000in}{-0.000000in}}{\pgfqpoint{1.543815in}{1.715350in}}%
\pgfusepath{clip}%
\pgfsys@transformshift{-0.000000in}{0.000000in}%
\pgftext[left,bottom]{\includegraphics[interpolate=true,width=1.545000in,height=1.717500in]{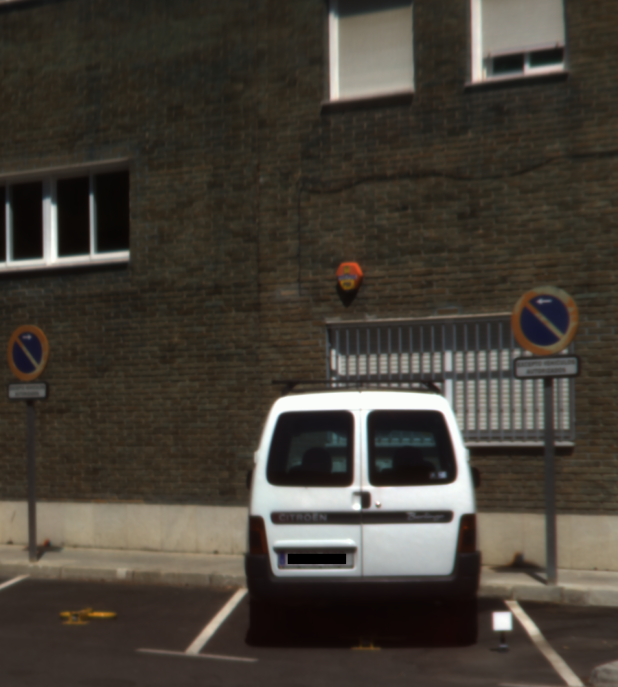}}%
\end{pgfscope}%
\begin{pgfscope}%
\pgfpathrectangle{\pgfqpoint{1.665935in}{-0.000000in}}{\pgfqpoint{1.543815in}{1.715350in}}%
\pgfusepath{clip}%
\pgfsys@transformshift{1.665935in}{0.000000in}%
\pgftext[left,bottom]{\includegraphics[interpolate=true,width=1.545000in,height=1.717500in]{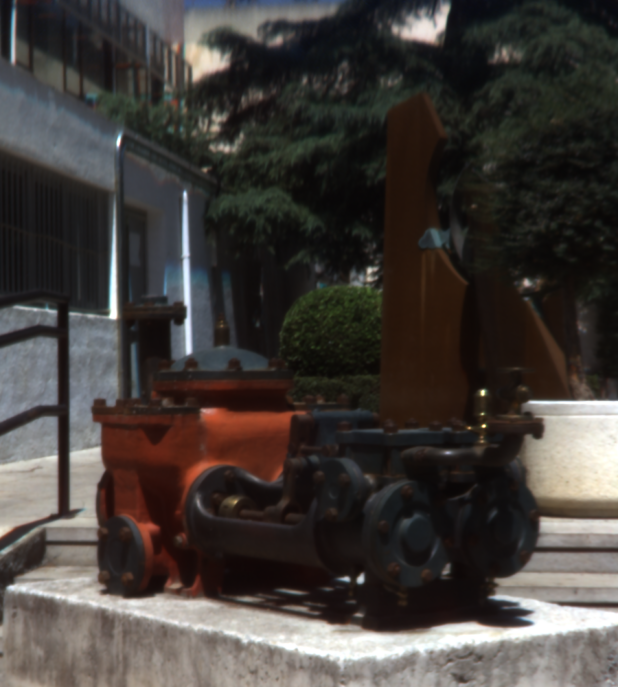}}%
\end{pgfscope}%
\end{pgfpicture}%
\makeatother%
\endgroup%

%% file: figures/example_reconstruction_images_2.pgf
%% Creator: Matplotlib, PGF backend
%%
%% To include the figure in your LaTeX document, write
%%   \input{<filename>.pgf}
%%
%% Make sure the required packages are loaded in your preamble
%%   \usepackage{pgf}
%%
%% and, on pdftex
%%   \usepackage[utf8]{inputenc}\DeclareUnicodeCharacter{2212}{-}
%%
%% or, on luatex and xetex
%%   \usepackage{unicode-math}
%%
%% Figures using additional raster images can only be included by \input if
%% they are in the same directory as the main LaTeX file. For loading figures
%% from other directories you can use the `import` package
%%   \usepackage{import}
%%
%% and then include the figures with
%%   \import{<path to file>}{<filename>.pgf}
%%
%% Matplotlib used the following preamble
%%   \usepackage{fontspec}
%%
\begingroup%
\makeatletter%
\begin{pgfpicture}%
\pgfpathrectangle{\pgfpointorigin}{\pgfqpoint{6.908195in}{8.334597in}}%
\pgfusepath{use as bounding box, clip}%
\begin{pgfscope}%
\pgfsetbuttcap%
\pgfsetmiterjoin%
\definecolor{currentfill}{rgb}{1.000000,1.000000,1.000000}%
\pgfsetfillcolor{currentfill}%
\pgfsetlinewidth{0.000000pt}%
\definecolor{currentstroke}{rgb}{1.000000,1.000000,1.000000}%
\pgfsetstrokecolor{currentstroke}%
\pgfsetdash{}{0pt}%
\pgfpathmoveto{\pgfqpoint{0.000000in}{0.000000in}}%
\pgfpathlineto{\pgfqpoint{6.908195in}{0.000000in}}%
\pgfpathlineto{\pgfqpoint{6.908195in}{8.334597in}}%
\pgfpathlineto{\pgfqpoint{0.000000in}{8.334597in}}%
\pgfpathclose%
\pgfusepath{fill}%
\end{pgfscope}%
\begin{pgfscope}%
\pgfpathrectangle{\pgfqpoint{0.134180in}{6.273583in}}{\pgfqpoint{1.936014in}{1.936014in}}%
\pgfusepath{clip}%
\pgfsys@transformshift{0.134180in}{6.273583in}%
\pgftext[left,bottom]{\includegraphics[interpolate=true,width=1.937500in,height=1.937500in]{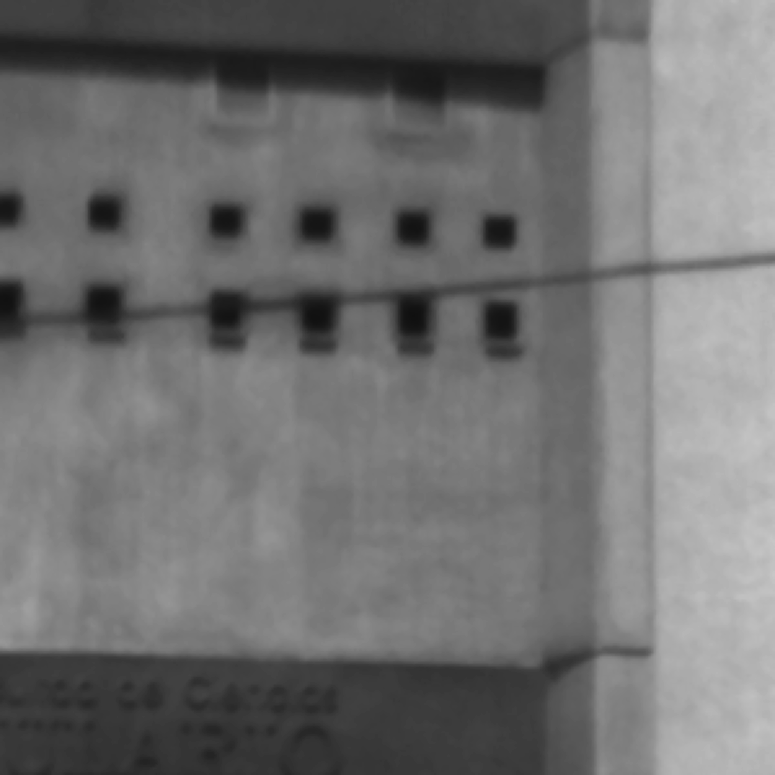}}%
\end{pgfscope}%
\begin{pgfscope}%
\definecolor{textcolor}{rgb}{0.000000,0.000000,0.000000}%
\pgfsetstrokecolor{textcolor}%
\pgfsetfillcolor{textcolor}%
\pgftext[x=1.102187in,y=8.233847in,,base]{\color{textcolor}\rmfamily\fontsize{9.000000}{10.800000}\selectfont Reference}%
\end{pgfscope}%
\begin{pgfscope}%
\pgfpathrectangle{\pgfqpoint{2.553180in}{6.273583in}}{\pgfqpoint{1.936014in}{1.936014in}}%
\pgfusepath{clip}%
\pgfsys@transformshift{2.553180in}{6.273583in}%
\pgftext[left,bottom]{\includegraphics[interpolate=true,width=1.937500in,height=1.937500in]{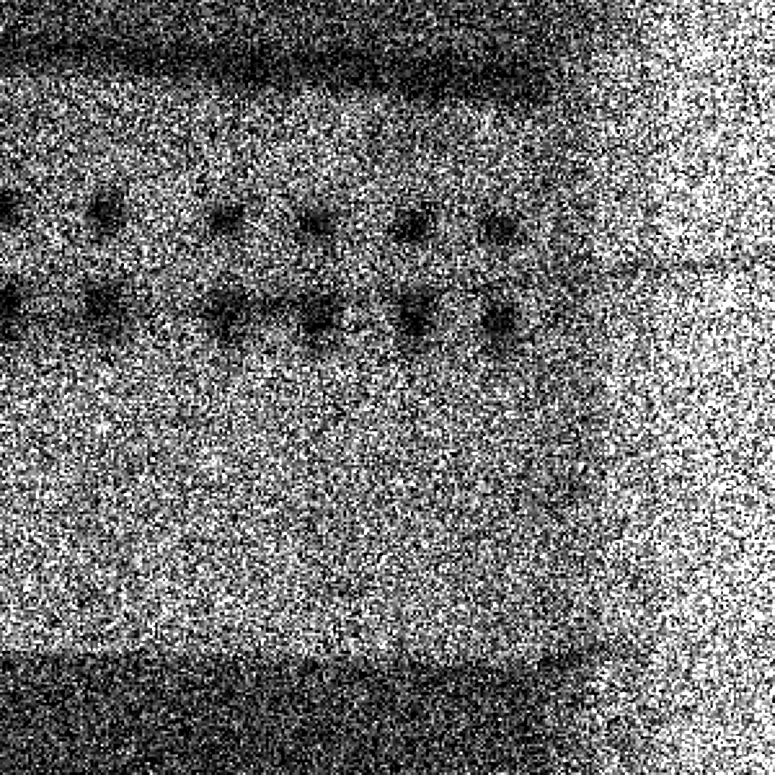}}%
\end{pgfscope}%
\begin{pgfscope}%
\definecolor{textcolor}{rgb}{0.000000,0.000000,0.000000}%
\pgfsetstrokecolor{textcolor}%
\pgfsetfillcolor{textcolor}%
\pgftext[x=3.521187in,y=8.240847in,,base]{\color{textcolor}\rmfamily\fontsize{9.000000}{10.800000}\selectfont SP\cite{pratt-spectral-1976}}%
\end{pgfscope}%
\begin{pgfscope}%
\pgfpathrectangle{\pgfqpoint{4.972180in}{6.273583in}}{\pgfqpoint{1.936014in}{1.936014in}}%
\pgfusepath{clip}%
\pgfsys@transformshift{4.972180in}{6.273583in}%
\pgftext[left,bottom]{\includegraphics[interpolate=true,width=1.937500in,height=1.937500in]{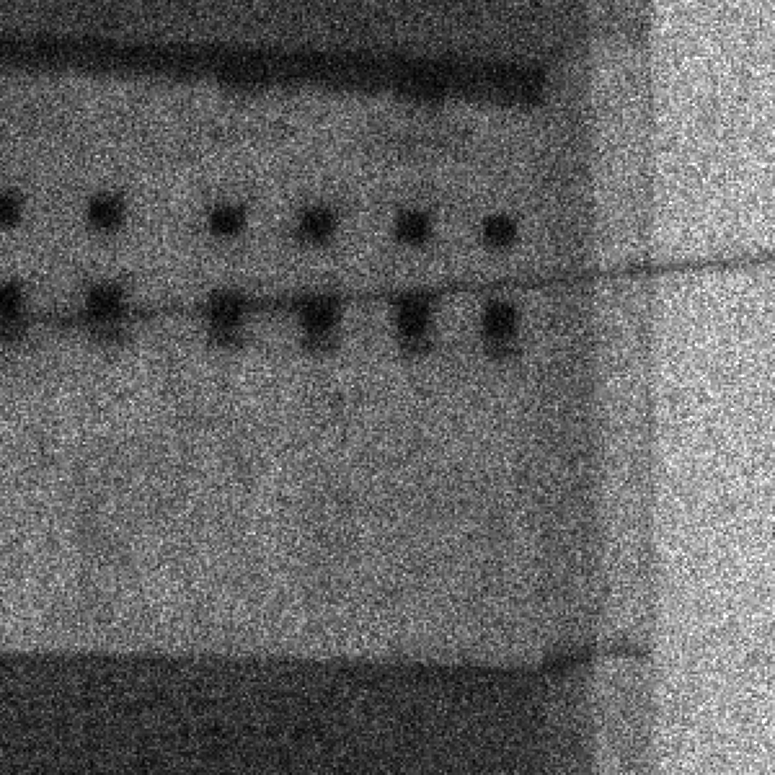}}%
\end{pgfscope}%
\begin{pgfscope}%
\definecolor{textcolor}{rgb}{0.000000,0.000000,0.000000}%
\pgfsetstrokecolor{textcolor}%
\pgfsetfillcolor{textcolor}%
\pgftext[x=5.940187in,y=8.240847in,,base]{\color{textcolor}\rmfamily\fontsize{9.000000}{10.800000}\selectfont WF\cite{pratt-spectral-1976}}%
\end{pgfscope}%
\begin{pgfscope}%
\pgfpathrectangle{\pgfqpoint{0.134180in}{4.182389in}}{\pgfqpoint{1.936014in}{1.936014in}}%
\pgfusepath{clip}%
\pgfsys@transformshift{0.134180in}{4.182389in}%
\pgftext[left,bottom]{\includegraphics[interpolate=true,width=1.937500in,height=1.937500in]{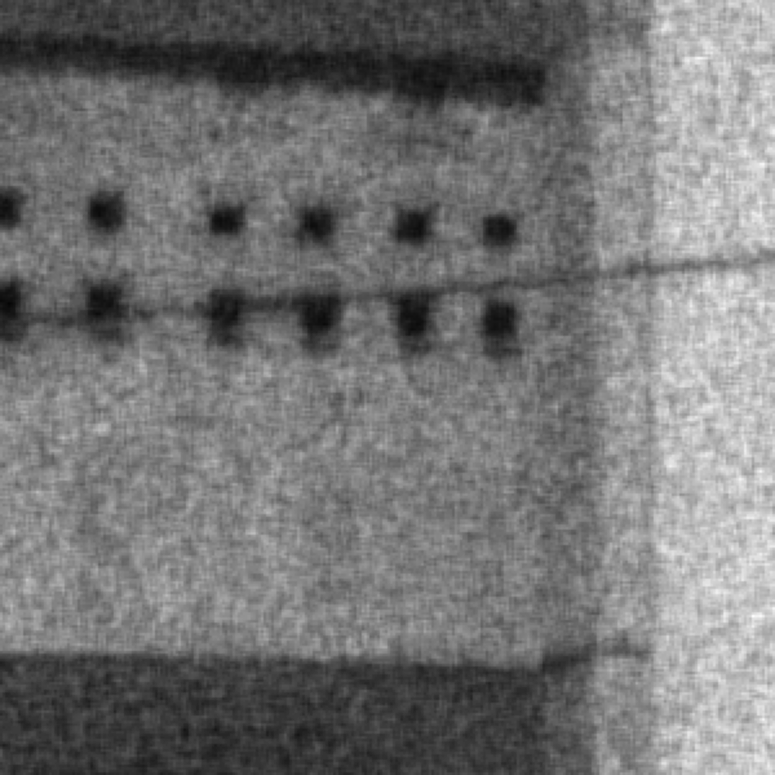}}%
\end{pgfscope}%
\begin{pgfscope}%
\definecolor{textcolor}{rgb}{0.000000,0.000000,0.000000}%
\pgfsetstrokecolor{textcolor}%
\pgfsetfillcolor{textcolor}%
\pgftext[x=1.102187in,y=6.149653in,,base]{\color{textcolor}\rmfamily\fontsize{9.000000}{10.800000}\selectfont SSW\cite{murakami-color-2008}}%
\end{pgfscope}%
\begin{pgfscope}%
\pgfpathrectangle{\pgfqpoint{2.553180in}{4.182389in}}{\pgfqpoint{1.936014in}{1.936014in}}%
\pgfusepath{clip}%
\pgfsys@transformshift{2.553180in}{4.182389in}%
\pgftext[left,bottom]{\includegraphics[interpolate=true,width=1.937500in,height=1.937500in]{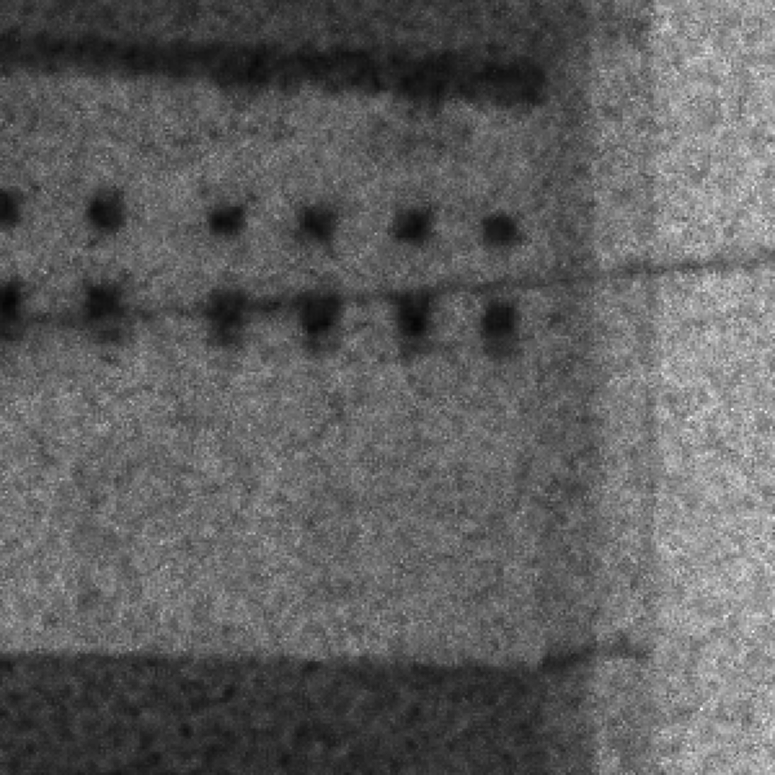}}%
\end{pgfscope}%
\begin{pgfscope}%
\definecolor{textcolor}{rgb}{0.000000,0.000000,0.000000}%
\pgfsetstrokecolor{textcolor}%
\pgfsetfillcolor{textcolor}%
\pgftext[x=3.521187in,y=6.149653in,,base]{\color{textcolor}\rmfamily\fontsize{9.000000}{10.800000}\selectfont EPSSW\cite{urban-spectral-2009}}%
\end{pgfscope}%
\begin{pgfscope}%
\pgfpathrectangle{\pgfqpoint{4.972180in}{4.182389in}}{\pgfqpoint{1.936014in}{1.936014in}}%
\pgfusepath{clip}%
\pgfsys@transformshift{4.972180in}{4.182389in}%
\pgftext[left,bottom]{\includegraphics[interpolate=true,width=1.937500in,height=1.937500in]{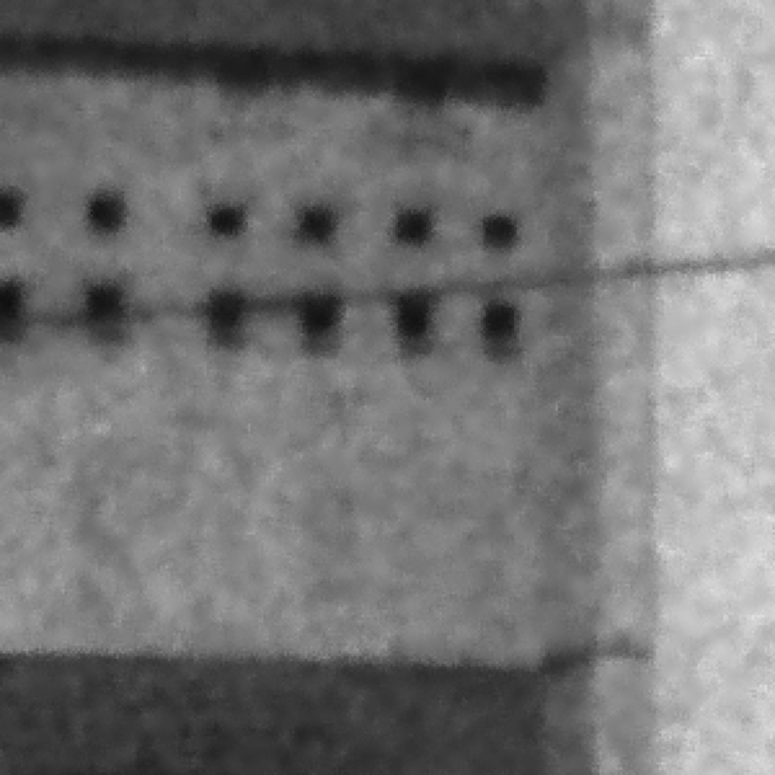}}%
\end{pgfscope}%
\begin{pgfscope}%
\definecolor{textcolor}{rgb}{0.000000,0.000000,0.000000}%
\pgfsetstrokecolor{textcolor}%
\pgfsetfillcolor{textcolor}%
\pgftext[x=5.940187in,y=6.142653in,,base]{\color{textcolor}\rmfamily\fontsize{9.000000}{10.800000}\selectfont SPRE}%
\end{pgfscope}%
\begin{pgfscope}%
\pgfpathrectangle{\pgfqpoint{0.134180in}{2.091194in}}{\pgfqpoint{1.936014in}{1.936014in}}%
\pgfusepath{clip}%
\pgfsys@transformshift{0.134180in}{2.091194in}%
\pgftext[left,bottom]{\includegraphics[interpolate=true,width=1.937500in,height=1.937500in]{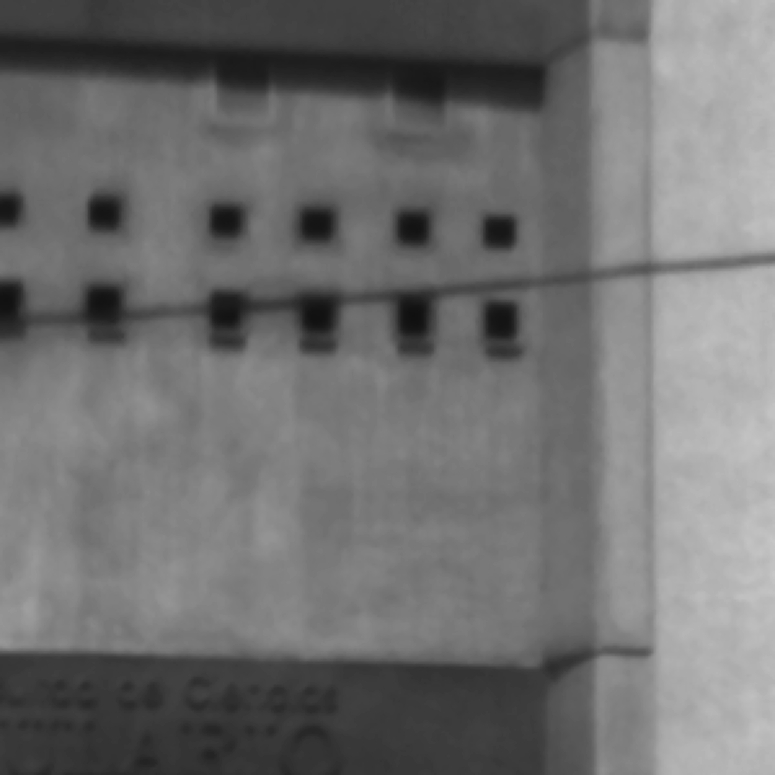}}%
\end{pgfscope}%
\begin{pgfscope}%
\definecolor{textcolor}{rgb}{0.000000,0.000000,0.000000}%
\pgfsetstrokecolor{textcolor}%
\pgfsetfillcolor{textcolor}%
\pgftext[x=1.102187in,y=4.051459in,,base]{\color{textcolor}\rmfamily\fontsize{9.000000}{10.800000}\selectfont Reference}%
\end{pgfscope}%
\begin{pgfscope}%
\pgfpathrectangle{\pgfqpoint{2.553180in}{2.091194in}}{\pgfqpoint{1.936014in}{1.936014in}}%
\pgfusepath{clip}%
\pgfsys@transformshift{2.553180in}{2.091194in}%
\pgftext[left,bottom]{\includegraphics[interpolate=true,width=1.937500in,height=1.937500in]{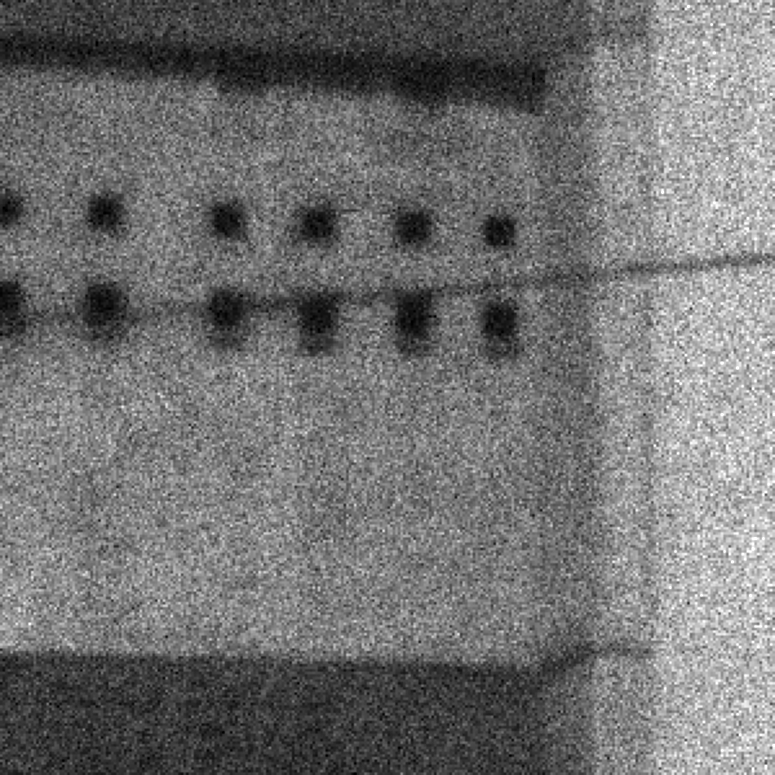}}%
\end{pgfscope}%
\begin{pgfscope}%
\definecolor{textcolor}{rgb}{0.000000,0.000000,0.000000}%
\pgfsetstrokecolor{textcolor}%
\pgfsetfillcolor{textcolor}%
\pgftext[x=3.521187in,y=4.058458in,,base]{\color{textcolor}\rmfamily\fontsize{9.000000}{10.800000}\selectfont SP\cite{pratt-spectral-1976}}%
\end{pgfscope}%
\begin{pgfscope}%
\pgfpathrectangle{\pgfqpoint{4.972180in}{2.091194in}}{\pgfqpoint{1.936014in}{1.936014in}}%
\pgfusepath{clip}%
\pgfsys@transformshift{4.972180in}{2.091194in}%
\pgftext[left,bottom]{\includegraphics[interpolate=true,width=1.937500in,height=1.937500in]{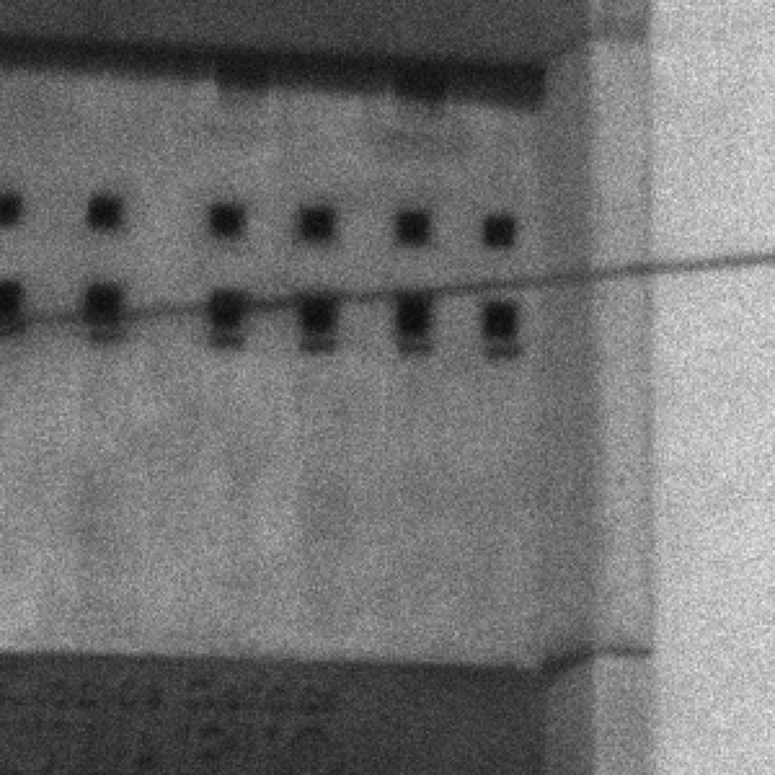}}%
\end{pgfscope}%
\begin{pgfscope}%
\definecolor{textcolor}{rgb}{0.000000,0.000000,0.000000}%
\pgfsetstrokecolor{textcolor}%
\pgfsetfillcolor{textcolor}%
\pgftext[x=5.940187in,y=4.058458in,,base]{\color{textcolor}\rmfamily\fontsize{9.000000}{10.800000}\selectfont WF\cite{pratt-spectral-1976}}%
\end{pgfscope}%
\begin{pgfscope}%
\pgfpathrectangle{\pgfqpoint{0.134180in}{-0.000000in}}{\pgfqpoint{1.936014in}{1.936014in}}%
\pgfusepath{clip}%
\pgfsys@transformshift{0.134180in}{0.000000in}%
\pgftext[left,bottom]{\includegraphics[interpolate=true,width=1.937500in,height=1.937500in]{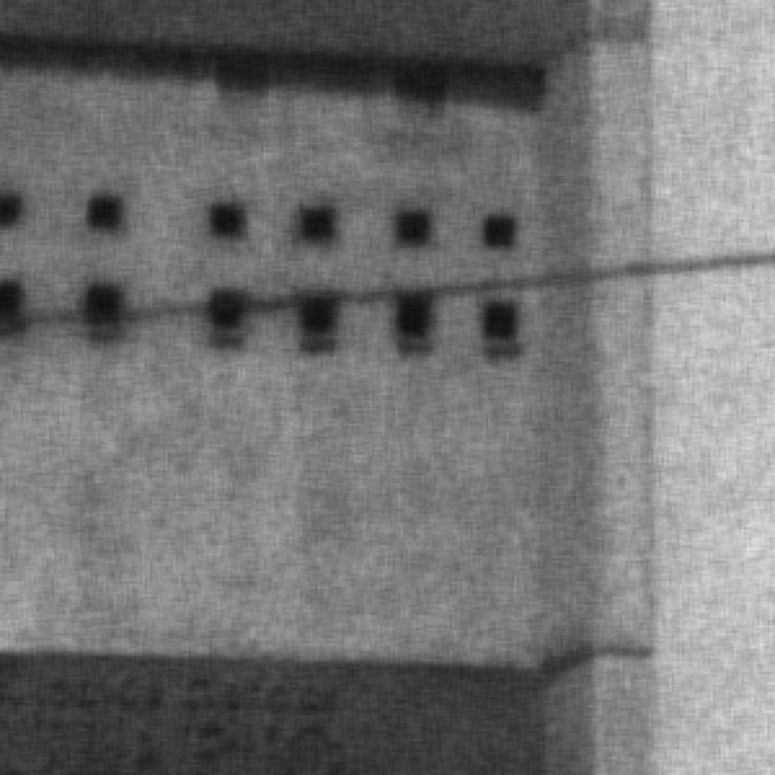}}%
\end{pgfscope}%
\begin{pgfscope}%
\definecolor{textcolor}{rgb}{0.000000,0.000000,0.000000}%
\pgfsetstrokecolor{textcolor}%
\pgfsetfillcolor{textcolor}%
\pgftext[x=1.102187in,y=1.967264in,,base]{\color{textcolor}\rmfamily\fontsize{9.000000}{10.800000}\selectfont SSW\cite{murakami-color-2008}}%
\end{pgfscope}%
\begin{pgfscope}%
\pgfpathrectangle{\pgfqpoint{2.553180in}{-0.000000in}}{\pgfqpoint{1.936014in}{1.936014in}}%
\pgfusepath{clip}%
\pgfsys@transformshift{2.553180in}{0.000000in}%
\pgftext[left,bottom]{\includegraphics[interpolate=true,width=1.937500in,height=1.937500in]{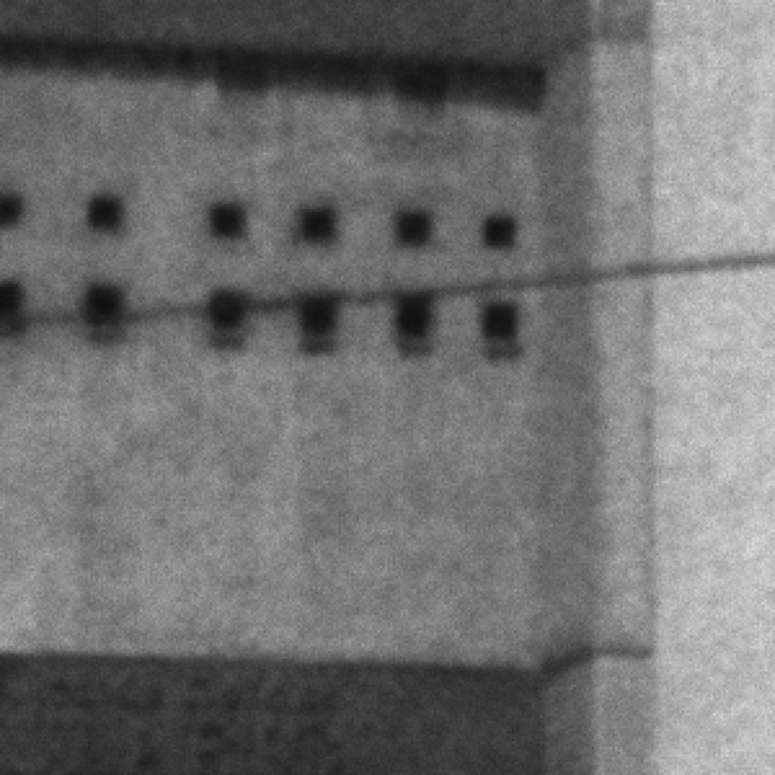}}%
\end{pgfscope}%
\begin{pgfscope}%
\definecolor{textcolor}{rgb}{0.000000,0.000000,0.000000}%
\pgfsetstrokecolor{textcolor}%
\pgfsetfillcolor{textcolor}%
\pgftext[x=3.521187in,y=1.967264in,,base]{\color{textcolor}\rmfamily\fontsize{9.000000}{10.800000}\selectfont EPSSW\cite{urban-spectral-2009}}%
\end{pgfscope}%
\begin{pgfscope}%
\pgfpathrectangle{\pgfqpoint{4.972180in}{-0.000000in}}{\pgfqpoint{1.936014in}{1.936014in}}%
\pgfusepath{clip}%
\pgfsys@transformshift{4.972180in}{0.000000in}%
\pgftext[left,bottom]{\includegraphics[interpolate=true,width=1.937500in,height=1.937500in]{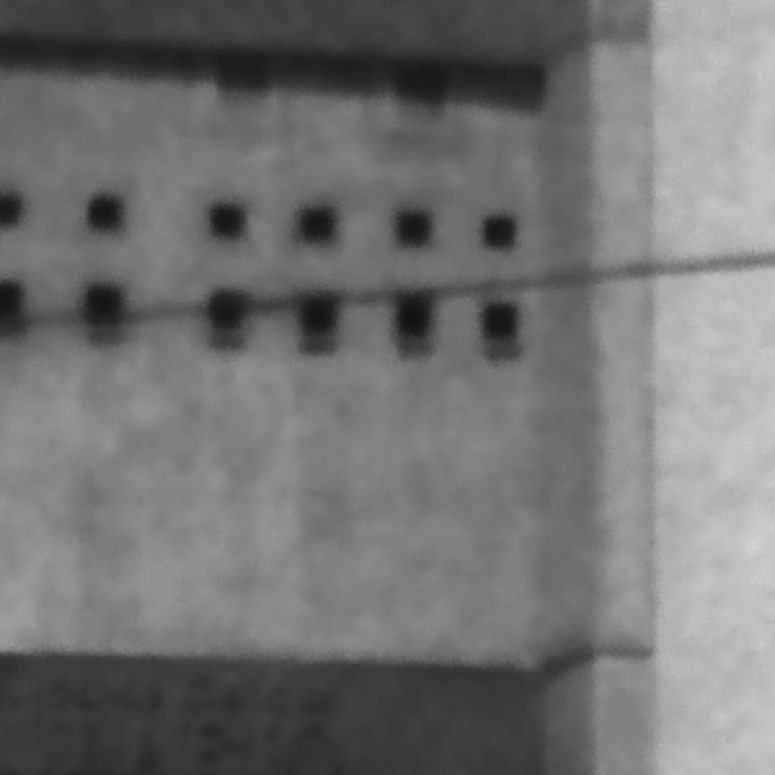}}%
\end{pgfscope}%
\begin{pgfscope}%
\definecolor{textcolor}{rgb}{0.000000,0.000000,0.000000}%
\pgfsetstrokecolor{textcolor}%
\pgfsetfillcolor{textcolor}%
\pgftext[x=5.940187in,y=1.960264in,,base]{\color{textcolor}\rmfamily\fontsize{9.000000}{10.800000}\selectfont SPRE}%
\end{pgfscope}%
\begin{pgfscope}%
\definecolor{textcolor}{rgb}{0.000000,0.000000,0.000000}%
\pgfsetstrokecolor{textcolor}%
\pgfsetfillcolor{textcolor}%
\pgftext[x=0.086750in, y=1.549375in, left, base,rotate=90.000000]{\color{textcolor}\rmfamily\fontsize{9.000000}{10.800000}\selectfont Intensity level 100}%
\end{pgfscope}%
\begin{pgfscope}%
\definecolor{textcolor}{rgb}{0.000000,0.000000,0.000000}%
\pgfsetstrokecolor{textcolor}%
\pgfsetfillcolor{textcolor}%
\pgftext[x=0.086750in, y=5.738875in, left, base,rotate=90.000000]{\color{textcolor}\rmfamily\fontsize{9.000000}{10.800000}\selectfont Intensity level 10}%
\end{pgfscope}%
\end{pgfpicture}%
\makeatother%
\endgroup%

%% file: figures/runtime_evaluation.pgf
%% Creator: Matplotlib, PGF backend
%%
%% To include the figure in your LaTeX document, write
%%   \input{<filename>.pgf}
%%
%% Make sure the required packages are loaded in your preamble
%%   \usepackage{pgf}
%%
%% and, on pdftex
%%   \usepackage[utf8]{inputenc}\DeclareUnicodeCharacter{2212}{-}
%%
%% or, on luatex and xetex
%%   \usepackage{unicode-math}
%%
%% Figures using additional raster images can only be included by \input if
%% they are in the same directory as the main LaTeX file. For loading figures
%% from other directories you can use the `import` package
%%   \usepackage{import}
%%
%% and then include the figures with
%%   \import{<path to file>}{<filename>.pgf}
%%
%% Matplotlib used the following preamble
%%   \usepackage{fontspec}
%%
\begingroup%
\makeatletter%
\begin{pgfpicture}%
\pgfpathrectangle{\pgfpointorigin}{\pgfqpoint{3.217750in}{1.848070in}}%
\pgfusepath{use as bounding box, clip}%
\begin{pgfscope}%
\pgfsetbuttcap%
\pgfsetmiterjoin%
\definecolor{currentfill}{rgb}{1.000000,1.000000,1.000000}%
\pgfsetfillcolor{currentfill}%
\pgfsetlinewidth{0.000000pt}%
\definecolor{currentstroke}{rgb}{1.000000,1.000000,1.000000}%
\pgfsetstrokecolor{currentstroke}%
\pgfsetdash{}{0pt}%
\pgfpathmoveto{\pgfqpoint{0.000000in}{-0.000000in}}%
\pgfpathlineto{\pgfqpoint{3.217750in}{-0.000000in}}%
\pgfpathlineto{\pgfqpoint{3.217750in}{1.848070in}}%
\pgfpathlineto{\pgfqpoint{0.000000in}{1.848070in}}%
\pgfpathclose%
\pgfusepath{fill}%
\end{pgfscope}%
\begin{pgfscope}%
\pgfsetbuttcap%
\pgfsetmiterjoin%
\definecolor{currentfill}{rgb}{1.000000,1.000000,1.000000}%
\pgfsetfillcolor{currentfill}%
\pgfsetlinewidth{0.000000pt}%
\definecolor{currentstroke}{rgb}{0.000000,0.000000,0.000000}%
\pgfsetstrokecolor{currentstroke}%
\pgfsetstrokeopacity{0.000000}%
\pgfsetdash{}{0pt}%
\pgfpathmoveto{\pgfqpoint{0.492278in}{0.208222in}}%
\pgfpathlineto{\pgfqpoint{3.217750in}{0.208222in}}%
\pgfpathlineto{\pgfqpoint{3.217750in}{1.848070in}}%
\pgfpathlineto{\pgfqpoint{0.492278in}{1.848070in}}%
\pgfpathclose%
\pgfusepath{fill}%
\end{pgfscope}%
\begin{pgfscope}%
\pgfpathrectangle{\pgfqpoint{0.492278in}{0.208222in}}{\pgfqpoint{2.725472in}{1.639848in}}%
\pgfusepath{clip}%
\pgfsetbuttcap%
\pgfsetmiterjoin%
\definecolor{currentfill}{rgb}{0.121569,0.466667,0.705882}%
\pgfsetfillcolor{currentfill}%
\pgfsetlinewidth{1.003750pt}%
\definecolor{currentstroke}{rgb}{0.000000,0.000000,0.000000}%
\pgfsetstrokecolor{currentstroke}%
\pgfsetdash{{1.000000pt}{1.650000pt}}{0.000000pt}%
\pgfpathmoveto{\pgfqpoint{0.616163in}{0.208222in}}%
\pgfpathlineto{\pgfqpoint{0.957915in}{0.208222in}}%
\pgfpathlineto{\pgfqpoint{0.957915in}{0.410956in}}%
\pgfpathlineto{\pgfqpoint{0.616163in}{0.410956in}}%
\pgfpathclose%
\pgfusepath{stroke,fill}%
\end{pgfscope}%
\begin{pgfscope}%
\pgfpathrectangle{\pgfqpoint{0.492278in}{0.208222in}}{\pgfqpoint{2.725472in}{1.639848in}}%
\pgfusepath{clip}%
\pgfsetbuttcap%
\pgfsetmiterjoin%
\definecolor{currentfill}{rgb}{0.172549,0.627451,0.172549}%
\pgfsetfillcolor{currentfill}%
\pgfsetlinewidth{1.003750pt}%
\definecolor{currentstroke}{rgb}{0.000000,0.000000,0.000000}%
\pgfsetstrokecolor{currentstroke}%
\pgfsetdash{}{0pt}%
\pgfpathmoveto{\pgfqpoint{1.043353in}{0.208222in}}%
\pgfpathlineto{\pgfqpoint{1.385105in}{0.208222in}}%
\pgfpathlineto{\pgfqpoint{1.385105in}{0.441998in}}%
\pgfpathlineto{\pgfqpoint{1.043353in}{0.441998in}}%
\pgfpathclose%
\pgfusepath{stroke,fill}%
\end{pgfscope}%
\begin{pgfscope}%
\pgfpathrectangle{\pgfqpoint{0.492278in}{0.208222in}}{\pgfqpoint{2.725472in}{1.639848in}}%
\pgfusepath{clip}%
\pgfsetbuttcap%
\pgfsetmiterjoin%
\definecolor{currentfill}{rgb}{1.000000,0.498039,0.054902}%
\pgfsetfillcolor{currentfill}%
\pgfsetlinewidth{1.003750pt}%
\definecolor{currentstroke}{rgb}{0.000000,0.000000,0.000000}%
\pgfsetstrokecolor{currentstroke}%
\pgfsetdash{{6.400000pt}{1.600000pt}{1.000000pt}{1.600000pt}}{0.000000pt}%
\pgfpathmoveto{\pgfqpoint{1.470543in}{0.208222in}}%
\pgfpathlineto{\pgfqpoint{1.812295in}{0.208222in}}%
\pgfpathlineto{\pgfqpoint{1.812295in}{0.616625in}}%
\pgfpathlineto{\pgfqpoint{1.470543in}{0.616625in}}%
\pgfpathclose%
\pgfusepath{stroke,fill}%
\end{pgfscope}%
\begin{pgfscope}%
\pgfpathrectangle{\pgfqpoint{0.492278in}{0.208222in}}{\pgfqpoint{2.725472in}{1.639848in}}%
\pgfusepath{clip}%
\pgfsetbuttcap%
\pgfsetmiterjoin%
\definecolor{currentfill}{rgb}{0.839216,0.152941,0.156863}%
\pgfsetfillcolor{currentfill}%
\pgfsetlinewidth{1.003750pt}%
\definecolor{currentstroke}{rgb}{0.000000,0.000000,0.000000}%
\pgfsetstrokecolor{currentstroke}%
\pgfsetdash{{1.000000pt}{1.650000pt}}{0.000000pt}%
\pgfpathmoveto{\pgfqpoint{1.897733in}{0.208222in}}%
\pgfpathlineto{\pgfqpoint{2.239485in}{0.208222in}}%
\pgfpathlineto{\pgfqpoint{2.239485in}{0.844502in}}%
\pgfpathlineto{\pgfqpoint{1.897733in}{0.844502in}}%
\pgfpathclose%
\pgfusepath{stroke,fill}%
\end{pgfscope}%
\begin{pgfscope}%
\pgfpathrectangle{\pgfqpoint{0.492278in}{0.208222in}}{\pgfqpoint{2.725472in}{1.639848in}}%
\pgfusepath{clip}%
\pgfsetbuttcap%
\pgfsetmiterjoin%
\definecolor{currentfill}{rgb}{0.580392,0.403922,0.741176}%
\pgfsetfillcolor{currentfill}%
\pgfsetlinewidth{1.003750pt}%
\definecolor{currentstroke}{rgb}{0.000000,0.000000,0.000000}%
\pgfsetstrokecolor{currentstroke}%
\pgfsetdash{{3.700000pt}{1.600000pt}}{0.000000pt}%
\pgfpathmoveto{\pgfqpoint{2.324923in}{0.208222in}}%
\pgfpathlineto{\pgfqpoint{2.666675in}{0.208222in}}%
\pgfpathlineto{\pgfqpoint{2.666675in}{1.776711in}}%
\pgfpathlineto{\pgfqpoint{2.324923in}{1.776711in}}%
\pgfpathclose%
\pgfusepath{stroke,fill}%
\end{pgfscope}%
\begin{pgfscope}%
\pgfpathrectangle{\pgfqpoint{0.492278in}{0.208222in}}{\pgfqpoint{2.725472in}{1.639848in}}%
\pgfusepath{clip}%
\pgfsetbuttcap%
\pgfsetmiterjoin%
\definecolor{currentfill}{rgb}{0.549020,0.337255,0.294118}%
\pgfsetfillcolor{currentfill}%
\pgfsetlinewidth{1.003750pt}%
\definecolor{currentstroke}{rgb}{0.000000,0.000000,0.000000}%
\pgfsetstrokecolor{currentstroke}%
\pgfsetdash{{3.000000pt}{1.000000pt}{1.000000pt}{1.000000pt}{1.000000pt}{1.000000pt}}{0.000000pt}%
\pgfpathmoveto{\pgfqpoint{2.752113in}{0.208222in}}%
\pgfpathlineto{\pgfqpoint{3.093865in}{0.208222in}}%
\pgfpathlineto{\pgfqpoint{3.093865in}{1.328346in}}%
\pgfpathlineto{\pgfqpoint{2.752113in}{1.328346in}}%
\pgfpathclose%
\pgfusepath{stroke,fill}%
\end{pgfscope}%
\begin{pgfscope}%
\pgfpathrectangle{\pgfqpoint{0.492278in}{0.208222in}}{\pgfqpoint{2.725472in}{1.639848in}}%
\pgfusepath{clip}%
\pgfsetrectcap%
\pgfsetroundjoin%
\pgfsetlinewidth{0.803000pt}%
\definecolor{currentstroke}{rgb}{0.690196,0.690196,0.690196}%
\pgfsetstrokecolor{currentstroke}%
\pgfsetdash{}{0pt}%
\pgfpathmoveto{\pgfqpoint{0.787039in}{0.208222in}}%
\pgfpathlineto{\pgfqpoint{0.787039in}{1.848070in}}%
\pgfusepath{stroke}%
\end{pgfscope}%
\begin{pgfscope}%
\pgfsetbuttcap%
\pgfsetroundjoin%
\definecolor{currentfill}{rgb}{0.000000,0.000000,0.000000}%
\pgfsetfillcolor{currentfill}%
\pgfsetlinewidth{0.803000pt}%
\definecolor{currentstroke}{rgb}{0.000000,0.000000,0.000000}%
\pgfsetstrokecolor{currentstroke}%
\pgfsetdash{}{0pt}%
\pgfsys@defobject{currentmarker}{\pgfqpoint{0.000000in}{-0.048611in}}{\pgfqpoint{0.000000in}{0.000000in}}{%
\pgfpathmoveto{\pgfqpoint{0.000000in}{0.000000in}}%
\pgfpathlineto{\pgfqpoint{0.000000in}{-0.048611in}}%
\pgfusepath{stroke,fill}%
}%
\begin{pgfscope}%
\pgfsys@transformshift{0.787039in}{0.208222in}%
\pgfsys@useobject{currentmarker}{}%
\end{pgfscope}%
\end{pgfscope}%
\begin{pgfscope}%
\definecolor{textcolor}{rgb}{0.000000,0.000000,0.000000}%
\pgfsetstrokecolor{textcolor}%
\pgfsetfillcolor{textcolor}%
\pgftext[x=0.787039in,y=0.111000in,,top]{\color{textcolor}\rmfamily\fontsize{9.000000}{10.800000}\selectfont SP}%
\end{pgfscope}%
\begin{pgfscope}%
\pgfpathrectangle{\pgfqpoint{0.492278in}{0.208222in}}{\pgfqpoint{2.725472in}{1.639848in}}%
\pgfusepath{clip}%
\pgfsetrectcap%
\pgfsetroundjoin%
\pgfsetlinewidth{0.803000pt}%
\definecolor{currentstroke}{rgb}{0.690196,0.690196,0.690196}%
\pgfsetstrokecolor{currentstroke}%
\pgfsetdash{}{0pt}%
\pgfpathmoveto{\pgfqpoint{1.214229in}{0.208222in}}%
\pgfpathlineto{\pgfqpoint{1.214229in}{1.848070in}}%
\pgfusepath{stroke}%
\end{pgfscope}%
\begin{pgfscope}%
\pgfsetbuttcap%
\pgfsetroundjoin%
\definecolor{currentfill}{rgb}{0.000000,0.000000,0.000000}%
\pgfsetfillcolor{currentfill}%
\pgfsetlinewidth{0.803000pt}%
\definecolor{currentstroke}{rgb}{0.000000,0.000000,0.000000}%
\pgfsetstrokecolor{currentstroke}%
\pgfsetdash{}{0pt}%
\pgfsys@defobject{currentmarker}{\pgfqpoint{0.000000in}{-0.048611in}}{\pgfqpoint{0.000000in}{0.000000in}}{%
\pgfpathmoveto{\pgfqpoint{0.000000in}{0.000000in}}%
\pgfpathlineto{\pgfqpoint{0.000000in}{-0.048611in}}%
\pgfusepath{stroke,fill}%
}%
\begin{pgfscope}%
\pgfsys@transformshift{1.214229in}{0.208222in}%
\pgfsys@useobject{currentmarker}{}%
\end{pgfscope}%
\end{pgfscope}%
\begin{pgfscope}%
\definecolor{textcolor}{rgb}{0.000000,0.000000,0.000000}%
\pgfsetstrokecolor{textcolor}%
\pgfsetfillcolor{textcolor}%
\pgftext[x=1.214229in,y=0.111000in,,top]{\color{textcolor}\rmfamily\fontsize{9.000000}{10.800000}\selectfont BPE}%
\end{pgfscope}%
\begin{pgfscope}%
\pgfpathrectangle{\pgfqpoint{0.492278in}{0.208222in}}{\pgfqpoint{2.725472in}{1.639848in}}%
\pgfusepath{clip}%
\pgfsetrectcap%
\pgfsetroundjoin%
\pgfsetlinewidth{0.803000pt}%
\definecolor{currentstroke}{rgb}{0.690196,0.690196,0.690196}%
\pgfsetstrokecolor{currentstroke}%
\pgfsetdash{}{0pt}%
\pgfpathmoveto{\pgfqpoint{1.641419in}{0.208222in}}%
\pgfpathlineto{\pgfqpoint{1.641419in}{1.848070in}}%
\pgfusepath{stroke}%
\end{pgfscope}%
\begin{pgfscope}%
\pgfsetbuttcap%
\pgfsetroundjoin%
\definecolor{currentfill}{rgb}{0.000000,0.000000,0.000000}%
\pgfsetfillcolor{currentfill}%
\pgfsetlinewidth{0.803000pt}%
\definecolor{currentstroke}{rgb}{0.000000,0.000000,0.000000}%
\pgfsetstrokecolor{currentstroke}%
\pgfsetdash{}{0pt}%
\pgfsys@defobject{currentmarker}{\pgfqpoint{0.000000in}{-0.048611in}}{\pgfqpoint{0.000000in}{0.000000in}}{%
\pgfpathmoveto{\pgfqpoint{0.000000in}{0.000000in}}%
\pgfpathlineto{\pgfqpoint{0.000000in}{-0.048611in}}%
\pgfusepath{stroke,fill}%
}%
\begin{pgfscope}%
\pgfsys@transformshift{1.641419in}{0.208222in}%
\pgfsys@useobject{currentmarker}{}%
\end{pgfscope}%
\end{pgfscope}%
\begin{pgfscope}%
\definecolor{textcolor}{rgb}{0.000000,0.000000,0.000000}%
\pgfsetstrokecolor{textcolor}%
\pgfsetfillcolor{textcolor}%
\pgftext[x=1.641419in,y=0.111000in,,top]{\color{textcolor}\rmfamily\fontsize{9.000000}{10.800000}\selectfont WF}%
\end{pgfscope}%
\begin{pgfscope}%
\pgfpathrectangle{\pgfqpoint{0.492278in}{0.208222in}}{\pgfqpoint{2.725472in}{1.639848in}}%
\pgfusepath{clip}%
\pgfsetrectcap%
\pgfsetroundjoin%
\pgfsetlinewidth{0.803000pt}%
\definecolor{currentstroke}{rgb}{0.690196,0.690196,0.690196}%
\pgfsetstrokecolor{currentstroke}%
\pgfsetdash{}{0pt}%
\pgfpathmoveto{\pgfqpoint{2.068609in}{0.208222in}}%
\pgfpathlineto{\pgfqpoint{2.068609in}{1.848070in}}%
\pgfusepath{stroke}%
\end{pgfscope}%
\begin{pgfscope}%
\pgfsetbuttcap%
\pgfsetroundjoin%
\definecolor{currentfill}{rgb}{0.000000,0.000000,0.000000}%
\pgfsetfillcolor{currentfill}%
\pgfsetlinewidth{0.803000pt}%
\definecolor{currentstroke}{rgb}{0.000000,0.000000,0.000000}%
\pgfsetstrokecolor{currentstroke}%
\pgfsetdash{}{0pt}%
\pgfsys@defobject{currentmarker}{\pgfqpoint{0.000000in}{-0.048611in}}{\pgfqpoint{0.000000in}{0.000000in}}{%
\pgfpathmoveto{\pgfqpoint{0.000000in}{0.000000in}}%
\pgfpathlineto{\pgfqpoint{0.000000in}{-0.048611in}}%
\pgfusepath{stroke,fill}%
}%
\begin{pgfscope}%
\pgfsys@transformshift{2.068609in}{0.208222in}%
\pgfsys@useobject{currentmarker}{}%
\end{pgfscope}%
\end{pgfscope}%
\begin{pgfscope}%
\definecolor{textcolor}{rgb}{0.000000,0.000000,0.000000}%
\pgfsetstrokecolor{textcolor}%
\pgfsetfillcolor{textcolor}%
\pgftext[x=2.068609in,y=0.111000in,,top]{\color{textcolor}\rmfamily\fontsize{9.000000}{10.800000}\selectfont SSW}%
\end{pgfscope}%
\begin{pgfscope}%
\pgfpathrectangle{\pgfqpoint{0.492278in}{0.208222in}}{\pgfqpoint{2.725472in}{1.639848in}}%
\pgfusepath{clip}%
\pgfsetrectcap%
\pgfsetroundjoin%
\pgfsetlinewidth{0.803000pt}%
\definecolor{currentstroke}{rgb}{0.690196,0.690196,0.690196}%
\pgfsetstrokecolor{currentstroke}%
\pgfsetdash{}{0pt}%
\pgfpathmoveto{\pgfqpoint{2.495799in}{0.208222in}}%
\pgfpathlineto{\pgfqpoint{2.495799in}{1.848070in}}%
\pgfusepath{stroke}%
\end{pgfscope}%
\begin{pgfscope}%
\pgfsetbuttcap%
\pgfsetroundjoin%
\definecolor{currentfill}{rgb}{0.000000,0.000000,0.000000}%
\pgfsetfillcolor{currentfill}%
\pgfsetlinewidth{0.803000pt}%
\definecolor{currentstroke}{rgb}{0.000000,0.000000,0.000000}%
\pgfsetstrokecolor{currentstroke}%
\pgfsetdash{}{0pt}%
\pgfsys@defobject{currentmarker}{\pgfqpoint{0.000000in}{-0.048611in}}{\pgfqpoint{0.000000in}{0.000000in}}{%
\pgfpathmoveto{\pgfqpoint{0.000000in}{0.000000in}}%
\pgfpathlineto{\pgfqpoint{0.000000in}{-0.048611in}}%
\pgfusepath{stroke,fill}%
}%
\begin{pgfscope}%
\pgfsys@transformshift{2.495799in}{0.208222in}%
\pgfsys@useobject{currentmarker}{}%
\end{pgfscope}%
\end{pgfscope}%
\begin{pgfscope}%
\definecolor{textcolor}{rgb}{0.000000,0.000000,0.000000}%
\pgfsetstrokecolor{textcolor}%
\pgfsetfillcolor{textcolor}%
\pgftext[x=2.495799in,y=0.111000in,,top]{\color{textcolor}\rmfamily\fontsize{9.000000}{10.800000}\selectfont EPSSW}%
\end{pgfscope}%
\begin{pgfscope}%
\pgfpathrectangle{\pgfqpoint{0.492278in}{0.208222in}}{\pgfqpoint{2.725472in}{1.639848in}}%
\pgfusepath{clip}%
\pgfsetrectcap%
\pgfsetroundjoin%
\pgfsetlinewidth{0.803000pt}%
\definecolor{currentstroke}{rgb}{0.690196,0.690196,0.690196}%
\pgfsetstrokecolor{currentstroke}%
\pgfsetdash{}{0pt}%
\pgfpathmoveto{\pgfqpoint{2.922989in}{0.208222in}}%
\pgfpathlineto{\pgfqpoint{2.922989in}{1.848070in}}%
\pgfusepath{stroke}%
\end{pgfscope}%
\begin{pgfscope}%
\pgfsetbuttcap%
\pgfsetroundjoin%
\definecolor{currentfill}{rgb}{0.000000,0.000000,0.000000}%
\pgfsetfillcolor{currentfill}%
\pgfsetlinewidth{0.803000pt}%
\definecolor{currentstroke}{rgb}{0.000000,0.000000,0.000000}%
\pgfsetstrokecolor{currentstroke}%
\pgfsetdash{}{0pt}%
\pgfsys@defobject{currentmarker}{\pgfqpoint{0.000000in}{-0.048611in}}{\pgfqpoint{0.000000in}{0.000000in}}{%
\pgfpathmoveto{\pgfqpoint{0.000000in}{0.000000in}}%
\pgfpathlineto{\pgfqpoint{0.000000in}{-0.048611in}}%
\pgfusepath{stroke,fill}%
}%
\begin{pgfscope}%
\pgfsys@transformshift{2.922989in}{0.208222in}%
\pgfsys@useobject{currentmarker}{}%
\end{pgfscope}%
\end{pgfscope}%
\begin{pgfscope}%
\definecolor{textcolor}{rgb}{0.000000,0.000000,0.000000}%
\pgfsetstrokecolor{textcolor}%
\pgfsetfillcolor{textcolor}%
\pgftext[x=2.922989in,y=0.111000in,,top]{\color{textcolor}\rmfamily\fontsize{9.000000}{10.800000}\selectfont SPRE}%
\end{pgfscope}%
\begin{pgfscope}%
\pgfpathrectangle{\pgfqpoint{0.492278in}{0.208222in}}{\pgfqpoint{2.725472in}{1.639848in}}%
\pgfusepath{clip}%
\pgfsetrectcap%
\pgfsetroundjoin%
\pgfsetlinewidth{0.803000pt}%
\definecolor{currentstroke}{rgb}{0.690196,0.690196,0.690196}%
\pgfsetstrokecolor{currentstroke}%
\pgfsetdash{}{0pt}%
\pgfpathmoveto{\pgfqpoint{0.492278in}{0.208222in}}%
\pgfpathlineto{\pgfqpoint{3.217750in}{0.208222in}}%
\pgfusepath{stroke}%
\end{pgfscope}%
\begin{pgfscope}%
\pgfsetbuttcap%
\pgfsetroundjoin%
\definecolor{currentfill}{rgb}{0.000000,0.000000,0.000000}%
\pgfsetfillcolor{currentfill}%
\pgfsetlinewidth{0.803000pt}%
\definecolor{currentstroke}{rgb}{0.000000,0.000000,0.000000}%
\pgfsetstrokecolor{currentstroke}%
\pgfsetdash{}{0pt}%
\pgfsys@defobject{currentmarker}{\pgfqpoint{-0.048611in}{0.000000in}}{\pgfqpoint{0.000000in}{0.000000in}}{%
\pgfpathmoveto{\pgfqpoint{0.000000in}{0.000000in}}%
\pgfpathlineto{\pgfqpoint{-0.048611in}{0.000000in}}%
\pgfusepath{stroke,fill}%
}%
\begin{pgfscope}%
\pgfsys@transformshift{0.492278in}{0.208222in}%
\pgfsys@useobject{currentmarker}{}%
\end{pgfscope}%
\end{pgfscope}%
\begin{pgfscope}%
\definecolor{textcolor}{rgb}{0.000000,0.000000,0.000000}%
\pgfsetstrokecolor{textcolor}%
\pgfsetfillcolor{textcolor}%
\pgftext[x=0.230805in, y=0.164847in, left, base]{\color{textcolor}\rmfamily\fontsize{9.000000}{10.800000}\selectfont 0.1}%
\end{pgfscope}%
\begin{pgfscope}%
\pgfpathrectangle{\pgfqpoint{0.492278in}{0.208222in}}{\pgfqpoint{2.725472in}{1.639848in}}%
\pgfusepath{clip}%
\pgfsetrectcap%
\pgfsetroundjoin%
\pgfsetlinewidth{0.803000pt}%
\definecolor{currentstroke}{rgb}{0.690196,0.690196,0.690196}%
\pgfsetstrokecolor{currentstroke}%
\pgfsetdash{}{0pt}%
\pgfpathmoveto{\pgfqpoint{0.492278in}{0.920880in}}%
\pgfpathlineto{\pgfqpoint{3.217750in}{0.920880in}}%
\pgfusepath{stroke}%
\end{pgfscope}%
\begin{pgfscope}%
\pgfsetbuttcap%
\pgfsetroundjoin%
\definecolor{currentfill}{rgb}{0.000000,0.000000,0.000000}%
\pgfsetfillcolor{currentfill}%
\pgfsetlinewidth{0.803000pt}%
\definecolor{currentstroke}{rgb}{0.000000,0.000000,0.000000}%
\pgfsetstrokecolor{currentstroke}%
\pgfsetdash{}{0pt}%
\pgfsys@defobject{currentmarker}{\pgfqpoint{-0.048611in}{0.000000in}}{\pgfqpoint{0.000000in}{0.000000in}}{%
\pgfpathmoveto{\pgfqpoint{0.000000in}{0.000000in}}%
\pgfpathlineto{\pgfqpoint{-0.048611in}{0.000000in}}%
\pgfusepath{stroke,fill}%
}%
\begin{pgfscope}%
\pgfsys@transformshift{0.492278in}{0.920880in}%
\pgfsys@useobject{currentmarker}{}%
\end{pgfscope}%
\end{pgfscope}%
\begin{pgfscope}%
\definecolor{textcolor}{rgb}{0.000000,0.000000,0.000000}%
\pgfsetstrokecolor{textcolor}%
\pgfsetfillcolor{textcolor}%
\pgftext[x=0.230805in, y=0.877505in, left, base]{\color{textcolor}\rmfamily\fontsize{9.000000}{10.800000}\selectfont 1.0}%
\end{pgfscope}%
\begin{pgfscope}%
\pgfpathrectangle{\pgfqpoint{0.492278in}{0.208222in}}{\pgfqpoint{2.725472in}{1.639848in}}%
\pgfusepath{clip}%
\pgfsetrectcap%
\pgfsetroundjoin%
\pgfsetlinewidth{0.803000pt}%
\definecolor{currentstroke}{rgb}{0.690196,0.690196,0.690196}%
\pgfsetstrokecolor{currentstroke}%
\pgfsetdash{}{0pt}%
\pgfpathmoveto{\pgfqpoint{0.492278in}{1.633538in}}%
\pgfpathlineto{\pgfqpoint{3.217750in}{1.633538in}}%
\pgfusepath{stroke}%
\end{pgfscope}%
\begin{pgfscope}%
\pgfsetbuttcap%
\pgfsetroundjoin%
\definecolor{currentfill}{rgb}{0.000000,0.000000,0.000000}%
\pgfsetfillcolor{currentfill}%
\pgfsetlinewidth{0.803000pt}%
\definecolor{currentstroke}{rgb}{0.000000,0.000000,0.000000}%
\pgfsetstrokecolor{currentstroke}%
\pgfsetdash{}{0pt}%
\pgfsys@defobject{currentmarker}{\pgfqpoint{-0.048611in}{0.000000in}}{\pgfqpoint{0.000000in}{0.000000in}}{%
\pgfpathmoveto{\pgfqpoint{0.000000in}{0.000000in}}%
\pgfpathlineto{\pgfqpoint{-0.048611in}{0.000000in}}%
\pgfusepath{stroke,fill}%
}%
\begin{pgfscope}%
\pgfsys@transformshift{0.492278in}{1.633538in}%
\pgfsys@useobject{currentmarker}{}%
\end{pgfscope}%
\end{pgfscope}%
\begin{pgfscope}%
\definecolor{textcolor}{rgb}{0.000000,0.000000,0.000000}%
\pgfsetstrokecolor{textcolor}%
\pgfsetfillcolor{textcolor}%
\pgftext[x=0.166555in, y=1.590163in, left, base]{\color{textcolor}\rmfamily\fontsize{9.000000}{10.800000}\selectfont 10.0}%
\end{pgfscope}%
\begin{pgfscope}%
\pgfsetbuttcap%
\pgfsetroundjoin%
\definecolor{currentfill}{rgb}{0.000000,0.000000,0.000000}%
\pgfsetfillcolor{currentfill}%
\pgfsetlinewidth{0.602250pt}%
\definecolor{currentstroke}{rgb}{0.000000,0.000000,0.000000}%
\pgfsetstrokecolor{currentstroke}%
\pgfsetdash{}{0pt}%
\pgfsys@defobject{currentmarker}{\pgfqpoint{-0.027778in}{0.000000in}}{\pgfqpoint{0.000000in}{0.000000in}}{%
\pgfpathmoveto{\pgfqpoint{0.000000in}{0.000000in}}%
\pgfpathlineto{\pgfqpoint{-0.027778in}{0.000000in}}%
\pgfusepath{stroke,fill}%
}%
\begin{pgfscope}%
\pgfsys@transformshift{0.492278in}{0.422754in}%
\pgfsys@useobject{currentmarker}{}%
\end{pgfscope}%
\end{pgfscope}%
\begin{pgfscope}%
\pgfsetbuttcap%
\pgfsetroundjoin%
\definecolor{currentfill}{rgb}{0.000000,0.000000,0.000000}%
\pgfsetfillcolor{currentfill}%
\pgfsetlinewidth{0.602250pt}%
\definecolor{currentstroke}{rgb}{0.000000,0.000000,0.000000}%
\pgfsetstrokecolor{currentstroke}%
\pgfsetdash{}{0pt}%
\pgfsys@defobject{currentmarker}{\pgfqpoint{-0.027778in}{0.000000in}}{\pgfqpoint{0.000000in}{0.000000in}}{%
\pgfpathmoveto{\pgfqpoint{0.000000in}{0.000000in}}%
\pgfpathlineto{\pgfqpoint{-0.027778in}{0.000000in}}%
\pgfusepath{stroke,fill}%
}%
\begin{pgfscope}%
\pgfsys@transformshift{0.492278in}{0.548246in}%
\pgfsys@useobject{currentmarker}{}%
\end{pgfscope}%
\end{pgfscope}%
\begin{pgfscope}%
\pgfsetbuttcap%
\pgfsetroundjoin%
\definecolor{currentfill}{rgb}{0.000000,0.000000,0.000000}%
\pgfsetfillcolor{currentfill}%
\pgfsetlinewidth{0.602250pt}%
\definecolor{currentstroke}{rgb}{0.000000,0.000000,0.000000}%
\pgfsetstrokecolor{currentstroke}%
\pgfsetdash{}{0pt}%
\pgfsys@defobject{currentmarker}{\pgfqpoint{-0.027778in}{0.000000in}}{\pgfqpoint{0.000000in}{0.000000in}}{%
\pgfpathmoveto{\pgfqpoint{0.000000in}{0.000000in}}%
\pgfpathlineto{\pgfqpoint{-0.027778in}{0.000000in}}%
\pgfusepath{stroke,fill}%
}%
\begin{pgfscope}%
\pgfsys@transformshift{0.492278in}{0.637285in}%
\pgfsys@useobject{currentmarker}{}%
\end{pgfscope}%
\end{pgfscope}%
\begin{pgfscope}%
\pgfsetbuttcap%
\pgfsetroundjoin%
\definecolor{currentfill}{rgb}{0.000000,0.000000,0.000000}%
\pgfsetfillcolor{currentfill}%
\pgfsetlinewidth{0.602250pt}%
\definecolor{currentstroke}{rgb}{0.000000,0.000000,0.000000}%
\pgfsetstrokecolor{currentstroke}%
\pgfsetdash{}{0pt}%
\pgfsys@defobject{currentmarker}{\pgfqpoint{-0.027778in}{0.000000in}}{\pgfqpoint{0.000000in}{0.000000in}}{%
\pgfpathmoveto{\pgfqpoint{0.000000in}{0.000000in}}%
\pgfpathlineto{\pgfqpoint{-0.027778in}{0.000000in}}%
\pgfusepath{stroke,fill}%
}%
\begin{pgfscope}%
\pgfsys@transformshift{0.492278in}{0.706349in}%
\pgfsys@useobject{currentmarker}{}%
\end{pgfscope}%
\end{pgfscope}%
\begin{pgfscope}%
\pgfsetbuttcap%
\pgfsetroundjoin%
\definecolor{currentfill}{rgb}{0.000000,0.000000,0.000000}%
\pgfsetfillcolor{currentfill}%
\pgfsetlinewidth{0.602250pt}%
\definecolor{currentstroke}{rgb}{0.000000,0.000000,0.000000}%
\pgfsetstrokecolor{currentstroke}%
\pgfsetdash{}{0pt}%
\pgfsys@defobject{currentmarker}{\pgfqpoint{-0.027778in}{0.000000in}}{\pgfqpoint{0.000000in}{0.000000in}}{%
\pgfpathmoveto{\pgfqpoint{0.000000in}{0.000000in}}%
\pgfpathlineto{\pgfqpoint{-0.027778in}{0.000000in}}%
\pgfusepath{stroke,fill}%
}%
\begin{pgfscope}%
\pgfsys@transformshift{0.492278in}{0.762778in}%
\pgfsys@useobject{currentmarker}{}%
\end{pgfscope}%
\end{pgfscope}%
\begin{pgfscope}%
\pgfsetbuttcap%
\pgfsetroundjoin%
\definecolor{currentfill}{rgb}{0.000000,0.000000,0.000000}%
\pgfsetfillcolor{currentfill}%
\pgfsetlinewidth{0.602250pt}%
\definecolor{currentstroke}{rgb}{0.000000,0.000000,0.000000}%
\pgfsetstrokecolor{currentstroke}%
\pgfsetdash{}{0pt}%
\pgfsys@defobject{currentmarker}{\pgfqpoint{-0.027778in}{0.000000in}}{\pgfqpoint{0.000000in}{0.000000in}}{%
\pgfpathmoveto{\pgfqpoint{0.000000in}{0.000000in}}%
\pgfpathlineto{\pgfqpoint{-0.027778in}{0.000000in}}%
\pgfusepath{stroke,fill}%
}%
\begin{pgfscope}%
\pgfsys@transformshift{0.492278in}{0.810488in}%
\pgfsys@useobject{currentmarker}{}%
\end{pgfscope}%
\end{pgfscope}%
\begin{pgfscope}%
\pgfsetbuttcap%
\pgfsetroundjoin%
\definecolor{currentfill}{rgb}{0.000000,0.000000,0.000000}%
\pgfsetfillcolor{currentfill}%
\pgfsetlinewidth{0.602250pt}%
\definecolor{currentstroke}{rgb}{0.000000,0.000000,0.000000}%
\pgfsetstrokecolor{currentstroke}%
\pgfsetdash{}{0pt}%
\pgfsys@defobject{currentmarker}{\pgfqpoint{-0.027778in}{0.000000in}}{\pgfqpoint{0.000000in}{0.000000in}}{%
\pgfpathmoveto{\pgfqpoint{0.000000in}{0.000000in}}%
\pgfpathlineto{\pgfqpoint{-0.027778in}{0.000000in}}%
\pgfusepath{stroke,fill}%
}%
\begin{pgfscope}%
\pgfsys@transformshift{0.492278in}{0.851817in}%
\pgfsys@useobject{currentmarker}{}%
\end{pgfscope}%
\end{pgfscope}%
\begin{pgfscope}%
\pgfsetbuttcap%
\pgfsetroundjoin%
\definecolor{currentfill}{rgb}{0.000000,0.000000,0.000000}%
\pgfsetfillcolor{currentfill}%
\pgfsetlinewidth{0.602250pt}%
\definecolor{currentstroke}{rgb}{0.000000,0.000000,0.000000}%
\pgfsetstrokecolor{currentstroke}%
\pgfsetdash{}{0pt}%
\pgfsys@defobject{currentmarker}{\pgfqpoint{-0.027778in}{0.000000in}}{\pgfqpoint{0.000000in}{0.000000in}}{%
\pgfpathmoveto{\pgfqpoint{0.000000in}{0.000000in}}%
\pgfpathlineto{\pgfqpoint{-0.027778in}{0.000000in}}%
\pgfusepath{stroke,fill}%
}%
\begin{pgfscope}%
\pgfsys@transformshift{0.492278in}{0.888271in}%
\pgfsys@useobject{currentmarker}{}%
\end{pgfscope}%
\end{pgfscope}%
\begin{pgfscope}%
\pgfsetbuttcap%
\pgfsetroundjoin%
\definecolor{currentfill}{rgb}{0.000000,0.000000,0.000000}%
\pgfsetfillcolor{currentfill}%
\pgfsetlinewidth{0.602250pt}%
\definecolor{currentstroke}{rgb}{0.000000,0.000000,0.000000}%
\pgfsetstrokecolor{currentstroke}%
\pgfsetdash{}{0pt}%
\pgfsys@defobject{currentmarker}{\pgfqpoint{-0.027778in}{0.000000in}}{\pgfqpoint{0.000000in}{0.000000in}}{%
\pgfpathmoveto{\pgfqpoint{0.000000in}{0.000000in}}%
\pgfpathlineto{\pgfqpoint{-0.027778in}{0.000000in}}%
\pgfusepath{stroke,fill}%
}%
\begin{pgfscope}%
\pgfsys@transformshift{0.492278in}{1.135412in}%
\pgfsys@useobject{currentmarker}{}%
\end{pgfscope}%
\end{pgfscope}%
\begin{pgfscope}%
\pgfsetbuttcap%
\pgfsetroundjoin%
\definecolor{currentfill}{rgb}{0.000000,0.000000,0.000000}%
\pgfsetfillcolor{currentfill}%
\pgfsetlinewidth{0.602250pt}%
\definecolor{currentstroke}{rgb}{0.000000,0.000000,0.000000}%
\pgfsetstrokecolor{currentstroke}%
\pgfsetdash{}{0pt}%
\pgfsys@defobject{currentmarker}{\pgfqpoint{-0.027778in}{0.000000in}}{\pgfqpoint{0.000000in}{0.000000in}}{%
\pgfpathmoveto{\pgfqpoint{0.000000in}{0.000000in}}%
\pgfpathlineto{\pgfqpoint{-0.027778in}{0.000000in}}%
\pgfusepath{stroke,fill}%
}%
\begin{pgfscope}%
\pgfsys@transformshift{0.492278in}{1.260905in}%
\pgfsys@useobject{currentmarker}{}%
\end{pgfscope}%
\end{pgfscope}%
\begin{pgfscope}%
\pgfsetbuttcap%
\pgfsetroundjoin%
\definecolor{currentfill}{rgb}{0.000000,0.000000,0.000000}%
\pgfsetfillcolor{currentfill}%
\pgfsetlinewidth{0.602250pt}%
\definecolor{currentstroke}{rgb}{0.000000,0.000000,0.000000}%
\pgfsetstrokecolor{currentstroke}%
\pgfsetdash{}{0pt}%
\pgfsys@defobject{currentmarker}{\pgfqpoint{-0.027778in}{0.000000in}}{\pgfqpoint{0.000000in}{0.000000in}}{%
\pgfpathmoveto{\pgfqpoint{0.000000in}{0.000000in}}%
\pgfpathlineto{\pgfqpoint{-0.027778in}{0.000000in}}%
\pgfusepath{stroke,fill}%
}%
\begin{pgfscope}%
\pgfsys@transformshift{0.492278in}{1.349943in}%
\pgfsys@useobject{currentmarker}{}%
\end{pgfscope}%
\end{pgfscope}%
\begin{pgfscope}%
\pgfsetbuttcap%
\pgfsetroundjoin%
\definecolor{currentfill}{rgb}{0.000000,0.000000,0.000000}%
\pgfsetfillcolor{currentfill}%
\pgfsetlinewidth{0.602250pt}%
\definecolor{currentstroke}{rgb}{0.000000,0.000000,0.000000}%
\pgfsetstrokecolor{currentstroke}%
\pgfsetdash{}{0pt}%
\pgfsys@defobject{currentmarker}{\pgfqpoint{-0.027778in}{0.000000in}}{\pgfqpoint{0.000000in}{0.000000in}}{%
\pgfpathmoveto{\pgfqpoint{0.000000in}{0.000000in}}%
\pgfpathlineto{\pgfqpoint{-0.027778in}{0.000000in}}%
\pgfusepath{stroke,fill}%
}%
\begin{pgfscope}%
\pgfsys@transformshift{0.492278in}{1.419007in}%
\pgfsys@useobject{currentmarker}{}%
\end{pgfscope}%
\end{pgfscope}%
\begin{pgfscope}%
\pgfsetbuttcap%
\pgfsetroundjoin%
\definecolor{currentfill}{rgb}{0.000000,0.000000,0.000000}%
\pgfsetfillcolor{currentfill}%
\pgfsetlinewidth{0.602250pt}%
\definecolor{currentstroke}{rgb}{0.000000,0.000000,0.000000}%
\pgfsetstrokecolor{currentstroke}%
\pgfsetdash{}{0pt}%
\pgfsys@defobject{currentmarker}{\pgfqpoint{-0.027778in}{0.000000in}}{\pgfqpoint{0.000000in}{0.000000in}}{%
\pgfpathmoveto{\pgfqpoint{0.000000in}{0.000000in}}%
\pgfpathlineto{\pgfqpoint{-0.027778in}{0.000000in}}%
\pgfusepath{stroke,fill}%
}%
\begin{pgfscope}%
\pgfsys@transformshift{0.492278in}{1.475436in}%
\pgfsys@useobject{currentmarker}{}%
\end{pgfscope}%
\end{pgfscope}%
\begin{pgfscope}%
\pgfsetbuttcap%
\pgfsetroundjoin%
\definecolor{currentfill}{rgb}{0.000000,0.000000,0.000000}%
\pgfsetfillcolor{currentfill}%
\pgfsetlinewidth{0.602250pt}%
\definecolor{currentstroke}{rgb}{0.000000,0.000000,0.000000}%
\pgfsetstrokecolor{currentstroke}%
\pgfsetdash{}{0pt}%
\pgfsys@defobject{currentmarker}{\pgfqpoint{-0.027778in}{0.000000in}}{\pgfqpoint{0.000000in}{0.000000in}}{%
\pgfpathmoveto{\pgfqpoint{0.000000in}{0.000000in}}%
\pgfpathlineto{\pgfqpoint{-0.027778in}{0.000000in}}%
\pgfusepath{stroke,fill}%
}%
\begin{pgfscope}%
\pgfsys@transformshift{0.492278in}{1.523146in}%
\pgfsys@useobject{currentmarker}{}%
\end{pgfscope}%
\end{pgfscope}%
\begin{pgfscope}%
\pgfsetbuttcap%
\pgfsetroundjoin%
\definecolor{currentfill}{rgb}{0.000000,0.000000,0.000000}%
\pgfsetfillcolor{currentfill}%
\pgfsetlinewidth{0.602250pt}%
\definecolor{currentstroke}{rgb}{0.000000,0.000000,0.000000}%
\pgfsetstrokecolor{currentstroke}%
\pgfsetdash{}{0pt}%
\pgfsys@defobject{currentmarker}{\pgfqpoint{-0.027778in}{0.000000in}}{\pgfqpoint{0.000000in}{0.000000in}}{%
\pgfpathmoveto{\pgfqpoint{0.000000in}{0.000000in}}%
\pgfpathlineto{\pgfqpoint{-0.027778in}{0.000000in}}%
\pgfusepath{stroke,fill}%
}%
\begin{pgfscope}%
\pgfsys@transformshift{0.492278in}{1.564475in}%
\pgfsys@useobject{currentmarker}{}%
\end{pgfscope}%
\end{pgfscope}%
\begin{pgfscope}%
\pgfsetbuttcap%
\pgfsetroundjoin%
\definecolor{currentfill}{rgb}{0.000000,0.000000,0.000000}%
\pgfsetfillcolor{currentfill}%
\pgfsetlinewidth{0.602250pt}%
\definecolor{currentstroke}{rgb}{0.000000,0.000000,0.000000}%
\pgfsetstrokecolor{currentstroke}%
\pgfsetdash{}{0pt}%
\pgfsys@defobject{currentmarker}{\pgfqpoint{-0.027778in}{0.000000in}}{\pgfqpoint{0.000000in}{0.000000in}}{%
\pgfpathmoveto{\pgfqpoint{0.000000in}{0.000000in}}%
\pgfpathlineto{\pgfqpoint{-0.027778in}{0.000000in}}%
\pgfusepath{stroke,fill}%
}%
\begin{pgfscope}%
\pgfsys@transformshift{0.492278in}{1.600929in}%
\pgfsys@useobject{currentmarker}{}%
\end{pgfscope}%
\end{pgfscope}%
\begin{pgfscope}%
\pgfsetbuttcap%
\pgfsetroundjoin%
\definecolor{currentfill}{rgb}{0.000000,0.000000,0.000000}%
\pgfsetfillcolor{currentfill}%
\pgfsetlinewidth{0.602250pt}%
\definecolor{currentstroke}{rgb}{0.000000,0.000000,0.000000}%
\pgfsetstrokecolor{currentstroke}%
\pgfsetdash{}{0pt}%
\pgfsys@defobject{currentmarker}{\pgfqpoint{-0.027778in}{0.000000in}}{\pgfqpoint{0.000000in}{0.000000in}}{%
\pgfpathmoveto{\pgfqpoint{0.000000in}{0.000000in}}%
\pgfpathlineto{\pgfqpoint{-0.027778in}{0.000000in}}%
\pgfusepath{stroke,fill}%
}%
\begin{pgfscope}%
\pgfsys@transformshift{0.492278in}{1.848070in}%
\pgfsys@useobject{currentmarker}{}%
\end{pgfscope}%
\end{pgfscope}%
\begin{pgfscope}%
\definecolor{textcolor}{rgb}{0.000000,0.000000,0.000000}%
\pgfsetstrokecolor{textcolor}%
\pgfsetfillcolor{textcolor}%
\pgftext[x=0.111000in,y=1.028146in,,bottom,rotate=90.000000]{\color{textcolor}\rmfamily\fontsize{9.000000}{10.800000}\selectfont Seconds}%
\end{pgfscope}%
\begin{pgfscope}%
\pgfsetrectcap%
\pgfsetmiterjoin%
\pgfsetlinewidth{0.803000pt}%
\definecolor{currentstroke}{rgb}{0.000000,0.000000,0.000000}%
\pgfsetstrokecolor{currentstroke}%
\pgfsetdash{}{0pt}%
\pgfpathmoveto{\pgfqpoint{0.492278in}{0.208222in}}%
\pgfpathlineto{\pgfqpoint{0.492278in}{1.848070in}}%
\pgfusepath{stroke}%
\end{pgfscope}%
\begin{pgfscope}%
\pgfsetrectcap%
\pgfsetmiterjoin%
\pgfsetlinewidth{0.803000pt}%
\definecolor{currentstroke}{rgb}{0.000000,0.000000,0.000000}%
\pgfsetstrokecolor{currentstroke}%
\pgfsetdash{}{0pt}%
\pgfpathmoveto{\pgfqpoint{3.217750in}{0.208222in}}%
\pgfpathlineto{\pgfqpoint{3.217750in}{1.848070in}}%
\pgfusepath{stroke}%
\end{pgfscope}%
\begin{pgfscope}%
\pgfsetrectcap%
\pgfsetmiterjoin%
\pgfsetlinewidth{0.803000pt}%
\definecolor{currentstroke}{rgb}{0.000000,0.000000,0.000000}%
\pgfsetstrokecolor{currentstroke}%
\pgfsetdash{}{0pt}%
\pgfpathmoveto{\pgfqpoint{0.492278in}{0.208222in}}%
\pgfpathlineto{\pgfqpoint{3.217750in}{0.208222in}}%
\pgfusepath{stroke}%
\end{pgfscope}%
\begin{pgfscope}%
\pgfsetrectcap%
\pgfsetmiterjoin%
\pgfsetlinewidth{0.803000pt}%
\definecolor{currentstroke}{rgb}{0.000000,0.000000,0.000000}%
\pgfsetstrokecolor{currentstroke}%
\pgfsetdash{}{0pt}%
\pgfpathmoveto{\pgfqpoint{0.492278in}{1.848070in}}%
\pgfpathlineto{\pgfqpoint{3.217750in}{1.848070in}}%
\pgfusepath{stroke}%
\end{pgfscope}%
\end{pgfpicture}%
\makeatother%
\endgroup%

%% file: figures/real_filters.pgf
%% Creator: Matplotlib, PGF backend
%%
%% To include the figure in your LaTeX document, write
%%   \input{<filename>.pgf}
%%
%% Make sure the required packages are loaded in your preamble
%%   \usepackage{pgf}
%%
%% and, on pdftex
%%   \usepackage[utf8]{inputenc}\DeclareUnicodeCharacter{2212}{-}
%%
%% or, on luatex and xetex
%%   \usepackage{unicode-math}
%%
%% Figures using additional raster images can only be included by \input if
%% they are in the same directory as the main LaTeX file. For loading figures
%% from other directories you can use the `import` package
%%   \usepackage{import}
%%
%% and then include the figures with
%%   \import{<path to file>}{<filename>.pgf}
%%
%% Matplotlib used the following preamble
%%   \usepackage{fontspec}
%%
\begingroup%
\makeatletter%
\begin{pgfpicture}%
\pgfpathrectangle{\pgfpointorigin}{\pgfqpoint{3.240125in}{1.813325in}}%
\pgfusepath{use as bounding box, clip}%
\begin{pgfscope}%
\pgfsetbuttcap%
\pgfsetmiterjoin%
\definecolor{currentfill}{rgb}{1.000000,1.000000,1.000000}%
\pgfsetfillcolor{currentfill}%
\pgfsetlinewidth{0.000000pt}%
\definecolor{currentstroke}{rgb}{1.000000,1.000000,1.000000}%
\pgfsetstrokecolor{currentstroke}%
\pgfsetdash{}{0pt}%
\pgfpathmoveto{\pgfqpoint{0.000000in}{0.000000in}}%
\pgfpathlineto{\pgfqpoint{3.240125in}{0.000000in}}%
\pgfpathlineto{\pgfqpoint{3.240125in}{1.813325in}}%
\pgfpathlineto{\pgfqpoint{0.000000in}{1.813325in}}%
\pgfpathclose%
\pgfusepath{fill}%
\end{pgfscope}%
\begin{pgfscope}%
\pgfsetbuttcap%
\pgfsetmiterjoin%
\definecolor{currentfill}{rgb}{1.000000,1.000000,1.000000}%
\pgfsetfillcolor{currentfill}%
\pgfsetlinewidth{0.000000pt}%
\definecolor{currentstroke}{rgb}{0.000000,0.000000,0.000000}%
\pgfsetstrokecolor{currentstroke}%
\pgfsetstrokeopacity{0.000000}%
\pgfsetdash{}{0pt}%
\pgfpathmoveto{\pgfqpoint{0.440778in}{0.386152in}}%
\pgfpathlineto{\pgfqpoint{3.133750in}{0.386152in}}%
\pgfpathlineto{\pgfqpoint{3.133750in}{1.452797in}}%
\pgfpathlineto{\pgfqpoint{0.440778in}{1.452797in}}%
\pgfpathclose%
\pgfusepath{fill}%
\end{pgfscope}%
\begin{pgfscope}%
\pgfsetbuttcap%
\pgfsetroundjoin%
\definecolor{currentfill}{rgb}{0.000000,0.000000,0.000000}%
\pgfsetfillcolor{currentfill}%
\pgfsetlinewidth{0.803000pt}%
\definecolor{currentstroke}{rgb}{0.000000,0.000000,0.000000}%
\pgfsetstrokecolor{currentstroke}%
\pgfsetdash{}{0pt}%
\pgfsys@defobject{currentmarker}{\pgfqpoint{0.000000in}{-0.048611in}}{\pgfqpoint{0.000000in}{0.000000in}}{%
\pgfpathmoveto{\pgfqpoint{0.000000in}{0.000000in}}%
\pgfpathlineto{\pgfqpoint{0.000000in}{-0.048611in}}%
\pgfusepath{stroke,fill}%
}%
\begin{pgfscope}%
\pgfsys@transformshift{0.440778in}{0.386152in}%
\pgfsys@useobject{currentmarker}{}%
\end{pgfscope}%
\end{pgfscope}%
\begin{pgfscope}%
\definecolor{textcolor}{rgb}{0.000000,0.000000,0.000000}%
\pgfsetstrokecolor{textcolor}%
\pgfsetfillcolor{textcolor}%
\pgftext[x=0.440778in,y=0.288930in,,top]{\color{textcolor}\rmfamily\fontsize{9.000000}{10.800000}\selectfont 1}%
\end{pgfscope}%
\begin{pgfscope}%
\pgfsetbuttcap%
\pgfsetroundjoin%
\definecolor{currentfill}{rgb}{0.000000,0.000000,0.000000}%
\pgfsetfillcolor{currentfill}%
\pgfsetlinewidth{0.803000pt}%
\definecolor{currentstroke}{rgb}{0.000000,0.000000,0.000000}%
\pgfsetstrokecolor{currentstroke}%
\pgfsetdash{}{0pt}%
\pgfsys@defobject{currentmarker}{\pgfqpoint{0.000000in}{-0.048611in}}{\pgfqpoint{0.000000in}{0.000000in}}{%
\pgfpathmoveto{\pgfqpoint{0.000000in}{0.000000in}}%
\pgfpathlineto{\pgfqpoint{0.000000in}{-0.048611in}}%
\pgfusepath{stroke,fill}%
}%
\begin{pgfscope}%
\pgfsys@transformshift{0.945710in}{0.386152in}%
\pgfsys@useobject{currentmarker}{}%
\end{pgfscope}%
\end{pgfscope}%
\begin{pgfscope}%
\definecolor{textcolor}{rgb}{0.000000,0.000000,0.000000}%
\pgfsetstrokecolor{textcolor}%
\pgfsetfillcolor{textcolor}%
\pgftext[x=0.945710in,y=0.288930in,,top]{\color{textcolor}\rmfamily\fontsize{9.000000}{10.800000}\selectfont 10}%
\end{pgfscope}%
\begin{pgfscope}%
\pgfsetbuttcap%
\pgfsetroundjoin%
\definecolor{currentfill}{rgb}{0.000000,0.000000,0.000000}%
\pgfsetfillcolor{currentfill}%
\pgfsetlinewidth{0.803000pt}%
\definecolor{currentstroke}{rgb}{0.000000,0.000000,0.000000}%
\pgfsetstrokecolor{currentstroke}%
\pgfsetdash{}{0pt}%
\pgfsys@defobject{currentmarker}{\pgfqpoint{0.000000in}{-0.048611in}}{\pgfqpoint{0.000000in}{0.000000in}}{%
\pgfpathmoveto{\pgfqpoint{0.000000in}{0.000000in}}%
\pgfpathlineto{\pgfqpoint{0.000000in}{-0.048611in}}%
\pgfusepath{stroke,fill}%
}%
\begin{pgfscope}%
\pgfsys@transformshift{1.506746in}{0.386152in}%
\pgfsys@useobject{currentmarker}{}%
\end{pgfscope}%
\end{pgfscope}%
\begin{pgfscope}%
\definecolor{textcolor}{rgb}{0.000000,0.000000,0.000000}%
\pgfsetstrokecolor{textcolor}%
\pgfsetfillcolor{textcolor}%
\pgftext[x=1.506746in,y=0.288930in,,top]{\color{textcolor}\rmfamily\fontsize{9.000000}{10.800000}\selectfont 20}%
\end{pgfscope}%
\begin{pgfscope}%
\pgfsetbuttcap%
\pgfsetroundjoin%
\definecolor{currentfill}{rgb}{0.000000,0.000000,0.000000}%
\pgfsetfillcolor{currentfill}%
\pgfsetlinewidth{0.803000pt}%
\definecolor{currentstroke}{rgb}{0.000000,0.000000,0.000000}%
\pgfsetstrokecolor{currentstroke}%
\pgfsetdash{}{0pt}%
\pgfsys@defobject{currentmarker}{\pgfqpoint{0.000000in}{-0.048611in}}{\pgfqpoint{0.000000in}{0.000000in}}{%
\pgfpathmoveto{\pgfqpoint{0.000000in}{0.000000in}}%
\pgfpathlineto{\pgfqpoint{0.000000in}{-0.048611in}}%
\pgfusepath{stroke,fill}%
}%
\begin{pgfscope}%
\pgfsys@transformshift{2.011678in}{0.386152in}%
\pgfsys@useobject{currentmarker}{}%
\end{pgfscope}%
\end{pgfscope}%
\begin{pgfscope}%
\definecolor{textcolor}{rgb}{0.000000,0.000000,0.000000}%
\pgfsetstrokecolor{textcolor}%
\pgfsetfillcolor{textcolor}%
\pgftext[x=2.011678in,y=0.288930in,,top]{\color{textcolor}\rmfamily\fontsize{9.000000}{10.800000}\selectfont 29}%
\end{pgfscope}%
\begin{pgfscope}%
\pgfsetbuttcap%
\pgfsetroundjoin%
\definecolor{currentfill}{rgb}{0.000000,0.000000,0.000000}%
\pgfsetfillcolor{currentfill}%
\pgfsetlinewidth{0.803000pt}%
\definecolor{currentstroke}{rgb}{0.000000,0.000000,0.000000}%
\pgfsetstrokecolor{currentstroke}%
\pgfsetdash{}{0pt}%
\pgfsys@defobject{currentmarker}{\pgfqpoint{0.000000in}{-0.048611in}}{\pgfqpoint{0.000000in}{0.000000in}}{%
\pgfpathmoveto{\pgfqpoint{0.000000in}{0.000000in}}%
\pgfpathlineto{\pgfqpoint{0.000000in}{-0.048611in}}%
\pgfusepath{stroke,fill}%
}%
\begin{pgfscope}%
\pgfsys@transformshift{2.572714in}{0.386152in}%
\pgfsys@useobject{currentmarker}{}%
\end{pgfscope}%
\end{pgfscope}%
\begin{pgfscope}%
\definecolor{textcolor}{rgb}{0.000000,0.000000,0.000000}%
\pgfsetstrokecolor{textcolor}%
\pgfsetfillcolor{textcolor}%
\pgftext[x=2.572714in,y=0.288930in,,top]{\color{textcolor}\rmfamily\fontsize{9.000000}{10.800000}\selectfont 39}%
\end{pgfscope}%
\begin{pgfscope}%
\pgfsetbuttcap%
\pgfsetroundjoin%
\definecolor{currentfill}{rgb}{0.000000,0.000000,0.000000}%
\pgfsetfillcolor{currentfill}%
\pgfsetlinewidth{0.803000pt}%
\definecolor{currentstroke}{rgb}{0.000000,0.000000,0.000000}%
\pgfsetstrokecolor{currentstroke}%
\pgfsetdash{}{0pt}%
\pgfsys@defobject{currentmarker}{\pgfqpoint{0.000000in}{-0.048611in}}{\pgfqpoint{0.000000in}{0.000000in}}{%
\pgfpathmoveto{\pgfqpoint{0.000000in}{0.000000in}}%
\pgfpathlineto{\pgfqpoint{0.000000in}{-0.048611in}}%
\pgfusepath{stroke,fill}%
}%
\begin{pgfscope}%
\pgfsys@transformshift{3.133750in}{0.386152in}%
\pgfsys@useobject{currentmarker}{}%
\end{pgfscope}%
\end{pgfscope}%
\begin{pgfscope}%
\definecolor{textcolor}{rgb}{0.000000,0.000000,0.000000}%
\pgfsetstrokecolor{textcolor}%
\pgfsetfillcolor{textcolor}%
\pgftext[x=3.133750in,y=0.288930in,,top]{\color{textcolor}\rmfamily\fontsize{9.000000}{10.800000}\selectfont 49}%
\end{pgfscope}%
\begin{pgfscope}%
\definecolor{textcolor}{rgb}{0.000000,0.000000,0.000000}%
\pgfsetstrokecolor{textcolor}%
\pgfsetfillcolor{textcolor}%
\pgftext[x=1.787264in,y=0.122375in,,top]{\color{textcolor}\rmfamily\fontsize{9.000000}{10.800000}\selectfont Hyperspectral channel}%
\end{pgfscope}%
\begin{pgfscope}%
\pgfsetbuttcap%
\pgfsetroundjoin%
\definecolor{currentfill}{rgb}{0.000000,0.000000,0.000000}%
\pgfsetfillcolor{currentfill}%
\pgfsetlinewidth{0.803000pt}%
\definecolor{currentstroke}{rgb}{0.000000,0.000000,0.000000}%
\pgfsetstrokecolor{currentstroke}%
\pgfsetdash{}{0pt}%
\pgfsys@defobject{currentmarker}{\pgfqpoint{-0.048611in}{0.000000in}}{\pgfqpoint{0.000000in}{0.000000in}}{%
\pgfpathmoveto{\pgfqpoint{0.000000in}{0.000000in}}%
\pgfpathlineto{\pgfqpoint{-0.048611in}{0.000000in}}%
\pgfusepath{stroke,fill}%
}%
\begin{pgfscope}%
\pgfsys@transformshift{0.440778in}{0.386152in}%
\pgfsys@useobject{currentmarker}{}%
\end{pgfscope}%
\end{pgfscope}%
\begin{pgfscope}%
\definecolor{textcolor}{rgb}{0.000000,0.000000,0.000000}%
\pgfsetstrokecolor{textcolor}%
\pgfsetfillcolor{textcolor}%
\pgftext[x=0.179305in, y=0.342777in, left, base]{\color{textcolor}\rmfamily\fontsize{9.000000}{10.800000}\selectfont 0.0}%
\end{pgfscope}%
\begin{pgfscope}%
\pgfsetbuttcap%
\pgfsetroundjoin%
\definecolor{currentfill}{rgb}{0.000000,0.000000,0.000000}%
\pgfsetfillcolor{currentfill}%
\pgfsetlinewidth{0.803000pt}%
\definecolor{currentstroke}{rgb}{0.000000,0.000000,0.000000}%
\pgfsetstrokecolor{currentstroke}%
\pgfsetdash{}{0pt}%
\pgfsys@defobject{currentmarker}{\pgfqpoint{-0.048611in}{0.000000in}}{\pgfqpoint{0.000000in}{0.000000in}}{%
\pgfpathmoveto{\pgfqpoint{0.000000in}{0.000000in}}%
\pgfpathlineto{\pgfqpoint{-0.048611in}{0.000000in}}%
\pgfusepath{stroke,fill}%
}%
\begin{pgfscope}%
\pgfsys@transformshift{0.440778in}{0.909017in}%
\pgfsys@useobject{currentmarker}{}%
\end{pgfscope}%
\end{pgfscope}%
\begin{pgfscope}%
\definecolor{textcolor}{rgb}{0.000000,0.000000,0.000000}%
\pgfsetstrokecolor{textcolor}%
\pgfsetfillcolor{textcolor}%
\pgftext[x=0.179305in, y=0.865642in, left, base]{\color{textcolor}\rmfamily\fontsize{9.000000}{10.800000}\selectfont 0.5}%
\end{pgfscope}%
\begin{pgfscope}%
\pgfsetbuttcap%
\pgfsetroundjoin%
\definecolor{currentfill}{rgb}{0.000000,0.000000,0.000000}%
\pgfsetfillcolor{currentfill}%
\pgfsetlinewidth{0.803000pt}%
\definecolor{currentstroke}{rgb}{0.000000,0.000000,0.000000}%
\pgfsetstrokecolor{currentstroke}%
\pgfsetdash{}{0pt}%
\pgfsys@defobject{currentmarker}{\pgfqpoint{-0.048611in}{0.000000in}}{\pgfqpoint{0.000000in}{0.000000in}}{%
\pgfpathmoveto{\pgfqpoint{0.000000in}{0.000000in}}%
\pgfpathlineto{\pgfqpoint{-0.048611in}{0.000000in}}%
\pgfusepath{stroke,fill}%
}%
\begin{pgfscope}%
\pgfsys@transformshift{0.440778in}{1.431882in}%
\pgfsys@useobject{currentmarker}{}%
\end{pgfscope}%
\end{pgfscope}%
\begin{pgfscope}%
\definecolor{textcolor}{rgb}{0.000000,0.000000,0.000000}%
\pgfsetstrokecolor{textcolor}%
\pgfsetfillcolor{textcolor}%
\pgftext[x=0.179305in, y=1.388507in, left, base]{\color{textcolor}\rmfamily\fontsize{9.000000}{10.800000}\selectfont 1.0}%
\end{pgfscope}%
\begin{pgfscope}%
\definecolor{textcolor}{rgb}{0.000000,0.000000,0.000000}%
\pgfsetstrokecolor{textcolor}%
\pgfsetfillcolor{textcolor}%
\pgftext[x=0.123750in,y=0.919475in,,bottom,rotate=90.000000]{\color{textcolor}\rmfamily\fontsize{9.000000}{10.800000}\selectfont Spectral sensitivity}%
\end{pgfscope}%
\begin{pgfscope}%
\pgfpathrectangle{\pgfqpoint{0.440778in}{0.386152in}}{\pgfqpoint{2.692972in}{1.066644in}}%
\pgfusepath{clip}%
\pgfsetrectcap%
\pgfsetroundjoin%
\pgfsetlinewidth{1.505625pt}%
\definecolor{currentstroke}{rgb}{0.000000,0.816000,1.000000}%
\pgfsetstrokecolor{currentstroke}%
\pgfsetdash{}{0pt}%
\pgfpathmoveto{\pgfqpoint{0.440778in}{0.386152in}}%
\pgfpathlineto{\pgfqpoint{0.496881in}{0.386152in}}%
\pgfpathlineto{\pgfqpoint{0.552985in}{0.386152in}}%
\pgfpathlineto{\pgfqpoint{0.609088in}{0.386152in}}%
\pgfpathlineto{\pgfqpoint{0.665192in}{0.386152in}}%
\pgfpathlineto{\pgfqpoint{0.721296in}{1.027533in}}%
\pgfpathlineto{\pgfqpoint{0.777399in}{1.348224in}}%
\pgfpathlineto{\pgfqpoint{0.833503in}{1.348224in}}%
\pgfpathlineto{\pgfqpoint{0.889606in}{1.348224in}}%
\pgfpathlineto{\pgfqpoint{0.945710in}{1.348224in}}%
\pgfpathlineto{\pgfqpoint{1.001814in}{1.348224in}}%
\pgfpathlineto{\pgfqpoint{1.057917in}{1.348224in}}%
\pgfpathlineto{\pgfqpoint{1.114021in}{1.348224in}}%
\pgfpathlineto{\pgfqpoint{1.170124in}{1.348224in}}%
\pgfpathlineto{\pgfqpoint{1.226228in}{1.348224in}}%
\pgfpathlineto{\pgfqpoint{1.282331in}{0.706843in}}%
\pgfpathlineto{\pgfqpoint{1.338435in}{0.386152in}}%
\pgfpathlineto{\pgfqpoint{1.394539in}{0.386152in}}%
\pgfpathlineto{\pgfqpoint{1.450642in}{0.386152in}}%
\pgfpathlineto{\pgfqpoint{1.506746in}{0.386152in}}%
\pgfpathlineto{\pgfqpoint{1.562849in}{0.386152in}}%
\pgfpathlineto{\pgfqpoint{1.618953in}{0.386152in}}%
\pgfpathlineto{\pgfqpoint{1.675057in}{0.386152in}}%
\pgfpathlineto{\pgfqpoint{1.731160in}{0.386152in}}%
\pgfpathlineto{\pgfqpoint{1.787264in}{0.386152in}}%
\pgfpathlineto{\pgfqpoint{1.843367in}{0.386152in}}%
\pgfpathlineto{\pgfqpoint{1.899471in}{0.386152in}}%
\pgfpathlineto{\pgfqpoint{1.955575in}{0.386152in}}%
\pgfpathlineto{\pgfqpoint{2.011678in}{0.386152in}}%
\pgfpathlineto{\pgfqpoint{2.067782in}{0.386152in}}%
\pgfpathlineto{\pgfqpoint{2.123885in}{0.386152in}}%
\pgfpathlineto{\pgfqpoint{2.179989in}{0.386152in}}%
\pgfpathlineto{\pgfqpoint{2.236092in}{0.386152in}}%
\pgfpathlineto{\pgfqpoint{2.292196in}{0.386152in}}%
\pgfpathlineto{\pgfqpoint{2.348300in}{0.386152in}}%
\pgfpathlineto{\pgfqpoint{2.404403in}{0.386152in}}%
\pgfpathlineto{\pgfqpoint{2.460507in}{0.386152in}}%
\pgfpathlineto{\pgfqpoint{2.516610in}{0.386152in}}%
\pgfpathlineto{\pgfqpoint{2.572714in}{0.386152in}}%
\pgfpathlineto{\pgfqpoint{2.628818in}{0.386152in}}%
\pgfpathlineto{\pgfqpoint{2.684921in}{0.386152in}}%
\pgfpathlineto{\pgfqpoint{2.741025in}{0.386152in}}%
\pgfpathlineto{\pgfqpoint{2.797128in}{0.386152in}}%
\pgfpathlineto{\pgfqpoint{2.853232in}{0.386152in}}%
\pgfpathlineto{\pgfqpoint{2.909336in}{0.386152in}}%
\pgfpathlineto{\pgfqpoint{2.965439in}{0.386152in}}%
\pgfpathlineto{\pgfqpoint{3.021543in}{0.386152in}}%
\pgfpathlineto{\pgfqpoint{3.077646in}{0.386152in}}%
\pgfpathlineto{\pgfqpoint{3.133750in}{0.386152in}}%
\pgfusepath{stroke}%
\end{pgfscope}%
\begin{pgfscope}%
\pgfpathrectangle{\pgfqpoint{0.440778in}{0.386152in}}{\pgfqpoint{2.692972in}{1.066644in}}%
\pgfusepath{clip}%
\pgfsetrectcap%
\pgfsetroundjoin%
\pgfsetlinewidth{1.505625pt}%
\definecolor{currentstroke}{rgb}{0.000000,1.000000,0.880000}%
\pgfsetstrokecolor{currentstroke}%
\pgfsetdash{}{0pt}%
\pgfpathmoveto{\pgfqpoint{0.440778in}{0.386152in}}%
\pgfpathlineto{\pgfqpoint{0.496881in}{0.386152in}}%
\pgfpathlineto{\pgfqpoint{0.552985in}{0.386152in}}%
\pgfpathlineto{\pgfqpoint{0.609088in}{0.386152in}}%
\pgfpathlineto{\pgfqpoint{0.665192in}{0.386152in}}%
\pgfpathlineto{\pgfqpoint{0.721296in}{0.727758in}}%
\pgfpathlineto{\pgfqpoint{0.777399in}{1.410968in}}%
\pgfpathlineto{\pgfqpoint{0.833503in}{1.410968in}}%
\pgfpathlineto{\pgfqpoint{0.889606in}{1.410968in}}%
\pgfpathlineto{\pgfqpoint{0.945710in}{1.410968in}}%
\pgfpathlineto{\pgfqpoint{1.001814in}{1.410968in}}%
\pgfpathlineto{\pgfqpoint{1.057917in}{1.410968in}}%
\pgfpathlineto{\pgfqpoint{1.114021in}{1.410968in}}%
\pgfpathlineto{\pgfqpoint{1.170124in}{1.410968in}}%
\pgfpathlineto{\pgfqpoint{1.226228in}{1.410968in}}%
\pgfpathlineto{\pgfqpoint{1.282331in}{1.410968in}}%
\pgfpathlineto{\pgfqpoint{1.338435in}{1.410968in}}%
\pgfpathlineto{\pgfqpoint{1.394539in}{1.410968in}}%
\pgfpathlineto{\pgfqpoint{1.450642in}{1.410968in}}%
\pgfpathlineto{\pgfqpoint{1.506746in}{1.410968in}}%
\pgfpathlineto{\pgfqpoint{1.562849in}{1.410968in}}%
\pgfpathlineto{\pgfqpoint{1.618953in}{0.841626in}}%
\pgfpathlineto{\pgfqpoint{1.675057in}{0.386152in}}%
\pgfpathlineto{\pgfqpoint{1.731160in}{0.386152in}}%
\pgfpathlineto{\pgfqpoint{1.787264in}{0.386152in}}%
\pgfpathlineto{\pgfqpoint{1.843367in}{0.386152in}}%
\pgfpathlineto{\pgfqpoint{1.899471in}{0.386152in}}%
\pgfpathlineto{\pgfqpoint{1.955575in}{0.386152in}}%
\pgfpathlineto{\pgfqpoint{2.011678in}{0.386152in}}%
\pgfpathlineto{\pgfqpoint{2.067782in}{0.386152in}}%
\pgfpathlineto{\pgfqpoint{2.123885in}{0.386152in}}%
\pgfpathlineto{\pgfqpoint{2.179989in}{0.386152in}}%
\pgfpathlineto{\pgfqpoint{2.236092in}{0.386152in}}%
\pgfpathlineto{\pgfqpoint{2.292196in}{0.386152in}}%
\pgfpathlineto{\pgfqpoint{2.348300in}{0.386152in}}%
\pgfpathlineto{\pgfqpoint{2.404403in}{0.386152in}}%
\pgfpathlineto{\pgfqpoint{2.460507in}{0.386152in}}%
\pgfpathlineto{\pgfqpoint{2.516610in}{0.386152in}}%
\pgfpathlineto{\pgfqpoint{2.572714in}{0.386152in}}%
\pgfpathlineto{\pgfqpoint{2.628818in}{0.386152in}}%
\pgfpathlineto{\pgfqpoint{2.684921in}{0.386152in}}%
\pgfpathlineto{\pgfqpoint{2.741025in}{0.386152in}}%
\pgfpathlineto{\pgfqpoint{2.797128in}{0.386152in}}%
\pgfpathlineto{\pgfqpoint{2.853232in}{0.386152in}}%
\pgfpathlineto{\pgfqpoint{2.909336in}{0.386152in}}%
\pgfpathlineto{\pgfqpoint{2.965439in}{0.386152in}}%
\pgfpathlineto{\pgfqpoint{3.021543in}{0.386152in}}%
\pgfpathlineto{\pgfqpoint{3.077646in}{0.386152in}}%
\pgfpathlineto{\pgfqpoint{3.133750in}{0.386152in}}%
\pgfusepath{stroke}%
\end{pgfscope}%
\begin{pgfscope}%
\pgfpathrectangle{\pgfqpoint{0.440778in}{0.386152in}}{\pgfqpoint{2.692972in}{1.066644in}}%
\pgfusepath{clip}%
\pgfsetrectcap%
\pgfsetroundjoin%
\pgfsetlinewidth{1.505625pt}%
\definecolor{currentstroke}{rgb}{0.000000,0.432000,1.000000}%
\pgfsetstrokecolor{currentstroke}%
\pgfsetdash{}{0pt}%
\pgfpathmoveto{\pgfqpoint{0.440778in}{0.497697in}}%
\pgfpathlineto{\pgfqpoint{0.496881in}{1.166964in}}%
\pgfpathlineto{\pgfqpoint{0.552985in}{1.390053in}}%
\pgfpathlineto{\pgfqpoint{0.609088in}{1.390053in}}%
\pgfpathlineto{\pgfqpoint{0.665192in}{1.390053in}}%
\pgfpathlineto{\pgfqpoint{0.721296in}{1.390053in}}%
\pgfpathlineto{\pgfqpoint{0.777399in}{1.390053in}}%
\pgfpathlineto{\pgfqpoint{0.833503in}{1.390053in}}%
\pgfpathlineto{\pgfqpoint{0.889606in}{1.390053in}}%
\pgfpathlineto{\pgfqpoint{0.945710in}{1.390053in}}%
\pgfpathlineto{\pgfqpoint{1.001814in}{1.278509in}}%
\pgfpathlineto{\pgfqpoint{1.057917in}{0.609242in}}%
\pgfpathlineto{\pgfqpoint{1.114021in}{0.386152in}}%
\pgfpathlineto{\pgfqpoint{1.170124in}{0.386152in}}%
\pgfpathlineto{\pgfqpoint{1.226228in}{0.386152in}}%
\pgfpathlineto{\pgfqpoint{1.282331in}{0.386152in}}%
\pgfpathlineto{\pgfqpoint{1.338435in}{0.386152in}}%
\pgfpathlineto{\pgfqpoint{1.394539in}{0.386152in}}%
\pgfpathlineto{\pgfqpoint{1.450642in}{0.386152in}}%
\pgfpathlineto{\pgfqpoint{1.506746in}{0.386152in}}%
\pgfpathlineto{\pgfqpoint{1.562849in}{0.386152in}}%
\pgfpathlineto{\pgfqpoint{1.618953in}{0.386152in}}%
\pgfpathlineto{\pgfqpoint{1.675057in}{0.386152in}}%
\pgfpathlineto{\pgfqpoint{1.731160in}{0.386152in}}%
\pgfpathlineto{\pgfqpoint{1.787264in}{0.386152in}}%
\pgfpathlineto{\pgfqpoint{1.843367in}{0.386152in}}%
\pgfpathlineto{\pgfqpoint{1.899471in}{0.386152in}}%
\pgfpathlineto{\pgfqpoint{1.955575in}{0.386152in}}%
\pgfpathlineto{\pgfqpoint{2.011678in}{0.386152in}}%
\pgfpathlineto{\pgfqpoint{2.067782in}{0.386152in}}%
\pgfpathlineto{\pgfqpoint{2.123885in}{0.386152in}}%
\pgfpathlineto{\pgfqpoint{2.179989in}{0.386152in}}%
\pgfpathlineto{\pgfqpoint{2.236092in}{0.386152in}}%
\pgfpathlineto{\pgfqpoint{2.292196in}{0.386152in}}%
\pgfpathlineto{\pgfqpoint{2.348300in}{0.386152in}}%
\pgfpathlineto{\pgfqpoint{2.404403in}{0.386152in}}%
\pgfpathlineto{\pgfqpoint{2.460507in}{0.386152in}}%
\pgfpathlineto{\pgfqpoint{2.516610in}{0.386152in}}%
\pgfpathlineto{\pgfqpoint{2.572714in}{0.386152in}}%
\pgfpathlineto{\pgfqpoint{2.628818in}{0.386152in}}%
\pgfpathlineto{\pgfqpoint{2.684921in}{0.386152in}}%
\pgfpathlineto{\pgfqpoint{2.741025in}{0.386152in}}%
\pgfpathlineto{\pgfqpoint{2.797128in}{0.386152in}}%
\pgfpathlineto{\pgfqpoint{2.853232in}{0.386152in}}%
\pgfpathlineto{\pgfqpoint{2.909336in}{0.386152in}}%
\pgfpathlineto{\pgfqpoint{2.965439in}{0.386152in}}%
\pgfpathlineto{\pgfqpoint{3.021543in}{0.386152in}}%
\pgfpathlineto{\pgfqpoint{3.077646in}{0.386152in}}%
\pgfpathlineto{\pgfqpoint{3.133750in}{0.386152in}}%
\pgfusepath{stroke}%
\end{pgfscope}%
\begin{pgfscope}%
\pgfpathrectangle{\pgfqpoint{0.440778in}{0.386152in}}{\pgfqpoint{2.692972in}{1.066644in}}%
\pgfusepath{clip}%
\pgfsetrectcap%
\pgfsetroundjoin%
\pgfsetlinewidth{1.505625pt}%
\definecolor{currentstroke}{rgb}{1.000000,0.160000,0.000000}%
\pgfsetstrokecolor{currentstroke}%
\pgfsetdash{}{0pt}%
\pgfpathmoveto{\pgfqpoint{0.440778in}{0.386152in}}%
\pgfpathlineto{\pgfqpoint{0.496881in}{0.386152in}}%
\pgfpathlineto{\pgfqpoint{0.552985in}{0.386152in}}%
\pgfpathlineto{\pgfqpoint{0.609088in}{0.386152in}}%
\pgfpathlineto{\pgfqpoint{0.665192in}{0.386152in}}%
\pgfpathlineto{\pgfqpoint{0.721296in}{0.386152in}}%
\pgfpathlineto{\pgfqpoint{0.777399in}{0.386152in}}%
\pgfpathlineto{\pgfqpoint{0.833503in}{0.386152in}}%
\pgfpathlineto{\pgfqpoint{0.889606in}{0.386152in}}%
\pgfpathlineto{\pgfqpoint{0.945710in}{0.386152in}}%
\pgfpathlineto{\pgfqpoint{1.001814in}{0.386152in}}%
\pgfpathlineto{\pgfqpoint{1.057917in}{0.386152in}}%
\pgfpathlineto{\pgfqpoint{1.114021in}{0.386152in}}%
\pgfpathlineto{\pgfqpoint{1.170124in}{0.386152in}}%
\pgfpathlineto{\pgfqpoint{1.226228in}{0.386152in}}%
\pgfpathlineto{\pgfqpoint{1.282331in}{0.386152in}}%
\pgfpathlineto{\pgfqpoint{1.338435in}{0.386152in}}%
\pgfpathlineto{\pgfqpoint{1.394539in}{0.386152in}}%
\pgfpathlineto{\pgfqpoint{1.450642in}{0.386152in}}%
\pgfpathlineto{\pgfqpoint{1.506746in}{0.386152in}}%
\pgfpathlineto{\pgfqpoint{1.562849in}{0.386152in}}%
\pgfpathlineto{\pgfqpoint{1.618953in}{0.386152in}}%
\pgfpathlineto{\pgfqpoint{1.675057in}{0.386152in}}%
\pgfpathlineto{\pgfqpoint{1.731160in}{0.386152in}}%
\pgfpathlineto{\pgfqpoint{1.787264in}{0.386152in}}%
\pgfpathlineto{\pgfqpoint{1.843367in}{0.386152in}}%
\pgfpathlineto{\pgfqpoint{1.899471in}{0.386152in}}%
\pgfpathlineto{\pgfqpoint{1.955575in}{0.386152in}}%
\pgfpathlineto{\pgfqpoint{2.011678in}{0.386152in}}%
\pgfpathlineto{\pgfqpoint{2.067782in}{0.386152in}}%
\pgfpathlineto{\pgfqpoint{2.123885in}{0.386152in}}%
\pgfpathlineto{\pgfqpoint{2.179989in}{0.386152in}}%
\pgfpathlineto{\pgfqpoint{2.236092in}{0.386152in}}%
\pgfpathlineto{\pgfqpoint{2.292196in}{0.613889in}}%
\pgfpathlineto{\pgfqpoint{2.348300in}{1.297099in}}%
\pgfpathlineto{\pgfqpoint{2.404403in}{1.410968in}}%
\pgfpathlineto{\pgfqpoint{2.460507in}{1.410968in}}%
\pgfpathlineto{\pgfqpoint{2.516610in}{1.410968in}}%
\pgfpathlineto{\pgfqpoint{2.572714in}{1.410968in}}%
\pgfpathlineto{\pgfqpoint{2.628818in}{1.410968in}}%
\pgfpathlineto{\pgfqpoint{2.684921in}{1.410968in}}%
\pgfpathlineto{\pgfqpoint{2.741025in}{1.410968in}}%
\pgfpathlineto{\pgfqpoint{2.797128in}{1.410968in}}%
\pgfpathlineto{\pgfqpoint{2.853232in}{1.410968in}}%
\pgfpathlineto{\pgfqpoint{2.909336in}{1.410968in}}%
\pgfpathlineto{\pgfqpoint{2.965439in}{1.410968in}}%
\pgfpathlineto{\pgfqpoint{3.021543in}{1.410968in}}%
\pgfpathlineto{\pgfqpoint{3.077646in}{0.727758in}}%
\pgfpathlineto{\pgfqpoint{3.133750in}{0.386152in}}%
\pgfusepath{stroke}%
\end{pgfscope}%
\begin{pgfscope}%
\pgfpathrectangle{\pgfqpoint{0.440778in}{0.386152in}}{\pgfqpoint{2.692972in}{1.066644in}}%
\pgfusepath{clip}%
\pgfsetrectcap%
\pgfsetroundjoin%
\pgfsetlinewidth{1.505625pt}%
\definecolor{currentstroke}{rgb}{0.384000,1.000000,0.000000}%
\pgfsetstrokecolor{currentstroke}%
\pgfsetdash{}{0pt}%
\pgfpathmoveto{\pgfqpoint{0.440778in}{0.386152in}}%
\pgfpathlineto{\pgfqpoint{0.496881in}{0.386152in}}%
\pgfpathlineto{\pgfqpoint{0.552985in}{0.386152in}}%
\pgfpathlineto{\pgfqpoint{0.609088in}{0.386152in}}%
\pgfpathlineto{\pgfqpoint{0.665192in}{0.386152in}}%
\pgfpathlineto{\pgfqpoint{0.721296in}{0.386152in}}%
\pgfpathlineto{\pgfqpoint{0.777399in}{0.386152in}}%
\pgfpathlineto{\pgfqpoint{0.833503in}{0.386152in}}%
\pgfpathlineto{\pgfqpoint{0.889606in}{0.386152in}}%
\pgfpathlineto{\pgfqpoint{0.945710in}{0.386152in}}%
\pgfpathlineto{\pgfqpoint{1.001814in}{0.386152in}}%
\pgfpathlineto{\pgfqpoint{1.057917in}{0.386152in}}%
\pgfpathlineto{\pgfqpoint{1.114021in}{0.386152in}}%
\pgfpathlineto{\pgfqpoint{1.170124in}{0.386152in}}%
\pgfpathlineto{\pgfqpoint{1.226228in}{0.386152in}}%
\pgfpathlineto{\pgfqpoint{1.282331in}{0.386152in}}%
\pgfpathlineto{\pgfqpoint{1.338435in}{0.386152in}}%
\pgfpathlineto{\pgfqpoint{1.394539in}{0.386152in}}%
\pgfpathlineto{\pgfqpoint{1.450642in}{0.386152in}}%
\pgfpathlineto{\pgfqpoint{1.506746in}{0.386152in}}%
\pgfpathlineto{\pgfqpoint{1.562849in}{0.386152in}}%
\pgfpathlineto{\pgfqpoint{1.618953in}{0.717300in}}%
\pgfpathlineto{\pgfqpoint{1.675057in}{1.379596in}}%
\pgfpathlineto{\pgfqpoint{1.731160in}{1.379596in}}%
\pgfpathlineto{\pgfqpoint{1.787264in}{1.379596in}}%
\pgfpathlineto{\pgfqpoint{1.843367in}{1.379596in}}%
\pgfpathlineto{\pgfqpoint{1.899471in}{1.379596in}}%
\pgfpathlineto{\pgfqpoint{1.955575in}{1.379596in}}%
\pgfpathlineto{\pgfqpoint{2.011678in}{1.379596in}}%
\pgfpathlineto{\pgfqpoint{2.067782in}{1.379596in}}%
\pgfpathlineto{\pgfqpoint{2.123885in}{1.379596in}}%
\pgfpathlineto{\pgfqpoint{2.179989in}{1.048448in}}%
\pgfpathlineto{\pgfqpoint{2.236092in}{0.386152in}}%
\pgfpathlineto{\pgfqpoint{2.292196in}{0.386152in}}%
\pgfpathlineto{\pgfqpoint{2.348300in}{0.386152in}}%
\pgfpathlineto{\pgfqpoint{2.404403in}{0.386152in}}%
\pgfpathlineto{\pgfqpoint{2.460507in}{0.386152in}}%
\pgfpathlineto{\pgfqpoint{2.516610in}{0.386152in}}%
\pgfpathlineto{\pgfqpoint{2.572714in}{0.386152in}}%
\pgfpathlineto{\pgfqpoint{2.628818in}{0.386152in}}%
\pgfpathlineto{\pgfqpoint{2.684921in}{0.386152in}}%
\pgfpathlineto{\pgfqpoint{2.741025in}{0.386152in}}%
\pgfpathlineto{\pgfqpoint{2.797128in}{0.386152in}}%
\pgfpathlineto{\pgfqpoint{2.853232in}{0.386152in}}%
\pgfpathlineto{\pgfqpoint{2.909336in}{0.386152in}}%
\pgfpathlineto{\pgfqpoint{2.965439in}{0.386152in}}%
\pgfpathlineto{\pgfqpoint{3.021543in}{0.386152in}}%
\pgfpathlineto{\pgfqpoint{3.077646in}{0.386152in}}%
\pgfpathlineto{\pgfqpoint{3.133750in}{0.386152in}}%
\pgfusepath{stroke}%
\end{pgfscope}%
\begin{pgfscope}%
\pgfpathrectangle{\pgfqpoint{0.440778in}{0.386152in}}{\pgfqpoint{2.692972in}{1.066644in}}%
\pgfusepath{clip}%
\pgfsetrectcap%
\pgfsetroundjoin%
\pgfsetlinewidth{1.505625pt}%
\definecolor{currentstroke}{rgb}{1.000000,0.752000,0.000000}%
\pgfsetstrokecolor{currentstroke}%
\pgfsetdash{}{0pt}%
\pgfpathmoveto{\pgfqpoint{0.440778in}{0.386152in}}%
\pgfpathlineto{\pgfqpoint{0.496881in}{0.386152in}}%
\pgfpathlineto{\pgfqpoint{0.552985in}{0.386152in}}%
\pgfpathlineto{\pgfqpoint{0.609088in}{0.386152in}}%
\pgfpathlineto{\pgfqpoint{0.665192in}{0.386152in}}%
\pgfpathlineto{\pgfqpoint{0.721296in}{0.386152in}}%
\pgfpathlineto{\pgfqpoint{0.777399in}{0.386152in}}%
\pgfpathlineto{\pgfqpoint{0.833503in}{0.386152in}}%
\pgfpathlineto{\pgfqpoint{0.889606in}{0.386152in}}%
\pgfpathlineto{\pgfqpoint{0.945710in}{0.386152in}}%
\pgfpathlineto{\pgfqpoint{1.001814in}{0.386152in}}%
\pgfpathlineto{\pgfqpoint{1.057917in}{0.386152in}}%
\pgfpathlineto{\pgfqpoint{1.114021in}{0.386152in}}%
\pgfpathlineto{\pgfqpoint{1.170124in}{0.386152in}}%
\pgfpathlineto{\pgfqpoint{1.226228in}{0.386152in}}%
\pgfpathlineto{\pgfqpoint{1.282331in}{0.386152in}}%
\pgfpathlineto{\pgfqpoint{1.338435in}{0.386152in}}%
\pgfpathlineto{\pgfqpoint{1.394539in}{0.386152in}}%
\pgfpathlineto{\pgfqpoint{1.450642in}{0.386152in}}%
\pgfpathlineto{\pgfqpoint{1.506746in}{0.386152in}}%
\pgfpathlineto{\pgfqpoint{1.562849in}{0.386152in}}%
\pgfpathlineto{\pgfqpoint{1.618953in}{0.386152in}}%
\pgfpathlineto{\pgfqpoint{1.675057in}{0.386152in}}%
\pgfpathlineto{\pgfqpoint{1.731160in}{0.386152in}}%
\pgfpathlineto{\pgfqpoint{1.787264in}{0.386152in}}%
\pgfpathlineto{\pgfqpoint{1.843367in}{0.386152in}}%
\pgfpathlineto{\pgfqpoint{1.899471in}{0.386152in}}%
\pgfpathlineto{\pgfqpoint{1.955575in}{0.386152in}}%
\pgfpathlineto{\pgfqpoint{2.011678in}{0.386152in}}%
\pgfpathlineto{\pgfqpoint{2.067782in}{0.495373in}}%
\pgfpathlineto{\pgfqpoint{2.123885in}{1.150697in}}%
\pgfpathlineto{\pgfqpoint{2.179989in}{1.369139in}}%
\pgfpathlineto{\pgfqpoint{2.236092in}{1.369139in}}%
\pgfpathlineto{\pgfqpoint{2.292196in}{1.369139in}}%
\pgfpathlineto{\pgfqpoint{2.348300in}{1.369139in}}%
\pgfpathlineto{\pgfqpoint{2.404403in}{1.369139in}}%
\pgfpathlineto{\pgfqpoint{2.460507in}{1.369139in}}%
\pgfpathlineto{\pgfqpoint{2.516610in}{1.369139in}}%
\pgfpathlineto{\pgfqpoint{2.572714in}{1.369139in}}%
\pgfpathlineto{\pgfqpoint{2.628818in}{1.150697in}}%
\pgfpathlineto{\pgfqpoint{2.684921in}{0.495373in}}%
\pgfpathlineto{\pgfqpoint{2.741025in}{0.386152in}}%
\pgfpathlineto{\pgfqpoint{2.797128in}{0.386152in}}%
\pgfpathlineto{\pgfqpoint{2.853232in}{0.386152in}}%
\pgfpathlineto{\pgfqpoint{2.909336in}{0.386152in}}%
\pgfpathlineto{\pgfqpoint{2.965439in}{0.386152in}}%
\pgfpathlineto{\pgfqpoint{3.021543in}{0.386152in}}%
\pgfpathlineto{\pgfqpoint{3.077646in}{0.386152in}}%
\pgfpathlineto{\pgfqpoint{3.133750in}{0.386152in}}%
\pgfusepath{stroke}%
\end{pgfscope}%
\begin{pgfscope}%
\pgfpathrectangle{\pgfqpoint{0.440778in}{0.386152in}}{\pgfqpoint{2.692972in}{1.066644in}}%
\pgfusepath{clip}%
\pgfsetrectcap%
\pgfsetroundjoin%
\pgfsetlinewidth{1.505625pt}%
\definecolor{currentstroke}{rgb}{0.000000,1.000000,0.400000}%
\pgfsetstrokecolor{currentstroke}%
\pgfsetdash{}{0pt}%
\pgfpathmoveto{\pgfqpoint{0.440778in}{0.386152in}}%
\pgfpathlineto{\pgfqpoint{0.496881in}{0.386152in}}%
\pgfpathlineto{\pgfqpoint{0.552985in}{0.386152in}}%
\pgfpathlineto{\pgfqpoint{0.609088in}{0.386152in}}%
\pgfpathlineto{\pgfqpoint{0.665192in}{0.386152in}}%
\pgfpathlineto{\pgfqpoint{0.721296in}{0.386152in}}%
\pgfpathlineto{\pgfqpoint{0.777399in}{0.386152in}}%
\pgfpathlineto{\pgfqpoint{0.833503in}{0.386152in}}%
\pgfpathlineto{\pgfqpoint{0.889606in}{0.386152in}}%
\pgfpathlineto{\pgfqpoint{0.945710in}{0.386152in}}%
\pgfpathlineto{\pgfqpoint{1.001814in}{0.386152in}}%
\pgfpathlineto{\pgfqpoint{1.057917in}{0.386152in}}%
\pgfpathlineto{\pgfqpoint{1.114021in}{0.386152in}}%
\pgfpathlineto{\pgfqpoint{1.170124in}{0.813740in}}%
\pgfpathlineto{\pgfqpoint{1.226228in}{1.348224in}}%
\pgfpathlineto{\pgfqpoint{1.282331in}{1.348224in}}%
\pgfpathlineto{\pgfqpoint{1.338435in}{1.348224in}}%
\pgfpathlineto{\pgfqpoint{1.394539in}{1.348224in}}%
\pgfpathlineto{\pgfqpoint{1.450642in}{1.348224in}}%
\pgfpathlineto{\pgfqpoint{1.506746in}{1.348224in}}%
\pgfpathlineto{\pgfqpoint{1.562849in}{1.348224in}}%
\pgfpathlineto{\pgfqpoint{1.618953in}{1.348224in}}%
\pgfpathlineto{\pgfqpoint{1.675057in}{1.348224in}}%
\pgfpathlineto{\pgfqpoint{1.731160in}{0.813740in}}%
\pgfpathlineto{\pgfqpoint{1.787264in}{0.386152in}}%
\pgfpathlineto{\pgfqpoint{1.843367in}{0.386152in}}%
\pgfpathlineto{\pgfqpoint{1.899471in}{0.386152in}}%
\pgfpathlineto{\pgfqpoint{1.955575in}{0.386152in}}%
\pgfpathlineto{\pgfqpoint{2.011678in}{0.386152in}}%
\pgfpathlineto{\pgfqpoint{2.067782in}{0.386152in}}%
\pgfpathlineto{\pgfqpoint{2.123885in}{0.386152in}}%
\pgfpathlineto{\pgfqpoint{2.179989in}{0.386152in}}%
\pgfpathlineto{\pgfqpoint{2.236092in}{0.386152in}}%
\pgfpathlineto{\pgfqpoint{2.292196in}{0.386152in}}%
\pgfpathlineto{\pgfqpoint{2.348300in}{0.386152in}}%
\pgfpathlineto{\pgfqpoint{2.404403in}{0.386152in}}%
\pgfpathlineto{\pgfqpoint{2.460507in}{0.386152in}}%
\pgfpathlineto{\pgfqpoint{2.516610in}{0.386152in}}%
\pgfpathlineto{\pgfqpoint{2.572714in}{0.386152in}}%
\pgfpathlineto{\pgfqpoint{2.628818in}{0.386152in}}%
\pgfpathlineto{\pgfqpoint{2.684921in}{0.386152in}}%
\pgfpathlineto{\pgfqpoint{2.741025in}{0.386152in}}%
\pgfpathlineto{\pgfqpoint{2.797128in}{0.386152in}}%
\pgfpathlineto{\pgfqpoint{2.853232in}{0.386152in}}%
\pgfpathlineto{\pgfqpoint{2.909336in}{0.386152in}}%
\pgfpathlineto{\pgfqpoint{2.965439in}{0.386152in}}%
\pgfpathlineto{\pgfqpoint{3.021543in}{0.386152in}}%
\pgfpathlineto{\pgfqpoint{3.077646in}{0.386152in}}%
\pgfpathlineto{\pgfqpoint{3.133750in}{0.386152in}}%
\pgfusepath{stroke}%
\end{pgfscope}%
\begin{pgfscope}%
\pgfpathrectangle{\pgfqpoint{0.440778in}{0.386152in}}{\pgfqpoint{2.692972in}{1.066644in}}%
\pgfusepath{clip}%
\pgfsetrectcap%
\pgfsetroundjoin%
\pgfsetlinewidth{1.505625pt}%
\definecolor{currentstroke}{rgb}{1.000000,0.000000,0.000000}%
\pgfsetstrokecolor{currentstroke}%
\pgfsetdash{}{0pt}%
\pgfpathmoveto{\pgfqpoint{0.440778in}{0.386152in}}%
\pgfpathlineto{\pgfqpoint{0.496881in}{0.386152in}}%
\pgfpathlineto{\pgfqpoint{0.552985in}{0.386152in}}%
\pgfpathlineto{\pgfqpoint{0.609088in}{0.386152in}}%
\pgfpathlineto{\pgfqpoint{0.665192in}{0.386152in}}%
\pgfpathlineto{\pgfqpoint{0.721296in}{0.386152in}}%
\pgfpathlineto{\pgfqpoint{0.777399in}{0.386152in}}%
\pgfpathlineto{\pgfqpoint{0.833503in}{0.386152in}}%
\pgfpathlineto{\pgfqpoint{0.889606in}{0.386152in}}%
\pgfpathlineto{\pgfqpoint{0.945710in}{0.386152in}}%
\pgfpathlineto{\pgfqpoint{1.001814in}{0.386152in}}%
\pgfpathlineto{\pgfqpoint{1.057917in}{0.386152in}}%
\pgfpathlineto{\pgfqpoint{1.114021in}{0.386152in}}%
\pgfpathlineto{\pgfqpoint{1.170124in}{0.386152in}}%
\pgfpathlineto{\pgfqpoint{1.226228in}{0.386152in}}%
\pgfpathlineto{\pgfqpoint{1.282331in}{0.386152in}}%
\pgfpathlineto{\pgfqpoint{1.338435in}{0.386152in}}%
\pgfpathlineto{\pgfqpoint{1.394539in}{0.386152in}}%
\pgfpathlineto{\pgfqpoint{1.450642in}{0.386152in}}%
\pgfpathlineto{\pgfqpoint{1.506746in}{0.386152in}}%
\pgfpathlineto{\pgfqpoint{1.562849in}{0.386152in}}%
\pgfpathlineto{\pgfqpoint{1.618953in}{0.386152in}}%
\pgfpathlineto{\pgfqpoint{1.675057in}{0.386152in}}%
\pgfpathlineto{\pgfqpoint{1.731160in}{0.386152in}}%
\pgfpathlineto{\pgfqpoint{1.787264in}{0.386152in}}%
\pgfpathlineto{\pgfqpoint{1.843367in}{0.386152in}}%
\pgfpathlineto{\pgfqpoint{1.899471in}{0.386152in}}%
\pgfpathlineto{\pgfqpoint{1.955575in}{0.386152in}}%
\pgfpathlineto{\pgfqpoint{2.011678in}{0.386152in}}%
\pgfpathlineto{\pgfqpoint{2.067782in}{0.386152in}}%
\pgfpathlineto{\pgfqpoint{2.123885in}{0.386152in}}%
\pgfpathlineto{\pgfqpoint{2.179989in}{0.386152in}}%
\pgfpathlineto{\pgfqpoint{2.236092in}{0.386152in}}%
\pgfpathlineto{\pgfqpoint{2.292196in}{0.386152in}}%
\pgfpathlineto{\pgfqpoint{2.348300in}{0.386152in}}%
\pgfpathlineto{\pgfqpoint{2.404403in}{0.386152in}}%
\pgfpathlineto{\pgfqpoint{2.460507in}{0.386152in}}%
\pgfpathlineto{\pgfqpoint{2.516610in}{0.386152in}}%
\pgfpathlineto{\pgfqpoint{2.572714in}{1.069363in}}%
\pgfpathlineto{\pgfqpoint{2.628818in}{1.410968in}}%
\pgfpathlineto{\pgfqpoint{2.684921in}{1.410968in}}%
\pgfpathlineto{\pgfqpoint{2.741025in}{1.410968in}}%
\pgfpathlineto{\pgfqpoint{2.797128in}{1.410968in}}%
\pgfpathlineto{\pgfqpoint{2.853232in}{1.410968in}}%
\pgfpathlineto{\pgfqpoint{2.909336in}{1.410968in}}%
\pgfpathlineto{\pgfqpoint{2.965439in}{1.410968in}}%
\pgfpathlineto{\pgfqpoint{3.021543in}{1.410968in}}%
\pgfpathlineto{\pgfqpoint{3.077646in}{1.410968in}}%
\pgfpathlineto{\pgfqpoint{3.133750in}{0.613889in}}%
\pgfusepath{stroke}%
\end{pgfscope}%
\begin{pgfscope}%
\pgfpathrectangle{\pgfqpoint{0.440778in}{0.386152in}}{\pgfqpoint{2.692972in}{1.066644in}}%
\pgfusepath{clip}%
\pgfsetrectcap%
\pgfsetroundjoin%
\pgfsetlinewidth{1.505625pt}%
\definecolor{currentstroke}{rgb}{0.000000,1.000000,0.016000}%
\pgfsetstrokecolor{currentstroke}%
\pgfsetdash{}{0pt}%
\pgfpathmoveto{\pgfqpoint{0.440778in}{0.386152in}}%
\pgfpathlineto{\pgfqpoint{0.496881in}{0.386152in}}%
\pgfpathlineto{\pgfqpoint{0.552985in}{0.386152in}}%
\pgfpathlineto{\pgfqpoint{0.609088in}{0.386152in}}%
\pgfpathlineto{\pgfqpoint{0.665192in}{0.386152in}}%
\pgfpathlineto{\pgfqpoint{0.721296in}{0.386152in}}%
\pgfpathlineto{\pgfqpoint{0.777399in}{0.386152in}}%
\pgfpathlineto{\pgfqpoint{0.833503in}{0.386152in}}%
\pgfpathlineto{\pgfqpoint{0.889606in}{0.386152in}}%
\pgfpathlineto{\pgfqpoint{0.945710in}{0.386152in}}%
\pgfpathlineto{\pgfqpoint{1.001814in}{0.386152in}}%
\pgfpathlineto{\pgfqpoint{1.057917in}{0.386152in}}%
\pgfpathlineto{\pgfqpoint{1.114021in}{0.386152in}}%
\pgfpathlineto{\pgfqpoint{1.170124in}{0.500021in}}%
\pgfpathlineto{\pgfqpoint{1.226228in}{1.183231in}}%
\pgfpathlineto{\pgfqpoint{1.282331in}{1.410968in}}%
\pgfpathlineto{\pgfqpoint{1.338435in}{1.410968in}}%
\pgfpathlineto{\pgfqpoint{1.394539in}{1.410968in}}%
\pgfpathlineto{\pgfqpoint{1.450642in}{1.410968in}}%
\pgfpathlineto{\pgfqpoint{1.506746in}{1.410968in}}%
\pgfpathlineto{\pgfqpoint{1.562849in}{1.410968in}}%
\pgfpathlineto{\pgfqpoint{1.618953in}{1.410968in}}%
\pgfpathlineto{\pgfqpoint{1.675057in}{1.410968in}}%
\pgfpathlineto{\pgfqpoint{1.731160in}{1.410968in}}%
\pgfpathlineto{\pgfqpoint{1.787264in}{1.410968in}}%
\pgfpathlineto{\pgfqpoint{1.843367in}{1.410968in}}%
\pgfpathlineto{\pgfqpoint{1.899471in}{1.410968in}}%
\pgfpathlineto{\pgfqpoint{1.955575in}{1.410968in}}%
\pgfpathlineto{\pgfqpoint{2.011678in}{1.410968in}}%
\pgfpathlineto{\pgfqpoint{2.067782in}{1.410968in}}%
\pgfpathlineto{\pgfqpoint{2.123885in}{1.183231in}}%
\pgfpathlineto{\pgfqpoint{2.179989in}{0.500021in}}%
\pgfpathlineto{\pgfqpoint{2.236092in}{0.386152in}}%
\pgfpathlineto{\pgfqpoint{2.292196in}{0.386152in}}%
\pgfpathlineto{\pgfqpoint{2.348300in}{0.386152in}}%
\pgfpathlineto{\pgfqpoint{2.404403in}{0.386152in}}%
\pgfpathlineto{\pgfqpoint{2.460507in}{0.386152in}}%
\pgfpathlineto{\pgfqpoint{2.516610in}{0.386152in}}%
\pgfpathlineto{\pgfqpoint{2.572714in}{0.386152in}}%
\pgfpathlineto{\pgfqpoint{2.628818in}{0.386152in}}%
\pgfpathlineto{\pgfqpoint{2.684921in}{0.386152in}}%
\pgfpathlineto{\pgfqpoint{2.741025in}{0.386152in}}%
\pgfpathlineto{\pgfqpoint{2.797128in}{0.386152in}}%
\pgfpathlineto{\pgfqpoint{2.853232in}{0.386152in}}%
\pgfpathlineto{\pgfqpoint{2.909336in}{0.386152in}}%
\pgfpathlineto{\pgfqpoint{2.965439in}{0.386152in}}%
\pgfpathlineto{\pgfqpoint{3.021543in}{0.386152in}}%
\pgfpathlineto{\pgfqpoint{3.077646in}{0.386152in}}%
\pgfpathlineto{\pgfqpoint{3.133750in}{0.386152in}}%
\pgfusepath{stroke}%
\end{pgfscope}%
\begin{pgfscope}%
\pgfsetrectcap%
\pgfsetmiterjoin%
\pgfsetlinewidth{0.803000pt}%
\definecolor{currentstroke}{rgb}{0.000000,0.000000,0.000000}%
\pgfsetstrokecolor{currentstroke}%
\pgfsetdash{}{0pt}%
\pgfpathmoveto{\pgfqpoint{0.440778in}{0.386152in}}%
\pgfpathlineto{\pgfqpoint{0.440778in}{1.452797in}}%
\pgfusepath{stroke}%
\end{pgfscope}%
\begin{pgfscope}%
\pgfsetrectcap%
\pgfsetmiterjoin%
\pgfsetlinewidth{0.803000pt}%
\definecolor{currentstroke}{rgb}{0.000000,0.000000,0.000000}%
\pgfsetstrokecolor{currentstroke}%
\pgfsetdash{}{0pt}%
\pgfpathmoveto{\pgfqpoint{3.133750in}{0.386152in}}%
\pgfpathlineto{\pgfqpoint{3.133750in}{1.452797in}}%
\pgfusepath{stroke}%
\end{pgfscope}%
\begin{pgfscope}%
\pgfsetrectcap%
\pgfsetmiterjoin%
\pgfsetlinewidth{0.803000pt}%
\definecolor{currentstroke}{rgb}{0.000000,0.000000,0.000000}%
\pgfsetstrokecolor{currentstroke}%
\pgfsetdash{}{0pt}%
\pgfpathmoveto{\pgfqpoint{0.440778in}{0.386152in}}%
\pgfpathlineto{\pgfqpoint{3.133750in}{0.386152in}}%
\pgfusepath{stroke}%
\end{pgfscope}%
\begin{pgfscope}%
\pgfsetrectcap%
\pgfsetmiterjoin%
\pgfsetlinewidth{0.803000pt}%
\definecolor{currentstroke}{rgb}{0.000000,0.000000,0.000000}%
\pgfsetstrokecolor{currentstroke}%
\pgfsetdash{}{0pt}%
\pgfpathmoveto{\pgfqpoint{0.440778in}{1.452797in}}%
\pgfpathlineto{\pgfqpoint{3.133750in}{1.452797in}}%
\pgfusepath{stroke}%
\end{pgfscope}%
\begin{pgfscope}%
\pgfsetbuttcap%
\pgfsetroundjoin%
\definecolor{currentfill}{rgb}{0.000000,0.000000,0.000000}%
\pgfsetfillcolor{currentfill}%
\pgfsetlinewidth{0.803000pt}%
\definecolor{currentstroke}{rgb}{0.000000,0.000000,0.000000}%
\pgfsetstrokecolor{currentstroke}%
\pgfsetdash{}{0pt}%
\pgfsys@defobject{currentmarker}{\pgfqpoint{0.000000in}{0.000000in}}{\pgfqpoint{0.000000in}{0.048611in}}{%
\pgfpathmoveto{\pgfqpoint{0.000000in}{0.000000in}}%
\pgfpathlineto{\pgfqpoint{0.000000in}{0.048611in}}%
\pgfusepath{stroke,fill}%
}%
\begin{pgfscope}%
\pgfsys@transformshift{0.440778in}{1.452797in}%
\pgfsys@useobject{currentmarker}{}%
\end{pgfscope}%
\end{pgfscope}%
\begin{pgfscope}%
\definecolor{textcolor}{rgb}{0.000000,0.000000,0.000000}%
\pgfsetstrokecolor{textcolor}%
\pgfsetfillcolor{textcolor}%
\pgftext[x=0.440778in,y=1.550019in,,bottom]{\color{textcolor}\rmfamily\fontsize{9.000000}{10.800000}\selectfont 391}%
\end{pgfscope}%
\begin{pgfscope}%
\pgfsetbuttcap%
\pgfsetroundjoin%
\definecolor{currentfill}{rgb}{0.000000,0.000000,0.000000}%
\pgfsetfillcolor{currentfill}%
\pgfsetlinewidth{0.803000pt}%
\definecolor{currentstroke}{rgb}{0.000000,0.000000,0.000000}%
\pgfsetstrokecolor{currentstroke}%
\pgfsetdash{}{0pt}%
\pgfsys@defobject{currentmarker}{\pgfqpoint{0.000000in}{0.000000in}}{\pgfqpoint{0.000000in}{0.048611in}}{%
\pgfpathmoveto{\pgfqpoint{0.000000in}{0.000000in}}%
\pgfpathlineto{\pgfqpoint{0.000000in}{0.048611in}}%
\pgfusepath{stroke,fill}%
}%
\begin{pgfscope}%
\pgfsys@transformshift{0.945710in}{1.452797in}%
\pgfsys@useobject{currentmarker}{}%
\end{pgfscope}%
\end{pgfscope}%
\begin{pgfscope}%
\definecolor{textcolor}{rgb}{0.000000,0.000000,0.000000}%
\pgfsetstrokecolor{textcolor}%
\pgfsetfillcolor{textcolor}%
\pgftext[x=0.945710in,y=1.550019in,,bottom]{\color{textcolor}\rmfamily\fontsize{9.000000}{10.800000}\selectfont 445}%
\end{pgfscope}%
\begin{pgfscope}%
\pgfsetbuttcap%
\pgfsetroundjoin%
\definecolor{currentfill}{rgb}{0.000000,0.000000,0.000000}%
\pgfsetfillcolor{currentfill}%
\pgfsetlinewidth{0.803000pt}%
\definecolor{currentstroke}{rgb}{0.000000,0.000000,0.000000}%
\pgfsetstrokecolor{currentstroke}%
\pgfsetdash{}{0pt}%
\pgfsys@defobject{currentmarker}{\pgfqpoint{0.000000in}{0.000000in}}{\pgfqpoint{0.000000in}{0.048611in}}{%
\pgfpathmoveto{\pgfqpoint{0.000000in}{0.000000in}}%
\pgfpathlineto{\pgfqpoint{0.000000in}{0.048611in}}%
\pgfusepath{stroke,fill}%
}%
\begin{pgfscope}%
\pgfsys@transformshift{1.506746in}{1.452797in}%
\pgfsys@useobject{currentmarker}{}%
\end{pgfscope}%
\end{pgfscope}%
\begin{pgfscope}%
\definecolor{textcolor}{rgb}{0.000000,0.000000,0.000000}%
\pgfsetstrokecolor{textcolor}%
\pgfsetfillcolor{textcolor}%
\pgftext[x=1.506746in,y=1.550019in,,bottom]{\color{textcolor}\rmfamily\fontsize{9.000000}{10.800000}\selectfont 506}%
\end{pgfscope}%
\begin{pgfscope}%
\pgfsetbuttcap%
\pgfsetroundjoin%
\definecolor{currentfill}{rgb}{0.000000,0.000000,0.000000}%
\pgfsetfillcolor{currentfill}%
\pgfsetlinewidth{0.803000pt}%
\definecolor{currentstroke}{rgb}{0.000000,0.000000,0.000000}%
\pgfsetstrokecolor{currentstroke}%
\pgfsetdash{}{0pt}%
\pgfsys@defobject{currentmarker}{\pgfqpoint{0.000000in}{0.000000in}}{\pgfqpoint{0.000000in}{0.048611in}}{%
\pgfpathmoveto{\pgfqpoint{0.000000in}{0.000000in}}%
\pgfpathlineto{\pgfqpoint{0.000000in}{0.048611in}}%
\pgfusepath{stroke,fill}%
}%
\begin{pgfscope}%
\pgfsys@transformshift{2.011678in}{1.452797in}%
\pgfsys@useobject{currentmarker}{}%
\end{pgfscope}%
\end{pgfscope}%
\begin{pgfscope}%
\definecolor{textcolor}{rgb}{0.000000,0.000000,0.000000}%
\pgfsetstrokecolor{textcolor}%
\pgfsetfillcolor{textcolor}%
\pgftext[x=2.011678in,y=1.550019in,,bottom]{\color{textcolor}\rmfamily\fontsize{9.000000}{10.800000}\selectfont 561}%
\end{pgfscope}%
\begin{pgfscope}%
\pgfsetbuttcap%
\pgfsetroundjoin%
\definecolor{currentfill}{rgb}{0.000000,0.000000,0.000000}%
\pgfsetfillcolor{currentfill}%
\pgfsetlinewidth{0.803000pt}%
\definecolor{currentstroke}{rgb}{0.000000,0.000000,0.000000}%
\pgfsetstrokecolor{currentstroke}%
\pgfsetdash{}{0pt}%
\pgfsys@defobject{currentmarker}{\pgfqpoint{0.000000in}{0.000000in}}{\pgfqpoint{0.000000in}{0.048611in}}{%
\pgfpathmoveto{\pgfqpoint{0.000000in}{0.000000in}}%
\pgfpathlineto{\pgfqpoint{0.000000in}{0.048611in}}%
\pgfusepath{stroke,fill}%
}%
\begin{pgfscope}%
\pgfsys@transformshift{2.572714in}{1.452797in}%
\pgfsys@useobject{currentmarker}{}%
\end{pgfscope}%
\end{pgfscope}%
\begin{pgfscope}%
\definecolor{textcolor}{rgb}{0.000000,0.000000,0.000000}%
\pgfsetstrokecolor{textcolor}%
\pgfsetfillcolor{textcolor}%
\pgftext[x=2.572714in,y=1.550019in,,bottom]{\color{textcolor}\rmfamily\fontsize{9.000000}{10.800000}\selectfont 622}%
\end{pgfscope}%
\begin{pgfscope}%
\pgfsetbuttcap%
\pgfsetroundjoin%
\definecolor{currentfill}{rgb}{0.000000,0.000000,0.000000}%
\pgfsetfillcolor{currentfill}%
\pgfsetlinewidth{0.803000pt}%
\definecolor{currentstroke}{rgb}{0.000000,0.000000,0.000000}%
\pgfsetstrokecolor{currentstroke}%
\pgfsetdash{}{0pt}%
\pgfsys@defobject{currentmarker}{\pgfqpoint{0.000000in}{0.000000in}}{\pgfqpoint{0.000000in}{0.048611in}}{%
\pgfpathmoveto{\pgfqpoint{0.000000in}{0.000000in}}%
\pgfpathlineto{\pgfqpoint{0.000000in}{0.048611in}}%
\pgfusepath{stroke,fill}%
}%
\begin{pgfscope}%
\pgfsys@transformshift{3.133750in}{1.452797in}%
\pgfsys@useobject{currentmarker}{}%
\end{pgfscope}%
\end{pgfscope}%
\begin{pgfscope}%
\definecolor{textcolor}{rgb}{0.000000,0.000000,0.000000}%
\pgfsetstrokecolor{textcolor}%
\pgfsetfillcolor{textcolor}%
\pgftext[x=3.133750in,y=1.550019in,,bottom]{\color{textcolor}\rmfamily\fontsize{9.000000}{10.800000}\selectfont 683}%
\end{pgfscope}%
\begin{pgfscope}%
\definecolor{textcolor}{rgb}{0.000000,0.000000,0.000000}%
\pgfsetstrokecolor{textcolor}%
\pgfsetfillcolor{textcolor}%
\pgftext[x=1.787264in,y=1.716575in,,base]{\color{textcolor}\rmfamily\fontsize{9.000000}{10.800000}\selectfont Wavelength in nm}%
\end{pgfscope}%
\begin{pgfscope}%
\pgfsetrectcap%
\pgfsetmiterjoin%
\pgfsetlinewidth{0.803000pt}%
\definecolor{currentstroke}{rgb}{0.000000,0.000000,0.000000}%
\pgfsetstrokecolor{currentstroke}%
\pgfsetdash{}{0pt}%
\pgfpathmoveto{\pgfqpoint{0.440778in}{0.386152in}}%
\pgfpathlineto{\pgfqpoint{0.440778in}{1.452797in}}%
\pgfusepath{stroke}%
\end{pgfscope}%
\begin{pgfscope}%
\pgfsetrectcap%
\pgfsetmiterjoin%
\pgfsetlinewidth{0.803000pt}%
\definecolor{currentstroke}{rgb}{0.000000,0.000000,0.000000}%
\pgfsetstrokecolor{currentstroke}%
\pgfsetdash{}{0pt}%
\pgfpathmoveto{\pgfqpoint{3.133750in}{0.386152in}}%
\pgfpathlineto{\pgfqpoint{3.133750in}{1.452797in}}%
\pgfusepath{stroke}%
\end{pgfscope}%
\begin{pgfscope}%
\pgfsetrectcap%
\pgfsetmiterjoin%
\pgfsetlinewidth{0.803000pt}%
\definecolor{currentstroke}{rgb}{0.000000,0.000000,0.000000}%
\pgfsetstrokecolor{currentstroke}%
\pgfsetdash{}{0pt}%
\pgfpathmoveto{\pgfqpoint{0.440778in}{0.386152in}}%
\pgfpathlineto{\pgfqpoint{3.133750in}{0.386152in}}%
\pgfusepath{stroke}%
\end{pgfscope}%
\begin{pgfscope}%
\pgfsetrectcap%
\pgfsetmiterjoin%
\pgfsetlinewidth{0.803000pt}%
\definecolor{currentstroke}{rgb}{0.000000,0.000000,0.000000}%
\pgfsetstrokecolor{currentstroke}%
\pgfsetdash{}{0pt}%
\pgfpathmoveto{\pgfqpoint{0.440778in}{1.452797in}}%
\pgfpathlineto{\pgfqpoint{3.133750in}{1.452797in}}%
\pgfusepath{stroke}%
\end{pgfscope}%
\end{pgfpicture}%
\makeatother%
\endgroup%

%% file: figures/real_evaluation.pgf
%% Creator: Matplotlib, PGF backend
%%
%% To include the figure in your LaTeX document, write
%%   \input{<filename>.pgf}
%%
%% Make sure the required packages are loaded in your preamble
%%   \usepackage{pgf}
%%
%% and, on pdftex
%%   \usepackage[utf8]{inputenc}\DeclareUnicodeCharacter{2212}{-}
%%
%% or, on luatex and xetex
%%   \usepackage{unicode-math}
%%
%% Figures using additional raster images can only be included by \input if
%% they are in the same directory as the main LaTeX file. For loading figures
%% from other directories you can use the `import` package
%%   \usepackage{import}
%%
%% and then include the figures with
%%   \import{<path to file>}{<filename>.pgf}
%%
%% Matplotlib used the following preamble
%%   \usepackage{fontspec}
%%
\begingroup%
\makeatletter%
\begin{pgfpicture}%
\pgfpathrectangle{\pgfpointorigin}{\pgfqpoint{3.417376in}{1.887325in}}%
\pgfusepath{use as bounding box, clip}%
\begin{pgfscope}%
\pgfsetbuttcap%
\pgfsetmiterjoin%
\definecolor{currentfill}{rgb}{1.000000,1.000000,1.000000}%
\pgfsetfillcolor{currentfill}%
\pgfsetlinewidth{0.000000pt}%
\definecolor{currentstroke}{rgb}{1.000000,1.000000,1.000000}%
\pgfsetstrokecolor{currentstroke}%
\pgfsetdash{}{0pt}%
\pgfpathmoveto{\pgfqpoint{-0.000000in}{0.000000in}}%
\pgfpathlineto{\pgfqpoint{3.417376in}{0.000000in}}%
\pgfpathlineto{\pgfqpoint{3.417376in}{1.887325in}}%
\pgfpathlineto{\pgfqpoint{-0.000000in}{1.887325in}}%
\pgfpathclose%
\pgfusepath{fill}%
\end{pgfscope}%
\begin{pgfscope}%
\pgfsetbuttcap%
\pgfsetmiterjoin%
\definecolor{currentfill}{rgb}{1.000000,1.000000,1.000000}%
\pgfsetfillcolor{currentfill}%
\pgfsetlinewidth{0.000000pt}%
\definecolor{currentstroke}{rgb}{0.000000,0.000000,0.000000}%
\pgfsetstrokecolor{currentstroke}%
\pgfsetstrokeopacity{0.000000}%
\pgfsetdash{}{0pt}%
\pgfpathmoveto{\pgfqpoint{0.592966in}{0.067625in}}%
\pgfpathlineto{\pgfqpoint{1.579938in}{0.067625in}}%
\pgfpathlineto{\pgfqpoint{1.579938in}{1.563761in}}%
\pgfpathlineto{\pgfqpoint{0.592966in}{1.563761in}}%
\pgfpathclose%
\pgfusepath{fill}%
\end{pgfscope}%
\begin{pgfscope}%
\pgfpathrectangle{\pgfqpoint{0.592966in}{0.067625in}}{\pgfqpoint{0.986972in}{1.496136in}}%
\pgfusepath{clip}%
\pgfsetbuttcap%
\pgfsetmiterjoin%
\definecolor{currentfill}{rgb}{1.000000,0.498039,0.054902}%
\pgfsetfillcolor{currentfill}%
\pgfsetlinewidth{1.003750pt}%
\definecolor{currentstroke}{rgb}{0.000000,0.000000,0.000000}%
\pgfsetstrokecolor{currentstroke}%
\pgfsetdash{{6.400000pt}{1.600000pt}{1.000000pt}{1.600000pt}}{0.000000pt}%
\pgfpathmoveto{\pgfqpoint{0.637828in}{0.067625in}}%
\pgfpathlineto{\pgfqpoint{0.862140in}{0.067625in}}%
\pgfpathlineto{\pgfqpoint{0.862140in}{1.493154in}}%
\pgfpathlineto{\pgfqpoint{0.637828in}{1.493154in}}%
\pgfpathclose%
\pgfusepath{stroke,fill}%
\end{pgfscope}%
\begin{pgfscope}%
\pgfpathrectangle{\pgfqpoint{0.592966in}{0.067625in}}{\pgfqpoint{0.986972in}{1.496136in}}%
\pgfusepath{clip}%
\pgfsetbuttcap%
\pgfsetmiterjoin%
\definecolor{currentfill}{rgb}{0.839216,0.152941,0.156863}%
\pgfsetfillcolor{currentfill}%
\pgfsetlinewidth{1.003750pt}%
\definecolor{currentstroke}{rgb}{0.000000,0.000000,0.000000}%
\pgfsetstrokecolor{currentstroke}%
\pgfsetdash{{1.000000pt}{1.650000pt}}{0.000000pt}%
\pgfpathmoveto{\pgfqpoint{0.862140in}{0.067625in}}%
\pgfpathlineto{\pgfqpoint{1.086452in}{0.067625in}}%
\pgfpathlineto{\pgfqpoint{1.086452in}{0.699064in}}%
\pgfpathlineto{\pgfqpoint{0.862140in}{0.699064in}}%
\pgfpathclose%
\pgfusepath{stroke,fill}%
\end{pgfscope}%
\begin{pgfscope}%
\pgfpathrectangle{\pgfqpoint{0.592966in}{0.067625in}}{\pgfqpoint{0.986972in}{1.496136in}}%
\pgfusepath{clip}%
\pgfsetbuttcap%
\pgfsetmiterjoin%
\definecolor{currentfill}{rgb}{0.580392,0.403922,0.741176}%
\pgfsetfillcolor{currentfill}%
\pgfsetlinewidth{1.003750pt}%
\definecolor{currentstroke}{rgb}{0.000000,0.000000,0.000000}%
\pgfsetstrokecolor{currentstroke}%
\pgfsetdash{{3.700000pt}{1.600000pt}}{0.000000pt}%
\pgfpathmoveto{\pgfqpoint{1.086452in}{0.067625in}}%
\pgfpathlineto{\pgfqpoint{1.310764in}{0.067625in}}%
\pgfpathlineto{\pgfqpoint{1.310764in}{0.977086in}}%
\pgfpathlineto{\pgfqpoint{1.086452in}{0.977086in}}%
\pgfpathclose%
\pgfusepath{stroke,fill}%
\end{pgfscope}%
\begin{pgfscope}%
\pgfpathrectangle{\pgfqpoint{0.592966in}{0.067625in}}{\pgfqpoint{0.986972in}{1.496136in}}%
\pgfusepath{clip}%
\pgfsetbuttcap%
\pgfsetmiterjoin%
\definecolor{currentfill}{rgb}{0.549020,0.337255,0.294118}%
\pgfsetfillcolor{currentfill}%
\pgfsetlinewidth{1.003750pt}%
\definecolor{currentstroke}{rgb}{0.000000,0.000000,0.000000}%
\pgfsetstrokecolor{currentstroke}%
\pgfsetdash{{3.000000pt}{1.000000pt}{1.000000pt}{1.000000pt}{1.000000pt}{1.000000pt}}{0.000000pt}%
\pgfpathmoveto{\pgfqpoint{1.310764in}{0.067625in}}%
\pgfpathlineto{\pgfqpoint{1.535075in}{0.067625in}}%
\pgfpathlineto{\pgfqpoint{1.535075in}{0.614648in}}%
\pgfpathlineto{\pgfqpoint{1.310764in}{0.614648in}}%
\pgfpathclose%
\pgfusepath{stroke,fill}%
\end{pgfscope}%
\begin{pgfscope}%
\pgfpathrectangle{\pgfqpoint{0.592966in}{0.067625in}}{\pgfqpoint{0.986972in}{1.496136in}}%
\pgfusepath{clip}%
\pgfsetrectcap%
\pgfsetroundjoin%
\pgfsetlinewidth{0.803000pt}%
\definecolor{currentstroke}{rgb}{0.690196,0.690196,0.690196}%
\pgfsetstrokecolor{currentstroke}%
\pgfsetdash{}{0pt}%
\pgfpathmoveto{\pgfqpoint{0.592966in}{0.067625in}}%
\pgfpathlineto{\pgfqpoint{1.579938in}{0.067625in}}%
\pgfusepath{stroke}%
\end{pgfscope}%
\begin{pgfscope}%
\pgfsetbuttcap%
\pgfsetroundjoin%
\definecolor{currentfill}{rgb}{0.000000,0.000000,0.000000}%
\pgfsetfillcolor{currentfill}%
\pgfsetlinewidth{0.803000pt}%
\definecolor{currentstroke}{rgb}{0.000000,0.000000,0.000000}%
\pgfsetstrokecolor{currentstroke}%
\pgfsetdash{}{0pt}%
\pgfsys@defobject{currentmarker}{\pgfqpoint{-0.048611in}{0.000000in}}{\pgfqpoint{0.000000in}{0.000000in}}{%
\pgfpathmoveto{\pgfqpoint{0.000000in}{0.000000in}}%
\pgfpathlineto{\pgfqpoint{-0.048611in}{0.000000in}}%
\pgfusepath{stroke,fill}%
}%
\begin{pgfscope}%
\pgfsys@transformshift{0.592966in}{0.067625in}%
\pgfsys@useobject{currentmarker}{}%
\end{pgfscope}%
\end{pgfscope}%
\begin{pgfscope}%
\definecolor{textcolor}{rgb}{0.000000,0.000000,0.000000}%
\pgfsetstrokecolor{textcolor}%
\pgfsetfillcolor{textcolor}%
\pgftext[x=0.267243in, y=0.024250in, left, base]{\color{textcolor}\rmfamily\fontsize{9.000000}{10.800000}\selectfont 0.00}%
\end{pgfscope}%
\begin{pgfscope}%
\pgfpathrectangle{\pgfqpoint{0.592966in}{0.067625in}}{\pgfqpoint{0.986972in}{1.496136in}}%
\pgfusepath{clip}%
\pgfsetrectcap%
\pgfsetroundjoin%
\pgfsetlinewidth{0.803000pt}%
\definecolor{currentstroke}{rgb}{0.690196,0.690196,0.690196}%
\pgfsetstrokecolor{currentstroke}%
\pgfsetdash{}{0pt}%
\pgfpathmoveto{\pgfqpoint{0.592966in}{0.407656in}}%
\pgfpathlineto{\pgfqpoint{1.579938in}{0.407656in}}%
\pgfusepath{stroke}%
\end{pgfscope}%
\begin{pgfscope}%
\pgfsetbuttcap%
\pgfsetroundjoin%
\definecolor{currentfill}{rgb}{0.000000,0.000000,0.000000}%
\pgfsetfillcolor{currentfill}%
\pgfsetlinewidth{0.803000pt}%
\definecolor{currentstroke}{rgb}{0.000000,0.000000,0.000000}%
\pgfsetstrokecolor{currentstroke}%
\pgfsetdash{}{0pt}%
\pgfsys@defobject{currentmarker}{\pgfqpoint{-0.048611in}{0.000000in}}{\pgfqpoint{0.000000in}{0.000000in}}{%
\pgfpathmoveto{\pgfqpoint{0.000000in}{0.000000in}}%
\pgfpathlineto{\pgfqpoint{-0.048611in}{0.000000in}}%
\pgfusepath{stroke,fill}%
}%
\begin{pgfscope}%
\pgfsys@transformshift{0.592966in}{0.407656in}%
\pgfsys@useobject{currentmarker}{}%
\end{pgfscope}%
\end{pgfscope}%
\begin{pgfscope}%
\definecolor{textcolor}{rgb}{0.000000,0.000000,0.000000}%
\pgfsetstrokecolor{textcolor}%
\pgfsetfillcolor{textcolor}%
\pgftext[x=0.267243in, y=0.364281in, left, base]{\color{textcolor}\rmfamily\fontsize{9.000000}{10.800000}\selectfont 0.05}%
\end{pgfscope}%
\begin{pgfscope}%
\pgfpathrectangle{\pgfqpoint{0.592966in}{0.067625in}}{\pgfqpoint{0.986972in}{1.496136in}}%
\pgfusepath{clip}%
\pgfsetrectcap%
\pgfsetroundjoin%
\pgfsetlinewidth{0.803000pt}%
\definecolor{currentstroke}{rgb}{0.690196,0.690196,0.690196}%
\pgfsetstrokecolor{currentstroke}%
\pgfsetdash{}{0pt}%
\pgfpathmoveto{\pgfqpoint{0.592966in}{0.747687in}}%
\pgfpathlineto{\pgfqpoint{1.579938in}{0.747687in}}%
\pgfusepath{stroke}%
\end{pgfscope}%
\begin{pgfscope}%
\pgfsetbuttcap%
\pgfsetroundjoin%
\definecolor{currentfill}{rgb}{0.000000,0.000000,0.000000}%
\pgfsetfillcolor{currentfill}%
\pgfsetlinewidth{0.803000pt}%
\definecolor{currentstroke}{rgb}{0.000000,0.000000,0.000000}%
\pgfsetstrokecolor{currentstroke}%
\pgfsetdash{}{0pt}%
\pgfsys@defobject{currentmarker}{\pgfqpoint{-0.048611in}{0.000000in}}{\pgfqpoint{0.000000in}{0.000000in}}{%
\pgfpathmoveto{\pgfqpoint{0.000000in}{0.000000in}}%
\pgfpathlineto{\pgfqpoint{-0.048611in}{0.000000in}}%
\pgfusepath{stroke,fill}%
}%
\begin{pgfscope}%
\pgfsys@transformshift{0.592966in}{0.747687in}%
\pgfsys@useobject{currentmarker}{}%
\end{pgfscope}%
\end{pgfscope}%
\begin{pgfscope}%
\definecolor{textcolor}{rgb}{0.000000,0.000000,0.000000}%
\pgfsetstrokecolor{textcolor}%
\pgfsetfillcolor{textcolor}%
\pgftext[x=0.267243in, y=0.704312in, left, base]{\color{textcolor}\rmfamily\fontsize{9.000000}{10.800000}\selectfont 0.10}%
\end{pgfscope}%
\begin{pgfscope}%
\pgfpathrectangle{\pgfqpoint{0.592966in}{0.067625in}}{\pgfqpoint{0.986972in}{1.496136in}}%
\pgfusepath{clip}%
\pgfsetrectcap%
\pgfsetroundjoin%
\pgfsetlinewidth{0.803000pt}%
\definecolor{currentstroke}{rgb}{0.690196,0.690196,0.690196}%
\pgfsetstrokecolor{currentstroke}%
\pgfsetdash{}{0pt}%
\pgfpathmoveto{\pgfqpoint{0.592966in}{1.087718in}}%
\pgfpathlineto{\pgfqpoint{1.579938in}{1.087718in}}%
\pgfusepath{stroke}%
\end{pgfscope}%
\begin{pgfscope}%
\pgfsetbuttcap%
\pgfsetroundjoin%
\definecolor{currentfill}{rgb}{0.000000,0.000000,0.000000}%
\pgfsetfillcolor{currentfill}%
\pgfsetlinewidth{0.803000pt}%
\definecolor{currentstroke}{rgb}{0.000000,0.000000,0.000000}%
\pgfsetstrokecolor{currentstroke}%
\pgfsetdash{}{0pt}%
\pgfsys@defobject{currentmarker}{\pgfqpoint{-0.048611in}{0.000000in}}{\pgfqpoint{0.000000in}{0.000000in}}{%
\pgfpathmoveto{\pgfqpoint{0.000000in}{0.000000in}}%
\pgfpathlineto{\pgfqpoint{-0.048611in}{0.000000in}}%
\pgfusepath{stroke,fill}%
}%
\begin{pgfscope}%
\pgfsys@transformshift{0.592966in}{1.087718in}%
\pgfsys@useobject{currentmarker}{}%
\end{pgfscope}%
\end{pgfscope}%
\begin{pgfscope}%
\definecolor{textcolor}{rgb}{0.000000,0.000000,0.000000}%
\pgfsetstrokecolor{textcolor}%
\pgfsetfillcolor{textcolor}%
\pgftext[x=0.267243in, y=1.044343in, left, base]{\color{textcolor}\rmfamily\fontsize{9.000000}{10.800000}\selectfont 0.15}%
\end{pgfscope}%
\begin{pgfscope}%
\pgfpathrectangle{\pgfqpoint{0.592966in}{0.067625in}}{\pgfqpoint{0.986972in}{1.496136in}}%
\pgfusepath{clip}%
\pgfsetrectcap%
\pgfsetroundjoin%
\pgfsetlinewidth{0.803000pt}%
\definecolor{currentstroke}{rgb}{0.690196,0.690196,0.690196}%
\pgfsetstrokecolor{currentstroke}%
\pgfsetdash{}{0pt}%
\pgfpathmoveto{\pgfqpoint{0.592966in}{1.427748in}}%
\pgfpathlineto{\pgfqpoint{1.579938in}{1.427748in}}%
\pgfusepath{stroke}%
\end{pgfscope}%
\begin{pgfscope}%
\pgfsetbuttcap%
\pgfsetroundjoin%
\definecolor{currentfill}{rgb}{0.000000,0.000000,0.000000}%
\pgfsetfillcolor{currentfill}%
\pgfsetlinewidth{0.803000pt}%
\definecolor{currentstroke}{rgb}{0.000000,0.000000,0.000000}%
\pgfsetstrokecolor{currentstroke}%
\pgfsetdash{}{0pt}%
\pgfsys@defobject{currentmarker}{\pgfqpoint{-0.048611in}{0.000000in}}{\pgfqpoint{0.000000in}{0.000000in}}{%
\pgfpathmoveto{\pgfqpoint{0.000000in}{0.000000in}}%
\pgfpathlineto{\pgfqpoint{-0.048611in}{0.000000in}}%
\pgfusepath{stroke,fill}%
}%
\begin{pgfscope}%
\pgfsys@transformshift{0.592966in}{1.427748in}%
\pgfsys@useobject{currentmarker}{}%
\end{pgfscope}%
\end{pgfscope}%
\begin{pgfscope}%
\definecolor{textcolor}{rgb}{0.000000,0.000000,0.000000}%
\pgfsetstrokecolor{textcolor}%
\pgfsetfillcolor{textcolor}%
\pgftext[x=0.267243in, y=1.384373in, left, base]{\color{textcolor}\rmfamily\fontsize{9.000000}{10.800000}\selectfont 0.20}%
\end{pgfscope}%
\begin{pgfscope}%
\definecolor{textcolor}{rgb}{0.000000,0.000000,0.000000}%
\pgfsetstrokecolor{textcolor}%
\pgfsetfillcolor{textcolor}%
\pgftext[x=0.211688in,y=0.815693in,,bottom,rotate=90.000000]{\color{textcolor}\rmfamily\fontsize{9.000000}{10.800000}\selectfont SA}%
\end{pgfscope}%
\begin{pgfscope}%
\pgfsetrectcap%
\pgfsetmiterjoin%
\pgfsetlinewidth{0.803000pt}%
\definecolor{currentstroke}{rgb}{0.000000,0.000000,0.000000}%
\pgfsetstrokecolor{currentstroke}%
\pgfsetdash{}{0pt}%
\pgfpathmoveto{\pgfqpoint{0.592966in}{0.067625in}}%
\pgfpathlineto{\pgfqpoint{0.592966in}{1.563761in}}%
\pgfusepath{stroke}%
\end{pgfscope}%
\begin{pgfscope}%
\pgfsetrectcap%
\pgfsetmiterjoin%
\pgfsetlinewidth{0.803000pt}%
\definecolor{currentstroke}{rgb}{0.000000,0.000000,0.000000}%
\pgfsetstrokecolor{currentstroke}%
\pgfsetdash{}{0pt}%
\pgfpathmoveto{\pgfqpoint{1.579938in}{0.067625in}}%
\pgfpathlineto{\pgfqpoint{1.579938in}{1.563761in}}%
\pgfusepath{stroke}%
\end{pgfscope}%
\begin{pgfscope}%
\pgfsetrectcap%
\pgfsetmiterjoin%
\pgfsetlinewidth{0.803000pt}%
\definecolor{currentstroke}{rgb}{0.000000,0.000000,0.000000}%
\pgfsetstrokecolor{currentstroke}%
\pgfsetdash{}{0pt}%
\pgfpathmoveto{\pgfqpoint{0.592966in}{0.067625in}}%
\pgfpathlineto{\pgfqpoint{1.579938in}{0.067625in}}%
\pgfusepath{stroke}%
\end{pgfscope}%
\begin{pgfscope}%
\pgfsetrectcap%
\pgfsetmiterjoin%
\pgfsetlinewidth{0.803000pt}%
\definecolor{currentstroke}{rgb}{0.000000,0.000000,0.000000}%
\pgfsetstrokecolor{currentstroke}%
\pgfsetdash{}{0pt}%
\pgfpathmoveto{\pgfqpoint{0.592966in}{1.563761in}}%
\pgfpathlineto{\pgfqpoint{1.579938in}{1.563761in}}%
\pgfusepath{stroke}%
\end{pgfscope}%
\begin{pgfscope}%
\pgfsetbuttcap%
\pgfsetmiterjoin%
\definecolor{currentfill}{rgb}{1.000000,1.000000,1.000000}%
\pgfsetfillcolor{currentfill}%
\pgfsetlinewidth{0.000000pt}%
\definecolor{currentstroke}{rgb}{0.000000,0.000000,0.000000}%
\pgfsetstrokecolor{currentstroke}%
\pgfsetstrokeopacity{0.000000}%
\pgfsetdash{}{0pt}%
\pgfpathmoveto{\pgfqpoint{1.714938in}{0.067625in}}%
\pgfpathlineto{\pgfqpoint{2.701910in}{0.067625in}}%
\pgfpathlineto{\pgfqpoint{2.701910in}{1.563761in}}%
\pgfpathlineto{\pgfqpoint{1.714938in}{1.563761in}}%
\pgfpathclose%
\pgfusepath{fill}%
\end{pgfscope}%
\begin{pgfscope}%
\pgfpathrectangle{\pgfqpoint{1.714938in}{0.067625in}}{\pgfqpoint{0.986972in}{1.496136in}}%
\pgfusepath{clip}%
\pgfsetbuttcap%
\pgfsetmiterjoin%
\definecolor{currentfill}{rgb}{1.000000,0.498039,0.054902}%
\pgfsetfillcolor{currentfill}%
\pgfsetlinewidth{1.003750pt}%
\definecolor{currentstroke}{rgb}{0.000000,0.000000,0.000000}%
\pgfsetstrokecolor{currentstroke}%
\pgfsetdash{{6.400000pt}{1.600000pt}{1.000000pt}{1.600000pt}}{0.000000pt}%
\pgfpathmoveto{\pgfqpoint{1.759800in}{0.067625in}}%
\pgfpathlineto{\pgfqpoint{1.984112in}{0.067625in}}%
\pgfpathlineto{\pgfqpoint{1.984112in}{1.489787in}}%
\pgfpathlineto{\pgfqpoint{1.759800in}{1.489787in}}%
\pgfpathclose%
\pgfusepath{stroke,fill}%
\end{pgfscope}%
\begin{pgfscope}%
\pgfpathrectangle{\pgfqpoint{1.714938in}{0.067625in}}{\pgfqpoint{0.986972in}{1.496136in}}%
\pgfusepath{clip}%
\pgfsetbuttcap%
\pgfsetmiterjoin%
\definecolor{currentfill}{rgb}{0.839216,0.152941,0.156863}%
\pgfsetfillcolor{currentfill}%
\pgfsetlinewidth{1.003750pt}%
\definecolor{currentstroke}{rgb}{0.000000,0.000000,0.000000}%
\pgfsetstrokecolor{currentstroke}%
\pgfsetdash{{1.000000pt}{1.650000pt}}{0.000000pt}%
\pgfpathmoveto{\pgfqpoint{1.984112in}{0.067625in}}%
\pgfpathlineto{\pgfqpoint{2.208424in}{0.067625in}}%
\pgfpathlineto{\pgfqpoint{2.208424in}{0.382639in}}%
\pgfpathlineto{\pgfqpoint{1.984112in}{0.382639in}}%
\pgfpathclose%
\pgfusepath{stroke,fill}%
\end{pgfscope}%
\begin{pgfscope}%
\pgfpathrectangle{\pgfqpoint{1.714938in}{0.067625in}}{\pgfqpoint{0.986972in}{1.496136in}}%
\pgfusepath{clip}%
\pgfsetbuttcap%
\pgfsetmiterjoin%
\definecolor{currentfill}{rgb}{0.580392,0.403922,0.741176}%
\pgfsetfillcolor{currentfill}%
\pgfsetlinewidth{1.003750pt}%
\definecolor{currentstroke}{rgb}{0.000000,0.000000,0.000000}%
\pgfsetstrokecolor{currentstroke}%
\pgfsetdash{{3.700000pt}{1.600000pt}}{0.000000pt}%
\pgfpathmoveto{\pgfqpoint{2.208424in}{0.067625in}}%
\pgfpathlineto{\pgfqpoint{2.432736in}{0.067625in}}%
\pgfpathlineto{\pgfqpoint{2.432736in}{1.240567in}}%
\pgfpathlineto{\pgfqpoint{2.208424in}{1.240567in}}%
\pgfpathclose%
\pgfusepath{stroke,fill}%
\end{pgfscope}%
\begin{pgfscope}%
\pgfpathrectangle{\pgfqpoint{1.714938in}{0.067625in}}{\pgfqpoint{0.986972in}{1.496136in}}%
\pgfusepath{clip}%
\pgfsetbuttcap%
\pgfsetmiterjoin%
\definecolor{currentfill}{rgb}{0.549020,0.337255,0.294118}%
\pgfsetfillcolor{currentfill}%
\pgfsetlinewidth{1.003750pt}%
\definecolor{currentstroke}{rgb}{0.000000,0.000000,0.000000}%
\pgfsetstrokecolor{currentstroke}%
\pgfsetdash{{3.000000pt}{1.000000pt}{1.000000pt}{1.000000pt}{1.000000pt}{1.000000pt}}{0.000000pt}%
\pgfpathmoveto{\pgfqpoint{2.432736in}{0.067625in}}%
\pgfpathlineto{\pgfqpoint{2.657048in}{0.067625in}}%
\pgfpathlineto{\pgfqpoint{2.657048in}{0.346868in}}%
\pgfpathlineto{\pgfqpoint{2.432736in}{0.346868in}}%
\pgfpathclose%
\pgfusepath{stroke,fill}%
\end{pgfscope}%
\begin{pgfscope}%
\pgfpathrectangle{\pgfqpoint{1.714938in}{0.067625in}}{\pgfqpoint{0.986972in}{1.496136in}}%
\pgfusepath{clip}%
\pgfsetrectcap%
\pgfsetroundjoin%
\pgfsetlinewidth{0.803000pt}%
\definecolor{currentstroke}{rgb}{0.690196,0.690196,0.690196}%
\pgfsetstrokecolor{currentstroke}%
\pgfsetdash{}{0pt}%
\pgfpathmoveto{\pgfqpoint{1.714938in}{0.067625in}}%
\pgfpathlineto{\pgfqpoint{2.701910in}{0.067625in}}%
\pgfusepath{stroke}%
\end{pgfscope}%
\begin{pgfscope}%
\pgfsetbuttcap%
\pgfsetroundjoin%
\definecolor{currentfill}{rgb}{0.000000,0.000000,0.000000}%
\pgfsetfillcolor{currentfill}%
\pgfsetlinewidth{0.803000pt}%
\definecolor{currentstroke}{rgb}{0.000000,0.000000,0.000000}%
\pgfsetstrokecolor{currentstroke}%
\pgfsetdash{}{0pt}%
\pgfsys@defobject{currentmarker}{\pgfqpoint{0.000000in}{0.000000in}}{\pgfqpoint{0.048611in}{0.000000in}}{%
\pgfpathmoveto{\pgfqpoint{0.000000in}{0.000000in}}%
\pgfpathlineto{\pgfqpoint{0.048611in}{0.000000in}}%
\pgfusepath{stroke,fill}%
}%
\begin{pgfscope}%
\pgfsys@transformshift{2.701910in}{0.067625in}%
\pgfsys@useobject{currentmarker}{}%
\end{pgfscope}%
\end{pgfscope}%
\begin{pgfscope}%
\definecolor{textcolor}{rgb}{0.000000,0.000000,0.000000}%
\pgfsetstrokecolor{textcolor}%
\pgfsetfillcolor{textcolor}%
\pgftext[x=2.799132in, y=0.024250in, left, base]{\color{textcolor}\rmfamily\fontsize{9.000000}{10.800000}\selectfont 0.0000}%
\end{pgfscope}%
\begin{pgfscope}%
\pgfpathrectangle{\pgfqpoint{1.714938in}{0.067625in}}{\pgfqpoint{0.986972in}{1.496136in}}%
\pgfusepath{clip}%
\pgfsetrectcap%
\pgfsetroundjoin%
\pgfsetlinewidth{0.803000pt}%
\definecolor{currentstroke}{rgb}{0.690196,0.690196,0.690196}%
\pgfsetstrokecolor{currentstroke}%
\pgfsetdash{}{0pt}%
\pgfpathmoveto{\pgfqpoint{1.714938in}{0.407656in}}%
\pgfpathlineto{\pgfqpoint{2.701910in}{0.407656in}}%
\pgfusepath{stroke}%
\end{pgfscope}%
\begin{pgfscope}%
\pgfsetbuttcap%
\pgfsetroundjoin%
\definecolor{currentfill}{rgb}{0.000000,0.000000,0.000000}%
\pgfsetfillcolor{currentfill}%
\pgfsetlinewidth{0.803000pt}%
\definecolor{currentstroke}{rgb}{0.000000,0.000000,0.000000}%
\pgfsetstrokecolor{currentstroke}%
\pgfsetdash{}{0pt}%
\pgfsys@defobject{currentmarker}{\pgfqpoint{0.000000in}{0.000000in}}{\pgfqpoint{0.048611in}{0.000000in}}{%
\pgfpathmoveto{\pgfqpoint{0.000000in}{0.000000in}}%
\pgfpathlineto{\pgfqpoint{0.048611in}{0.000000in}}%
\pgfusepath{stroke,fill}%
}%
\begin{pgfscope}%
\pgfsys@transformshift{2.701910in}{0.407656in}%
\pgfsys@useobject{currentmarker}{}%
\end{pgfscope}%
\end{pgfscope}%
\begin{pgfscope}%
\definecolor{textcolor}{rgb}{0.000000,0.000000,0.000000}%
\pgfsetstrokecolor{textcolor}%
\pgfsetfillcolor{textcolor}%
\pgftext[x=2.799132in, y=0.364281in, left, base]{\color{textcolor}\rmfamily\fontsize{9.000000}{10.800000}\selectfont 0.0005}%
\end{pgfscope}%
\begin{pgfscope}%
\pgfpathrectangle{\pgfqpoint{1.714938in}{0.067625in}}{\pgfqpoint{0.986972in}{1.496136in}}%
\pgfusepath{clip}%
\pgfsetrectcap%
\pgfsetroundjoin%
\pgfsetlinewidth{0.803000pt}%
\definecolor{currentstroke}{rgb}{0.690196,0.690196,0.690196}%
\pgfsetstrokecolor{currentstroke}%
\pgfsetdash{}{0pt}%
\pgfpathmoveto{\pgfqpoint{1.714938in}{0.747687in}}%
\pgfpathlineto{\pgfqpoint{2.701910in}{0.747687in}}%
\pgfusepath{stroke}%
\end{pgfscope}%
\begin{pgfscope}%
\pgfsetbuttcap%
\pgfsetroundjoin%
\definecolor{currentfill}{rgb}{0.000000,0.000000,0.000000}%
\pgfsetfillcolor{currentfill}%
\pgfsetlinewidth{0.803000pt}%
\definecolor{currentstroke}{rgb}{0.000000,0.000000,0.000000}%
\pgfsetstrokecolor{currentstroke}%
\pgfsetdash{}{0pt}%
\pgfsys@defobject{currentmarker}{\pgfqpoint{0.000000in}{0.000000in}}{\pgfqpoint{0.048611in}{0.000000in}}{%
\pgfpathmoveto{\pgfqpoint{0.000000in}{0.000000in}}%
\pgfpathlineto{\pgfqpoint{0.048611in}{0.000000in}}%
\pgfusepath{stroke,fill}%
}%
\begin{pgfscope}%
\pgfsys@transformshift{2.701910in}{0.747687in}%
\pgfsys@useobject{currentmarker}{}%
\end{pgfscope}%
\end{pgfscope}%
\begin{pgfscope}%
\definecolor{textcolor}{rgb}{0.000000,0.000000,0.000000}%
\pgfsetstrokecolor{textcolor}%
\pgfsetfillcolor{textcolor}%
\pgftext[x=2.799132in, y=0.704312in, left, base]{\color{textcolor}\rmfamily\fontsize{9.000000}{10.800000}\selectfont 0.0010}%
\end{pgfscope}%
\begin{pgfscope}%
\pgfpathrectangle{\pgfqpoint{1.714938in}{0.067625in}}{\pgfqpoint{0.986972in}{1.496136in}}%
\pgfusepath{clip}%
\pgfsetrectcap%
\pgfsetroundjoin%
\pgfsetlinewidth{0.803000pt}%
\definecolor{currentstroke}{rgb}{0.690196,0.690196,0.690196}%
\pgfsetstrokecolor{currentstroke}%
\pgfsetdash{}{0pt}%
\pgfpathmoveto{\pgfqpoint{1.714938in}{1.087718in}}%
\pgfpathlineto{\pgfqpoint{2.701910in}{1.087718in}}%
\pgfusepath{stroke}%
\end{pgfscope}%
\begin{pgfscope}%
\pgfsetbuttcap%
\pgfsetroundjoin%
\definecolor{currentfill}{rgb}{0.000000,0.000000,0.000000}%
\pgfsetfillcolor{currentfill}%
\pgfsetlinewidth{0.803000pt}%
\definecolor{currentstroke}{rgb}{0.000000,0.000000,0.000000}%
\pgfsetstrokecolor{currentstroke}%
\pgfsetdash{}{0pt}%
\pgfsys@defobject{currentmarker}{\pgfqpoint{0.000000in}{0.000000in}}{\pgfqpoint{0.048611in}{0.000000in}}{%
\pgfpathmoveto{\pgfqpoint{0.000000in}{0.000000in}}%
\pgfpathlineto{\pgfqpoint{0.048611in}{0.000000in}}%
\pgfusepath{stroke,fill}%
}%
\begin{pgfscope}%
\pgfsys@transformshift{2.701910in}{1.087718in}%
\pgfsys@useobject{currentmarker}{}%
\end{pgfscope}%
\end{pgfscope}%
\begin{pgfscope}%
\definecolor{textcolor}{rgb}{0.000000,0.000000,0.000000}%
\pgfsetstrokecolor{textcolor}%
\pgfsetfillcolor{textcolor}%
\pgftext[x=2.799132in, y=1.044343in, left, base]{\color{textcolor}\rmfamily\fontsize{9.000000}{10.800000}\selectfont 0.0015}%
\end{pgfscope}%
\begin{pgfscope}%
\pgfpathrectangle{\pgfqpoint{1.714938in}{0.067625in}}{\pgfqpoint{0.986972in}{1.496136in}}%
\pgfusepath{clip}%
\pgfsetrectcap%
\pgfsetroundjoin%
\pgfsetlinewidth{0.803000pt}%
\definecolor{currentstroke}{rgb}{0.690196,0.690196,0.690196}%
\pgfsetstrokecolor{currentstroke}%
\pgfsetdash{}{0pt}%
\pgfpathmoveto{\pgfqpoint{1.714938in}{1.427748in}}%
\pgfpathlineto{\pgfqpoint{2.701910in}{1.427748in}}%
\pgfusepath{stroke}%
\end{pgfscope}%
\begin{pgfscope}%
\pgfsetbuttcap%
\pgfsetroundjoin%
\definecolor{currentfill}{rgb}{0.000000,0.000000,0.000000}%
\pgfsetfillcolor{currentfill}%
\pgfsetlinewidth{0.803000pt}%
\definecolor{currentstroke}{rgb}{0.000000,0.000000,0.000000}%
\pgfsetstrokecolor{currentstroke}%
\pgfsetdash{}{0pt}%
\pgfsys@defobject{currentmarker}{\pgfqpoint{0.000000in}{0.000000in}}{\pgfqpoint{0.048611in}{0.000000in}}{%
\pgfpathmoveto{\pgfqpoint{0.000000in}{0.000000in}}%
\pgfpathlineto{\pgfqpoint{0.048611in}{0.000000in}}%
\pgfusepath{stroke,fill}%
}%
\begin{pgfscope}%
\pgfsys@transformshift{2.701910in}{1.427748in}%
\pgfsys@useobject{currentmarker}{}%
\end{pgfscope}%
\end{pgfscope}%
\begin{pgfscope}%
\definecolor{textcolor}{rgb}{0.000000,0.000000,0.000000}%
\pgfsetstrokecolor{textcolor}%
\pgfsetfillcolor{textcolor}%
\pgftext[x=2.799132in, y=1.384373in, left, base]{\color{textcolor}\rmfamily\fontsize{9.000000}{10.800000}\selectfont 0.0020}%
\end{pgfscope}%
\begin{pgfscope}%
\definecolor{textcolor}{rgb}{0.000000,0.000000,0.000000}%
\pgfsetstrokecolor{textcolor}%
\pgfsetfillcolor{textcolor}%
\pgftext[x=3.211688in,y=0.815693in,,top,rotate=90.000000]{\color{textcolor}\rmfamily\fontsize{9.000000}{10.800000}\selectfont MSE}%
\end{pgfscope}%
\begin{pgfscope}%
\pgfsetrectcap%
\pgfsetmiterjoin%
\pgfsetlinewidth{0.803000pt}%
\definecolor{currentstroke}{rgb}{0.000000,0.000000,0.000000}%
\pgfsetstrokecolor{currentstroke}%
\pgfsetdash{}{0pt}%
\pgfpathmoveto{\pgfqpoint{1.714938in}{0.067625in}}%
\pgfpathlineto{\pgfqpoint{1.714938in}{1.563761in}}%
\pgfusepath{stroke}%
\end{pgfscope}%
\begin{pgfscope}%
\pgfsetrectcap%
\pgfsetmiterjoin%
\pgfsetlinewidth{0.803000pt}%
\definecolor{currentstroke}{rgb}{0.000000,0.000000,0.000000}%
\pgfsetstrokecolor{currentstroke}%
\pgfsetdash{}{0pt}%
\pgfpathmoveto{\pgfqpoint{2.701910in}{0.067625in}}%
\pgfpathlineto{\pgfqpoint{2.701910in}{1.563761in}}%
\pgfusepath{stroke}%
\end{pgfscope}%
\begin{pgfscope}%
\pgfsetrectcap%
\pgfsetmiterjoin%
\pgfsetlinewidth{0.803000pt}%
\definecolor{currentstroke}{rgb}{0.000000,0.000000,0.000000}%
\pgfsetstrokecolor{currentstroke}%
\pgfsetdash{}{0pt}%
\pgfpathmoveto{\pgfqpoint{1.714938in}{0.067625in}}%
\pgfpathlineto{\pgfqpoint{2.701910in}{0.067625in}}%
\pgfusepath{stroke}%
\end{pgfscope}%
\begin{pgfscope}%
\pgfsetrectcap%
\pgfsetmiterjoin%
\pgfsetlinewidth{0.803000pt}%
\definecolor{currentstroke}{rgb}{0.000000,0.000000,0.000000}%
\pgfsetstrokecolor{currentstroke}%
\pgfsetdash{}{0pt}%
\pgfpathmoveto{\pgfqpoint{1.714938in}{1.563761in}}%
\pgfpathlineto{\pgfqpoint{2.701910in}{1.563761in}}%
\pgfusepath{stroke}%
\end{pgfscope}%
\begin{pgfscope}%
\pgfsetbuttcap%
\pgfsetmiterjoin%
\definecolor{currentfill}{rgb}{1.000000,1.000000,1.000000}%
\pgfsetfillcolor{currentfill}%
\pgfsetfillopacity{0.800000}%
\pgfsetlinewidth{1.003750pt}%
\definecolor{currentstroke}{rgb}{0.800000,0.800000,0.800000}%
\pgfsetstrokecolor{currentstroke}%
\pgfsetstrokeopacity{0.800000}%
\pgfsetdash{}{0pt}%
\pgfpathmoveto{\pgfqpoint{0.025000in}{1.599825in}}%
\pgfpathlineto{\pgfqpoint{3.392376in}{1.599825in}}%
\pgfpathquadraticcurveto{\pgfqpoint{3.417376in}{1.599825in}}{\pgfqpoint{3.417376in}{1.624825in}}%
\pgfpathlineto{\pgfqpoint{3.417376in}{1.799825in}}%
\pgfpathquadraticcurveto{\pgfqpoint{3.417376in}{1.824825in}}{\pgfqpoint{3.392376in}{1.824825in}}%
\pgfpathlineto{\pgfqpoint{0.025000in}{1.824825in}}%
\pgfpathquadraticcurveto{\pgfqpoint{0.000000in}{1.824825in}}{\pgfqpoint{0.000000in}{1.799825in}}%
\pgfpathlineto{\pgfqpoint{0.000000in}{1.624825in}}%
\pgfpathquadraticcurveto{\pgfqpoint{0.000000in}{1.599825in}}{\pgfqpoint{0.025000in}{1.599825in}}%
\pgfpathclose%
\pgfusepath{stroke,fill}%
\end{pgfscope}%
\begin{pgfscope}%
\pgfsetbuttcap%
\pgfsetmiterjoin%
\definecolor{currentfill}{rgb}{1.000000,0.498039,0.054902}%
\pgfsetfillcolor{currentfill}%
\pgfsetlinewidth{1.003750pt}%
\definecolor{currentstroke}{rgb}{0.000000,0.000000,0.000000}%
\pgfsetstrokecolor{currentstroke}%
\pgfsetdash{{6.400000pt}{1.600000pt}{1.000000pt}{1.600000pt}}{0.000000pt}%
\pgfpathmoveto{\pgfqpoint{0.050000in}{1.681075in}}%
\pgfpathlineto{\pgfqpoint{0.300000in}{1.681075in}}%
\pgfpathlineto{\pgfqpoint{0.300000in}{1.768575in}}%
\pgfpathlineto{\pgfqpoint{0.050000in}{1.768575in}}%
\pgfpathclose%
\pgfusepath{stroke,fill}%
\end{pgfscope}%
\begin{pgfscope}%
\definecolor{textcolor}{rgb}{0.000000,0.000000,0.000000}%
\pgfsetstrokecolor{textcolor}%
\pgfsetfillcolor{textcolor}%
\pgftext[x=0.312500in,y=1.681075in,left,base]{\color{textcolor}\rmfamily\fontsize{9.000000}{10.800000}\selectfont WF\cite{pratt-spectral-1976}}%
\end{pgfscope}%
\begin{pgfscope}%
\pgfsetbuttcap%
\pgfsetmiterjoin%
\definecolor{currentfill}{rgb}{0.839216,0.152941,0.156863}%
\pgfsetfillcolor{currentfill}%
\pgfsetlinewidth{1.003750pt}%
\definecolor{currentstroke}{rgb}{0.000000,0.000000,0.000000}%
\pgfsetstrokecolor{currentstroke}%
\pgfsetdash{{1.000000pt}{1.650000pt}}{0.000000pt}%
\pgfpathmoveto{\pgfqpoint{0.857250in}{1.681075in}}%
\pgfpathlineto{\pgfqpoint{1.107250in}{1.681075in}}%
\pgfpathlineto{\pgfqpoint{1.107250in}{1.768575in}}%
\pgfpathlineto{\pgfqpoint{0.857250in}{1.768575in}}%
\pgfpathclose%
\pgfusepath{stroke,fill}%
\end{pgfscope}%
\begin{pgfscope}%
\definecolor{textcolor}{rgb}{0.000000,0.000000,0.000000}%
\pgfsetstrokecolor{textcolor}%
\pgfsetfillcolor{textcolor}%
\pgftext[x=1.119750in,y=1.681075in,left,base]{\color{textcolor}\rmfamily\fontsize{9.000000}{10.800000}\selectfont SSW\cite{murakami-color-2008}}%
\end{pgfscope}%
\begin{pgfscope}%
\pgfsetbuttcap%
\pgfsetmiterjoin%
\definecolor{currentfill}{rgb}{0.580392,0.403922,0.741176}%
\pgfsetfillcolor{currentfill}%
\pgfsetlinewidth{1.003750pt}%
\definecolor{currentstroke}{rgb}{0.000000,0.000000,0.000000}%
\pgfsetstrokecolor{currentstroke}%
\pgfsetdash{{3.700000pt}{1.600000pt}}{0.000000pt}%
\pgfpathmoveto{\pgfqpoint{1.723375in}{1.681075in}}%
\pgfpathlineto{\pgfqpoint{1.973375in}{1.681075in}}%
\pgfpathlineto{\pgfqpoint{1.973375in}{1.768575in}}%
\pgfpathlineto{\pgfqpoint{1.723375in}{1.768575in}}%
\pgfpathclose%
\pgfusepath{stroke,fill}%
\end{pgfscope}%
\begin{pgfscope}%
\definecolor{textcolor}{rgb}{0.000000,0.000000,0.000000}%
\pgfsetstrokecolor{textcolor}%
\pgfsetfillcolor{textcolor}%
\pgftext[x=1.985875in,y=1.681075in,left,base]{\color{textcolor}\rmfamily\fontsize{9.000000}{10.800000}\selectfont EPSSW\cite{urban-spectral-2009}}%
\end{pgfscope}%
\begin{pgfscope}%
\pgfsetbuttcap%
\pgfsetmiterjoin%
\definecolor{currentfill}{rgb}{0.549020,0.337255,0.294118}%
\pgfsetfillcolor{currentfill}%
\pgfsetlinewidth{1.003750pt}%
\definecolor{currentstroke}{rgb}{0.000000,0.000000,0.000000}%
\pgfsetstrokecolor{currentstroke}%
\pgfsetdash{{3.000000pt}{1.000000pt}{1.000000pt}{1.000000pt}{1.000000pt}{1.000000pt}}{0.000000pt}%
\pgfpathmoveto{\pgfqpoint{2.764251in}{1.681075in}}%
\pgfpathlineto{\pgfqpoint{3.014251in}{1.681075in}}%
\pgfpathlineto{\pgfqpoint{3.014251in}{1.768575in}}%
\pgfpathlineto{\pgfqpoint{2.764251in}{1.768575in}}%
\pgfpathclose%
\pgfusepath{stroke,fill}%
\end{pgfscope}%
\begin{pgfscope}%
\definecolor{textcolor}{rgb}{0.000000,0.000000,0.000000}%
\pgfsetstrokecolor{textcolor}%
\pgfsetfillcolor{textcolor}%
\pgftext[x=3.026751in,y=1.681075in,left,base]{\color{textcolor}\rmfamily\fontsize{9.000000}{10.800000}\selectfont SPRE}%
\end{pgfscope}%
\end{pgfpicture}%
\makeatother%
\endgroup%

%% file: OSA-journal-template.bbl
\begin{thebibliography}{10}
\newcommand{\enquote}[1]{``#1''}

\bibitem{moroni-pet-2015}
M.~Moroni, A.~Mei, A.~Leonardi, E.~Lupo, and F.~Marca, \enquote{{PET} and {PVC}
  {Separation} with {Hyperspectral} {Imagery},}
  {\protect\JournalTitle{Sensors}} \textbf{15}, 2205--2227 (2015).

\bibitem{sowa-classification-2006}
M.~G. Sowa, L.~Leonardi, J.~R. Payette, K.~M. Cross, M.~Gomez, and J.~S. Fish,
  \enquote{Classification of burn injuries using near-infrared spectroscopy,}
  {\protect\JournalTitle{Journal of Biomedical Optics}} \textbf{11}, 054002
  (2006).

\bibitem{degardin-near-2016}
K.~Dégardin, A.~Guillemain, N.~V. Guerreiro, and Y.~Roggo, \enquote{Near
  infrared spectroscopy for counterfeit detection using a large database of
  pharmaceutical tablets,} {\protect\JournalTitle{Journal of Pharmaceutical and
  Biomedical Analysis}} \textbf{128}, 89--97 (2016).

\bibitem{dowell-automated-1998}
F.~E. Dowell, \enquote{Automated {Color} {Classification} of {Single} {Wheat}
  {Kernels} {Using} {Visible} and {Near}-{Infrared} {Reflectance},}
  {\protect\JournalTitle{Cereal Chemistry Journal}} \textbf{75}, 142--144
  (1998).

\bibitem{balabin-gasoline-2010}
R.~M. Balabin, R.~Z. Safieva, and E.~I. Lomakina, \enquote{Gasoline
  classification using near infrared ({NIR}) spectroscopy data: {Comparison} of
  multivariate techniques,} {\protect\JournalTitle{Analytica Chimica Acta}}
  \textbf{671}, 27--35 (2010).

\bibitem{lapray-multispectral-2014}
P.-J. Lapray, X.~Wang, J.-B. Thomas, and P.~Gouton, \enquote{Multispectral
  {Filter} {Arrays}: {Recent} {Advances} and {Practical} {Implementation},}
  {\protect\JournalTitle{Sensors}} \textbf{14}, 21626--21659 (2014).

\bibitem{lambrechts-cmos-compatible-2014}
A.~Lambrechts, P.~Gonzalez, B.~Geelen, P.~Soussan, K.~Tack, and M.~Jayapala,
  \enquote{A {CMOS}-compatible, integrated approach to hyper- and multispectral
  imaging,} in \emph{2014 {IEEE} {International} {Electron} {Devices}
  {Meeting},}  (IEEE, San Francisco, CA, USA, 2014), pp. 10.5.1--10.5.4.

\bibitem{brauers-multispectral-2008}
J.~Brauers, N.~Schulte, and T.~Aach, \enquote{Multispectral {Filter}-{Wheel}
  {Cameras}: {Geometric} {Distortion} {Model} and {Compensation} {Algorithms},}
  {\protect\JournalTitle{IEEE Transactions on Image Processing}} \textbf{17},
  2368--2380 (2008).

\bibitem{genser-multispectral-2019}
N.~Genser, J.~Seiler, and A.~Kaup, \enquote{Camera {Array} for
  {Multi}-{Spectral} {Imaging},} {\protect\JournalTitle{IEEE Transactions on
  Image Processing}}  (2020).

\bibitem{genser-deep-2020}
N.~Genser, A.~Spruck, J.~Seiler, and A.~Kaup, \enquote{Deep {Learning} based
  {Cross}-{Spectral} {Disparity} {Estimation} for {Stereo} {Imaging},}
  {\protect\JournalTitle{IEEE International Conference on Image Processing}}
  (2020).

\bibitem{genser-joint-2020}
N.~Genser, J.~Seiler, and A.~Kaup, \enquote{Joint {Content}-{Adaptive}
  {Dictionary}-{Learning} and {Sparse} {Selective} {Extrapolation} for
  {Cross}-{Spectral} {Image} {Reconstruction},} {\protect\JournalTitle{IEEE
  International Conference on Image Processing}}  (2020).

\bibitem{cortes-multipectral-2003}
A.~R. Cortés, \enquote{Multispectral {Analysis} and spectral {Reflectance}
  {Reconstruction} of {Art} {Paintings},} Ph.D. thesis, École Nationale
  Supérieure des Télécommunications (2003).

\bibitem{schoberl-photometric-2012}
M.~Schöberl, \enquote{Photometric limits for digital camera systems,}
  {\protect\JournalTitle{Journal of Electronic Imaging}} \textbf{21}, 020501
  (2012).

\bibitem{gow-comprehensive-2007}
R.~Gow, D.~Renshaw, K.~Findlater, L.~Grant, S.~McLeod, J.~Hart, and R.~Nicol,
  \enquote{A {Comprehensive} {Tool} for {Modeling} {CMOS}
  {Image}-{Sensor}-{Noise} {Performance},} {\protect\JournalTitle{IEEE
  Transactions on Electron Devices}} \textbf{54}, 1321--1329 (2007).

\bibitem{pratt-spectral-1976}
W.~K. Pratt and C.~E. Mancill, \enquote{Spectral estimation techniques for the
  spectral calibration of a color image scanner,}
  {\protect\JournalTitle{Applied Optics}} \textbf{15}, 73 (1976).

\bibitem{immerkaer-fast-1996}
J.~Immerkær, \enquote{Fast {Noise} {Variance} {Estimation},}
  {\protect\JournalTitle{Computer Vision and Image Understanding}} \textbf{64},
  300--302 (1996).

\bibitem{zhang-study-2015}
L.~Zhang, D.~Liang, Z.~Pan, and X.~Ma, \enquote{Study on the key technology of
  reconstruction spectral reflectance based on the algorithm of compressive
  sensing,} {\protect\JournalTitle{Optical and Quantum Electronics}}
  \textbf{47}, 1679--1692 (2015).

\bibitem{mansouri-representation-2008}
A.~Mansouri, T.~Sliwa, J.~Y. Hardeberg, and Y.~Voisin, \enquote{Representation
  and estimation of spectral reflectances using projection on {PCA} and wavelet
  bases,} {\protect\JournalTitle{Color Research \& Application}} \textbf{33},
  485--493 (2008).

\bibitem{eckhard-evaluating-2014}
T.~Eckhard, E.~M. Valero, J.~Hernández-Andrés, and V.~Heikkinen,
  \enquote{Evaluating logarithmic kernel for spectral reflectance
  estimation—effects on model parametrization, training set size, and number
  of sensor spectral channels,} {\protect\JournalTitle{Journal of the Optical
  Society of America A}} \textbf{31}, 541 (2014).

\bibitem{shen-spectral-2006}
H.-L. Shen and J.~H. Xin, \enquote{Spectral characterization of a color scanner
  based on optimized adaptive estimation,} {\protect\JournalTitle{Journal of
  the Optical Society of America A}} \textbf{23}, 1566 (2006).

\bibitem{zhang-spectral-2008}
W.-F. Zhang and D.-Q. Dai, \enquote{Spectral reflectance estimation from camera
  responses by support vector regression and a composite model,}
  {\protect\JournalTitle{Journal of the Optical Society of America A}}
  \textbf{25}, 2286 (2008).

\bibitem{shen-reflectance-2007}
H.-L. Shen, P.-Q. Cai, S.-J. Shao, and J.~H. Xin, \enquote{Reflectance
  reconstruction for multispectral imaging by adaptive {Wiener} estimation,}
  {\protect\JournalTitle{Optics Express}} \textbf{15}, 15545 (2007).

\bibitem{murakami-color-2008}
Y.~Murakami, K.~Fukura, M.~Yamaguchi, and N.~Ohyama, \enquote{Color
  reproduction from low-{SNR} multispectral images using spatio-spectral
  {Wiener} estimation,} {\protect\JournalTitle{Optics Express}} \textbf{16},
  4106 (2008).

\bibitem{urban-spectral-2009}
P.~Urban, M.~R. Rosen, and R.~S. Berns, \enquote{Spectral image reconstruction
  using an edge preserving spatio-spectral {Wiener} estimation,}
  {\protect\JournalTitle{Journal of the Optical Society of America A}}
  \textbf{26}, 1865 (2009).

\bibitem{he-guided-2013}
K.~He, J.~Sun, and X.~Tang, \enquote{Guided {Image} {Filtering},}
  {\protect\JournalTitle{IEEE Transactions on Pattern Analysis and Machine
  Intelligence}} \textbf{35}, 1397--1409 (2013).

\bibitem{therrien-discrete-1992}
C.~W. Therrien, \emph{Discrete random signals and statistical signal
  processing} (Prentice Hall, Englewood Cliffs, NJ, 1992).

\bibitem{kruse-spectral-1993}
F.~Kruse, A.~Lefkoff, J.~Boardman, K.~Heidebrecht, A.~Shapiro, P.~Barloon, and
  A.~Goetz, \enquote{The spectral image processing system
  ({SIPS})—interactive visualization and analysis of imaging spectrometer
  data,} {\protect\JournalTitle{Remote Sensing of Environment}} \textbf{44},
  145--163 (1993).

\bibitem{leibe-sparse-2016}
B.~Arad and O.~Ben-Shahar, \enquote{Sparse {Recovery} of {Hyperspectral}
  {Signal} from {Natural} {RGB} {Images},} in \emph{Computer {Vision} –
  {ECCV} 2016,}  vol. 9911 B.~Leibe, J.~Matas, N.~Sebe, and M.~Welling, eds.
  (Springer International Publishing, Cham, 2016), pp. 19--34. Series Title:
  Lecture Notes in Computer Science.

\bibitem{eckhard-outdoor-2015}
J.~Eckhard, T.~Eckhard, E.~M. Valero, J.~L. Nieves, and E.~G. Contreras,
  \enquote{Outdoor scene reflectance measurements using a {Bragg}-grating-based
  hyperspectral imager,} {\protect\JournalTitle{Applied Optics}} \textbf{54},
  D15 (2015).

\bibitem{yasuma-generalized-2010}
F.~Yasuma, T.~Mitsunaga, D.~Iso, and S.~K. Nayar, \enquote{Generalized
  {Assorted} {Pixel} {Camera}: {Postcapture} {Control} of {Resolution},
  {Dynamic} {Range}, and {Spectrum},} {\protect\JournalTitle{IEEE Transactions
  on Image Processing}} \textbf{19}, 2241--2253 (2010).

\bibitem{petersen-matrix}
K.~B. Petersen, M.~S. Pedersen, J.~Larsen, K.~Strimmer, L.~Christiansen,
  K.~Hansen, L.~He, L.~Thibaut, M.~Barão, S.~Hattinger, V.~Sima, and W.~The,
  \enquote{The matrix cookbook,} Tech. rep. (2006).

\end{thebibliography}
